\documentclass[a4paper,11pt]{article}
\usepackage{color,xcolor,ucs}
\usepackage[top=0.3in, bottom=0.5in, left = 0.65in, right = 0.65in]{geometry}
\usepackage[linkcolor=black,colorlinks=true,urlcolor=blue]{hyperref}
\usepackage{mathtools}

\usepackage{color,xcolor,ucs}
\usepackage{mathtools}   \usepackage{tikz} 
\usepackage{ amssymb }
\usepackage{extarrows} 
\usepackage{pgf,tikz}
\usepackage{float}
\usetikzlibrary{positioning}
\usetikzlibrary{shapes.geometric}
\usetikzlibrary{shapes.misc}
\usetikzlibrary{arrows}
\usepackage{caption}
\usepackage{mathrsfs}
\usetikzlibrary{arrows,shapes,automata,backgrounds,petri,positioning}
\usetikzlibrary{decorations.pathmorphing}
\usetikzlibrary{decorations.shapes}
\usetikzlibrary{decorations.text}
\usetikzlibrary{decorations.fractals}
\usetikzlibrary{decorations.footprints}
\usetikzlibrary{shadows}
\usetikzlibrary{calc}
\usetikzlibrary{spy}
\usepackage{amsmath}
\usepackage{array}
\usepackage{ amssymb }
\usepackage{braket}
\usepackage{qcircuit}
\usepackage{soul}
\usepackage{braket} 
\usepackage{relsize}

\usepackage{amsmath}
\usepackage{ amssymb }
\usepackage{braket}
\usepackage{qcircuit}
\usepackage{soul}
\usepackage{braket}

\title{{\LARGE Poisson structures for the 4-vertex model, and the higher-spin XXX chain, and Yang-Baxter algebras}}
\author{Pete Rigas}
\date{}

\begin{document}

\maketitle

\begin{abstract}
            We implement the quantum inverse scattering method for the 4-vertex model. In comparison to previous works of the author which examined the 6-vertex, and 20-vertex, models, the 4-vertex model exhibits different characteristics, ranging from L-operators expressed in terms of projectors and Pauli matrices to algebraic and combinatorial properties, including Poisson structure and boxed plane partitions. We derive a system of relations from the structure of operators that can be leveraged for studying characteristics of the higher-spin XXX chain in the weak finite volume limit. From explicit relations for operators of the 4-vertex transfer matrix, we conclude by discussing corresponding aspects of the Yang-Baxter algebra, which is closely related to the operators obtained from products of L-operators for approximating the transfer, and quantum monodromy, matrices. The structure of computations from L-operators of the 4-vertex model directly transfers to L-operators of the higher-spin XXX chain, revealing a similar structure of another Yang-Baxter algebra of interest. \footnote{\textbf{MSC Class}: 34L25; 82B23; 82C10; 81U02; 60K35}
\end{abstract}

\textbf{Keywords}: Statistical physics, square ice, ice rule, triangular ice, six-vertex model, twenty-vertex model, quantum inverse scattering, Poisson structure, crossing probabilities, integrability, action-angle variables

\section{Introduction}

\subsection{Overview}

The 4-vertex model, as a degeneration of the 6, and 20, vertex models, is a model of Statistical Mechanics related to one originally introduced by Pauling, {\color{blue}[35]}, that has garnered interest for connections to determinantal representations {\color{blue}[8},{\color{blue}16},{\color{blue}34},{\color{blue}47]} symmetric functions {\color{blue}[19},{\color{blue}33]}, delocalization behavior under sufficiently flat, and even sloped, boundary conditions {\color{blue}[11},{\color{blue}37]}, integrability of inhomogeneous, or homogeneous, limit shapes {\color{blue}[25},{\color{blue}36]}, R, and K, matrices {\color{blue}[6},{\color{blue}7},{\color{blue}14},{\color{blue}15},{\color{blue}21},{\color{blue}27},{\color{blue}43]}, the Bethe ansatz {\color{blue}[12},{\color{blue}13},{\color{blue}21},{\color{blue}23]}, and correlation functions {\color{blue}[27},{\color{blue}28]}, amongst several other topics, ranging from integrability, exact solvability, quantum integrability, and the like {\color{blue}[1},{\color{blue}2},{\color{blue}3},{\color{blue}5},{\color{blue}9},{\color{blue}10},{\color{blue}18},{\color{blue}22},{\color{blue}24},{\color{blue}29},{\color{blue}30},{\color{blue}32]}. Besides aspects of the 6-vertex model which were examined in seminal computations for the residual free entropy over the torus, {\color{blue}[31]}, this vertex model has emerged as an intense object of study with several applications to Mathematical and Statistical Physics, two examples of which were examined in previous work from the author, {\color{blue}[42},{\color{blue}46]}, in which notions of integrability, and exact solvability, were explored through two and three-dimensional Poisson structures. Besides more algebraic characteristics of vertex models which become apparent when studying quantum inverse scattering type methods, several questions and longstanding conjectures remain open, whether related to scaling limits, {\color{blue}[9},{\color{blue}44]}, that can be easily computed for other models of Statistical Mechanics and Markov processes, along with potential connections with behavior of the phase transition for the Gaussian free field, {\color{blue}[39]}, and the dependence of boundary conditions, {\color{blue}[20]}, in which domain-wall boundary conditions have been of great interest, and thermodynamical properties, {\color{blue}[26]}, to name a few. The primary mechanism through which interactions for the 6-vertex model are mediated is through the ice rule, which as a conservation rule states that, about each vertex, there should be two ingoing, and two outgoing, arrows. Other families of vertex models, such as the 8-vertex, and odd 8-vertex, models, arise when other constraints are placed on the ice rule, such as an even number of ingoing, or outgoing, arrows at each vertex, which can not only impact local behaviors of parafermionic type observables that have been studied for other models, {\color{blue}[41]}, which can be leveraged for studying parafermions of vertex models, but also for computing transfer, and quantum monodromy, matrices that have been formalized for open boundary conditions of spin chains {\color{blue}[43]}.

To continue making contact with all such characteristics of vertex models mentioned previously, we initiate a similar program for studying quantum inverse scattering type methods for the 4-vertex model. In comparison to previous work of the author that closely studied the quantum inverse scattering method for the 6-vertex, and 20-vertex, models, {\color{blue}[42},{\color{blue}47]}, the quantum inverse scattering type method, by virtue of the 4-vertex model having a smaller sample space than the 6-vertex, and 20-vertex, models, enables one to view more simple combinatorial behaviors, and characteristics, of determinantal processes, and related objects, over the square lattice. Moreover, from a previous application of the quantum inverse scattering method to the 4-vertex model, {\color{blue}[4]}, correlation functions were computed, which share connections with boxed plane partitions and other combinatorial objects. Despite the fact that determinantal representations have already been investigated for boxed plane partitions, and other objects, for the 6-vertex and 20-vertex models in references from the previous paragraph, counterparts of two and three-dimensional Poisson structures have not. In two, and three, dimensions, learning about characteristics of Poisson structures is not only informative for determining whether various integrability properties of vertex models are expected to hold, {\color{blue}[42]}, but also for quantifying how the structure of L-operators appears in the partition function, and correlations.

In three dimensions, over $\textbf{T}$, in comparison to in two dimensions, over $\textbf{Z}^2$, Poisson structure that was investigated by the author for the 20-vertex model, {\color{blue}[46]}, exhibited several unexpected connections. The three-dimensional Poisson structure has many intriguing characteristics, several of which stem from the fact that the larger set of relations for the 20-vertex model over $\textbf{T}$, $81$, in comparison to the smaller set of relations for the 6-vertex model over $\textbf{Z}^2$, $16$, arise from operators from the block representations $A \big( \underline{u} \big),\cdots, I \big( \underline{u}\big)$ for the three-dimensional transfer matrix. Such operators which constitute the block representation of the transfer matrix, and quantum monodromy matrix, for the 20-vertex model share in similar interpretations as the collection of operators $A \big( \underline{u}\big), B \big(\underline{u}\big), C \big(\underline{u}\big), D \big(\underline{u}\big)$ for the transfer matrix of the 6-vertex model; depending upon the spectral parameters that are enforced in the vertical and horizontal directions of $\textbf{Z}^2$ before a homogenizing limit is taken, the fact that $\textbf{T}$ has an additional degree of freedom not only impacts how spectral parameters interact at each vertex of a configuration, but also the approximation for each Poisson bracket within the three-dimensional set of $81$ relations. For the 4-vertex model, the corresponding Poisson structure is comprised of a smaller set of relations than the $16$ of the 6-vertex model. Despite the fact that there is a dramatic reduction in complexity of the number of Poisson brackets that one must approximate from products of L-operators that are used to obtain the transfer matrix in the 4-vertex model, versus the number of Poisson brackets in the structure of the 6, and 20, vertex models, studying Poisson structure of the 4-vertex model can unveil more simplistic structures of the partition function, correlations, and boxed plane partitions.

With such an overview in mind, beginning in the next section we define several objects associated with the quantum inverse scattering framework from previous works of the author, namely in {\color{blue}[42},{\color{blue}46]}, which will share connections with combinatorial aspects of the 4-vertex model studied in {\color{blue}[4]}.

\subsection{Objects for the 4-vertex model, from the 6, and 20, vertex models}

\noindent We introduce several objects from closely related vertex models to the 4-vertex model, from previous works of the author {\color{blue}[42},{\color{blue}46]}. To define the forthcoming objects, we will make significant use of the Poisson bracket, which not only plays a pivotal role in characterizing integrability under the presence of inhomogeneities, from integrability of inhomogeneous limit shapes, {\color{blue}[25]}, but also for further probing the Poisson structure of vertex models. The Poisson bracket $\big\{ \cdot , \cdot \big\}$ satisfies the following set of properties, given test functions $f$, $g$ and $h$,

\begin{itemize}
    \item [$\bullet$] \textit{Anticommutativity}. $\big\{ f, g \big\}  =  - \big\{ g , f \big\} $

    \item[$\bullet$] \textit{Bilinearity}. For real $a,b$, $\big\{ af + bg , h \big\} = a \big\{ f ,h \big\} + b \big\{ g , h \big\},$ and $\big\{ h , af + bg \big\} = a \big\{ h , f \big\} + b \big\{ h , g \big\} $

    \item[$\bullet$] \textit{Leibniz' rule}. $\big\{ fg , h \big\} = \big\{ f , h \big\} g + f \big\{ g , h \big\}$

    \item[$\bullet$] \textit{Jacobi identity}. $\big\{ f , \big\{ g , h \big\} \big\} + \big\{ g , \big\{ h , f \big\} \big\} + \big\{ h , \big\{ f , g \big\} \big\} = 0$ \end{itemize}

\noindent For the 6-vertex model, under a wide variety of boundary conditions discussed across various works in the literature, {\color{blue}[11},{\color{blue}12},{\color{blue}18},{\color{blue}37},{\color{blue}41]}, the weight for each vertex over the square lattice takes the form,

\begin{align*}
  w_{6V}(\omega) \equiv w(\omega) =        a_1^{n_1} a_2^{n_2} b_1^{n_3} b_2^{n_4} c_1^{n_5} c_2^{n_6}  \underset{c_1 \equiv c_2 \equiv c}{\underset{b_1 \equiv b_2 \equiv b}{\underset{a_1 \equiv a_2 \equiv a}{\overset{\mathrm{Isotropic}}{\Longleftrightarrow}} }}          a^{n_1+n_2}  b^{n_3+n_4}  c^{n_5+n_6}    \text{, }   
\end{align*}

\noindent (see Figure 1 and Figure 2 for arrow, and line, depictions, respectively, of each possible configuration) which can be used to define the probability measure,

\begin{align*}
  \textbf{P}_{\textbf{T}_N}[      \omega         ]   \equiv \textbf{P}[   \omega     ]     =  \frac{w_{6V}(\omega)}{Z_{\textbf{T}_N}}   \equiv \frac{w(\omega)}{Z_{\textbf{T}_N}}   \text{, }  
\end{align*}

\noindent supported over the torus. Enforcing an isotropic parameter choice in the vertex weight function can have the effect of homogenizing the weight distribution, ie the total number of vertices of the finite volume which are of type $1$, $2$, or $3$, over all vertices over finite volume. For the quantum inverse scattering method, in the 6-vertex model the Poisson bracket that is significantly manipulated for obtaining asymptotic approximations of brackets within the two-dimensional Poisson structure takes the form, for two test functions $F$ and $G$,

\begin{align*}
 \big\{ F ,  G \big\}   \equiv    i \int_{[-L,L]} \bigg[          \frac{\delta F}{\delta \psi} \frac{\delta G}{\delta \bar{\psi}} - \frac{\delta F}{\delta \bar{\psi}} \frac{\delta G}{\delta \psi }             \bigg]  \text{ } \mathrm{d} x      \text{. } 
\end{align*}

\noindent To determine the total number of relations appearing with the bracket defined above, within either a two, or three, dimensional Poisson structure, one can introduce the operation of taking the tensor product of the Poisson bracket, which takes the form,

\begin{align*}
 \big\{  A \overset{\bigotimes}{,} B \big\} \equiv     i \int_{[-L,L]} \bigg[  \frac{\delta A}{\delta \psi} \bigotimes \frac{\delta B}{\delta \bar{\psi}} - \frac{\delta A}{\delta \bar{\psi}} \bigotimes  \frac{\delta B}{\delta \psi }                     \bigg]    \text{ } \mathrm{d} x     \text{, } 
\end{align*}

\noindent where $A$ and $B$ denote any possible choice of operators, which in the two-dimensional case takes the form from the representation,

\[ \begin{bmatrix}
       A \big( u \big)   & B \big( u \big)   \\
    C \big( u \big)  & D \big( u \big)  \text{ }  
  \end{bmatrix}  \text{. }
\]

\noindent In {\color{blue}[42]}, the author made significant use of an L-operator taken under boundary conditions for the inhomogeneous 6-vertex model, which takes the form, for spectral parameters $\lambda_{\alpha}$, and $v_k$,

 \[
       L_{\alpha , k   } \big( \lambda_{\alpha} , v_{k} \big)    \equiv 
  \begin{bmatrix}
     \mathrm{sin} \big( \lambda_{\alpha} - v_k + \eta \sigma^z_k \big)       &    \mathrm{sin} \big( 2 \eta \big) \sigma^{-}_k    \\
      \mathrm{sin} \big( 2 \eta \big) \sigma^{+}_k     &   \mathrm{sin}  \big( \lambda_{\alpha} - v_k - \eta \sigma^z_k \big)     
  \end{bmatrix} \text{, } 
\]

\begin{figure}
\begin{align*}
\includegraphics[width=0.6\columnwidth]{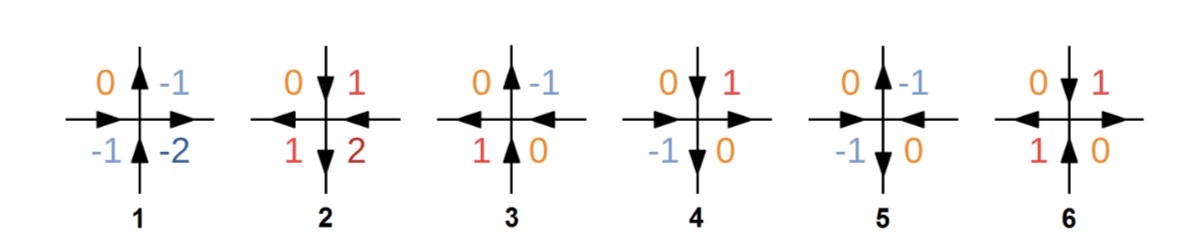}
\end{align*}
\caption{One depiction of each possible vertex for the six-vertex model, adapted from {\color{blue}[11]}.}
\end{figure}

\begin{figure}
\begin{align*}
\includegraphics[width=0.55\columnwidth]{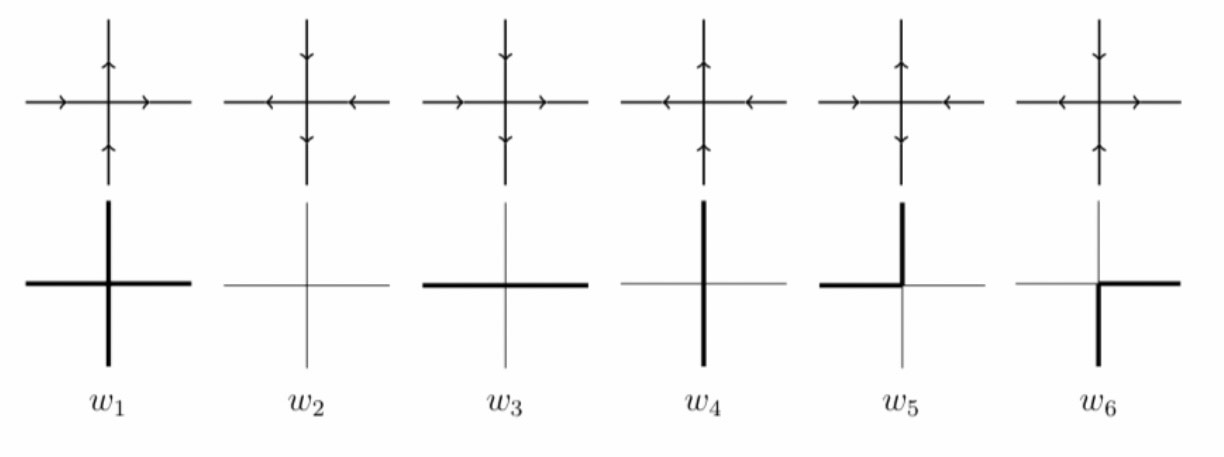}
\end{align*}
\caption{Another depiction of each possible vertex for the six-vertex model, adapted from {\color{blue}[25]}.}
\end{figure}

\noindent in which $\sigma^z_k,\sigma^{+}_k$, and $\sigma^{-}_k$ are Pauli basis elements. With such a representation for the operator above, one can perform several computations with respect to the Poisson bracket for characterizing the two-dimensional Poisson structure. The associated transfer matrix of the inhomogeneous 6-vertex model under such boundary conditions takes the form,

\begin{align*}
 T_a \big( u , \big\{ v_k \big\} , H , 0 \big) : \textbf{C}^2 \otimes \big( \textbf{C}^2 \big)^{\otimes |N|} \longrightarrow \textbf{C}^2 \otimes \big( \textbf{C}^2 \big)^{\otimes |N|}      \mapsto    \overset{-N}{\underset{i=1}{\prod}}   \bigg[ \mathrm{diag} \big( \mathrm{exp} \big( 2H \big) ,  \mathrm{exp} \big( 2 H \big)  \big)       R_{ia} \big( u - v_i \big)      \bigg]   \text{, } 
\end{align*}

\noindent where the R-matrix takes the form,

\[
R \equiv R \big( u , H , V \big) \equiv 
  \begin{bmatrix}
      a \text{ }  \mathrm{exp} \big(  H + V \big)    & 0 & 0 & 0  \\
    0 & b \text{ } \mathrm{exp} \big( H - V \big) & c & 0  \\0 & c & b \text{ }  \mathrm{exp} \big( - H + V \big) & 0 \\ 0 & 0 & 0 & a \text{ }  \mathrm{exp} \big( - H - V \big) \\ 
  \end{bmatrix}  \text{, }
\]

\noindent from the parametrization of 6-vertex weights,

\begin{align*}
   a_1 \equiv      a \text{ }  \mathrm{exp} \big(  H + V \big)  \text{, } \\ a_2 \equiv   a  \text{ }  \mathrm{exp} \big( - H - V \big)  \text{, } \\  b_1 \equiv  \text{ }  \mathrm{exp} \big( H - V \big)  \text{, } \\ b_2 \equiv \text{ }  \mathrm{exp} \big( - H + V \big)  \text{, } \\ c_1 \equiv  c \lambda  \text{, } \\ c_2 \equiv c \lambda^{-1} \text{, } 
\end{align*}

\noindent under the influence of two external fields $H$ and $V$, and some $\lambda >0$. For the 6-vertex model, in comparison to the 20-vertex model that will be introduced next, there exists two-dimensional action-angle coordinates, $\Phi$, which satisfies,

\begin{align*}
        \big\{  \Phi \big( \underline{\lambda} \big)  ,   \bar{\Phi \big( \underline{\lambda} \big) } \big\} = 0       \text{, } 
\end{align*}

\noindent with respect to the complex conjugate action-angle coordinates $\bar{\Phi}$. Over two dimensions, as is also the case that will be performed over three dimensions, a weak finite volume limit of the transfer matrix is taken, with,

\begin{align*}
   T_a \big( u , H , V \big)  =   \underset{v_k \longrightarrow - \infty} {\mathrm{lim}}  T_a \big(    u ,  \big\{ v_k \big\} , H , V \big)  =    \underset{v_i \longrightarrow - \infty} {\mathrm{lim}} \text{ } \bigg[  \overset{N}{\underset{i=1}{\prod}}   \big( 
    \mathrm{diag} \big( \mathrm{exp} \big( 2H \big) ,  \mathrm{exp} \big( 2 H \big)  \big)   \big)     R_{ia} \big( u - v_i \big) \bigg]   \text{. } 
   \end{align*}

\noindent With such a representation for the transfer matrix, for another choice of spectral parameters $u^{\prime}$, the relations with respect to the Poisson bracket which appear in the two-dimensional Poisson structure takes the form,

\[
\big\{ T_a \big( u , \big\{ v_k \big\} \big) \overset{\bigotimes}{,}   T_a \big( u^{\prime} , \big\{ v^{\prime}_k \big\} \big)   \big\} = \bigg\{    \begin{bmatrix} 
A \big( u \big)  & B \big( u \big)  \\ C \big( u \big) & D \big( u \big)   
\end{bmatrix}\overset{\bigotimes}{,}   \begin{bmatrix} 
A \big( u^{\prime} \big)  & B \big( u^{\prime} \big)  \\ C \big( u^{\prime} \big) & D \big( u^{\prime} \big)   
\end{bmatrix}  \bigg\}  \text{. } 
\]

\noindent Equipped with the two representations which have operators $A \big( u \big), \cdots, D \big( u \big), A \big( u^{\prime} \big), \cdots, D \big( u^{\prime}\big)$, one can study extensions of seminal computations with the Poisson bracket provided through Hamiltonian methods in {\color{blue}[17]}, from a suggestion of research interest provided by Keating, Reshetikhin, and Sridhar in {\color{blue}[25]}. For the 20-vertex model, in the place of spectral parameters $u$, or $u^{\prime}$, which are introduced into the operators for the block representation of the two-dimensional transfer matrix, for the three-dimensional transfer matrix the corresponding representation takes the form,

\[
\begin{bmatrix}
 A \big( \underline{u} \big) & D \big( \underline{u} \big)  & G \big( \underline{u} \big) \\ B \big( \underline{u} \big) & E \big( \underline{u} \big) & H \big( \underline{u} \big)  \\ C \big( \underline{u} \big)  &  F \big( \underline{u} \big) & I \big( \underline{u} \big) 
\end{bmatrix}  \text{. } 
\]

\noindent To make use of similar structures that were exploited for the quantum inverse scattering method in two dimensions under the presence of inhomogeneities for the 6-vertex model, {\color{blue}[42]}, for the 20-vertex model the L-operator,

\[
\hat{L} \big( \xi \big) \equiv L^{3D}_1 =  \mathrm{exp} \big( \lambda_3 ( q^{-2 } \xi^s ) \big)    \bigg[ \begin{smallmatrix}
        q^{D_1}       &    q^{-2} a_1 q^{-D_1-D_2} \xi^{s-s_1}        &   a_1 a_2 q^{-D_1 - 3D_2} \xi^{s - s_1 - s_2}  \\ a^{\dagger}_1 q^{D_1} \xi^{s_1} 
             &      q^{-D_1 + D_2} - q^{-2} q^{D_1 -D_2} \xi^{s}     &     - a_2 q^{D_1 - 3D_2} \xi^{s-s_2}  \\ 0  &    a^{\dagger}_2 q^{D_2} \xi^{s_2} &  q^{-D_2} \\   
  \end{smallmatrix} \bigg] \text{, }         
\]

\noindent is manipulated for obtaining explicit forms for the operators $A \big(\underline{u}\big), \cdots, I \big(\underline{u}\big)$ of the product representation for the three-dimensional transfer matrix. Under a relabeling of terms in the L-operator above for the 20-vertex model, through,

\[  \bigg[ \begin{smallmatrix}     q^{D^j_k}       &    q^{-2} a^j_k q^{-D^j_k -D^j_{k+1}} \xi^{s-s^k_j}        &  a^j_k a^j_{k+1} q^{-D^j_k - 3D^j_{k+1}} \xi^{s - s^j_k - s^j_{k+1}}  \\ \big( a^j_k \big)^{\dagger} q^{D^j_k} \xi^{s^j_k} 
             &      q^{-D^j_k + D^j_{k+1}} - q^{-2} q^{D^j_k -D^j_{k+1}} \xi^{s}     &     - a^j_k q^{D^j_k - 3D^j_{k+1}} \xi^{s-s^j_k}  \\ 0  &    a^{\dagger}_j q^{D^j_k} \xi^{s^j_k} &  q^{-D^j_k} \\    \end{smallmatrix} \bigg]\]

\noindent where, given some position vector $\textbf{r}$, and basis vectors $e_k$ and $e_{k+1}$ of $\textbf{T}$,

\begin{align*}
D^j_k  \equiv \big(  D      \otimes \textbf{1} \big) \textbf{1}_{\{\textbf{r} \equiv e_k\}}  \text{, } D^j_{k+1}\equiv \big(  \textbf{1} \otimes D \big) \textbf{1}_{\{\textbf{r} \equiv e_{k+1}\}}  \text{. }  
\end{align*}

\noindent To define a probability measure for the 20-vertex model, as is the case for the weight function for each vertex in a 6-vertex configuration, introduce,

\begin{align*}
     w_0 \equiv   a_1 a_2 a_3   \text{, } \\ 
    w_1 \equiv   b_1 a_2 b_3   \text{, } \\  w_2    \equiv   b_1 a_2 c_3    \text{, } \\ w_3 \equiv       a_1 b_2 b_3 + c_1 c_2 c_3  \text{, } \\   w_4 \equiv    c_1 a_2 a_3        \text{, } \\               w_5 \equiv   b_1 c_2 a_3   \text{, } \\  w_6 \equiv  b_1 b_2 a_3    \text{. } 
\end{align*}

\noindent With the weight function $w_{20V} \big( \cdot \big)$ (see Figure 3 and Figure 4 for a depiction of each possible configuration), the probability measure for the 20-vertex model takes the form,

\begin{align*}
   \textbf{P}^{20V}_{\textbf{T}}[      \omega         ]   \equiv \textbf{P}^{20V}[   \omega     ]     =  \frac{w_{20V}(\omega)}{Z^{20V}_{\textbf{T}}} \equiv \frac{w(\omega)}{Z_{\textbf{T}}} \text{, } 
\end{align*}

\noindent for some $\omega \in \Omega^{20V}$, supported over the torus. The three-dimensional transfer matrix takes the form,

\begin{align*}
 T^{3D}_{a,b} \big(    \big\{ u_i \big\} , \big\{ v^{\prime}_j  \big\} , \big\{ w^{\prime\prime}_k \big\}     \big) :   \textbf{C}^3 \otimes \big( \textbf{C}^3 \big)^{\otimes ( |N| + ||M||_1 )}  \longrightarrow   \textbf{C}^3 \otimes \big( \textbf{C}^3 \big)^{\otimes ( |N| + ||M||_1 )}   \mapsto  \overset{-N}{\underset{j=0}{\prod}} \text{ }  \overset{\underline{M}}{\underset{k=0}{\prod}} \bigg[  \mathrm{diag} \big( \mathrm{exp} \big(  \alpha \big(i, j,k \big)  \big) \\  , \mathrm{exp} \big(  \alpha \big( i,j,k\big)  \big)   , \mathrm{exp} \big(  \alpha \big( i, j,k \big)  \big) \big)  R_{ia,jb,kc} \big( u - u_i ,  u^{\prime} - v^{\prime}_j , w-w^{\prime\prime}_k \big) \bigg]  \text{, } 
\end{align*}

\noindent where the R-matrix in the product is taken to be the universal R-matrix, which has the factorization,

\begin{align*}
  R = R_{\leq \delta} R_{\sim \delta} R_{\geq \delta} K   \text{, }
\end{align*}

\noindent into q-exponentials and several related objects. The factorization above is one of many in the literature that have also been examined by the author for further developing the Bethe ansatz, and quantum inverse scattering, frameworks, for several models in Statistical Mechanics, with possible extensions of future interest for closely related models {\color{blue}[5},{\color{blue}8},{\color{blue}13},{\color{blue}14},{\color{blue}15},{\color{blue}18},{\color{blue}20},{\color{blue}22},{\color{blue}23},{\color{blue}24},{\color{blue}43]}. Fundamentally, as extensions of the quantum inverse scattering method of Hamiltonian systems, {\color{blue}[17]}, one must identify how sets of relations with the Poisson structure are related to products of L-operators, which determine, explicitly, the structure of operators in the product representation for the transfer matrix.

\begin{figure}
\begin{align*}
\includegraphics[width=0.6\columnwidth]{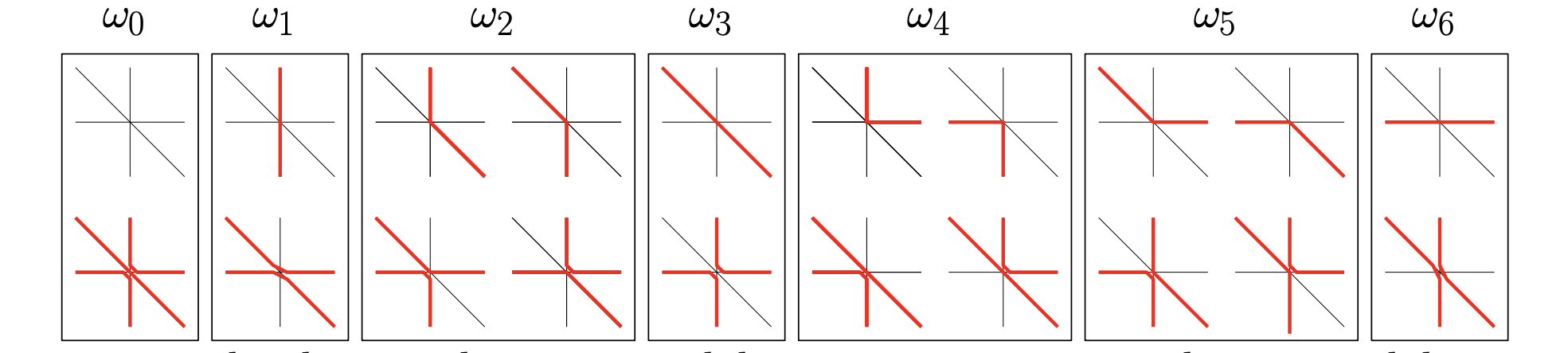}
\end{align*}
\caption{A depiction of each possible vertex for the triangular, or three dimensional, six-vertex model, adapted from {\color{blue}[16]}.}
\end{figure}

\begin{figure}
\begin{align*}
\includegraphics[width=0.65\columnwidth]{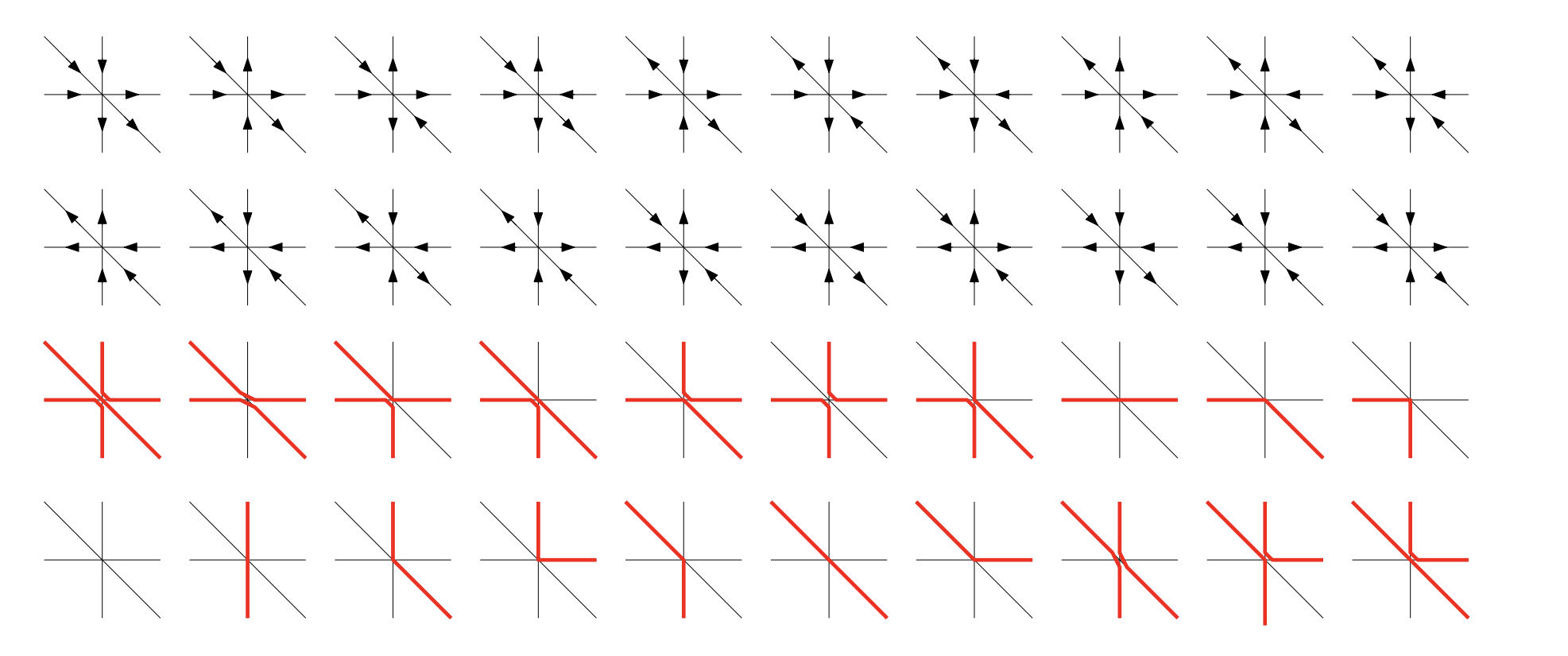}
\end{align*}
\caption{A depiction of each Boltzman weight for the triangular, or three dimensional, six-vertex model, also adapted from {\color{blue}[16]}.}
\end{figure}

\noindent Unlike the 6-vertex model under domain-wall boundary conditions, for the 20-vertex model there do not exist three-dimensional analogs for action angle coordinates. If such a three-dimensional counterpart to the two-dimensional object would exist, then the Poisson bracket,

\begin{align*}
        \big\{  \Phi^{3D} \big( \underline{\lambda} \big)  ,   \bar{\Phi^{3D} \big( \underline{\lambda} \big) } \big\}         \text{, } 
\end{align*}

\noindent between the three-dimensional action-angle variables, with its complex conjugate, would vanish. Irrespective of whether three-dimensional action angle coordinates exist, a weak finite volume limit of the three-dimensional transfer matrix can be taken, in which,

\begin{align*}
 \textbf{T}^{3D} \big( \underline{\lambda} \big)   \equiv    \underset{ N \longrightarrow -  \infty}{\underset{\underline{M} \longrightarrow + \infty}{\mathrm{lim}}} \mathrm{tr} \bigg[     \overset{\underline{M}}{\underset{j=0}{\prod}}  \text{ }  \overset{-N}{\underset{k=0}{\prod}} \mathrm{exp} \big( \lambda_3 ( q^{-2} \xi^{s^j_k} ) \big)     \bigg[ \begin{smallmatrix}     q^{D^j_k}       &    q^{-2} a^j_k q^{-D^j_k -D^j_{k+1}} \xi^{s-s^k_j}        &  a^j_k a^j_{k+1} q^{-D^j_k - 3D^j_{k+1}} \xi^{s - s^j_k - s^j_{k+1}}  \\ \big( a^j_k \big)^{\dagger} q^{D^j_k} \xi^{s^j_k} 
             &      q^{-D^j_k + D^j_{k+1}} - q^{-2} q^{D^j_k -D^j_{k+1}} \xi^{s}     &     - a^j_k q^{D^j_k - 3D^j_{k+1}} \xi^{s-s^j_k}  \\ 0  &    a^{\dagger}_j q^{D^j_k} \xi^{s^j_k} &  q^{-D^j_k} \\    \end{smallmatrix} \bigg]      \bigg] 
\text{. }
\end{align*}

\noindent As is the case for the 6-vertex model, one can take tensor products of the Poisson bracket for two 20-vertex transfer matrices, which takes the form,

\[
\big\{ \textbf{T} \big( \underline{u}  , \big\{ u_i \big\} ,  \big\{ v_j \big\} , \big\{ w_k \big\}  \big) \overset{\bigotimes}{,}  \textbf{T} \big( \underline{u^{\prime}} , \big\{ u^{\prime}_i \big\} ,  \big\{ v^{\prime}_j \big\} , \big\{ w^{\prime}_k \big\}   \big)   \big\} = \bigg\{    \begin{bmatrix}
 A \big( \underline{u} 
 \big) & D \big( \underline{u} \big)  & G \big( \underline{u}  \big) \\ B \big( \underline{u}  \big) & E \big( \underline{u}  \big) & H \big( \underline{u}  \big)  \\ C \big( \underline{u}  \big)  &  F \big( \underline{u}  \big) & I \big( \underline{u}  \big) 
\end{bmatrix}\overset{\bigotimes}{,}\begin{bmatrix}
 A \big( \underline{u^{\prime}} \big) & D \big( \underline{u^{\prime}} \big)  & G \big( \underline{u^{\prime}} \big) \\ B \big( \underline{u^{\prime}}\big) & E \big( \underline{u^{\prime}} \big) & H \big( \underline{u^{\prime}} \big)  \\ C \big( \underline{u^{\prime}} \big)  &  F \big( \underline{u^{\prime}} \big) & I \big( \underline{u^{\prime}} \big) 
\end{bmatrix} \bigg\} \text{. } 
\]

\noindent Besides the fact that two, and three, dimensional Poisson structure, for the 6-vertex, and 20-vertex, models, respectively, can be studied within the quantum inverse scattering framework results from the fact that the transfer matrix of each vertex model satisfies a series of relations with respect to the tensor product of the Poisson bracket. Fix some $\chi>0$. For the collection of such relations with respect to the tensor product of the Poisson bracket, denote a solution to the nonlinear Schrodinger's equation,

\begin{align*}
   i \frac{\partial \psi}{\partial t} = - \frac{\partial^2 \psi}{\partial x^2} + 2 \chi \big| \psi \big|^2 \psi      \text{, } 
\end{align*}

\noindent with $\psi$, under well posed initial data. For the 6-vertex model, such relations take the form,

\begin{align*}
      \big\{  T_{-} \big( x , \underline{\lambda} \big) \overset{\bigotimes}{,} T_{-} \big( x , \underline{\mu} \big)           \big\} = r \big( \underline{\lambda} - \underline{\mu} \big) T_{-} \big( x , \underline{\lambda} \big)\bigotimes T_{-} \big( x , \underline{\mu} \big) - T_{-} \big( x , \underline{\lambda} \big) \bigotimes T_{-} \big( x , \underline{\mu} \big) r_{-} \big( \underline{\lambda} - \underline{\mu} \big)         \text{, } \\     \big\{  T_{+} \big( x , \underline{\lambda} \big) \overset{\bigotimes}{,} T_{+} \big( x , \underline{\mu} \big)           \big\} = T_{+} \big( x , \underline{\lambda} \big) \bigotimes T_{+} \big( x , \underline{\mu} \big) r_{+} \big( \underline{\lambda} - \underline{\mu} \big) - r \big( \underline{\lambda} - \underline{\mu} \big) T_{+} \big( x , \underline{\lambda} \big) \bigotimes T_{+} \big( x , \underline{\mu} \big)   \text{, } 
\end{align*}

\noindent where,

\begin{align*}
  r_{\pm} \big( \lambda - \mu \big) = \underset{y \longrightarrow \pm \infty}{\mathrm{lim}}   E \big( y , \mu - \lambda \big)   \bigotimes E \big( y , \lambda - \mu \big) r \big( \lambda - \mu \big)         \text{, } 
\end{align*}

\noindent for,

\begin{align*}
  E \big( x , \lambda \big) \equiv \mathrm{exp} \big( \lambda x U_1 
 \big) \text{, } 
\end{align*}

\noindent and,

\[
U_0  \equiv \sqrt{\chi}
  \begin{bmatrix}
       0  &  \psi \\
    \bar{\psi}  & 0 \text{ }  
  \end{bmatrix} \text{, } 
\]

\[
U_1  \equiv 
 \frac{1}{2i} \begin{bmatrix}
   1     &  0  \\
   0  &  -1 \text{ }  
  \end{bmatrix} \text{. } 
\]

\noindent As was the case for the 6-vertex model, for the 20-vertex model denote $\psi$ to be a solution of the three-dimensional nonlinear Schrodinger's equation, which can be obtained from the two-dimensional PDE from the 6-vertex model setting, also taken under well posed initial data. For the 20-vertex model, the three-dimensional counterpart of the two-dimensional relations above take the form,

\begin{align*}
      \big\{  \textbf{T}_{-} \big( x , \underline{\lambda} \big) \overset{\bigotimes}{,} \textbf{T}_{-} \big( x , \underline{\mu} \big)           \big\} = r^{3D} \big( \underline{\lambda} - \underline{\mu} \big) \textbf{T}_{-} \big( x , \underline{\lambda} \big)\bigotimes \textbf{T}_{-} \big( x , \underline{\mu} \big) - \textbf{T}_{-} \big( x , \underline{\lambda} \big) \bigotimes \textbf{T}_{-} \big( x , \underline{\mu} \big) r^{3D}_{-} \big( \underline{\lambda} - \underline{\mu} \big)     \text{, } \\     \big\{  \textbf{T}_{+} \big( x , \underline{\lambda} \big) \overset{\bigotimes}{,} \textbf{T}_{+} \big( x , \underline{\mu} \big)           \big\} = \textbf{T}_{+} \big( x , \underline{\lambda} \big) \bigotimes \textbf{T}_{+} \big( x , \underline{\mu} \big) r^{3D}_{+} \big( \underline{\lambda} - \underline{\mu} \big) - r^{3D} \big( \underline{\lambda} - \underline{\mu}  \big) \textbf{T}_{+} \big( x , \underline{\lambda} \big) \bigotimes \textbf{T}_{+} \big( x , \underline{\mu} \big)        \text{, } 
\end{align*}

\noindent where,

\begin{align*}
  r^{3D}_{\pm} \big( \underline{\lambda} - \underline{\mu} \big) = \underset{x \longrightarrow + \infty}{\underset{y \longrightarrow \pm \infty}{\mathrm{lim}}} \bigg[   E^{3D} \big( x,y , \underline{\mu} - \underline{\lambda} \big)   \bigotimes \bigg[ E^{3D} \big( x,y , \underline{\lambda} - \underline{\mu} \big) r^{3D} \big( \underline{\lambda} - \underline{\mu} \big)  \bigg] \bigg]    \text{, } 
\end{align*}

\noindent for,

\begin{align*}
   E^{3D,\mathrm{6V}} \big(   \underline{u^{\prime}}  ,  v^{\prime}_k - v_k ,  u^{\prime}_k - u_k  , w^{\prime}_k - w_k \big)  \equiv    \mathrm{exp} \bigg(  \frac{1}{2i} \begin{bmatrix} 1 & 0 & 0 \\ 0 & -1 & 0 \\ 0 & 0 & 0 \end{bmatrix}   +  \begin{bmatrix} 0 & 0 & \psi  \\ 0 & \bar{\psi}   & 0 \\ 0 & 0 & 0 \end{bmatrix}        \bigg)       \text{. } 
\end{align*}

\noindent With objects of the 6-vertex, and 20-vertex, models, we now provide counterpart objects for the 4-vertex model. As alluded to in the introduction, manipulating L-operators, the transfer matrix, and the quantum monodromy matrix, for the 4-vertex model within the quantum inverse scattering framework. Despite the fact that previous works of the author have investigated completely different facets of the quantum inverse scattering framework, {\color{blue}[42},{\color{blue}46]}, focusing primarily upon notions of integrability and Poisson structure, performing similar computations with L-operators of the 4-vertex model can provide new insights into various Poisson structures for vertex models with smaller state spaces than the 6-vertex, and 20-vertex, models. To this end, the first object for the 4-vertex model, the weight function, takes the form,

\begin{align*}
 w^{4V}_{\textbf{Z}^2} \big( \omega \big) \equiv  a^{n_1+n_2} c^{n_5 + n_6}  \text{, }
\end{align*}

\noindent for a configuration $\omega \in \Omega^{4V}$. The parameters $a$ and $c$ appearing in the expression for the 4-vertex weight above coincide with the isotropic parameters $a,c$ introduced for the weight function of the 6-vertex model over vertices of $\textbf{Z}^2$. From the weight function, the corresponding probability measure over the state space consisting of $4$ configurations takes the form,

\begin{align*}
 \textbf{P}^{4V,\xi}_{\textbf{Z}^2} \big[ \omega \big] \equiv \textbf{P}^{\xi}_{\textbf{Z}^2} \big[ \omega \big]   =  \frac{w^{4V}_{\textbf{Z}^2} \big( \omega \big) }{Z^{4V}_{\textbf{Z}^2} } \text{, }
\end{align*}

\noindent under fixed boundary conditions $\xi$, which can be taken to slightly differ from domain-walls enforced for the 6-vertex and 20-vertex models. The measure is normalized by the partition function of the 4-vertex model so that the resulting ratio is indeed a probability measure. In the presence of such boundary conditions, the partition function for the 4-vertex model, within the context of the quantum inverse scattering method has several representations, one of which is, {\color{blue}[4]},

\begin{align*}
     Z \big( \big\{ w_a \big\} , \big\{ w_b \big\} , \big\{ w_c \big\} \big) =   \underset{\omega \in \Omega^{4V}}{\sum}  \bigg[      \underset{-N \leq k \leq -1}{\prod}  \big( w_a \big)^{l^a_k}_k  \big( w_b \big)^{l^b_k}_k  \big( w_c \big)^{l^c_k}_k              \underset{1 \leq j \leq N}{\prod}    \big( w_a \big)^{l^a_j}_j \big( w_b \big)^{l^b_j}_j  \big( w_c \big)^{l^c_j}_j     \bigg]        \text{, }
\end{align*}

\noindent over the finite volume $N \times N \times N$, for the number of vertices of types $a,b,c$, which are given by $l^a,l^b,l^c$, respectively, corresponding to a product over weights,

\begin{align*}
  w_{a_1} \equiv w_{a_2} \equiv w_a  \text{, } \\      w_{b_1} \equiv w_{b_2} \equiv w_b      \text{, } \\      w_{c_1} \equiv w_{c_2} \equiv w_c  \text{. }
\end{align*}

\noindent From representations of the probability measure and normalizing constant, the partition function, so that the probability over all configurations in the 4-vertex model equals $1$, for defining objects within the quantum inverse scattering framework, as has been done for the 6-vertex model, and 20-vertex model, introduce,

\begin{align*}
    T^{4V} \big( \underline{u} \big) \equiv T \big( u \big) =     \underset{0 \leq j \leq M}{\prod}  L \big( j | u \big) \equiv    \begin{bmatrix}
        A \big( \underline{u} \big) & B \big( \underline{u} \big) \\ C \big( \underline{u} \big) & D \big( \underline{u} \big) 
    \end{bmatrix}     = \begin{bmatrix}
        A \big( u \big) & B \big( u \big) \\ C \big( u \big) & D \big( u \big) 
    \end{bmatrix} \equiv \underset{0 \leq j \leq M}{\prod}   \begin{bmatrix}
        \underline{\mathcal{I}^j_1}  & \underline{\mathcal{I}^j_2}  \\ \underline{\mathcal{I}^j_3}  & \underline{\mathcal{I}^j_4}  
    \end{bmatrix} \text{, }
\end{align*}

\noindent corresponding to quantum monodromy matrix, taken under fixed boundary conditions, for the L-operator,

\begin{align*}
    L^{4V} \big(  n | u \big) \equiv L \big( n | u \big)    \equiv  \begin{bmatrix}
         L_{11} \big( n | u \big) & L_{12} \big( n | u \big) \\ L_{21} \big( n | u \big) & L_{22} \big( n | u \big) 
    \end{bmatrix}  = \begin{bmatrix}
       - u e_n & \sigma^{-}_n \\ \sigma^{+}_n & u^{-1} e_n 
    \end{bmatrix}
    \text{, }
\end{align*}

\noindent given some $n>0$, where $\sigma^{\pm}$ denote the Pauli matrices, and $e$ denotes the projector $\frac{1}{2} \big( \sigma^{\pm} + 1 \big)$. When performing computations with the L-operator, each degree of freedom takes the form as the union of each Pauli basis element, in addition to the union over all projects, each of which respectively take the form,

\begin{align*}
 \mathrm{DOF\text{ }  1} \equiv \underset{n \in \textbf{Z}}{\bigcup}  \sigma^-_n     \text{, } \\ \mathrm{DOF\text{ }  2} \equiv  \underset{n \in \textbf{Z}}{\bigcup}  \sigma^+_n    \text{, } \\ \mathrm{DOF\text{ }  3} \equiv  \underset{n \in \textbf{Z}}{\bigcup}  e_n  \text{. }
\end{align*}

\noindent Furthermore, also fix the basis elements,

\begin{align*}
 \ket{\Uparrow} =  \begin{bmatrix}0 \\ 1
 \end{bmatrix} \text{, } \\ \ket{\Downarrow } = \begin{bmatrix} 1 \\ 0
 \end{bmatrix}  \text{. }
\end{align*}

\noindent The quantum monodromy matrix can be used to define the transfer matrix, as,

\begin{align*}
 t^{4V} \big( u \big) =  \mathrm{tr} \big( T \big( u \big) \big)   \text{. }
\end{align*}

\noindent In the entries of the L-operator above for the 4-vertex model, as in other representations for the L-operator for the 6-vertex model in the presence of inhomogeneities, {\color{blue}[8]}, $\sigma^{\pm}_n$ denote Pauli matrices, while,

\begin{align*}
  e_n = \textbf{I} \otimes \underset{n-2}{\underbrace{\cdots}} \otimes e \otimes \underset{n-2}{\underbrace{\cdots}} \otimes \textbf{I}  \text{. }
\end{align*}

\noindent From the transfer matrix, one can study solutions to the Bethe equations, analytical expressions for the maximal and minimal eigenvalues of the spectrum of the transfer matrix, {\color{blue}[25]}, in addition to several other qualities of objects within the quantum inverse scattering method that relate to integrability and action-angle coordinates {\color{blue}[25},{\color{blue}42]}.

\bigskip

\noindent As will be discussed in later sections, computations with L-operators for the 4-vertex model can be transferred to the higher-spin XXX chain. For this vertex model that one more configuration in the state space than the 4-vertex model, the L-operator takes a very similar form as the L-operator does for the 4-vertex model, which takes the form, {\color{blue}[28]},

\begin{align*}
 L^{XXX} \big( n | \lambda \big)  \equiv L \big( n | \lambda \big)  \equiv      \begin{bmatrix}
         L_{11} \big( n | \lambda \big) & L_{12} \big( n | \lambda \big) \\ L_{21} \big( n | \lambda \big) & L_{22} \big( n | \lambda \big) 
    \end{bmatrix} =   \begin{bmatrix}
  \lambda I_n + i S^3_n & i S^-_n \\ i S^+_n & \lambda I_n - i S^3_n
    \end{bmatrix}
    \text{, }
\end{align*}

\noindent where $\lambda$ denotes the spectral parameter, and $n$, as in the previous definition of the L-operator for the 4-vertex model, denotes a quantum parameter. To rigorously analyze the transfer, and quantum monodromy, matrices, of the 4-vertex model requires knowledge of the Yang-Baxter algebra. Within the quantum inverse scattering framework, in previous work of the author a three-dimensional analog of such an algebra was characterized for being able to compute determinantal, and contour integral, representations of nonlocal correlations for the 20-vertex model {\color{blue}[47]}. As expected, the transfer matrix for the 5-vertex model, from the transfer matrix for the higher-spin XXX chain takes the form,

\begin{align*}
    T^{XXX} \big( \underline{u} \big) \equiv T \big( u \big) =     \underset{0 \leq j \leq M}{\prod}  L \big( n | \lambda \big)\equiv \begin{bmatrix}
        A^{XXX} \big( \underline{u} \big) & B^{XXX} \big( \underline{u} \big) \\ C^{XXX} \big( \underline{u} \big) & D^{XXX} \big( \underline{u} \big) 
    \end{bmatrix} = \begin{bmatrix}
        A^{XXX} \big( u \big) & B^{XXX} \big( u \big) \\ C^{XXX} \big( u \big) & D^{XXX} \big( u \big) 
    \end{bmatrix} \text{. }
\end{align*}

\noindent For the 4-vertex model, the fact that the model has a smaller state space than the 6-vertex model, and the 20-vertex model, allows for one to study combinatorial implications of the quantum inverse scattering method, not only through boxed plane partitions of two and three dimensional spaces, but also through classes of polynomials that can be placed into correspondence with the partition function. Beginning in the next section, we demonstrate how computations with the L-operator provide information on properties of the transfer and quantum monodromy matrices.

\subsection{Paper organization}

\subsubsection{General description of objectives}

\noindent With the quantum inverse scattering type objects defined for the 4-vertex, 5-vertex, 6-vertex, and 20-vertex, models, we proceed to explicitly characterize the asymptotic representation of the transfer matrix for the 4-vertex model in the next section from the corresponding L-operator. In performing such computations, we further streamline manipulation of objects associated with the quantum inverse scattering method, which was originally suggested as a direction of interest from integrability of inhomogeneous limit shapes of the 6-vertex model {\color{blue}[25]}. In comparison to the primary result of applying the quantum inverse scattering method for characterizing integrability of a Hamiltonian flow for the 6-vertex model in the presence of inhomogeneities {\color{blue}[42]}, applying the quantum inverse scattering method in higher dimensions reflects on difficulties of being able to conclude that vertex models with higher, or even lower, dimensional sample spaces are exactly solvable, or integrable, which is the case for the 20-vertex model which satisfies a weakened version of integrability from that of the 6-vertex model {\color{blue}[46]}. From information on the asymptotic expansions of the transfer and quantum monodromy matrices that we gather for the 4-vertex model, analagous properties can be deduced for the 5-vertex model, whose L-operator representation differs from that of the 5-vertex model by one entry.

In the asymptotic expansions for each of the four operators of the transfer matrix for the 4-vertex model, we make use of the the sets,

\begin{align*}
 \mathrm{support} \big( A \big( u \big) \big)  =              \mathrm{span}  \big\{   A_{n-1} \big( u \big)         \big\}     \text{, } \\  \mathrm{support} \big( B \big( u \big) \big)  =  \mathrm{span}  \big\{   B_{n-1} \big( u \big)         \big\}    \text{, } \\  \mathrm{support} \big( C \big( u \big) \big)  =  \mathrm{span}  \big\{   C_{n-1} \big( u \big)         \big\}   \text{, } \\            \mathrm{support} \big( D \big( u \big) \big)  =  \mathrm{span}  \big\{   D_{n-1} \big( u \big)         \big\}   \text{. }
\end{align*}

\noindent corresponding to the supports of each operator which are obtained from representations of products of L-operators. Taking the union over supports of the operators of the transfer matrix representation satisfies,

\begin{align*}
   \big[  \mathrm{support} \big( A \big( u \big) \big) \cup \mathrm{support} \big( B \big( u \big) \big)  \cup \mathrm{support} \big( C \big( u \big) \big)  \cup \mathrm{support} \big( D \big( u \big) \big) \big]  \subsetneq \mathrm{span} \big( \textbf{Z}^2 \big)   \text{. }
\end{align*}

\subsubsection{Computations for obtaining an asymptotic approximation of the transfer matrix representation}

\noindent From computations with the L-operator in previous works of the author for the 6-vertex model, and the 20-vertex model, for the 4-vertex model, one can also obtain a product representation for the transfer matrix through the following approach. First, one works out a base case in which four entries of the product representation are obtained; such entries of the first product representation are used to build entries of higher order product representation for the transfer matrix by repeatedly multiplying L-operators together. Second, from the entries of the product representation obtained in the first result in the next section, entries of the desired asymptotic representation for the transfer matrix are obtained by making use of the support sets introduced above for each of the operators $A,B,C,D$. Equipped with such support sets for each operator, entries of higher order product representations for the transfer matrix are expressed in terms of indicator functions and linear combinations of Pauli basis elements appearing in the original definition of the L-operator for the 4-vertex model. Third, with one possible expansion for the transfer matrix of the 4-vertex model from products of L-operators, approximations of relations from the Poisson structure of the 4-vertex model can be obtained. As is the case from computations of the L-operator for the 6-vertex model, in addition to those for L-operators of the 20-vertex model, repeated applications of (AC), and (LR), can be used to decompose Poisson brackets that are dependent upon several parameters into a superposition of Poisson brackets. In the final superposition that is obtained for each Poisson bracket, a desired approximation for every term within the superposition can be obtained because each bracket is dependent upon one term in each argument. Fourth, after having obtained approximations for each bracket within the Poisson structure, we discuss how Poisson structure, and several accompanying computations with the Poisson bracket, can also be obtained for the higher-spin XXX chain. Despite the fact that spin chains and vertex models are defined on completely different state spaces, in addition to being defined in terms of different local interactions (ie, the ice rule for vertex models on the square lattice versus nearest neighbor interactions for spin chains on the real line), L-operators for vertex models can be more or less be immediately leveraged for study transfer matrices, and quantum monodromy matrices, of the higher-spin XXX chain. Fifth, after having performed computations with the Poisson bracket from asymptotic approximations of the operators $A,B,C,D$ of the 4-vertex transfer matrix, with several straightforward computations one can approximate products of operators from the transfer matrix representation that was also used for approximating brackets in the Poisson structure. Given the fact that such computations for relations of the Yang-Baxter algebra for the 4-vertex model, and relatedly, for the higher-spin XXX chain are obtained from direct manipulation of asymptotic representations for $A,B,C,D$, they are included in the Appendix.

\section{L-operator computations}

\noindent We obtain expressions for products of L-operators, first beginning with two terms. In a similar way that entries of L-operators, such as those given in three-dimensional L-operators from the set of expressions on Page \textit{29} of {\color{blue}[46]}, originally developed from a construction of the Universal R-matrix, {\color{blue}[7]}, determine entries of the product representations for L-operators that are given in results on Pages \textit{32}, \textit{35}, \textit{37}, and \textit{39}, all of which lead to the one portion of the generalized product representation for the first row of the transfer matrix on Page \textit{51}. The fact that the L-operator for the 4-vertex, and 5-vertex, models introduced in the previous section is lower dimensional significantly reduces the amount of terms that must be computed explicitly in the product representation.

\bigskip

\noindent \textbf{Lemma} \textit{0} (\textit{product representation for two terms}, \textbf{Lemma} \textit{1}, {\color{blue}[42]}, \textbf{Lemma} \textit{2}, {\color{blue}[45]}). The first representation corresponding to the product of two L-operators for the 4-vertex model equals,

\[
\begin{bmatrix}
  \underline{\mathcal{I}_1}  & \underline{\mathcal{I}_2} \\ \underline{\mathcal{I}_3} &  \underline{\mathcal{I}_4} 
\end{bmatrix}
\]

\noindent \textit{Proof of Lemma 0}. By inspection, explicitly, each entry takes the form,

\begin{align*}
 \underline{\mathcal{I}_1} \equiv              i u \sigma^+_0 \sigma^-_0 i u \sigma^+_1 \sigma^-_1 +  \sigma^-_0 \sigma^+_1    \text{, }  \\ \\ 
 \underline{\mathcal{I}_2} \equiv   i u \sigma^+_0 \sigma^-_0 \sigma^-_1  + \sigma^-_0 i u \sigma^+_1 \sigma^-_1      \text{, } \\ \\ \underline{\mathcal{I}_3} \equiv      i u \sigma^+_0 \sigma^-_0 \sigma^+_1 + \sigma^+_0 i u \sigma^+_1 \sigma^-_1      \text{, } \\ \\   \underline{\mathcal{I}_4} \equiv            \sigma^+_0 \sigma^-_1 + i u \sigma^+_0 \sigma^-_0 i u \sigma^+_1 \sigma^-_1  \text{, }
\end{align*}

\noindent from which we conclude the argument. \boxed{}

\bigskip

\noindent We generalize the entries of the product representation obtained above by obtaining higher order terms of a system of recrusive relations that are used to obtain entries of the L-operator product representation. When taking products of L-operators, as a proxy for the asymptotic expansion of the transfer matrix, we denote entries of the product representation with $\underline{\mathcal{I}^{\prime}_1}, \cdots, \underline{\mathcal{I}^{\prime}_4}$, for the entries of the product representation with two L-operators. Asymptotically, the operators $\underline{\mathcal{I}^{\prime \cdots\prime}_1},\cdots, \underline{\mathcal{I}^{\prime\cdots\prime}_4}$ are the entries of the desired product representation.

\bigskip

\noindent \textbf{Lemma} \textit{1} (\textit{adding on another term to the product representation of L-operators}, \textbf{Lemma} \textit{1}, {\color{blue}[42]}, \textbf{Lemma} \textit{2}, {\color{blue}[45]}, \textbf{Lemma} \textit{3}, {\color{blue}[45]}). The second representation corresponding to the product of the previous relation obtained in \textbf{Lemma} \textit{1}, with the next L-operator, can be expressed as,

\[
\begin{bmatrix}
 \underline{\mathcal{I}^{\prime}_1}   & \underline{\mathcal{I}^{\prime}_2}  \\ \underline{\mathcal{I}^{\prime}_3}   &  \underline{\mathcal{I}^{\prime}_4} 
\end{bmatrix}
\]

\noindent which is equivalent to the union of the set of linear combinations of,

\[
\begin{bmatrix}
 i u \sigma^+_0 \sigma^-_0    i u \sigma^+_1 \sigma^-_1 i u \sigma^+_1 \sigma^-_1 i u \sigma^+_2 \sigma^-_2 + \sigma^-_0 \sigma^+_1 i u \sigma^+_1 \sigma^-_1 i u \sigma^+_2 \sigma^-_2 + i u \sigma^+_0 \sigma^-_0 i u \sigma^+_1  \sigma^-_1 \sigma^-_1 \sigma^+_2 + \sigma^-_0 \sigma^+_1 \sigma^-_1 \sigma^+_2 \\    \\            i u \sigma^+_0 \sigma^-_0 i u \sigma^+_1 \sigma^-_1 i u \sigma^+_1 \sigma^-_1 \sigma^+_2  + \sigma^-_0 \sigma^+_1 i u \sigma^+_1 \sigma^-_1 \sigma^+_2 - i u \sigma^+_0 \sigma^-_0 i u \sigma^+_1 \sigma^-_1 \sigma^+_1 i u \sigma^+_2 \sigma^-_2 + \sigma^-_0 \sigma^+_1 \sigma^+_1 i u \sigma^+_2 \sigma^-_2 \\ + i u \sigma^+_0 \sigma^+_1 \sigma^+_1 \sigma^-_2 + i u  \sigma^+_0 \sigma^+_1 i u \sigma^+_1 \sigma^-_1 i u \sigma^+_2 \sigma^-_2 +  \sigma^+_0 i u \sigma^+_1 \sigma^-_1 \sigma^+_1 \sigma^-_2 +    \sigma^+_0 i u \sigma^+_1 \sigma^-_1 i u \sigma^+_1 \sigma^-_1 \\ \times i u \sigma^+_2 \sigma^-_2          
\end{bmatrix} \text{, }
\]

\noindent corresponding to the first row of the product representation, and of,

\[
\begin{bmatrix}
 i u \sigma^+_0 \sigma^-_0 i u \sigma^+_1 \sigma^-_1 i u \sigma^+_1 \sigma^-_1 \sigma^-_2 + i u \sigma^+_0 \sigma^-_0 i u \sigma^+_1 \sigma^-_1 \sigma^-_1 i u \sigma^+_2 \sigma^-_2  + \sigma^-_0 \sigma^+_1 i u \sigma^+_1 \sigma^-_1 \sigma^-_2 +          \sigma^-_0 \sigma^+_1 \sigma^-_1 i u \sigma^+_2 \sigma^-_2 \\ +     i u \sigma^+_0 \sigma^-_0 \sigma^-_1 \sigma^+_2 \sigma^-_2 + i u      \sigma^+_0 \sigma^-_0 \sigma^-_1 i u \sigma^+_1 \sigma^-_1 i u \sigma^+_2 \sigma^-_2 + \sigma^-_0 i u \sigma^+_1 \sigma^-_1 \sigma^+_1 \sigma^-_2 + \sigma^-_0 i u \sigma^+_1 \sigma^-_1 i u \\ \times \sigma^+_1 \sigma^-_1 i u \sigma^+_2 \sigma^-_2          \\  \\                                      i u \sigma^+_0 \sigma^-_0 i u \sigma^+_1 \sigma^-_1 i u \sigma^+_1 \sigma^-_1 \sigma^-_2 + i u \sigma^+_0       \sigma^-_0 i u \sigma^+_1 \sigma^-_1 \sigma^-_1 i u \sigma^+_2 \sigma^-_2    + \sigma^-_0 \sigma^+_1 i u \sigma^+_1 \sigma^-_1 \sigma^-_2 + \sigma^-_0 \sigma^+_1 \sigma^-_1 i u \sigma^+_2 \sigma^-_2   \\ +                i u \sigma^+_0 \sigma^-_0 \sigma^-_1 \sigma^+_1 \sigma^-_2 + i u \sigma^+_0 \sigma^-_0 \sigma^-_1 i u \sigma^+_1 \sigma^-_1 i u \sigma^+_2 \sigma^-_2 + \sigma^-_0 i u \sigma^+_1 \sigma^-_1 \sigma^+_1 \sigma^-_2 + \sigma^-_0 i u \sigma^+_1 \sigma^-_1 i u \\ \times \sigma^+_1 \sigma^-_1          i u \sigma^+_2 \sigma^-_2        
\end{bmatrix} \text{, }
\]

\noindent corresponding to the second row of the product representation.

\bigskip

\noindent We also obtain a system of expressions for the product representation of L-operators by multiplying the expression from $\textbf{Lemma}$ \textit{1} by more than one L-operator at a time.

\bigskip

\noindent \textbf{Lemma} \textit{2} (\textit{adding an even number of terms to the product representation of L-operators}, \textbf{Lemma} \textit{1}, {\color{blue}[42]}, \textbf{Lemma} \textit{2}, {\color{blue}[45]}, \textbf{Lemma} \textit{3}, {\color{blue}[45]}). There exists functions $\mathscr{I}^1_1 \big(u, u^{-1} \big) \equiv \mathscr{I}^1_1$, $\mathscr{I}^2_1 \big(u, u^{-1} \big) \equiv \mathscr{I}^2_1$, $\cdots$, $\mathscr{I}^4 \big(u, u^{-1} \big) \equiv \mathscr{I}^4$, for which the general form for the product representation of L-operators, from the product representation in the previous result, can be expressed as the span of the two following representations, the first of which equals,

\[
\begin{bmatrix}
\mathscr{I}^1_1 + \mathscr{I}^2_1 + \mathscr{I}^3_1  \\ \\ \mathscr{I}^1_2 + \mathscr{I}^2_2 + \mathscr{I}^2_3  \\ 
\end{bmatrix} \equiv \begin{bmatrix}
\mathscr{I}^1_1 \big( \underline{u} \big) + \mathscr{I}^2_1 \big( \underline{u} \big) + \mathscr{I}^3_1  \big( \underline{u} \big) \\ \\ \mathscr{I}^1_2  \big( \underline{u} \big) + \mathscr{I}^2_2  \big( \underline{u} \big) + \mathscr{I}^2_3 \big( \underline{u} \big) \\ 
\end{bmatrix} \text{, }
\]

\noindent corresponding to the first column of the product representation for the transfer matrix, and the second of which equals,

\[
\begin{bmatrix}
 \mathscr{I}^1_3 + \mathscr{I}^2_3 + \mathscr{I}^3_3  \\ \\ \mathscr{I}^4_1 + \mathscr{I}^4_2 + \mathscr{I}^4_3  \\ 
\end{bmatrix}  \equiv \begin{bmatrix}
 \mathscr{I}^1_3 \big( \underline{u} \big)  + \mathscr{I}^2_3 \big( \underline{u} \big) + \mathscr{I}^3_3 \big( \underline{u} \big)  \\ \\ \mathscr{I}^4_1 \big( \underline{u} \big) + \mathscr{I}^4_2 \big( \underline{u} \big) + \mathscr{I}^4_3 \big( \underline{u} \big)  \\ 
\end{bmatrix} \text{. }
\]

\noindent From the entries of the transfer matrix, one can introduce the Poisson structure $\mathscr{P}$ of the 4-vertex model below. Such a structure is given by the following collection of Poisson brackets. Equipped with another spectral parameter $u^{\prime}$, one would like to approximate the collection of Poisson brackets, $\mathscr{C}_1, \mathscr{C}_2, \mathscr{C}_3, \mathscr{C}_4$, given by,

\[
\mathscr{P} \equiv \left\{\!\begin{array}{ll@{}>{{}}l}      
\mathscr{C}_1 \equiv  \left\{\!\begin{array}{ll@{}>{{}}l}          \big\{ \underline{\mathcal{I}^{\prime\cdots\prime}_1 \big( u , u^{-1} \big)}, \underline{\mathcal{I}^{\prime\cdots\prime}_1 \big( u^{\prime} , \big( u^{\prime} \big)^{-1} \big)}\big\} 
\text{, } \\  
  \big\{ \underline{\mathcal{I}^{\prime\cdots\prime}_1 \big( u , u^{-1} \big)},  \underline{\mathcal{I}^{\prime\cdots\prime}_2 \big( u^{\prime} , \big( u^{\prime} \big)^{-1} \big)}  \big\} \text{, }  \\      \big\{ \underline{\mathcal{I}^{\prime\cdots\prime}_1 \big( u , u^{-1} \big)},  \underline{\mathcal{I}^{\prime\cdots\prime}_3 \big( u^{\prime} , \big( u^{\prime} \big)^{-1} \big)} \big\}    \text{, } \\  \big\{ \underline{\mathcal{I}^{\prime\cdots\prime}_1 \big( u , u^{-1} \big)}, \underline{\mathcal{I}^{\prime\cdots\prime}_4 \big( u^{\prime} , \big( u^{\prime} \big)^{-1} \big)}\big\}   \text{, } \end{array}\right.  \\  \mathscr{C}_2 \equiv  \left\{\!\begin{array}{ll@{}>{{}}l}          \big\{ \underline{\mathcal{I}^{\prime\cdots\prime}_2 \big( u , u^{-1} \big)}, \underline{\mathcal{I}^{\prime\cdots\prime}_1 \big( u^{\prime} , \big( u^{\prime} \big)^{-1} \big)}\big\} 
\text{, } \\  
  \big\{ \underline{\mathcal{I}^{\prime\cdots\prime}_2 \big( u , u^{-1} \big)},  \underline{\mathcal{I}^{\prime\cdots\prime}_2 \big( u^{\prime} , \big( u^{\prime} \big)^{-1} \big)}  \big\} \text{, }  \\      \big\{ \underline{\mathcal{I}^{\prime\cdots\prime}_2 \big( u , u^{-1} \big)},  \underline{\mathcal{I}^{\prime\cdots\prime}_3 \big( u^{\prime} , \big( u^{\prime} \big)^{-1} \big)} \big\}    \text{, } \\  \big\{ \underline{\mathcal{I}^{\prime\cdots\prime}_2 \big( u , u^{-1} \big)}, \underline{\mathcal{I}^{\prime\cdots\prime}_4 \big( u^{\prime} , \big( u^{\prime} \big)^{-1} \big)}\big\}   \text{, } \end{array}\right.  \\  \mathscr{C}_3 \equiv  \left\{\!\begin{array}{ll@{}>{{}}l}         \big\{ \underline{\mathcal{I}^{\prime\cdots\prime}_3 \big( u , u^{-1} \big)}, \underline{\mathcal{I}^{\prime\cdots\prime}_1 \big( u^{\prime} , \big( u^{\prime} \big)^{-1} \big)}\big\} 
\text{, } \\  
  \big\{ \underline{\mathcal{I}^{\prime\cdots\prime}_3 \big( u , u^{-1} \big)},  \underline{\mathcal{I}^{\prime\cdots\prime}_2 \big( u^{\prime} , \big( u^{\prime} \big)^{-1} \big)}  \big\} \text{, }  \\      \big\{ \underline{\mathcal{I}^{\prime\cdots\prime}_3 \big( u , u^{-1} \big)},  \underline{\mathcal{I}^{\prime\cdots\prime}_3 \big( u^{\prime} , \big( u^{\prime} \big)^{-1} \big)} \big\}    \text{, } \\  \big\{ \underline{\mathcal{I}^{\prime\cdots\prime}_3 \big( u , u^{-1} \big)}, \underline{\mathcal{I}^{\prime\cdots\prime}_4 \big( u^{\prime} , \big( u^{\prime} \big)^{-1} \big)}\big\}   \text{, } \end{array}\right. \\  \mathscr{C}_4 \equiv  \left\{\!\begin{array}{ll@{}>{{}}l}        \big\{ \underline{\mathcal{I}^{\prime\cdots\prime}_4 \big( u , u^{-1} \big)}, \underline{\mathcal{I}^{\prime\cdots\prime}_1 \big( u^{\prime} , \big( u^{\prime} \big)^{-1} \big)}\big\} 
\text{, } \\  
  \big\{ \underline{\mathcal{I}^{\prime\cdots\prime}_4 \big( u , u^{-1} \big)},  \underline{\mathcal{I}^{\prime\cdots\prime}_2 \big( u^{\prime} , \big( u^{\prime} \big)^{-1} \big)}  \big\} \text{, }  \\      \big\{ \underline{\mathcal{I}^{\prime\cdots\prime}_4 \big( u , u^{-1} \big)},  \underline{\mathcal{I}^{\prime\cdots\prime}_3 \big( u^{\prime} , \big( u^{\prime} \big)^{-1} \big)} \big\}    \text{, } \\  \big\{ \underline{\mathcal{I}^{\prime\cdots\prime}_4 \big( u , u^{-1} \big)}, \underline{\mathcal{I}^{\prime\cdots\prime}_4 \big( u^{\prime} , \big( u^{\prime} \big)^{-1} \big)}\big\}   \text{. }   \end{array}\right. 
\end{array}\right. 
\]

\noindent To approximate each one of the Poisson brackets within the collections of terms $\mathscr{C}_1, \cdots, \mathscr{C}_4$, in light of the decompositions for $\underline{\mathcal{I}_1}, \cdots, \underline{\mathcal{I}_4}$, the system above is equivalent to,

  \[ \mathscr{P} \equiv  \left\{\!\begin{array}{ll@{}>{{}}l} \mathscr{C}_1 \equiv  \left\{\!\begin{array}{ll@{}>{{}}l}          \big\{ \mathscr{I}^1_1 \big( u \big) + \mathscr{I}^1_2 \big( u \big) + \mathscr{I}^1_3 \big( u \big), \mathscr{I}^1_1 \big( u^{\prime} \big) + \mathscr{I}^1_2 \big( u^{\prime} \big) + \mathscr{I}^1_3 \big( u^{\prime} \big)          \big\}  \overset{(\mathrm{BL})}{=} \big\{ \mathscr{I}^1_1 \big( u \big), \mathscr{I}^1_1 \big( u^{\prime} \big) \big\} \\ + \big\{ \mathscr{I}^1_1 \big( u \big), \mathscr{I}^1_2 \big( u^{\prime} \big) \big\}  +  \big\{ \mathscr{I}^1_1 \big( u \big), \mathscr{I}^1_3 \big( u^{\prime} \big) \big\}  + \big\{ \mathscr{I}^1_2 \big( u \big), \mathscr{I}^1_1 \big( u^{\prime} \big) \big\} + \big\{ \mathscr{I}^1_2 \big( u \big), \mathscr{I}^1_2 \big( u^{\prime} \big) \big\} \\  + \big\{ \mathscr{I}^1_2 \big( u \big), \mathscr{I}^1_3 \big( u^{\prime} \big) \big\} + \big\{ \mathscr{I}^1_3 \big( u \big), \mathscr{I}^1_1 \big( u^{\prime} \big) \big\} + \big\{ \mathscr{I}^1_3 \big( u \big), \mathscr{I}^1_2 \big( u^{\prime} \big) \big\} + \big\{ \mathscr{I}^1_3 \big( u \big), \mathscr{I}^1_3 \big( u^{\prime} \big) \big\}
\text{, } \\ \\  
  \big\{ \mathscr{I}^1_1 \big( u \big) + \mathscr{I}^1_2 \big( u \big) + \mathscr{I}^1_3 \big( u \big),  \mathscr{I}^2_1 \big( u^{\prime} \big)   + \mathscr{I}^2_2 \big( u^{\prime} \big)  + \mathscr{I}^2_3 \big( u^{\prime} \big)  + \mathscr{I}^2_4 \big( u^{\prime} \big)  \big\}     \overset{(\mathrm{BL})}{=}    \big\{ \mathscr{I}^1_1 \big( u \big), \mathscr{I}^2_1 \big( u^{\prime} \big)  \big\} \\  + \big\{ \mathscr{I}^1_1 \big( u \big),  \mathscr{I}^2_2 \big( u^{\prime} \big)   \big\} + \big\{ \mathscr{I}^1_1 \big( u \big), \mathscr{I}^2_3 \big( u^{\prime} \big)   \big\} +    \big\{ \mathscr{I}^1_1 \big( u \big), \mathscr{I}^2_4 \big( u^{\prime} \big)   \big\}   + \big\{ \mathscr{I}^1_2 \big( u \big), \mathscr{I}^2_1 \big( u^{\prime} \big) \big\}  \\ +  \big\{ \mathscr{I}^1_2 \big( u \big), \mathscr{I}^2_2 \big( u^{\prime} \big) \big\}  + \big\{ \mathscr{I}^1_2 \big( u \big), \mathscr{I}^2_3 \big( u^{\prime} \big) \big\} + \big\{ \mathscr{I}^1_2 \big( u \big), \mathscr{I}^2_4 \big( u^{\prime} \big) \big\}   \text{, }  \\  \\     \big\{  \mathscr{I}^1_1 \big( \underline{u} \big) + \mathscr{I}^1_2 \big( \underline{u}  \big) + \mathscr{I}^1_3 \big( \underline{u}  \big),  \mathscr{I}^3_1 \big( \underline{u^{\prime}} \big)  + \mathscr{I}^3_2 \big( \underline{u^{\prime}} \big) + \mathscr{I}^3_3 \big( \underline{u^{\prime}} \big) + \mathscr{I}^3_4  \big( \underline{u^{\prime}} \big)  \big\}  \overset{(\mathrm{BL})}{=}        \big\{ \mathscr{I}^1_1 \big( \underline{u}  \big) ,  \mathscr{I}^3_1 \big( \underline{u^{\prime}}  \big) \big\}     \\  \big\{ \mathscr{I}^1_1 \big( \underline{u}  \big) ,  \mathscr{I}^3_2 \big( \underline{u^{\prime}}  \big) \big\}   +  \big\{ \mathscr{I}^1_1 \big( \underline{u}  \big) ,  \mathscr{I}^3_3 \big( \underline{u^{\prime}}  \big) \big\}   +  \big\{ \mathscr{I}^1_1 \big( \underline{u}  \big) ,  \mathscr{I}^3_4 \big( \underline{u^{\prime}}  \big) \big\}   +  \big\{ \mathscr{I}^1_2 \big( \underline{u}  \big) ,  \mathscr{I}^3_1 \big( \underline{u^{\prime}}  \big) \big\}  \end{array}\right. 
\end{array}\right.  \] \[ \mathscr{P} \equiv  \left\{\!\begin{array}{ll@{}>{{}}l} \mathscr{C}_1 \equiv  \left\{\!\begin{array}{ll@{}>{{}}l}    + \big\{ \mathscr{I}^1_2 \big( \underline{u}  \big) ,  \mathscr{I}^3_2 \big( \underline{u^{\prime}}  \big) \big\} +  \big\{ \mathscr{I}^1_2 \big( \underline{u} \big) , \mathscr{I}^3_3 \big( \underline{u^{\prime}} \big)  \big\} + \big\{ \mathscr{I}^1_2 \big( \underline{u} \big) , \mathscr{I}^3_4 \big( \underline{u^{\prime}} \big)    \big\} + \big\{ \mathscr{I}^1_3 \big( \underline{u} \big) , \mathscr{I}^3_1 \big( \underline{u^{\prime}} \big)    \big\} \\ + \big\{ \mathscr{I}^1_3 \big( \underline{u} \big) , \mathscr{I}^3_2 \big( \underline{u^{\prime}} \big)    \big\} + \big\{ \mathscr{I}^1_3 \big( \underline{u} \big) , \mathscr{I}^3_3 \big( \underline{u^{\prime}} \big)    \big\} + \big\{ \mathscr{I}^1_3 \big( \underline{u} \big) , \mathscr{I}^3_4 \big( \underline{u^{\prime}} \big)    \big\} \text{, }    \\  \big\{  \mathscr{I}^1_1 \big( u \big) + \mathscr{I}^1_2 \big( u \big) + \mathscr{I}^1_3 \big( u \big),  \mathscr{I}^4_1 \big( \underline{u^{\prime}} \big) +     \mathscr{I}^4_2 \big( \underline{u^{\prime}} \big) + \mathscr{I}^4_3 \big( \underline{u^{\prime}} \big) +     \mathscr{I}^4_4 \big( \underline{u^{\prime}} \big) +     \mathscr{I}^4_5 \big( \underline{u^{\prime}} \big)           \big\}  \\ \overset{(\mathrm{BL})}{=} \big\{  \mathscr{I}^1_1 \big( \underline{u} \big) ,         \mathscr{I}^4_1 \big( \underline{u^{\prime}}    \big\}   +  \big\{  \mathscr{I}^1_1 \big( \underline{u} \big) ,         \mathscr{I}^4_2 \big( \underline{u^{\prime}}    \big\} + \big\{  \mathscr{I}^1_1 \big( \underline{u} \big) ,         \mathscr{I}^4_3 \big( \underline{u^{\prime}}    \big\}  + \big\{  \mathscr{I}^1_1 \big( \underline{u} \big) ,         \mathscr{I}^4_4 \big( \underline{u^{\prime}}    \big\} \\  + \big\{  \mathscr{I}^1_1 \big( \underline{u} \big) ,         \mathscr{I}^4_5 \big( \underline{u^{\prime}} \big)    \big\}    +  \big\{ \mathscr{I}^1_2 \big( \underline{u} \big)  ,  \mathscr{I}^4_1 \big( \underline{u^{\prime}} \big)   \big\} + \big\{ \mathscr{I}^1_2 \big( \underline{u} \big)  ,  \mathscr{I}^4_2 \big( \underline{u^{\prime}} \big) \big\} + \big\{ \mathscr{I}^1_2 \big( \underline{u} \big)  ,  \mathscr{I}^4_3 \big( \underline{u^{\prime}} \big) \big\}   \\ + \big\{ \mathscr{I}^1_2 \big( \underline{u} \big)  ,  \mathscr{I}^4_4 \big( \underline{u^{\prime}} \big) \big\}  + \big\{ \mathscr{I}^1_2 \big( \underline{u} \big)  ,  \mathscr{I}^4_5 \big( \underline{u^{\prime}} \big) \big\} + \big\{ \mathscr{I}^1_3 \big( \underline{u} \big), \mathscr{I}^4_1 \big( \underline{u^{\prime}} \big)  \big\} +  \big\{ \mathscr{I}^1_3 \big( \underline{u} \big), \mathscr{I}^4_2 \big( \underline{u^{\prime}} \big)  \big\} \\ + \big\{ \mathscr{I}^1_3 \big( \underline{u} \big), \mathscr{I}^4_3 \big( \underline{u^{\prime}} \big)  \big\} + \big\{ \mathscr{I}^1_3 \big( \underline{u} \big), \mathscr{I}^4_4 \big( \underline{u^{\prime}} \big)  \big\} + \big\{ \mathscr{I}^1_3 \big( \underline{u} \big), \mathscr{I}^4_5 \big( \underline{u^{\prime}} \big)  \big\}  \text{, }  \end{array}\right.  \\ \\ \mathscr{C}_2 \equiv  \left\{\!\begin{array}{ll@{}>{{}}l}   \big\{ \mathscr{I}^2_1 \big( \underline{u} \big)  + \mathscr{I}^2_2 \big( \underline{u} \big)  + \mathscr{I}^2_3 \big( \underline{u} \big)  + \mathscr{I}^2_4 \big( \underline{u} \big)   ,                  \mathscr{I}^1_1 \big( \underline{u^{\prime}} \big) +       \mathscr{I}^1_2 \big( \underline{u^{\prime}} \big) +     \mathscr{I}^1_3 \big( \underline{u^{\prime}} \big)     \big\} \overset{(\mathrm{BL})}{=}         \big\{ \mathscr{I}^2_1 \big( \underline{u} \big) , \mathscr{I}^1_1 \big( \underline{u^{\prime}} \big) \big\} 
    \\  +   \big\{ \mathscr{I}^2_1 \big( \underline{u} \big) , \mathscr{I}^1_2 \big( \underline{u^{\prime}} \big) \big\}  +  \big\{ \mathscr{I}^2_1 \big( \underline{u} \big) , \mathscr{I}^1_3 \big( \underline{u^{\prime}} \big) \big\}   +  \big\{ \mathscr{I}^2_2 \big( \underline{u} \big) , \mathscr{I}^1_1 \big( \underline{u^{\prime}} \big) \big\}  +  \big\{ \mathscr{I}^2_2 \big( \underline{u} \big) , \mathscr{I}^1_2 \big( \underline{u^{\prime}} \big) \big\}    \\   +  \big\{ \mathscr{I}^2_2 \big( \underline{u} \big) , \mathscr{I}^1_3 \big( \underline{u^{\prime}} \big) \big\}  + \big\{ \mathscr{I}^2_3 \big( \underline{u} \big) , \mathscr{I}^1_1 \big( \underline{u^{\prime}} \big) \big\}  +  \big\{ \mathscr{I}^2_3 \big( \underline{u} \big) , \mathscr{I}^1_2 \big( \underline{u^{\prime}} \big) \big\}    +  \big\{ \mathscr{I}^2_3 \big( \underline{u} \big) , \mathscr{I}^1_3 \big( \underline{u^{\prime}} \big) \big\} \\  + \big\{ \mathscr{I}^2_4 \big( \underline{u} \big) ,     \mathscr{I}^1_1 \big( \underline{u^{\prime}} \big)    \big\}  + \big\{ \mathscr{I}^2_4 \big( \underline{u} \big) ,     \mathscr{I}^1_2 \big( \underline{u^{\prime}} \big)    \big\} + \big\{ \mathscr{I}^2_4 \big( \underline{u} \big) ,     \mathscr{I}^1_3 \big( \underline{u^{\prime}} \big)    \big\} + \big\{ \mathscr{I}^2_4 \big( \underline{u} \big) ,     \mathscr{I}^1_4 \big( \underline{u^{\prime}} \big)    \big\} 
\text{, } \\  \\ 
  \big\{ \mathscr{I}^2_1 \big( \underline{u} \big)  + \mathscr{I}^2_2 \big( \underline{u} \big)  + \mathscr{I}^2_3 \big( \underline{u} \big)  + \mathscr{I}^2_4 \big( \underline{u} \big)  ,  \mathscr{I}^2_1 \big( \underline{u^{\prime}} \big)  + \mathscr{I}^2_2 \big( \underline{u^{\prime}} \big)  + \mathscr{I}^2_3 \big( \underline{u^{\prime}} \big)  + \mathscr{I}^2_4 \big( \underline{u^{\prime}} \big) \big\}  \\ \overset{(\mathrm{BL})}{=} \big\{ \mathscr{I}^2_1 \big( \underline{u} \big)  ,           \mathscr{I}^2_1 \big( \underline{u^{\prime}} \big)     \big\}  + \big\{ \mathscr{I}^2_1 \big( \underline{u} \big)  ,           \mathscr{I}^2_2 \big( \underline{u^{\prime}} \big)     \big\}  + \big\{ \mathscr{I}^2_1 \big( \underline{u} \big)  ,           \mathscr{I}^2_3 \big( \underline{u^{\prime}} \big)     \big\} + \big\{ \mathscr{I}^2_1 \big( \underline{u} \big)  ,           \mathscr{I}^2_4  \big( \underline{u^{\prime}} \big)     \big\} \\ + \big\{ \mathscr{I}^2_2 \big( \underline{u} \big)  ,           \mathscr{I}^2_1 \big( \underline{u^{\prime}} \big)     \big\}  + \big\{ \mathscr{I}^2_2 \big( \underline{u} \big)  ,           \mathscr{I}^2_2 \big( \underline{u^{\prime}} \big)     \big\}  + \big\{ \mathscr{I}^2_2 \big( \underline{u} \big)  ,           \mathscr{I}^2_3 \big( \underline{u^{\prime}} \big)     \big\} + \big\{ \mathscr{I}^2_2 \big( \underline{u} \big)  ,           \mathscr{I}^2_4  \big( \underline{u^{\prime}} \big)     \big\} \\ +  \big\{ \mathscr{I}^2_3 \big( \underline{u} \big)  ,           \mathscr{I}^2_1 \big( \underline{u^{\prime}} \big)     \big\}  + \big\{ \mathscr{I}^2_3 \big( \underline{u} \big)  ,           \mathscr{I}^2_2 \big( \underline{u^{\prime}} \big)     \big\}  + \big\{ \mathscr{I}^2_3 \big( \underline{u} \big)  ,           \mathscr{I}^2_3 \big( \underline{u^{\prime}} \big)     \big\} + \big\{ \mathscr{I}^2_3 \big( \underline{u} \big)  ,           \mathscr{I}^2_4  \big( \underline{u^{\prime}} \big)     \big\} \\ +   \big\{ \mathscr{I}^2_4 \big( \underline{u} \big)  ,           \mathscr{I}^2_1 \big( \underline{u^{\prime}} \big)     \big\}  + \big\{ \mathscr{I}^2_4 \big( \underline{u} \big)  ,           \mathscr{I}^2_2 \big( \underline{u^{\prime}} \big)     \big\}  + \big\{ \mathscr{I}^2_4 \big( \underline{u} \big)  ,           \mathscr{I}^2_3 \big( \underline{u^{\prime}} \big)     \big\} + \big\{ \mathscr{I}^2_4 \big( \underline{u} \big)  ,           \mathscr{I}^2_4  \big( \underline{u^{\prime}} \big)     \big\} \text{, } \\  \\      \big\{ \mathscr{I}^2_1 \big( \underline{u} \big)  + \mathscr{I}^2_2 \big( \underline{u} \big)  + \mathscr{I}^2_3 \big( \underline{u} \big)  + \mathscr{I}^2_4 \big( \underline{u} \big) ,   \mathscr{I}^3_1 \big( \underline{u^{\prime}} \big) + \mathscr{I}^3_2 \big( \underline{u^{\prime}} \big) + \mathscr{I}^3_3 \big( \underline{u^{\prime}} \big) 
 + \mathscr{I}^3_4 \big( \underline{u^{\prime}} \big)  \big\} \\ \overset{(\mathrm{BL})}{=}   \big\{ \mathscr{I}^2_1 \big( \underline{u} \big)  , \mathscr{I}^3_1 \big( \underline{u^{\prime}} \big)  \big\}  + \big\{ \mathscr{I}^2_1 \big( \underline{u} \big)  , \mathscr{I}^3_2 \big( \underline{u^{\prime}} \big)   \big\} +   \big\{ \mathscr{I}^2_1 \big( \underline{u} \big)  , \mathscr{I}^3_3 \big( \underline{u^{\prime}} \big)   \big\}    + \big\{ \mathscr{I}^2_1 \big( \underline{u} \big)  , \mathscr{I}^3_4 \big( \underline{u^{\prime}} \big)   \big\}    \\ +  \big\{ \mathscr{I}^2_2 \big( \underline{u} \big)  , \mathscr{I}^3_1 \big( \underline{u^{\prime}} \big)  \big\}  + \big\{ \mathscr{I}^2_2 \big( \underline{u} \big)  , \mathscr{I}^3_2 \big( \underline{u^{\prime}} \big)   \big\} +   \big\{ \mathscr{I}^2_2 \big( \underline{u} \big)  , \mathscr{I}^3_3 \big( \underline{u^{\prime}} \big)   \big\}    + \big\{ \mathscr{I}^2_2 \big( \underline{u} \big)  , \mathscr{I}^3_4 \big( \underline{u^{\prime}} \big)   \big\} \\ + \big\{ \mathscr{I}^2_3 \big( \underline{u} \big)  , \mathscr{I}^3_1 \big( \underline{u^{\prime}} \big)  \big\}  + \big\{ \mathscr{I}^2_3 \big( \underline{u} \big)  , \mathscr{I}^3_2 \big( \underline{u^{\prime}} \big)   \big\} +   \big\{ \mathscr{I}^2_3 \big( \underline{u} \big)  , \mathscr{I}^3_3 \big( \underline{u^{\prime}} \big)   \big\}    + \big\{ \mathscr{I}^2_3 \big( \underline{u} \big)  , \mathscr{I}^3_4 \big( \underline{u^{\prime}} \big)   \big\}  \\ + \big\{ \mathscr{I}^2_4 \big( \underline{u} \big)  , \mathscr{I}^3_1 \big( \underline{u^{\prime}} \big)  \big\}  + \big\{ \mathscr{I}^2_4 \big( \underline{u} \big)  , \mathscr{I}^3_2 \big( \underline{u^{\prime}} \big)   \big\} +   \big\{ \mathscr{I}^2_4 \big( \underline{u} \big)  , \mathscr{I}^3_3 \big( \underline{u^{\prime}} \big)   \big\}    + \big\{ \mathscr{I}^2_4 \big( \underline{u} \big)  , \mathscr{I}^3_4 \big( \underline{u^{\prime}} \big)   \big\}    \text{, } \\  \\  \big\{ \mathscr{I}^2_1 \big( \underline{u} \big)  + \mathscr{I}^2_2 \big( \underline{u} \big)  + \mathscr{I}^2_3 \big( \underline{u} \big)  + \mathscr{I}^2_4 \big( \underline{u} \big) ,  \mathscr{I}^4_1 \big( \underline{u^{\prime}} \big) + \mathscr{I}^4_2 \big( \underline{u^{\prime}} \big)  + \mathscr{I}^4_3 \big( \underline{u^{\prime}} \big) + \mathscr{I}^4_4 \big( \underline{u^{\prime}} \big)   \big\} \\ \overset{(\mathrm{BL})}{=}         \big\{ \mathscr{I}^2_1 \big( \underline{u} \big) , \mathscr{I}^4_1 \big( \underline{u^{\prime}} \big) \big\}  +  \big\{ \mathscr{I}^2_1 \big( \underline{u} \big) , \mathscr{I}^4_2 \big( \underline{u^{\prime}} \big) \big\}  +  \big\{ \mathscr{I}^2_1 \big( \underline{u} \big) , \mathscr{I}^4_3 \big( \underline{u^{\prime}} \big) \big\}  +  \big\{ \mathscr{I}^2_1 \big( \underline{u} \big) , \mathscr{I}^4_4 \big( \underline{u^{\prime}} \big) \big\} \\    +    \big\{ \mathscr{I}^2_2 \big( \underline{u} \big) , \mathscr{I}^4_1 \big( \underline{u^{\prime}} \big) \big\}  +  \big\{ \mathscr{I}^2_2 \big( \underline{u} \big) , \mathscr{I}^4_2 \big( \underline{u^{\prime}} \big) \big\}  +  \big\{ \mathscr{I}^2_2 \big( \underline{u} \big) , \mathscr{I}^4_3 \big( \underline{u^{\prime}} \big) \big\}  +  \big\{ \mathscr{I}^2_2 \big( \underline{u} \big) , \mathscr{I}^4_4 \big( \underline{u^{\prime}} \big) \big\} \\  +    \big\{ \mathscr{I}^2_3 \big( \underline{u} \big) , \mathscr{I}^4_1 \big( \underline{u^{\prime}} \big) \big\}  +  \big\{ \mathscr{I}^2_3 \big( \underline{u} \big) , \mathscr{I}^4_2 \big( \underline{u^{\prime}} \big) \big\}  +  \big\{ \mathscr{I}^2_3 \big( \underline{u} \big) , \mathscr{I}^4_3 \big( \underline{u^{\prime}} \big) \big\}  +  \big\{ \mathscr{I}^2_3 \big( \underline{u} \big) , \mathscr{I}^4_4 \big( \underline{u^{\prime}} \big) \big\} \\  +    \big\{ \mathscr{I}^2_4 \big( \underline{u} \big) , \mathscr{I}^4_1 \big( \underline{u^{\prime}} \big) \big\}  +  \big\{ \mathscr{I}^2_4 \big( \underline{u} \big) , \mathscr{I}^4_2 \big( \underline{u^{\prime}} \big) \big\}  +  \big\{ \mathscr{I}^2_4 \big( \underline{u} \big) , \mathscr{I}^4_3 \big( \underline{u^{\prime}} \big) \big\}  +  \big\{ \mathscr{I}^2_4 \big( \underline{u} \big) , \mathscr{I}^4_4 \big( \underline{u^{\prime}} \big) \big\} \text{, } \end{array}\right.  \\ \\ 
 \mathscr{C}_3 \equiv  \left\{\!\begin{array}{ll@{}>{{}}l}         \big\{ \mathscr{I}^3_1 \big( \underline{u} \big) +     \mathscr{I}^3_2 \big( \underline{u} \big) + \mathscr{I}^3_3 \big( \underline{u} \big) + \mathscr{I}^3_4 \big( \underline{u} \big), \mathscr{I}^1_1 \big( \underline{u^{\prime}} \big) + \mathscr{I}^1_2 \big( \underline{u^{\prime}} \big) + \mathscr{I}^1_3 \big( \underline{u^{\prime}} \big) \big\} \overset{(\mathrm{BL})}{=} \big\{ \mathscr{I}^3_1 \big( \underline{u} \big) ,  \mathscr{I}^1_1 \big( \underline{u^{\prime}} \big) \big\} \\ + \big\{ \mathscr{I}^3_1 \big( \underline{u} \big) ,  \mathscr{I}^1_2 \big( \underline{u^{\prime}} \big) \big\}  + \big\{ \mathscr{I}^3_1 \big( \underline{u} \big) ,  \mathscr{I}^1_3 \big( \underline{u^{\prime}} \big) \big\} +   \big\{ \mathscr{I}^3_2 \big( \underline{u} \big) ,   \mathscr{I}^1_1 \big( \underline{u^{\prime}} \big)   \big\} +   \big\{ \mathscr{I}^3_2 \big( \underline{u} \big) ,   \mathscr{I}^1_2 \big( \underline{u^{\prime}} \big)   \big\}   \\  + \big\{ \mathscr{I}^3_2 \big( \underline{u} \big) ,   \mathscr{I}^1_3 \big( \underline{u^{\prime}} \big)   \big\}  + \big\{ \mathscr{I}^3_3 \big( \underline{u} \big) , \mathscr{I}^1_1 \big( \underline{u^{\prime}} \big) \big\} +  \big\{ \mathscr{I}^3_3 \big( \underline{u} \big) , \mathscr{I}^1_2 \big( \underline{u^{\prime}} \big) \big\} + \big\{ \mathscr{I}^3_3 \big( \underline{u} \big) , \mathscr{I}^1_3 \big( \underline{u^{\prime}} \big) \big\} \\ + \big\{ \mathscr{I}^3_4 \big( \underline{u} \big) , \mathscr{I}^1_1 \big( \underline{u^{\prime}} \big) \big\} +  \big\{ \mathscr{I}^3_4 \big( \underline{u} \big) , \mathscr{I}^1_2 \big( \underline{u^{\prime}} \big) \big\} + \big\{ \mathscr{I}^3_4 \big( \underline{u} \big) , \mathscr{I}^1_3 \big( \underline{u^{\prime}} \big) \big\}
\text{, } \\  \\  
  \big\{ \mathscr{I}^3_1 \big( \underline{u} \big) +     \mathscr{I}^3_2 \big( \underline{u} \big) + \mathscr{I}^3_3 \big( \underline{u} \big) + \mathscr{I}^3_4 \big( \underline{u} \big),                 \mathscr{I}^2_1 \big( \underline{u^{\prime}} \big) +    \mathscr{I}^2_2 \big( \underline{u^{\prime}} \big)   +  \mathscr{I}^2_3 \big( \underline{u^{\prime}} \big) +  \mathscr{I}^2_4 \big( \underline{u^{\prime}} \big)       \big\} \\ \overset{(\mathrm{BL})}{=}       \big\{ \mathscr{I}^3_1 \big( \underline{u} \big) , \mathscr{I}^2_1 \big( \underline{u^{\prime}} \big)  \big\}  +   \big\{ \mathscr{I}^3_1 \big( \underline{u} \big) , \mathscr{I}^2_2 \big( \underline{u^{\prime}} \big)  \big\}  +  \big\{ \mathscr{I}^3_1 \big( \underline{u} \big) , \mathscr{I}^2_3 \big( \underline{u^{\prime}} \big)  \big\}  +  \big\{ \mathscr{I}^3_1 \big( \underline{u} \big) , \mathscr{I}^2_4 \big( \underline{u^{\prime}} \big)  \big\} \\ +  \big\{ \mathscr{I}^3_2 \big( \underline{u} \big) , \mathscr{I}^2_1 \big( \underline{u^{\prime}} \big)  \big\}  +   \big\{ \mathscr{I}^3_2 \big( \underline{u} \big) , \mathscr{I}^2_2 \big( \underline{u^{\prime}} \big)  \big\}  +  \big\{ \mathscr{I}^3_2 \big( \underline{u} \big) , \mathscr{I}^2_3 \big( \underline{u^{\prime}} \big)  \big\}  +  \big\{ \mathscr{I}^3_2 \big( \underline{u} \big) , \mathscr{I}^2_4 \big( \underline{u^{\prime}} \big)  \big\}  \\  +   \big\{ \mathscr{I}^3_3 \big( \underline{u} \big) , \mathscr{I}^2_1 \big( \underline{u^{\prime}} \big)  \big\}  +   \big\{ \mathscr{I}^3_3 \big( \underline{u} \big) , \mathscr{I}^2_2 \big( \underline{u^{\prime}} \big)  \big\}  +  \big\{ \mathscr{I}^3_3 \big( \underline{u} \big) , \mathscr{I}^2_3 \big( \underline{u^{\prime}} \big)  \big\}  +  \big\{ \mathscr{I}^3_3 \big( \underline{u} \big) , \mathscr{I}^2_4 \big( \underline{u^{\prime}} \big)  \big\} \\ +   \big\{ \mathscr{I}^3_4 \big( \underline{u} \big) , \mathscr{I}^2_1 \big( \underline{u^{\prime}} \big)  \big\}  +   \big\{ \mathscr{I}^3_4 \big( \underline{u} \big) , \mathscr{I}^2_2 \big( \underline{u^{\prime}} \big)  \big\}  +  \big\{ \mathscr{I}^3_4 \big( \underline{u} \big) , \mathscr{I}^2_3 \big( \underline{u^{\prime}} \big)  \big\}  +  \big\{ \mathscr{I}^3_4 \big( \underline{u} \big) , \mathscr{I}^2_4 \big( \underline{u^{\prime}} \big)  \big\}     \\ \\   \big\{  \mathscr{I}^3_1 \big( \underline{u} \big) +     \mathscr{I}^3_2 \big( \underline{u} \big) + \mathscr{I}^3_3 \big( \underline{u} \big) + \mathscr{I}^3_4 \big( \underline{u} \big) ,  \mathscr{I}^3_1 \big( \underline{u^{\prime}} \big) + \mathscr{I}^3_2 \big( \underline{u^{\prime}} \big) + \mathscr{I}^3_3 \big( \underline{u^{\prime}} \big) + \mathscr{I}^3_4 \big( \underline{u^{\prime}} \big)  \big\}  \\  \overset{(\mathrm{BL})}{=}  \big\{  \mathscr{I}^3_1  \big( \underline{u} \big)  ,  \mathscr{I}^3_1  \big( \underline{u^{\prime}} \big)  \big\}  +  \big\{  \mathscr{I}^3_1  \big( \underline{u} \big)  ,  \mathscr{I}^3_2  \big( \underline{u^{\prime}} \big)  \big\}  + \big\{  \mathscr{I}^3_1  \big( \underline{u} \big)  ,  \mathscr{I}^3_3  \big( \underline{u^{\prime}} \big)  \big\}  + \big\{  \mathscr{I}^3_1  \big( \underline{u} \big)  ,  \mathscr{I}^3_4  \big( \underline{u^{\prime}} \big)  \big\} \\ + \big\{  \mathscr{I}^3_2  \big( \underline{u} \big)  ,  \mathscr{I}^3_1  \big( \underline{u^{\prime}} \big)  \big\}  +  \big\{  \mathscr{I}^3_2  \big( \underline{u} \big)  ,  \mathscr{I}^3_2  \big( \underline{u^{\prime}} \big)  \big\}  + \big\{  \mathscr{I}^3_2  \big( \underline{u} \big)  ,  \mathscr{I}^3_3  \big( \underline{u^{\prime}} \big)  \big\}  + \big\{  \mathscr{I}^3_2  \big( \underline{u} \big)  ,  \mathscr{I}^3_4  \big( \underline{u^{\prime}} \big)  \big\} \\ + \big\{  \mathscr{I}^3_3  \big( \underline{u} \big)  ,  \mathscr{I}^3_1  \big( \underline{u^{\prime}} \big)  \big\}  +  \big\{  \mathscr{I}^3_3  \big( \underline{u} \big)  ,  \mathscr{I}^3_2  \big( \underline{u^{\prime}} \big)  \big\}  + \big\{  \mathscr{I}^3_3  \big( \underline{u} \big)  ,  \mathscr{I}^3_3  \big( \underline{u^{\prime}} \big)  \big\}  + \big\{  \mathscr{I}^3_3  \big( \underline{u} \big)  ,  \mathscr{I}^3_4  \big( \underline{u^{\prime}} \big)  \big\} \\ + \big\{  \mathscr{I}^3_4  \big( \underline{u} \big)  ,  \mathscr{I}^3_1  \big( \underline{u^{\prime}} \big)  \big\}  +  \big\{  \mathscr{I}^3_4  \big( \underline{u} \big)  ,  \mathscr{I}^3_2  \big( \underline{u^{\prime}} \big)  \big\}  + \big\{  \mathscr{I}^3_4  \big( \underline{u} \big)  ,  \mathscr{I}^3_3  \big( \underline{u^{\prime}} \big)  \big\}  + \big\{  \mathscr{I}^3_4  \big( \underline{u} \big)  ,  \mathscr{I}^3_4  \big( \underline{u^{\prime}} \big)  \big\} \text{, } \\ \\  \big\{ 
\mathscr{I}^3_1 \big( \underline{u} \big) +     \mathscr{I}^3_2 \big( \underline{u} \big) + \mathscr{I}^3_3 \big( \underline{u} \big) + \mathscr{I}^3_4 \big( \underline{u} \big)  ,  \mathscr{I}^4_1 \big( \underline{u^{\prime}}             \big)  + \mathscr{I}^4_2 \big( \underline{u^{\prime}}             \big)  +\mathscr{I}^4_3 \big( \underline{u^{\prime}}             \big)  + \mathscr{I}^4_4 \big( \underline{u^{\prime}}       \big) + \mathscr{I}^4_5 \big( \underline{u^{\prime}}             \big)    \big\} \\ \overset{(\mathrm{BL})}{=}            \big\{ \mathscr{I}^3_1 \big( \underline{u} \big) , \mathscr{I}^4_1 \big( \underline{u^{\prime}} \big)  \big\}  +  \big\{ \mathscr{I}^3_1 \big( \underline{u} \big) , \mathscr{I}^4_2 \big( \underline{u^{\prime}} \big)  \big\} +  \big\{ \mathscr{I}^3_1 \big( \underline{u} \big) , \mathscr{I}^4_3 \big( \underline{u^{\prime}} \big)  \big\} +  \big\{ \mathscr{I}^3_1 \big( \underline{u} \big) , \mathscr{I}^4_4 \big( \underline{u^{\prime}} \big)  \big\} \\  +  \big\{ \mathscr{I}^3_2 \big( \underline{u} \big) , \mathscr{I}^4_1 \big( \underline{u^{\prime}} \big)  \big\}  +  \big\{ \mathscr{I}^3_2 \big( \underline{u} \big) , \mathscr{I}^4_2 \big( \underline{u^{\prime}} \big)  \big\} +  \big\{ \mathscr{I}^3_2 \big( \underline{u} \big) , \mathscr{I}^4_3 \big( \underline{u^{\prime}} \big)  \big\} +  \big\{ \mathscr{I}^3_2 \big( \underline{u} \big) , \mathscr{I}^4_4 \big( \underline{u^{\prime}} \big)  \big\}  \\ +  \big\{ \mathscr{I}^3_3 \big( \underline{u} \big) , \mathscr{I}^4_1 \big( \underline{u^{\prime}} \big)  \big\}  +  \big\{ \mathscr{I}^3_3 \big( \underline{u} \big) , \mathscr{I}^4_2 \big( \underline{u^{\prime}} \big)  \big\} +  \big\{ \mathscr{I}^3_3 \big( \underline{u} \big) , \mathscr{I}^4_3 \big( \underline{u^{\prime}} \big)  \big\} +  \big\{ \mathscr{I}^3_3 \big( \underline{u} \big) , \mathscr{I}^4_4 \big( \underline{u^{\prime}} \big)  \big\}   \text{, }  
    \end{array}\right.     \\ \\ 
 \mathscr{C}_4 \equiv  \left\{\!\begin{array}{ll@{}>{{}}l}            \big\{ \mathscr{I}^4_1 \big( \underline{u} \big) + \mathscr{I}^4_2 \big( \underline{u} \big) + \mathscr{I}^4_3 \big( \underline{u} \big) + \mathscr{I}^4_4 \big( \underline{u} \big)  + \mathscr{I}^4_5 \big( \underline{u} \big), \mathscr{I}^1_1 \big( \underline{u^{\prime}} \big) + \mathscr{I}^1_2 \big( \underline{u^{\prime}} \big) +     \mathscr{I}^1_3 \big( \underline{u^{\prime}} \big)          \big\} \\ \overset{(\mathrm{BL})}{=}          \big\{ \mathscr{I}^4_1 \big( \underline{u} \big) , \mathscr{I}^1_1 \big( \underline{u^{\prime}} \big) \big\} +  \big\{ \mathscr{I}^4_1 \big( \underline{u} \big) , \mathscr{I}^1_2 \big( \underline{u^{\prime}} \big) \big\} +  \big\{ \mathscr{I}^4_1 \big( \underline{u} \big) , \mathscr{I}^1_3 \big( \underline{u^{\prime}} \big) \big\}  +  \big\{ \mathscr{I}^4_2 \big( \underline{u}  \big) , \mathscr{I}^1_1 \big( \underline{u^{\prime}} \big) \big\}  \end{array}\right. 
\end{array}\right. 
\]

\[  \mathscr{P} \equiv     \left\{\!\begin{array}{ll@{}>{{}}l}   \mathscr{C}_3 \equiv  \left\{\!\begin{array}{ll@{}>{{}}l}   +   \big\{ \mathscr{I}^4_2 \big( \underline{u}  \big) , \mathscr{I}^1_2 \big( \underline{u^{\prime}} \big) \big\}   +  \big\{ \mathscr{I}^4_2 \big( \underline{u}  \big) , \mathscr{I}^1_3 \big( \underline{u^{\prime}} \big) \big\}   + \big\{ \mathscr{I}^4_3 \big( \underline{u} \big)  , \mathscr{I}^1_1 \big( \underline{u^{\prime}} \big) 
 \big\} +  \big\{ \mathscr{I}^4_3 \big( \underline{u} \big)  , \mathscr{I}^1_2 \big( \underline{u^{\prime}} \big) 
 \big\} \\   +  \big\{ \mathscr{I}^4_3 \big( \underline{u} \big)  , \mathscr{I}^1_3 \big( \underline{u^{\prime}} \big) 
 \big\} + \big\{ \mathscr{I}^4_4 \big( \underline{u} \big) , \mathscr{I}^1_1 \big( \underline{u^{\prime}} \big) \big\} +  \big\{ \mathscr{I}^4_4 \big( \underline{u} \big) , \mathscr{I}^1_2 \big( \underline{u^{\prime}} \big) \big\} +  \big\{ \mathscr{I}^4_4 \big( \underline{u} \big) , \mathscr{I}^1_3 \big( \underline{u^{\prime}} \big) \big\} \\ +  \big\{ \mathscr{I}^4_5 \big( \underline{u} \big) , \mathscr{I}^1_1 \big( \underline{u^{\prime}} \big)  \big\} +  \big\{ \mathscr{I}^4_5 \big( \underline{u} \big) , \mathscr{I}^1_2 \big( \underline{u^{\prime}} \big)  \big\} +  \big\{ \mathscr{I}^4_5 \big( \underline{u} \big) , \mathscr{I}^1_3 \big( \underline{u^{\prime}} \big)  \big\}
\text{, }        \\ \\ \big\{ \mathscr{I}^4_1 \big( \underline{u} \big) + \mathscr{I}^4_2 \big( \underline{u} \big) + \mathscr{I}^4_3 \big( \underline{u} \big) + \mathscr{I}^4_4 \big( \underline{u} \big)  + \mathscr{I}^4_5 \big( \underline{u} \big),  \mathscr{I}^3_1 \big( \underline{u^{\prime}} \big) + \mathscr{I}^3_2 \big( \underline{u^{\prime}} \big)  + \mathscr{I}^3_3 \big( \underline{u^{\prime}}\big) + \mathscr{I}^3_4 \big( \underline{u^{\prime}} \big) \big\}  \\  \overset{(\mathrm{BL})}{=}           \big\{ \mathscr{I}^4_1 \big( \underline{u} \big) , \mathscr{I}^3_1 \big( \underline{u^{\prime}} \big) \big\} +  \big\{ \mathscr{I}^4_1 \big( \underline{u} \big) , \mathscr{I}^3_2 \big( \underline{u^{\prime}} \big) \big\}  + \big\{ \mathscr{I}^4_1 \big( \underline{u} \big) , \mathscr{I}^3_3 \big( \underline{u^{\prime}} \big) \big\} + \big\{ \mathscr{I}^4_1 \big( \underline{u} \big) , \mathscr{I}^3_4 \big( \underline{u^{\prime}} \big) \big\}  \\ + \big\{ \mathscr{I}^4_2 \big( \underline{u} \big) , \mathscr{I}^3_1 \big( \underline{u^{\prime}} \big)  \big\} + \big\{  \mathscr{I}^4_2 \big( \underline{u} \big) , \mathscr{I}^3_2 \big( \underline{u^{\prime}} \big)   \big\} + \big\{  \mathscr{I}^4_2 \big( \underline{u} \big) , \mathscr{I}^3_3 \big( \underline{u^{\prime}} \big)   \big\} + \big\{  \mathscr{I}^4_2 \big( \underline{u} \big) , \mathscr{I}^3_4 \big( \underline{u^{\prime}} \big)   \big\} \\ + \big\{ \mathscr{I}^4_3 \big( \underline{u} \big) , \mathscr{I}^3_1  \big( \underline{u^{\prime}} \big)  \big\} + \big\{ \mathscr{I}^4_3 \big( \underline{u} \big) , \mathscr{I}^3_2  \big( \underline{u^{\prime}} \big)   \big\} + \big\{ \mathscr{I}^4_3 \big( \underline{u} \big) , \mathscr{I}^3_3  \big( \underline{u^{\prime}} \big)   \big\}  + \big\{ \mathscr{I}^4_3 \big( \underline{u} \big) , \mathscr{I}^3_4  \big( \underline{u^{\prime}} \big)   \big\}    \\  + \big\{ \mathscr{I}^4_4 \big( \underline{u} \big) ,  \mathscr{I}^3_1 \big( \underline{u^{\prime}} \big) \big\}   + \big\{ \mathscr{I}^4_4 \big( \underline{u} \big) ,  \mathscr{I}^3_2 \big( \underline{u^{\prime}} \big) \big\}  + \big\{ \mathscr{I}^4_4 \big( \underline{u} \big) ,  \mathscr{I}^3_3 \big( \underline{u^{\prime}} \big) \big\} +   \big\{ \mathscr{I}^4_4 \big( \underline{u} \big) ,  \mathscr{I}^3_4 \big( \underline{u^{\prime}} \big) \big\}  \\ + \big\{ \mathscr{I}^4_5 \big( \underline{u} \big) , \mathscr{I}^3_1 \big( \underline{u^{\prime}} \big)  \big\} +  \big\{ \mathscr{I}^4_5 \big( \underline{u} \big) , \mathscr{I}^3_2 \big( \underline{u^{\prime}} \big)  \big\}  + \big\{ \mathscr{I}^4_5 \big( \underline{u} \big) , \mathscr{I}^3_3 \big( \underline{u^{\prime}} \big)  \big\}  + \big\{ \mathscr{I}^4_5 \big( \underline{u} \big) , \mathscr{I}^3_4 \big( \underline{u^{\prime}} \big)  \big\}  \text{, } \\  \\      \big\{ \mathscr{I}^4_1 \big( \underline{u} \big) + \mathscr{I}^4_2 \big( \underline{u} \big) + \mathscr{I}^4_3 \big( \underline{u} \big) + \mathscr{I}^4_4 \big( \underline{u} \big)  + \mathscr{I}^4_5 \big( \underline{u} \big) ,  \mathscr{I}^3_1 \big( \underline{u}^{\prime} \big) + \mathscr{I}^3_2 \big( \underline{u^{\prime}} \big)   + \mathscr{I}^3_3 \big( \underline{u^{\prime}} \big\}   + \mathscr{I}^4_3 \big( \underline{u^{\prime}} \big) \big\} \\  \overset{(\mathrm{BL})}{=}  \big\{ \mathscr{I}^4_1 \big( \underline{u} \big) , \mathscr{I}^3_1 \big( \underline{u^{\prime}} \big) \big\} + \big\{  \mathscr{I}^4_1 \big( \underline{u} \big) , \mathscr{I}^3_2 \big( \underline{u^{\prime}} \big) \big\} +  \big\{  \mathscr{I}^4_1 \big( \underline{u} \big) , \mathscr{I}^3_3 \big( \underline{u^{\prime}} \big) \big\}   + \big\{  \mathscr{I}^4_1 \big( \underline{u} \big) , \mathscr{I}^3_4 \big( \underline{u^{\prime}} \big) \big\} \\ + \big\{  \mathscr{I}^4_2 \big( \underline{u} \big) , \mathscr{I}^3_1 \big( \underline{u^{\prime}} \big) \big\}  + \big\{  \mathscr{I}^4_2 \big( \underline{u} \big) , \mathscr{I}^3_2 \big( \underline{u^{\prime}} \big) \big\} + \big\{  \mathscr{I}^4_2 \big( \underline{u} \big) , \mathscr{I}^3_3 \big( \underline{u^{\prime}} \big) \big\}   + \big\{  \mathscr{I}^4_2 \big( \underline{u} \big) , \mathscr{I}^3_4 \big( \underline{u^{\prime}} \big) \big\} \\  + \big\{ \mathscr{I}^4_3 \big( \underline{u} \big) , \mathscr{I}^3_1 \big( \underline{u^{\prime}} \big) \big\} +    \big\{ \mathscr{I}^4_3 \big( \underline{u} \big) , \mathscr{I}^3_2 \big( \underline{u^{\prime}} \big) \big\} +  \big\{ \mathscr{I}^4_3 \big( \underline{u} \big) , \mathscr{I}^3_3 \big( \underline{u^{\prime}} \big) \big\} +  \big\{ \mathscr{I}^4_3 \big( \underline{u} \big) , \mathscr{I}^3_4 \big( \underline{u^{\prime}} \big) \big\} \\ + \big\{  \mathscr{I}^4_5 \big( \underline{u} \big) , \mathscr{I}^3_1 \big( \underline{u^{\prime}} \big)  \big\} + \big\{ \mathscr{I}^4_5 \big( \underline{u} \big) , \mathscr{I}^3_2 \big( \underline{u^{\prime}} \big)  \big\} + \big\{ \mathscr{I}^4_5 \big( \underline{u} \big) , \mathscr{I}^3_3 \big( \underline{u^{\prime}} \big)  \big\}  + \big\{ \mathscr{I}^4_5 \big( \underline{u} \big) , \mathscr{I}^3_4 \big( \underline{u^{\prime}} \big)  \big\}   \text{, } \\ \\   \big\{ \mathscr{I}^4_1 \big( \underline{u} \big) + \mathscr{I}^4_2 \big( \underline{u} \big) + \mathscr{I}^4_3 \big( \underline{u} \big) + \mathscr{I}^4_4 \big( \underline{u} \big)  + \mathscr{I}^4_5 \big( \underline{u} \big) , \mathscr{I}^4_1 \big( \underline{u^{\prime}} \big) + \mathscr{I}^4_2 \big( \underline{u^{\prime}} \big) + \mathscr{I}^4_3 \big( \underline{u^{\prime}} \big) + \mathscr{I}^4_4 \big( \underline{u^{\prime}} \big)  + \mathscr{I}^4_5 \big( \underline{u^{\prime}} \big)  \big\} \\  \overset{\mathrm{(BL)}}{=}         \big\{  \mathscr{I}^4_1 \big( \underline{u} \big) ,  \mathscr{I}^4_1 \big( \underline{u^{\prime}} \big)  \big\} +      \big\{ \mathscr{I}^4_1 \big( \underline{u} \big) ,  \mathscr{I}^4_2 \big( \underline{u^{\prime}} \big) \big\} +      \big\{ \mathscr{I}^4_1 \big( \underline{u} \big) ,  \mathscr{I}^4_3 \big( \underline{u^{\prime}} \big) \big\} +      \big\{ \mathscr{I}^4_1 \big( \underline{u} \big),  \mathscr{I}^4_4 \big( \underline{u^{\prime}} \big) 
 \big\} \\  +      \big\{ \mathscr{I}^4_1 \big( \underline{u} \big), \mathscr{I}^4_5 \big( \underline{u^{\prime}} \big) \big\} +   \big\{ \mathscr{I}^4_2 \big( \underline{u} \big), \mathscr{I}^4_1 \big( \underline{u^{\prime}} \big) \big\} +  \big\{ \mathscr{I}^4_2 \big( \underline{u} \big), \mathscr{I}^4_2 \big( \underline{u^{\prime}} \big) \big\}    +    \big\{ \mathscr{I}^4_2 \big( \underline{u} \big), \mathscr{I}^4_3 \big( \underline{u^{\prime}} \big) \big\}  \\   +  \big\{ \mathscr{I}^4_2 \big( \underline{u} \big), \mathscr{I}^4_4 \big( \underline{u^{\prime}} \big) \big\}   +  \big\{ \mathscr{I}^4_2 \big( \underline{u} \big), \mathscr{I}^4_5 \big( \underline{u^{\prime}} \big) \big\}    +                           \big\{ \mathscr{I}^4_3 \big( \underline{u} \big) , \mathscr{I}^4_1 \big( \underline{u^{\prime}} \big) \big\} + \big\{ \mathscr{I}^4_3 \big( \underline{u} \big) , \mathscr{I}^4_2 \big( \underline{u^{\prime}} \big) \big\}  \\ + \big\{ \mathscr{I}^4_3 \big( \underline{u} \big) , \mathscr{I}^4_3 \big( \underline{u^{\prime}} \big) \big\}     + \big\{ \mathscr{I}^4_3 \big( \underline{u} \big) , \mathscr{I}^4_4 \big( \underline{u^{\prime}} \big) \big\}    + \big\{ \mathscr{I}^4_3 \big( \underline{u} \big) , \mathscr{I}^4_5 \big( \underline{u^{\prime}} \big) \big\}  + \big\{ \mathscr{I}^4_4 \big( \underline{u} \big) , \mathscr{I}^4_1 \big( \underline{u^{\prime}} \big)    \big\}   \\  + \big\{  \mathscr{I}^4_4 \big( \underline{u} \big) , \mathscr{I}^4_2 \big( \underline{u^{\prime}} \big)    \big\} +  \big\{  \mathscr{I}^4_4 \big( \underline{u} \big) , \mathscr{I}^4_3 \big( \underline{u^{\prime}} \big)    \big\} + \big\{  \mathscr{I}^4_4 \big( \underline{u} \big) , \mathscr{I}^4_4 \big( \underline{u^{\prime}} \big)    \big\} + \big\{  \mathscr{I}^4_4 \big( \underline{u} \big) , \mathscr{I}^4_5 \big( \underline{u^{\prime}} \big)    \big\} \\ + \big\{  \mathscr{I}^4_5 \big( \underline{u} \big), \mathscr{I}^4_1 \big( \underline{u^{\prime}} \big) \big\} + \big\{  \mathscr{I}^4_5 \big( \underline{u} \big), \mathscr{I}^4_2 \big( \underline{u^{\prime}} \big)  \big\}  + \big\{  \mathscr{I}^4_5 \big( \underline{u} \big), \mathscr{I}^4_3 \big( \underline{u^{\prime}} \big)  \big\}  + \big\{  \mathscr{I}^4_5 \big( \underline{u} \big), \mathscr{I}^4_4 \big( \underline{u^{\prime}} \big)  \big\}  \\ + \big\{  \mathscr{I}^4_5 \big( \underline{u} \big), \mathscr{I}^4_5 \big( \underline{u^{\prime}} \big)  \big\} \text{, }
\end{array}\right. \end{array}\right.
\]

\noindent from applications of the bilinearity of the Poisson bracket in each argument. From the superposition for each Poisson bracket provided above, from previous computations with the bracket in the 6-vertex, and 20-vertex, models  {\color{blue}[42},{\color{blue}46]}, the following terms are approximately,

\begin{align*}
  \big\{ \mathscr{I}^1_1 \big( u \big) , \mathscr{I}^1_1 \big( u^{\prime} \big) \big\} \approx \frac{1}{u-u^{\prime}} \equiv C^1_1 \propto \mathscr{C}^1_1 \text{, } \\ \big\{ \mathscr{I}^1_2 \big( u \big), \mathscr{I}^1_2 \big( u^{\prime} \big) \big\} \approx \frac{1}{u-u^{\prime}}  \equiv C^1_2 \propto \mathscr{C}^1_2 \text{, }  \\ 
 \big\{ \mathscr{I}^1_3 \big( u \big), \mathscr{I}^1_3 \big( u^{\prime} \big) \big\} \approx \frac{1}{u-u^{\prime}} \equiv C^1_3  \propto \mathscr{C}^1_3 \text{, }  \\  \big\{ \mathscr{I}^2_1 \big( u \big), \mathscr{I}^2_1 \big( u^{\prime} \big) \big\} \approx \frac{1}{u-u^{\prime}} \equiv C^2_1 \propto \mathscr{C}^2_1 
 \text{, }  \\  \big\{  \mathscr{I}^2_2 \big( u \big), \mathscr{I}^2_2 \big( u^{\prime} \big) \big\} \approx \frac{1}{u-u^{\prime}}  \equiv C^2_2   \propto \mathscr{C}^2_2  \text{, } \\ \big\{ \mathscr{I}^2_3 \big( u \big), \mathscr{I}^2_3 \big( u^{\prime} \big) \big\} \approx \frac{1}{u-u^{\prime}}   \equiv C^2_3   \propto \mathscr{C}^2_3 \text{, }   \\  \big\{ \mathscr{I}^2_4 \big( u \big), \mathscr{I}^2_4  \big( u^{\prime} \big) \big\} \approx \frac{1}{u-u^{\prime}}  \equiv C^2_4   \propto \mathscr{C}^2_4 \text{, } \\  \big\{ \mathscr{I}^3_1 \big( u \big), \mathscr{I}^3_1 \big( u^{\prime} \big) \big\} \approx \frac{1}{u-u^{\prime}}  \equiv C^3_1  \propto \mathscr{C}^3_1  \text{, }  \\ \big\{ \mathscr{I}^3_2 \big( u \big), \mathscr{I}^3_2 \big( u^{\prime} \big) \big\} \approx \frac{1}{u-u^{\prime}}   \equiv C^3_2   \propto \mathscr{C}^3_2 
 \text{, } \\ \big\{ \mathscr{I}^3_3 \big( u \big), \mathscr{I}^3_3 \big( u^{\prime} \big) \big\} \approx \frac{1}{u-u^{\prime}}  \equiv C^3_3  \propto \mathscr{C}^3_3  \text{, } \\  \big\{ \mathscr{I}^3_4 \big( u \big), \mathscr{I}^3_4 \big( u^{\prime} \big) \big\} \approx \frac{1}{u-u^{\prime}}  \equiv C^3_4  \propto \mathscr{C}^3_4  \text{, }    \\ 
 \big\{ \mathscr{I}^4_1 \big( u \big) , \mathscr{I}^4_1 \big( u^{\prime} \big) \big\} \approx  \frac{1}{u-u^{\prime}}   \equiv C^4_1  \propto \mathscr{C}^4_1 \text{, } \\ \big\{ \mathscr{I}^4_2 \big( u \big) , \mathscr{I}^4_2 \big( u^{\prime} \big) \big\} \approx  \frac{1}{u-u^{\prime}}   \equiv C^4_2  \propto \mathscr{C}^4_2 \text{, } \end{align*}

 \begin{align*}  \big\{ \mathscr{I}^4_3 \big( u \big) , \mathscr{I}^4_3 \big( u^{\prime} \big) \big\} \approx  \frac{1}{u-u^{\prime}}  \equiv C^4_3 \propto \mathscr{C}^4_3  \text{, } \\ 
 \big\{ \mathscr{I}^4_4 \big( u \big) , \mathscr{I}^4_4 \big( u^{\prime} \big) \big\} \approx  \frac{1}{u-u^{\prime}}  \equiv C^4_4  \propto \mathscr{C}^4_4  \text{, } \\ \big\{ \mathscr{I}^4_5 \big( u \big) , \mathscr{I}^4_5 \big( u^{\prime} \big) \big\} \approx  \frac{1}{u-u^{\prime}}  \equiv C^4_5   \propto \mathscr{C}^4_5 \text{. }
\end{align*}

\noindent The constants $\mathscr{C}^1_1, \cdots$ are provided in the next section, \textit{3}. The sequence of approximations, along with each constant used to approximate the corresponding Poisson bracket, was first introduced in the study of Hamiltonian systems {\color{blue}[17]}. Despite the fact that seminal work from Fadeev and Takhtajan initially studied the behavior of solutions to the nonlinear Schrodinger's equation, vertex models can be studied within this framework. An inhomgeneous Hamiltonian flow was introduced for establishing that inhomogeneous limit shapes of the 6-vertex model are integrable, {\color{blue}[25]}, which has implications for integrability of the Hamiltonian flow studied by the author {\color{blue}[42]}. For the 20-vertex model, as a three-dimensional generalization of the 6-vertex model over the square lattice, in a later 2024 work the author studied a weakining of integrability properties for the 20-vertex model {\color{blue}[46]}. For the 20-vertex model, in comparison to the 6-vertex model, a counterpart of the stronger integrability condition does not hold due to the lack of existence for action-angle coordinates. With respect to the Poisson bracket, action-angle variables $\Phi$ which are used for demonstrating that the Hamiltonian flow of the 6-vertex model is integrable vanish, through the equality,

\begin{align*}
  \big\{ \Phi , \bar{\Phi} \big\} = 0   \text{. }
\end{align*}

\noindent For the 4-vertex model, as demonstrated through computations in previous sections with the L-operator, one can expect to be able to make use the quantum inverse scattering framework for studying the Poisson structure. Equipped with the Poisson bracket, repeatedly applying Leibniz' rule allows for one to isolate components, one at a time, appearing in products of terms in either the first, or second, argument of the bracket. In comparison to previous works of the author which have made significant use of the quantum inverse scattering framework, {\color{blue}[42},{\color{blue}46},{\color{blue}47]}, characteristics of the Yang-Baxter algebra continue to be of interest for further investigation. For the 20-vertex model, properties of the Yang-Baxter algebra were used for obtaining nonlocal correlations, in addition to contour integral expressions for evaluating the emptiness formation probability. In the following set of equalities, denote $\underline{u^{\prime}}$ and $\underline{u^{\prime\prime}}$ as two other spectral parameters over $\textbf{T}$ that are not equal to $\underline{u}$. For such an algebra that is used to further examine correlations of the 20-vertex model, the collection of relations for the Yang-Baxter algebra takes the form, {\color{blue}[47]},

\begin{align*}
         \underline{G \big( \underline{u} \big) E \big( \underline{u^{\prime}} \big) C \big( \underline{u^{\prime\prime}} \big) } =       f \big( \lambda_{\alpha} , \lambda_r , \lambda_{r^{\prime}} \big)   f \big( \lambda , \lambda^{\prime} \big)     C \big( \underline{u^{\prime\prime}} \big) E \big( \underline{u^{\prime}} \big)  G \big( \underline{u} \big)     +    f \big( \lambda_{\alpha} , \lambda_r , \lambda_{r^{\prime}} \big) g \big( \lambda^{\prime} , \lambda \big)     C \big( \underline{u^{\prime}}    \big) E \big( \underline{u^{\prime\prime}} \big)    G \big( \underline{u} \big)                        \\     +        g \big( \lambda_{\alpha} , \lambda_r , \lambda_{r^{\prime}} \big)      f \big(   \lambda , \lambda^{\prime} \big)  C \big( \underline{u^{\prime}} \big) E \big( \underline{u} \big)         G \big( \underline{u^{\prime\prime} } \big)              +  g \big( \lambda_{\alpha} , \lambda_r , \lambda_{r^{\prime}} \big)      g \big( \lambda^{\prime} , \lambda \big)   C \big(  \underline{u}     \big) E \big( \underline{u^{\prime}}       \big)                      G \big( \underline{u^{\prime\prime} } \big)                \text{, } \end{align*}

 \begin{align*}  \underline{ I \big( \underline{u} \big) H \big( \underline{u^{\prime}} \big) G \big( \underline{u^{\prime\prime}} \big)}  =           f \big( \lambda_{\alpha} , \lambda_r , \lambda_{r^{\prime}} \big)       f \big( \lambda , \lambda^{\prime} \big)                G \big( \underline{u^{\prime\prime}} \big) H \big( \underline{u^{\prime}} \big)        I \big( \underline{u} \big)     +   f \big( \lambda_{\alpha} , \lambda_r , \lambda_{r^{\prime}} \big)   g \big( \lambda^{\prime} , \lambda \big)    G \big( \underline{u^{\prime}} \big) H \big( \underline{u^{\prime\prime}} \big)                          I \big( \underline{u} \big)         \\    +  g \big( \lambda_{\alpha} , \lambda_r  , \lambda_{r^{\prime}} \big)    f \big( \lambda , \lambda^{\prime} \big) G \big( \underline{u^{\prime\prime}} \big) H \big(   \underline{u^{\prime}} \big) I \big( \underline{u^{\prime}}  \big)   +  g \big( \lambda_{\alpha} , \lambda_r  , \lambda_{r^{\prime}} \big)          g \big( \lambda^{\prime} , \lambda \big)                       G \big( \underline{u^{\prime}} \big)  H \big( \underline{u^{\prime\prime}} \big) I \big( \underline{u^{\prime}}  \big)        \text{, } \\   \\      \underline{  A \big( \underline{u }  \big) D \big( \underline{u^{\prime}} \big) G \big( \underline{u^{\prime\prime}} \big) }   =          f \big( \lambda_{\alpha}  , \lambda_r , \lambda_{r^{\prime}} \big)               f \big( \lambda , \lambda^{\prime} \big)                       G \big(    \underline{u^{\prime\prime}}   \big)    D \big( \underline{u^{\prime}} \big)       A \big( \underline{u} \big)   +  f \big( \lambda_{\alpha}  , \lambda_r , \lambda_{r^{\prime}} \big)    g \big( \lambda^{\prime} , \lambda \big)    G \big( \underline{u^{\prime}} \big) D \big( \underline{u^{\prime\prime}} \big)                               A \big( \underline{u} \big)  \\   +   g \big( \lambda_{\alpha} , \lambda_r , \lambda_{r^{\prime}} \big)           f \big( \lambda , \lambda^{\prime} \big)   G \big( \underline{u} \big)  D \big( \underline{u^{\prime\prime}} \big)      A \big( \underline{u} \big)      +  g \big( \lambda_{\alpha} , \lambda_r , \lambda_{r^{\prime}} \big)   g \big( \lambda^{\prime} , \lambda \big)  G \big( \underline{u^{\prime\prime}} D \big( \underline{u} \big)          A \big( \underline{u} \big)     \text{, } \\ \\ 
         \underline{A \big( \underline{u } \big) E \big( \underline{u^{\prime}} \big) I \big( \underline{u^{\prime\prime}} \big)}       =           f \big( \lambda_{\alpha}  , \lambda_r , \lambda_{r^{\prime}} \big)        f \big( \lambda , \lambda^{\prime} \big)     I \big( \underline{u^{\prime\prime}} \big) E \big( \underline{u^{\prime}} \big)  A \big( \underline{u} \big)    +    f \big( \lambda_{\alpha}  , \lambda_r , \lambda_{r^{\prime}} \big)    g \big( \lambda^{\prime} , \lambda \big)              I \big( \underline{u^{\prime}} \big) E \big( \underline{u^{\prime\prime}} \big)           A \big( \underline{u} \big)   \\  +   g \big( \lambda_{\alpha} , \lambda_r , \lambda_{r^{\prime}} \big)        f \big( \lambda , \lambda^{\prime} \big)     I \big( \underline{u}  \big) E \big( \underline{u^{\prime\prime}} \big)               A \big( \underline{u^{\prime}} \big)      + g \big( \lambda_{\alpha} , \lambda_r , \lambda_{r^{\prime}} \big)  g \big( \lambda^{\prime} , \lambda \big)                     I \big( \underline{u^{\prime\prime}} \big)           E \big( \underline{u}  \big)      A \big( \underline{u^{\prime}} \big)             \text{, } 
\end{align*}

\noindent where,

\begin{align*}
   f \big( \lambda_{\alpha} , \lambda_{r} , \lambda_{r^{\prime}} \big) \equiv     \frac{\mathrm{sin} \big( \lambda_{r^{\prime}} - \lambda_r - \lambda_{\alpha} + 2 \eta  \big)}{\mathrm{sin} \big( \lambda_{r^{\prime}} - \lambda_r - \lambda_{\alpha } \big) }                    \text{, } \\ 
g \big( \lambda_{\alpha} , \lambda_{r} , \lambda_{r^{\prime}} \big) \equiv   \frac{\mathrm{sin} \big( 2 \eta \big)}{\mathrm{sin} \big( \lambda_{r^{\prime}} - \lambda_r - \lambda_{\alpha} \big)}                  \text{, } 
\end{align*}

\noindent from blocks of the product representation for the three-dimensional transfer matrix, {\color{blue}[45]},

\[
     \begin{bmatrix}
 A^{20V} \big( \underline{u} \big) & D^{20V} \big( \underline{u} \big)  & G^{20V} \big( \underline{u} \big) \\ B^{20V} \big(\underline{u} \big) & E^{20V} \big( \underline{u} \big) & H^{20V} \big( \underline{u} \big)  \\ C^{20V} \big( \underline{u} \big)  &  F^{20V} \big( \underline{u} \big) & I^{20V} \big( \underline{u} \big) 
\end{bmatrix}   \equiv     \begin{bmatrix}
 A \big( \underline{u} \big) & D \big( \underline{u} \big)  & G \big( \underline{u} \big) \\ B \big(\underline{u} \big) & E \big( \underline{u} \big) & H \big( \underline{u} \big)  \\ C \big( \underline{u} \big)  &  F \big( \underline{u} \big) & I \big( \underline{u} \big) 
\end{bmatrix}   \text{. } 
\]

\noindent The Yang-Baxter algebra for the 4-vertex model, in absence of the product representation above, through the product representation,

\[
\begin{bmatrix}
        A^{4V} \big( \underline{u} \big) & B^{4V} \big( \underline{u} \big) \\ C^{4V} \big( \underline{u} \big) & D^{4V} \big( \underline{u} \big) 
    \end{bmatrix} \equiv \begin{bmatrix}
        A \big( \underline{u} \big) & B \big( \underline{u} \big) \\ C \big( \underline{u} \big) & D \big( \underline{u} \big) 
    \end{bmatrix}    \text{, }  
\]

\noindent implies that the corresponding Yang-Baxter algebra can be studied by taking products of operators from the asymptotic representation for the transfer matrix, and quantum monodromy matrix. In comparison to the structure of the Yang-Baxter algebra that one would expect to obtain from the 20-vertex model, characteristics of the Yang-Baxter algebra are dependent upon the terms,

\begin{align*}
 \underset{i \in \textbf{N}}{\bigcup}  i u \sigma^+_i \sigma^-_i  \text{, } \\ \underline{\mathcal{I}_1} \text{, } \\ \underline{\mathcal{I}_2}  \text{, } \\ \underline{\mathcal{I}_3}  \text{, } \\ \underline{\mathcal{I}_4}  \text{, } \\ \underline{\mathcal{I}^{\prime\cdots\prime}_1}    \text{, } \\ \underline{\mathcal{I}^{\prime\cdots\prime}_2}  \text{, } \\ \underline{\mathcal{I}^{\prime\cdots\prime}_3}  \text{, } \\  \underline{\mathcal{I}^{\prime\cdots\prime}_4}  \text{, } \\     A \big( \underline{u} \big)     \text{, } \\   B \big( \underline{u} \big)           \text{, } \\  C \big( \underline{u} \big)          \text{, } \\  D \big( \underline{u} \big) \text{, } \\ \textbf{1}_{\{ j^{\prime} > j \text{ } : \text{ } i u \sigma^+_{j^{\prime}} \sigma^-_{j^{\prime}} \in \mathrm{support} ( A ( u )) \}}    , \cdots,  \textbf{1}_{\{ j^{\prime} > j \text{ } : \text{ } i u \sigma^+_{j^{\prime}} \sigma^-_{j^{\prime}} \in \mathrm{support} ( D ( u )) \}}     \text{. } 
\end{align*}

\noindent In comparison to the embedded Poisson bracket,

\[\
\bigg\{    \begin{bmatrix}
 A \big( \underline{u} 
 \big) & D \big( \underline{u} \big)  & G \big( \underline{u}  \big) \\ B \big( \underline{u}  \big) & E \big( \underline{u}  \big) & H \big( \underline{u}  \big)  \\ C \big( \underline{u}  \big)  &  F \big( \underline{u}  \big) & I \big( \underline{u}  \big) 
\end{bmatrix}\overset{\bigotimes}{,} \bigg\{ \begin{bmatrix}
 A \big( \underline{u^{\prime}} \big) & D \big( \underline{u^{\prime}} \big)  & G \big( \underline{u^{\prime}} \big) \\ B \big( \underline{u^{\prime}}\big) & E \big( \underline{u^{\prime}} \big) & H \big( \underline{u^{\prime}} \big)  \\ C \big( \underline{u^{\prime}} \big)  &  F \big( \underline{u^{\prime}} \big) & I \big( \underline{u^{\prime}} \big) 
\end{bmatrix}  \overset{\bigotimes}{,} \begin{bmatrix}
 A \big( \underline{u^{\prime\prime}} \big) & D \big( \underline{u^{\prime\prime}} \big)  & G \big( \underline{u^{\prime\prime}} \big) \\ B \big( \underline{u^{\prime\prime}}\big) & E \big( \underline{u^{\prime\prime }} \big) & H \big( \underline{u^{\prime\prime}} \big)  \\ C \big( \underline{u^{\prime\prime}} \big)  &  F \big( \underline{u^{\prime\prime}} \big) & I \big( \underline{u^{\prime\prime}} \big) 
\end{bmatrix}  \bigg\}  \text{ } \bigg\} \text{. } 
\] 

\noindent of the 20-vertex model, one has the embedded Poisson bracket,

\[\
\bigg\{    \begin{bmatrix}
 A \big( \underline{u} 
 \big) & B \big( \underline{u} \big)  \\ C \big( \underline{u}  \big) & D \big( \underline{u}  \big) 
\end{bmatrix}\overset{\bigotimes}{,} \bigg\{  \begin{bmatrix}
 A \big( \underline{u^{\prime}} 
 \big) & B \big( \underline{u^{\prime}} \big)  \\ C \big( \underline{u^{\prime}}  \big) & D \big( \underline{u^{\prime}}  \big) 
\end{bmatrix} \overset{\bigotimes}{,} \begin{bmatrix}
 A \big( \underline{u^{\prime\prime}} 
 \big) & B \big( \underline{u^{\prime\prime}} \big)  \\ C \big( \underline{u^{\prime\prime}}  \big) & D \big( \underline{u^{\prime\prime}}  \big) 
\end{bmatrix}  \bigg\}  \text{ } \bigg\} \text{. } 
\]

\noindent for the 4-vertex model. Equipped with such information, in the next section we compute the Poisson brackets which have been rexpressed after three applications of the bilinearity property. From previous expressions obtained for $A \big( \underline{u} \big), B \big( \underline{u} \big), C \big( \underline{u} \big), D \big( \underline{u} \big)$, in the next section we obtain the sequence of desired approximations for each Poisson bracket through repeated applications of Leibniz' rule.

\section{Computations with the Poisson bracket} 

We make use of similarly minded computations with the bracket that have previously been investigated by the author in {\color{blue}[42},{\color{blue}46]}. First, we identify all of the terms which would asymptotically behave like $\frac{1}{u-u^{\prime}}$, and then approximate other remaining terms depending upon the arguments that appear for the bracket. Before proceeding to prove the desired result for the Poisson structure of the 4-vertex model, which can be immediately leveraged for the higher-spin XXX chain, from the product representation,

\begin{align*}
\begin{bmatrix}
        A^{XXX} \big( u \big) & B^{XXX} \big( u \big) \\ C^{XXX} \big( u \big) & D^{XXX} \big( u \big) 
    \end{bmatrix} \text{, }
\end{align*}

\noindent consider the following statement below. With the main point being to relate Poisson structures of a vertex model to that of a higher-spin chain, between the representations,

\[\begin{bmatrix}
        A^{4V} \big( \underline{u} \big) & B^{4V} \big( \underline{u} \big) \\ C^{4V} \big( \underline{u} \big) & D^{4V} \big( \underline{u} \big)  \text{, }
    \end{bmatrix}\]

\noindent and,

\[\begin{bmatrix}
        A^{XXX} \big( \underline{u} \big) & B^{XXX} \big( \underline{u} \big) \\ C^{XXX} \big( \underline{u} \big) & D^{XXX} \big( \underline{u} \big)  \text{, }
    \end{bmatrix}\]

    \noindent the mapping $\phi : \mathcal{R} \big( 4V \big)  \longrightarrow \mathcal{R} \big( XXX \big)$, into the target space of all possible XXX representations for the transfer matrix, from the codomain of all possible 4-vertex representations for the transfer matrix, acts on the generators,

    \begin{align*}
  \phi \big( \mathcal{I}_1 \big) \equiv   \phi \big( A^{4V} \big( \underline{u} \big)  \big) =   \lambda I_n + i S^3_n  = \mathscr{I}_1   \text{, } \\   \phi \big( \mathcal{I}_2 \big) \equiv   \phi \big( B^{4V} \big( \underline{u} \big)  \big) =   i S^-_n   = \mathscr{I}_2      \text{, } \\   \phi \big( \mathcal{I}_3 \big) \equiv  \phi \big( C^{4V} \big( \underline{u} \big)  \big) =  i S^+_n   = \mathscr{I}_3 \text{, } \\  \phi \big( \mathcal{I}_4 \big) \equiv   \phi \big( D^{4V} \big( \underline{u} \big)  \big) =    \lambda I_n - i S^3_n   = \mathscr{I}_4          \text{. } 
    \end{align*}

\noindent The following restriction of the spanning set for $N \equiv 2$ satisfies,

\begin{align*}
  \underset{ \underline{u} \in \textbf{Z}^2}{\underset{N \in \textbf{N}}{\mathrm{span}}} \big\{  A^{XXX} \big( \underline{u} \big) , B^{XXX} \big( \underline{u} \big), C^{XXX} \big( \underline{u} \big),  D^{XXX} \big( \underline{u} \big) \big\} \bigg|_{N \equiv 2} = \underset{ \underline{u} \in \textbf{Z}^2}{\underset{1 \leq N \leq 2}{\mathrm{span}}} \big\{ 
 \lambda I_n \big( \underline{u} \big)   + i S^3_n  \big( \underline{u} \big)  
 , i S^-_n \big( \underline{u} \big)  , i S^+_n \big( \underline{u} \big)  , \lambda I_n \big( \underline{u} \big)  
 \\ - i S^3_n \big( \underline{u} \big)   \big\}      \text{. }
\end{align*}

\noindent \textbf{Theorem} (\textit{Poisson structures of the 4-vertex model, and higher-spin XXX chain}). For the 4-vertex model, the \textit{Poisson structure},

\begin{align*}
  \mathscr{P}^{4V} \equiv \mathscr{C}^{4V}_1 \cup \mathscr{C}^{4V}_2 \cup \mathscr{C}^{4V}_3 \cup \mathscr{C}^{4V}_4   \text{, }
\end{align*}

\noindent is captured through the following sequence of approximations,

\[
\mathscr{C}^{4V}_1 \equiv \mathscr{C}_1 =   \left\{\!\begin{array}{ll@{}>{{}}l}          \big\{ \underline{\mathcal{I}^{\prime\cdots\prime}_1 \big( u , u^{-1} \big)}, \underline{\mathcal{I}^{\prime\cdots\prime}_1 \big( u^{\prime} , \big( u^{\prime} \big)^{-1} \big)}\big\} \approx  \mathscr{C}^1_1 
\text{, } \\  
  \big\{ \underline{\mathcal{I}^{\prime\cdots\prime}_1 \big( u , u^{-1} \big)},  \underline{\mathcal{I}^{\prime\cdots\prime}_2 \big( u^{\prime} , \big( u^{\prime} \big)^{-1} \big)}  \big\} \approx  \mathscr{C}^2_1   \text{, }  \\      \big\{ \underline{\mathcal{I}^{\prime\cdots\prime}_1 \big( u , u^{-1} \big)},  \underline{\mathcal{I}^{\prime\cdots\prime}_3 \big( u^{\prime} , \big( u^{\prime} \big)^{-1} \big)} \big\}  \approx  \mathscr{C}^3_1   \text{, } \\  \big\{ \underline{\mathcal{I}^{\prime\cdots\prime}_1 \big( u , u^{-1} \big)}, \underline{\mathcal{I}^{\prime\cdots\prime}_4 \big( u^{\prime} , \big( u^{\prime} \big)^{-1} \big)}\big\} \approx  \mathscr{C}^4_1   \text{, } \end{array}\right. \]   \[ \mathscr{C}^{4V}_2 \equiv 
 \mathscr{C}_2 =   \left\{\!\begin{array}{ll@{}>{{}}l}          \big\{ \underline{\mathcal{I}^{\prime\cdots\prime}_2 \big( u , u^{-1} \big)}, \underline{\mathcal{I}^{\prime\cdots\prime}_1 \big( u^{\prime} , \big( u^{\prime} \big)^{-1} \big)}\big\} \approx   \mathscr{C}^1_2 
\text{, } \\  
  \big\{ \underline{\mathcal{I}^{\prime\cdots\prime}_2 \big( u , u^{-1} \big)},  \underline{\mathcal{I}^{\prime\cdots\prime}_2 \big( u^{\prime} , \big( u^{\prime} \big)^{-1} \big)}  \big\} \approx  \mathscr{C}^2_2  \text{, }  \\      \big\{ \underline{\mathcal{I}^{\prime\cdots\prime}_2 \big( u , u^{-1} \big)},  \underline{\mathcal{I}^{\prime\cdots\prime}_3 \big( u^{\prime} , \big( u^{\prime} \big)^{-1} \big)} \big\}  \approx  \mathscr{C}^3_2   \text{, } \\  \big\{ \underline{\mathcal{I}^{\prime\cdots\prime}_2 \big( u , u^{-1} \big)}, \underline{\mathcal{I}^{\prime\cdots\prime}_4 \big( u^{\prime} , \big( u^{\prime} \big)^{-1} \big)}\big\}  \approx   \mathscr{C}^4_2  \text{, } \end{array}\right. \\  \mathscr{C}^{4V}_3 \equiv 
 \mathscr{C}_3 =   \left\{\!\begin{array}{ll@{}>{{}}l}         \big\{ \underline{\mathcal{I}^{\prime\cdots\prime}_3 \big( u , u^{-1} \big)}, \underline{\mathcal{I}^{\prime\cdots\prime}_1 \big( u^{\prime} , \big( u^{\prime} \big)^{-1} \big)}\big\}  \approx \mathscr{C}^1_3
\text{, } \\  
  \big\{ \underline{\mathcal{I}^{\prime\cdots\prime}_3 \big( u , u^{-1} \big)},  \underline{\mathcal{I}^{\prime\cdots\prime}_2 \big( u^{\prime} , \big( u^{\prime} \big)^{-1} \big)}  \big\} \approx \mathscr{C}^2_3 \text{, }  \\      \big\{ \underline{\mathcal{I}^{\prime\cdots\prime}_3 \big( u , u^{-1} \big)},  \underline{\mathcal{I}^{\prime\cdots\prime}_3 \big( u^{\prime} , \big( u^{\prime} \big)^{-1} \big)} \big\}  \approx  \mathscr{C}^3_3 \text{, } \\  \big\{ \underline{\mathcal{I}^{\prime\cdots\prime}_3 \big( u , u^{-1} \big)}, \underline{\mathcal{I}^{\prime\cdots\prime}_4 \big( u^{\prime} , \big( u^{\prime} \big)^{-1} \big)}\big\} \approx \mathscr{C}^4_3   \text{, } \end{array}\right. \]

  \[ \mathscr{C}^{4V}_4 \equiv 
 \mathscr{C}_4 =   \left\{\!\begin{array}{ll@{}>{{}}l}        \big\{ \underline{\mathcal{I}^{\prime\cdots\prime}_4 \big( u , u^{-1} \big)}, \underline{\mathcal{I}^{\prime\cdots\prime}_1 \big( u^{\prime} , \big( u^{\prime} \big)^{-1} \big)}\big\} \approx  \mathscr{C}^1_4 
\text{, } \\  
  \big\{ \underline{\mathcal{I}^{\prime\cdots\prime}_4 \big( u , u^{-1} \big)},  \underline{\mathcal{I}^{\prime\cdots\prime}_2 \big( u^{\prime} , \big( u^{\prime} \big)^{-1} \big)}  \big\} \approx  \mathscr{C}^2_4    \text{, }  \\      \big\{ \underline{\mathcal{I}^{\prime\cdots\prime}_4 \big( u , u^{-1} \big)},  \underline{\mathcal{I}^{\prime\cdots\prime}_3 \big( u^{\prime} , \big( u^{\prime} \big)^{-1} \big)} \big\}  \approx  \mathscr{C}^3_4  \text{, } \\  \big\{ \underline{\mathcal{I}^{\prime\cdots\prime}_4 \big( u , u^{-1} \big)}, \underline{\mathcal{I}^{\prime\cdots\prime}_4 \big( u^{\prime} , \big( u^{\prime} \big)^{-1} \big)}\big\}  \approx  \mathscr{C}^4_4   \text{. }   \end{array}\right. 
\]

\noindent From the Poisson structure of the 4-vertex model, that of the higher-spin XXX chain, 

\begin{align*}
  \mathscr{P}^{XXX} \equiv \mathcal{C}^{XXX}_1 \cup \mathcal{C}^{XXX}_2 \cup \mathcal{C}^{XXX}_3 \cup \mathcal{C}^{XXX}_4  \text{, }
\end{align*}

\noindent is captured through the following sequence of approximations,

\[
\mathcal{C}^{XXX}_1 \equiv \mathcal{C}_1 =   \left\{\!\begin{array}{ll@{}>{{}}l}          \big\{ \underline{\mathscr{I}^{\prime\cdots\prime}_1 \big( u , u^{-1} \big)} , \underline{\mathscr{I}^{\prime\cdots\prime}_1 \big( u^{\prime} , \big( u^{\prime} \big)^{-1} \big)} \big\} \approx \mathcal{C}^1_1 
\text{, } \\  
  \big\{ \underline{\mathscr{I}^{\prime\cdots\prime}_1 \big( u , u^{-1} \big)} , \underline{\mathscr{I}^{\prime\cdots\prime}_2 \big( u^{\prime} , \big( u^{\prime} \big)^{-1} \big)} \big\} \approx  \mathcal{C}^2_1 
\text{, }  \\     \big\{ \underline{\mathscr{I}^{\prime\cdots\prime}_1 \big( u , u^{-1} \big)} , \underline{\mathscr{I}^{\prime\cdots\prime}_3 \big( u^{\prime} , \big( u^{\prime} \big)^{-1} \big)} \big\} \approx \mathcal{C}^3_1 
\text{, } \\  \big\{ \underline{\mathscr{I}^{\prime\cdots\prime}_1 \big( u , u^{-1} \big)} , \underline{\mathscr{I}^{\prime\cdots\prime}_4 \big( u^{\prime} , \big( u^{\prime} \big)^{-1} \big)} \big\} \approx \mathcal{C}^4_1 
\text{, }  \end{array}\right. \] 

\[
\mathcal{C}^{XXX}_2 \equiv \mathcal{C}_2 =   \left\{\!\begin{array}{ll@{}>{{}}l}          \big\{ \underline{\mathscr{I}^{\prime\cdots\prime}_2 \big( u , \big( u \big)^{-1} \big)} ,  \underline{\mathscr{I}^{\prime\cdots\prime}_1 \big( u^{\prime} , \big( u^{\prime} \big)^{-1} \big)} \big\} \approx \mathcal{C}^1_2 
\text{, } \\  
  \big\{  \underline{\mathscr{I}^{\prime\cdots\prime}_2 \big( u , \big( u \big)^{-1} \big)} , \underline{\mathscr{I}^{\prime\cdots\prime}_2 \big( u^{\prime} , \big( u^{\prime} \big)^{-1} \big)} \big\} \approx \mathcal{C}^2_2 
\text{, }  \\     \big\{ \underline{\mathscr{I}^{\prime\cdots\prime}_2 \big( u , \big( u \big)^{-1} \big)} , \underline{\mathscr{I}^{\prime\cdots\prime}_3 \big( u^{\prime} , \big( u^{\prime} \big)^{-1} \big)}\big\} \approx \mathcal{C}^3_2 
\text{, } \\  \big\{ \underline{\mathscr{I}^{\prime\cdots\prime}_2 \big( u , \big( u \big)^{-1} \big)} ,  \underline{\mathscr{I}^{\prime\cdots\prime}_4 \big( u^{\prime} , \big( u^{\prime} \big)^{-1} \big)}\big\} \approx \mathcal{C}^4_2 
\text{, }  \end{array}\right.\\ 
\mathcal{C}^{XXX}_3 \equiv \mathcal{C}_3 =   \left\{\!\begin{array}{ll@{}>{{}}l}          \big\{ , \underline{\mathscr{I}^{\prime\cdots\prime}_1 \big( u^{\prime} , \big( u^{\prime} \big)^{-1} \big)} \big\} \approx  \mathcal{C}^1_3 
\text{, } \\  
  \big\{ , \underline{\mathscr{I}^{\prime\cdots\prime}_2 \big( u^{\prime} , \big( u^{\prime} \big)^{-1} \big)} \big\} \approx \mathcal{C}^2_3 
\text{, }  \\     \big\{ , \underline{\mathscr{I}^{\prime\cdots\prime}_3 \big( u^{\prime} , \big( u^{\prime} \big)^{-1} \big)} \big\} \approx \mathcal{C}^3_3 
\text{, } \\  \big\{ ,  \underline{\mathscr{I}^{\prime\cdots\prime}_4 \big( u^{\prime} , \big( u^{\prime} \big)^{-1} \big)}\big\} \approx \mathcal{C}^4_3 
\text{, }  \end{array}\right. \]

\[
\mathcal{C}^{XXX}_4 \equiv \mathcal{C}_4 =   \left\{\!\begin{array}{ll@{}>{{}}l}          \big\{ \underline{\mathscr{I}^{\prime\cdots\prime}_4 \big( u , \big( u \big)^{-1} \big)}  , \underline{\mathscr{I}^{\prime\cdots\prime}_1 \big( u^{\prime} , \big( u^{\prime} \big)^{-1} \big)} \big\} \approx \mathcal{C}^1_4 
\text{, } \\  
  \big\{ \underline{\mathscr{I}^{\prime\cdots\prime}_4 \big( u , \big( u \big)^{-1} \big)}  , \underline{\mathscr{I}^{\prime\cdots\prime}_2 \big( u^{\prime} , \big( u^{\prime} \big)^{-1} \big)} \big\} \approx \mathcal{C}^2_4 
\text{, }  \\     \big\{ \underline{\mathscr{I}^{\prime\cdots\prime}_4 \big( u , \big( u \big)^{-1} \big)}, \underline{\mathscr{I}^{\prime\cdots\prime}_3 \big( u^{\prime} , \big( u^{\prime} \big)^{-1} \big)}\big\} \approx \mathcal{C}^3_4 
\text{, } \\  \big\{ \underline{\mathscr{I}^{\prime\cdots\prime}_4 \big( u , \big( u \big)^{-1} \big)}, \underline{\mathscr{I}^{\prime\cdots\prime}_4 \big( u^{\prime} , \big( u^{\prime} \big)^{-1} \big)} \big\} \approx \mathcal{C}^4_4 
\text{. }  \end{array}\right. \]

\noindent In the sequence of approximations for each Poisson brackets from XXX L-operators, the representation from products of operators,

\begin{align*}
  \underset{\lambda \in \textbf{R}}{\underset{1 \leq n \leq N}{\prod}}   L^{XXX} \big( n | \lambda \big)             \text{. }
\end{align*}

\section{Appendix}

\subsection{Lemma 1}

\noindent \textit{Proof of Lemma 1}. By inspection, each one of the four entries, in order, of the union of the set of linear combinations of the two rows above can be obtained with,

\begin{align*}
\underline{\mathcal{I}^{\prime}_1} \equiv  \bigg[ i u \sigma^+_0 \sigma^-_0 i u \sigma^1_+ \sigma^-_1 + \sigma^-_0 \sigma^+_1 \bigg] \bigg[ i u \sigma^+_1 \sigma^-_1 i u \sigma^+_2 \sigma^-_2 + \sigma^-_1 \sigma^+_2 \bigg] + \bigg[ i u \sigma^+_0 \sigma^-_0 \sigma^+_1 + \sigma^+_0 i u \sigma^+_1 \sigma^-_1  \bigg] \bigg[ i u \sigma^+_1 \sigma^-_1 \sigma^-_2  \\  + \sigma^-_1 i u  \sigma^+_2 \sigma^-_2 \bigg] \text{, } \end{align*}

\begin{align*}
\underline{\mathcal{I}^{\prime}_2} \equiv            \bigg[ i u \sigma^+_0 \sigma^-_0 i u \sigma^+_1 \sigma^-_1 + \sigma^-_0 \sigma^+_0  \bigg]  \bigg[        i u \sigma^+_1 \sigma^-_1 \sigma^+_2  + \sigma^+_1 i u \sigma^+_2 \sigma^-_2     \bigg]  + \bigg[   i u \sigma^+_0 \sigma^+_1 + \sigma^+_0 i u \sigma^+_1 \sigma^-_1  \bigg]     \bigg[   \sigma^+_1 \sigma^-_2 + i u \sigma^+_1 \sigma^-_1  \\ \times  i u \sigma^+_2 \sigma^-_2  \bigg]    \text{, } \\ \\ 
\underline{\mathcal{I}^{\prime}_3} \equiv  
 \bigg[      i u \sigma^+_0 \sigma^-_0 i u \sigma^+_1 \sigma^-_1 + \sigma^-_0 \sigma^+_1      \bigg] \bigg[   i u \sigma^+_1 \sigma^-_1 \sigma^-_2 + \sigma^+_1 i u \sigma^+_2 \sigma^-_2     \bigg] +  \bigg[ i u \sigma^+_0    \sigma^-_0 \sigma^-_1 + \sigma^-_0 i u \sigma^+_1 \sigma^-_1     \bigg] \bigg[    \sigma^+_1 \sigma^-_2 + i u \sigma^+_1 \sigma^-_1 i u \\   \times  \sigma^+_2 \sigma^-_2   \bigg] \text{, } \\ \\ 
 \underline{\mathcal{I}^{\prime}_4} \equiv   \bigg[   i u \sigma^+_0 \sigma^-_0 i u \sigma^+_1 \sigma^-_1 + \sigma^-_0 \sigma^+_1    \bigg] \bigg[     i u \sigma^+_1 \sigma^-_1       \sigma^-_2 + \sigma^-_1 i u \sigma^+_2 \sigma^-_2       \bigg] + \bigg[   i u \sigma^+_0 \sigma^-_0 \sigma^-_1 + \sigma^-_0 i u \sigma^+_1 \sigma^-_1       \bigg] \bigg[ \sigma^+_1 \sigma^-_2 + i u \sigma^+_1 \sigma^-_1 i u \\ \times  \sigma^+_2 \sigma^-_2  \bigg] \text{, }
\end{align*}

\noindent from which we conclude the argument. \boxed{}

\subsection{Lemma 2}

\noindent \textit{Proof of Lemma 2}. We establish that the desired form of the product representation holds through induction over $n$. For the base case of the induction, by making use of the product representation of three L-operators for the 4-vertex model presented in the previous result, one would like to compute,

\[
\begin{bmatrix}
\underline{\mathcal{I}^{\prime}_1} &  \underline{\mathcal{I}^{\prime}_2} \\ \underline{\mathcal{I}^{\prime}_3} & \underline{\mathcal{I}^{\prime}_4} 
\end{bmatrix} \begin{bmatrix}
 i u \sigma^+_3 \sigma^-_3 & \sigma^-_3  \\ \sigma^+_3 &
 i u^{-1} \sigma^+_3 \sigma^-_3 \end{bmatrix} \text{. }
\]

\noindent To this end, the entries of the product representation obtained from the resultant expression above, which we denote with,

\[
\begin{bmatrix}
\underline{\mathcal{I}^{\prime\prime}_1} & \underline{\mathcal{I}^{\prime\prime}_2} \\ \underline{\mathcal{I}^{\prime\prime}_3} & \underline{\mathcal{I}^{\prime\prime}_4}
\end{bmatrix} \text{, }
\]

\noindent take the form,

\begin{align*}
\underline{\mathcal{I}^{\prime\prime}_1}  \equiv   i u \sigma^+_0 \sigma^-_0 i u \sigma^+_1 \sigma^-_1         i u \sigma^+_1 \sigma^-_1 i u \sigma^+_2 \sigma^-_2 i u \sigma^+_3 \sigma^-_3 + \sigma^-_0 \sigma^+_1   i u \sigma^+_1 \sigma^-_1 i u \sigma^+_2 \sigma^-_2 i u \sigma^+_3 \sigma^-_3 + i u \sigma^+_0 \sigma^-_0  i u \sigma^+_1 \sigma^-_1 \sigma^-_1 \sigma^+_2  i u \sigma^+_3 \sigma^-_3 \\  + \sigma^-_0 \sigma^+_1 \sigma^-_1 \sigma^+_2 i u \sigma^+_3 \sigma^-_3 + i u \sigma^+_0 \sigma^-_0 i u \sigma^+_1  \sigma^-_1 i u \sigma^+_1 \sigma^-_1 \sigma^-_2 \sigma^+_2 \sigma^-_3 + \sigma^-_0 \sigma^+_1 i u \sigma^+_1 \sigma^+_1 \sigma^-_3 + \sigma^-_0 \sigma^+_1 \sigma^+_1 i u  \\ \times \sigma^+_2 \sigma^-_2 \sigma^-_3  + i u \sigma^+_0 \sigma^+_1 \sigma^+_1 \sigma^-_2 \sigma^-_3 + i u \sigma^+_0 \sigma^+_1 i u \sigma^+_1 \sigma^-_1 i u \sigma^+_2 \sigma^-_2 \sigma^-_3 + \sigma^+_0 i u \sigma^+_1 \sigma^-_1 \sigma^+_1 \sigma^-_2   + \sigma^+_0 i u \sigma^+_1 \sigma^-_1 i u \sigma^+_1 \sigma^-_1    \\ \times i u \sigma^+_1 \sigma^-_1 i u \sigma^+_2 \sigma^-_2 \sigma^-_3           \text{, } \\ \\ \underline{\mathcal{I}^{\prime\prime}_2} \equiv      i u \sigma^+_0 \sigma^-_0 i u \sigma^+_1 \sigma^-_1    i u \sigma^+_1 \sigma^-_1        i u \sigma^+_2 \sigma^-_2 \sigma^+_3 + \sigma^-_0 \sigma^+_1 i u \sigma^+_1 \sigma^-_1 i u \sigma^+_2 \sigma^-_2 \sigma^+_3  + i u \sigma^+_0 \sigma^-_0 i u \sigma^+_1 \sigma^-_1 \sigma^+_2 \sigma^+_3  
    + \sigma^-_0 \sigma^+_1 \sigma^-_1 \sigma^+_2 \sigma^-_3 \\ + i u \sigma^+_0 \sigma^-_0 i u \sigma^+_1 \sigma^-_1 i u \sigma^+_1 \sigma^-_1 \sigma^-_2 i u^{-1} \sigma^+_3 \sigma^-_3  + i u \sigma^+_0 \sigma^-_0 i u \sigma^+_1 \sigma^-_1 \sigma^-_1 i u \sigma^+_2 \sigma^-_2 i u^{-1} \sigma^+_3 \sigma^-_3   + \sigma^-_0 \sigma^+_1 i u \sigma^+_1 \sigma^-_1  \sigma^-_2 i u^{-1} \sigma^+_3 \sigma^-_3 \\ + \sigma^-_0 \sigma^+_1 \sigma^-_1 i u \sigma^+_2 \sigma^-_2 i u^{-1} \sigma^+_3 \sigma^-_3 + i u \sigma^+_0 \sigma^-_0 \sigma^-_1 \sigma^+_1 \sigma^-_2 i u^{-1} \sigma^+_3 \sigma^-_3    +   i u \sigma^+_0 \sigma^-_0 \sigma^-_1 i u \sigma^+_1 \sigma^-_1 i u \sigma^+_2 \sigma^-_2 i u^{-1} \sigma^+_3 \sigma^-_3 \\   + \sigma^-_0 i u \sigma^+_1 \sigma^-_1 \sigma^+_1 \sigma^-_2 i u^{-1} \sigma^+_3 \sigma^-_3    + \sigma^-_0 i u \sigma^+_1 \sigma^-_1 i u \sigma^+_1 \sigma^-_1 i u \sigma^+_2 \sigma^-_2 i u^{-1} \sigma^+_3 \sigma^-_3 \text{, } \\ \\ \underline{\mathcal{I}^{\prime\prime}_3} \equiv    i u \sigma^+_0 \sigma^-_0 i u \sigma^+_1 \sigma^-_1  i u \sigma^+_1 \sigma^-_1 i u \sigma^+_2 \sigma^-_2 \sigma^-_3 + \sigma^-_0 \sigma^+_1 i u \sigma^+_1 \sigma^-_1 i u \sigma^+_2 \sigma^-_2 \sigma^-_3 +    i u \sigma^+_0 \sigma^-_0  i u \sigma^+_1 \sigma^-_1 \sigma^-_1 \sigma^+_2 \sigma^-_3  \\ +  \sigma^-_0 \sigma^+_1 \sigma^-_1 \sigma^+_2 \sigma^-_3 + i u \sigma^+_0 \sigma^-_0 i u \sigma^+_1 \sigma^-_1 i u \sigma^+_1 \sigma^-_1 \sigma^-_2 i u^{-1} \sigma^+_3 \sigma^-_3   + i u \sigma^+_0 \sigma^-_0 i u \sigma^+_1 \sigma^-_1 \sigma^-_1 i u \sigma^+_2 \sigma^-_2 i u^{-1} \sigma^+_3 \sigma^-_3 \\ + \sigma^-_0 \sigma^+_1 i u \sigma^+_1 \sigma^-_1 \sigma^-_2 i u^{-1}  \sigma^+_3 \sigma^-_3 + \sigma^-_0 \sigma^-_1 \sigma^-_1 i u \sigma^+_2 \sigma^-_2 i u^{-1} \sigma^+_3 \sigma^-_3 +  i u \sigma^+_0 \sigma^-_0 \sigma^-_1 \sigma^+_1 \sigma^-_2 i u^{-1} \sigma^+_3 \sigma^-_3        + i u \sigma^+_0 \sigma^-_0 \sigma^-_1 \\ \times       i u \sigma^+_1 \sigma^-_1 i u \sigma^+_2 \sigma^-_2 i u^{-1} \sigma^+_3 \sigma^-_3 + \sigma^-_0 i u \sigma^+_1 \sigma^-_1 \sigma^+_1 \sigma^-_2 i u \sigma^+_3 \sigma^-_3 +   \sigma^-_0 i u \sigma^+_1 \sigma^-_1 i u \sigma^+_1 \sigma^-_1 i u \sigma^+_2 \sigma^-_2 i u \sigma^+_3 \sigma^-_3          \text{, } \\ \\ \underline{\mathcal{I}^{\prime\prime}_4} \equiv       i u \sigma^+_0 \sigma^-_0 i u \sigma^+_1 \sigma^-_1 i u \sigma^+_1 \sigma^-_1 \sigma^+_2 \sigma^-_3 + \sigma^-_0 \sigma^+_1 i u \sigma^+_1 \sigma^-_1 \sigma^+_2 \sigma^-_3 - i u \sigma^+_0 \sigma^-_0 i u \sigma^+_1 \sigma^-_1 \sigma^+_1 i u \sigma^+_2 \sigma^-_2 \sigma^-_3 \\ +        \sigma^-_0 \sigma^+_1 \sigma^+_1 i u \sigma^+_2 \sigma^-_2 \sigma^-_3 + i u \sigma^+_0 \sigma^+_1 \sigma^+_1 \sigma^-_2 \sigma^-_3 + i u \sigma^+_0 \sigma^+_1 i u \sigma^+_1 \sigma^-_1 i u \sigma^+_2 \sigma^-_2 \sigma^-_3 + \sigma^+_0 i u \sigma^+_1 \sigma^-_1 \sigma^+_1 \sigma^-_2 \sigma^-_3 \\  + \sigma^+_0 i u \sigma^+_1 \sigma^-_1 i u \sigma^+_1 \sigma^-_1 i u \sigma^+_2 \sigma^-_2 \sigma^-_3 +   i u \sigma^+_0 \sigma^-_0 i u \sigma^+_1 \sigma^-_1 i u \sigma^+_1 \sigma^-_1 \sigma^-_2 i u^{-1}  \sigma^+_3 \sigma^-_3 + i u \sigma^+_0 \sigma^-_0 \\ \times i u \sigma^+_1 \sigma^-_1 \sigma^-_1 i u \sigma^+_2 \sigma^-_2 i u^{-1} \sigma^+_3 \sigma^-_3 + \sigma^-_0 \sigma^+_1 i u \sigma^+_1 \sigma^-_1 \sigma^-_2 i u^{-1} \sigma^+_3 \sigma^-_3 + \sigma^-_0 \sigma^+_1 \sigma^-_1 i u \sigma^+_2 \sigma^-_2 i u^{-1} \sigma^+_3 \sigma^-_3 \\     + i  u \sigma^+_0 \sigma^-_0 \sigma^-_1 \sigma^+_1 \sigma^-_2 i u^{-1 } \sigma^+_3 \sigma^-_3 + i u \sigma^+_0  \sigma^-_0 \sigma^-_1 i u \sigma^+_1 \sigma^-_1 i u \sigma^+_2 \sigma^-_2 i u^{-1} \sigma^+_3 \sigma^-_3 +    \sigma^-_0 i u \sigma^+_1 \sigma^-_1 \sigma^+_1 \\ \times \sigma^+_2 i u^{-1} \sigma^+_3 \sigma^-_3 + \sigma^-_0 i u \sigma^+_1 \sigma^-_1 i u \sigma^+_1 \sigma^-_1 i u \sigma^+_2 \sigma^-_2 i u^{-1} \sigma^+_3  \sigma^-_3 \text{. }       
\end{align*}

\noindent The computations for the base case of the induction produced above are left as exercises for the reader. With the set of expressions for the product representation of L-operators for the base case, performing the induction entails that the terms of the product representation would take the form,

\[
\begin{bmatrix}
\underline{\mathcal{I}^{\prime\prime\prime}_1} &  \underline{\mathcal{I}^{\prime\prime\prime}_2} \\ \underline{\mathcal{I}^{\prime\prime\prime}_3} & \underline{\mathcal{I}^{\prime\prime\prime}_4} 
\end{bmatrix} = \begin{bmatrix}
\underline{\mathcal{I}^{\prime}_1} &  \underline{\mathcal{I}^{\prime}_2} \\ \underline{\mathcal{I}^{\prime}_3} & \underline{\mathcal{I}^{\prime}_4} 
\end{bmatrix} \begin{bmatrix}
 i u \sigma^+_3 \sigma^-_3 & \sigma^-_3  \\ \sigma^+_3 &
 i u^{-1} \sigma^+_3 \sigma^-_3 \end{bmatrix}  \begin{bmatrix}
 i u \sigma^+_4 \sigma^-_4 & \sigma^-_4  \\ \sigma^+_4 &
 i u^{-1} \sigma^+_3 \sigma^-_4 \end{bmatrix}  \text{. }
\]

\noindent In this case, the entries of the product representation for the 4-vertex model take the form,

\begin{align*}
      \underline{\mathcal{I}^{\prime\prime\prime}_1} \equiv    i u \sigma^+_0 \sigma^-_0 i u \sigma^+_1 \sigma^-_1 i u \sigma^+_1 \sigma^-_1 i u \sigma^+_2 \sigma^-_2     i u \sigma^+_3 \sigma^-_3 i u \sigma^+_4 \sigma^-_4 + \sigma^-_0 \sigma^+_1   i u \sigma^+_1 \sigma^-_1 i u \sigma^+_2 \sigma^-_2 i u \sigma^+_3 i u \sigma^+_4 \sigma^-_4 + i u \sigma^+_0 \\ \sigma^-_0 i u \sigma^+_1 \sigma^-_1 \sigma^-_1 \sigma^+_2 i u \sigma^+_3 \sigma^-_3 i u \sigma^+_4 \sigma^-_4 + \sigma^-_0 \sigma^+_1 \sigma^-_1 \sigma^+_2 i u \sigma^+_3 \sigma^-_3 i u \sigma^+_4 \sigma^-_4    + i u \sigma^+_0 \sigma^-_0 i u \sigma^+_1 \sigma^-_1 i u \sigma^+_1 \sigma^-_1 \sigma^-_2 \sigma^+_2 \sigma^-_3 i u \sigma^+_4 \sigma^-_4 \\ + \sigma^-_0 \sigma^+_1    i u \sigma^+_1 \sigma^+_1 \sigma^-_3 i u \sigma^+_4 \sigma^-_4 + \sigma^-_0 \sigma^+_1 \sigma^+_1 i u \sigma^+_2 \sigma^-_2 \sigma^-_3 i u \sigma^+_4 \sigma^-_4   + i u \sigma^+_0 \sigma^+_1 i u \sigma^+_1 \sigma^-_1 i u \sigma^+_2 \sigma^-_2 \sigma^-_3 i u \sigma^+_4 \sigma^-_4   \end{align*}

      \begin{align*}
      + \sigma^+_0 i u \sigma^+_1 \sigma^-_1 i u \sigma^+_1 \sigma^-_1 i u \sigma^+_1 \sigma^-_1 i u \sigma^+_2 \sigma^-_2 \sigma^-_3 i u \sigma^+_4 \sigma^-_4  + i u \sigma^+_0 \sigma^-_0 i u \sigma^+_1 \sigma^-_1 i u \sigma^+_1 \sigma^-_1 i u \sigma^+_2 \sigma^-_2 \sigma^+_3 \sigma^+_4 \\ + \sigma^-_0 \sigma^+_1     i u \sigma^+_1 \sigma^-_1 i u \sigma^+_2 \sigma^-_2 \sigma^+_3 \sigma^+_4 +          i u\sigma^+_0 \sigma^-_0 i u \sigma^+_1 \sigma^-_1 \sigma^+_2 \sigma^+_3 \sigma^+_4           + \sigma^-_0             \sigma^+_1 \sigma^-_1 \sigma^+_2 \sigma^-_3 \sigma^+_4 \\ +         i u \sigma^+_0 \sigma^-_0 i u \sigma^+_1 i u \sigma^+_1 \sigma^-_1 \sigma^-_2 i u^{-1}  \sigma^+_3 \sigma^-_3 \sigma^+_4  +     i u \sigma^+_0 \sigma^-_0 i u \sigma^+_1 \sigma^-_1 \sigma^-_1 i u \sigma^+_2 \sigma^-_2 i u^{-1} \sigma^+_3 \sigma^-_3 \sigma^+_4 \\  +      \sigma^-_0 \sigma^+_1         i u \sigma^+_1 \sigma^-_1 \sigma^+_4 + \sigma^-_0 \sigma^+_1 \sigma^-_1 i u \sigma^+_2 \sigma^-_2  i u^{-1} \sigma^+_3 \sigma^-_3 \sigma^+_4 + i u   \sigma^+_0 \sigma^-_0       \sigma^-_1 \sigma^+_1 \sigma^-_2 i u^{-1} \sigma^+_3 \sigma^-_3 \sigma^+_4    \\      + i u \sigma^+_0 \sigma^-_0   \sigma^-_1 i u \sigma^+_1 \sigma^-_1 i u \sigma^+_2 \sigma^-_2 i u^{-1} \sigma^+_3 \sigma^-_3 \sigma^+_4      + \sigma^-_0 i u \sigma^+_1 \sigma^-_1 \sigma^+_1 \sigma^-_2 i u^{-1} \sigma^+_3 \sigma^-_3 \sigma^+_4 \\ + \sigma^-_0 i u \sigma^+_1 \sigma^-_1 i u \sigma^+_1 \sigma^-_1 i u \sigma^+_2      i u^{-1} \sigma^+_3 \sigma^-_3 \sigma^+_4                                                                        \text{, } \\ \\ \underline{\mathcal{I}^{\prime\prime\prime}_2} \equiv     i u \sigma^+_0 \sigma^-_0 i u \sigma^+_1 \sigma^-_1 i u \sigma^+_1 \sigma^-_1 i u \sigma^+_2 \sigma^-_2 i u \sigma^+_3 \sigma^-_3 \sigma^-_4 + \sigma^-_0 \sigma^+_1 i u \sigma^+_1 \sigma^-_1 i u \sigma^+_2 \sigma^-_2 i u \sigma^+_3 \sigma^-_3 \sigma^-_4   + i u \sigma^+_0 \sigma^-_0 i u \sigma^+_1 \sigma^-_1 \sigma^-_1 \sigma^+_2 \\ \times  i u \sigma^+_3 \sigma^-_3 \sigma^-_4     + i u \sigma^+_0 \sigma^-_0 i u \sigma^+_1 \sigma^-_1 i u \sigma^+_1 \sigma^-_1 \sigma^-_2 \sigma^+_2 \sigma^-_3 \sigma^-_4 + \sigma^-_0 \sigma^+_1 i u \sigma^+_1 \sigma^+_1   i u \sigma^+_1 \sigma^+_1 \sigma^-_3 \sigma^-_4 \\ + \sigma^-_0 \sigma^+_1 \sigma^+_1 i u \sigma^+_2 \sigma^-_2 \sigma^-_3 \sigma^-_4 + i u    \sigma^+_0 \sigma^+_1 \sigma^+_1 \sigma^-_2 \sigma^-_3 \sigma^-_4      + i \sigma^+_0 \sigma^+_1 i u \sigma^+_1 \sigma^-_1 i u \sigma^+_2 \sigma^-_2 \sigma^-_3 \sigma^-_4  \\  + \sigma^+_0 i u \sigma^+_1 \sigma^-_1 \sigma^+_1 \sigma^-_2 \sigma^-_3 \sigma^-_4       + \sigma^+_0 i u \sigma^+_1 \sigma^-_1 i u \sigma^+_1 \sigma^-_1 i u \sigma^+_2 \sigma^-_2 \sigma^-_3 \sigma^-_4     +                          i u \sigma^+_0 \sigma^-_0 i u \sigma^+_1 \sigma^-_1 i u \sigma^+_1 \sigma^-_1 i u \sigma^+_2 \sigma^-_2 \sigma^+_3 i u^{-1} 
 \\  \times   \sigma^+_4 \sigma^-_4 + i u \sigma^+_0 \sigma^-_0 i u \sigma^+_1 \sigma^-_1 i u \sigma^+_1 i u \sigma^+_1 \sigma^-_1 \sigma^-_2 i u^{-1} \sigma^+_3 \sigma^-_3 i u^{-1} \sigma^+_4 \sigma^-_4     + i u \sigma^+_0 \sigma^-_0 i u \sigma^+_1 \sigma^-_1 \sigma^-_1 \\ \times  i u \sigma^+_2 \sigma^-_2 i u^{-1} \sigma^+_3 \sigma^-_3 i u^{-1} \sigma^+_4 \sigma^-_4   + \sigma^-_0 \sigma^+_1 \sigma^-_1 i u \sigma^+_2 \sigma^-_2 i u^{-1} \sigma^+_3 \sigma^-_3 i u^{-1} \sigma^+_4 \sigma^-_4 + i u  \\ \times \sigma^+_0 \sigma^-_0 \sigma^-_1 \sigma^+_1 \sigma^-_2 i u^{-1} \sigma^+_3 \sigma^-_3 i u^{-1} \sigma^+_4 \sigma^-_4 + i u    \sigma^+_0 \sigma^-_0 \sigma^-_1 i u \sigma^+_1 \sigma^-_1 i u \sigma^+_2 \sigma^-_2 i u^{-1}   \sigma^+_3 \sigma^-_3 i u^{-1} \sigma^+_4 \sigma^-_4  \\ + \sigma^-_0           i u \sigma^+_1 \sigma^-_1 \sigma^-_2 i u^{-1} \sigma^+_3 \sigma^-_3 i u^{-1} \sigma^+_4 \sigma^-_4                + \sigma^-_0 i u \sigma^+_1 \sigma^-_1 i u \sigma^+_1 \sigma^-_1 i u \sigma^+_2 \sigma^-_2  i u^{-1} \sigma^+_3 \sigma^-_3 i u^{-1} \sigma^+_4 \sigma^-_4        \text{, }
\\ \\  
\underline{\mathcal{I}^{\prime\prime\prime}_3} \equiv       i u \sigma^+_0 \sigma^-_0 i  u \sigma^+_1 \sigma^-_1 i u \sigma^+_1 \sigma^-_1           i u \sigma^+_2 \sigma^-_2 i u \sigma^+_3 \sigma^-_3 \sigma^+_4 + \sigma^-_0 \sigma^+_1 i u \sigma^+_1 \sigma^-_1 i u \sigma^+_2 \sigma^-_2 i u \sigma^+_3 \sigma^-_3 \sigma^+_4               + i u  \sigma^+_0 \sigma^-_0 i u \sigma^+_1 \\ \times \sigma^-_1 \sigma^-_1 \sigma^+_2          i u \sigma^+_3 \sigma^-_3   \sigma^+_4 + \sigma^-_0 \sigma^+_1      \sigma^-_1 \sigma^+_2 i u \sigma^+_3    \sigma^-_3 \sigma^+_4 + i u \sigma^+_0 \sigma^-_0 i u \sigma^+_1 \sigma^-_1 \\  \times i u \sigma^+_1 \sigma^-_1 \sigma^-_2 \sigma^+_2 \sigma^-_3 \sigma^+_4     + \sigma^-_0 \sigma^+_1 i u \sigma^+_1 \sigma^-_1 i u \sigma^+_1 \sigma^-_1 \sigma^-_3 \sigma^+_4 + \sigma^-_0 \sigma^+_1 \sigma^+_1 i u \sigma^+_2 \sigma^-_2    \sigma^-_3 \sigma^+_4 \\ + i u \sigma^+_0 \sigma^+_1 \sigma^+_1 \sigma^-_2 \sigma^-_3 \sigma^+_4                          + \sigma^+_0 i u \sigma^+_1 \sigma^-_1 \sigma^-_2 \sigma^-_3 \sigma^+_4    +      \sigma^+_0 i u \sigma^+_1 \sigma^-_1 i u \sigma^+_1 \sigma^-_1 i u \sigma^+_2 \sigma^-_2 \sigma^-_3 \sigma^+_4 \\ + i u \sigma^+_0 \sigma^-_0 i u \sigma^+_1 \sigma^-_1 i u \sigma^+_1 \sigma^-_1 i u \sigma^+_2 \sigma^-_2 \sigma^-_3      i u^{-1} \sigma^+_4 \sigma^-_4 + \sigma^-_0 \sigma^+_1 \sigma^-_1 \sigma^+_2 \sigma^-_3 i u^{-1} \sigma^+_4 \sigma^-_4 \\  + i u \sigma^+_0 \sigma^-_0 i u \sigma^+_1 \sigma^-_1   i u \sigma^+_1 \sigma^-_1 \sigma^-_2 i u \sigma^+_3 \sigma^-_3 i u^{-1} \sigma^+_4 \sigma^-_4  + \sigma^-_0 \sigma^+_1 \sigma^-_1 \sigma^+_2 \sigma^-_3 i u^{-1} \sigma^+_4 \sigma^-_4    \\ + i u \sigma^+_0 \sigma^-_0 i u \sigma^+_1 \sigma^-_1 \sigma^-_1 i u \sigma^+_2 \sigma^-_2 i u^{-1} \sigma^+_3 \sigma^-_3 i u^{-1} \sigma^+_4 \sigma^-_4 + \sigma^-_0 \sigma^+_1 i u \sigma^+_1 \sigma^-_1 \sigma^-_2 i u^{-1} \sigma^+_3 \sigma^-_3 i u^{-1} \sigma^+_4 \sigma^-_4 \\   + i u \sigma^+_0 \sigma^-_0 \sigma^-_1 \sigma^+_1 \sigma^-_2 i u^{-1} \sigma^+_3 \sigma^-_3 i u^{-1} \sigma^+_4 \sigma^-_4       +  i u \sigma^+_0 \sigma^-_0 i u \sigma^+_1 \sigma^-_1 i u \sigma^+_2 \sigma^-_2 i u^{-1} \sigma^+_3 \sigma^-_3 i u^{-1} \sigma^+_4 \sigma^-_4 \\   + \sigma^-_0 \sigma^+_1 \sigma^-_1 i u \sigma^+_2 \sigma^-_2 i u^{-1} \sigma^+_3 \sigma^-_3 i u^{-1} \sigma^+_4 \sigma^-_4         + i u \sigma^+_0 \sigma^-_0 \sigma^-_1 \sigma^+_1 \sigma^-_2 i u^{-1} \sigma^+_3 \sigma^-_3 i u^{-1} \sigma^+_4 \sigma^-_4    \\     + i u \sigma^+_0 \sigma^-_0 \sigma^-_1 i u \sigma^+_1 \sigma^-_1 i u \sigma^+_2 \sigma^-_2 i u^{-1} \sigma^+_3 \sigma^-_3 i u^{-1} \sigma^+_4 \sigma^-_4  \\   + \sigma^-_0 i u \sigma^+_1 \sigma^-_1 \sigma^+_1 \sigma^-_2 i u \sigma^+_3 \sigma^-_3 i u^{-1} \sigma^+_4 \sigma^-_4 + \sigma^-_0 i u \sigma^+_1 \sigma^-_1 i u \sigma^+_1 \sigma^-_1 i u \sigma^+_2 \sigma^-_2 i u \sigma^+_3 \sigma^-_3 \\ \times i u^{-1} \sigma^+_4 \sigma^-_4                                                \text{, } \\ \\ \underline{\mathcal{I}^{\prime\prime\prime}_4} \equiv  i u \sigma^+_0 \sigma^-_0 i u \sigma^+_1 \sigma^-_1 i u \sigma^+_1 \sigma^-_1 i u \sigma^+_2 \sigma^-_2 \sigma^+_3 \sigma^+_4 + \sigma^-_0 \sigma^+_1 i u \sigma^+_1 \sigma^-_1 i u \sigma^+_2 \sigma^-_2      \sigma^+_3 \sigma^+_4      + i u \sigma^+_0 \sigma^-_0 i u \sigma^+_1 \sigma^-_1 \sigma^+_2 \sigma^+_3 \sigma^+_4    \\ +          \sigma^-_0 \sigma^+_1 \sigma^-_1 \sigma^+_2 \sigma^-_3 \sigma^+_4             + i u \sigma^+_0 \sigma^-_0 i u \sigma^+_1 \sigma^-_1 i  u \sigma^+_1 \sigma^-_1 \sigma^-_2 i u^{-1} \sigma^+_3 \sigma^-_3 \sigma^+_4 + i u \sigma^+_0                \sigma^-_0 i u \sigma^+_1 \sigma^-_1 \sigma^-_1 i u \sigma^+_2 \sigma^-_2 i u^{-1} \sigma^+_3 \sigma^-_3 \sigma^+_4 \\  + \sigma^-_0 \sigma^+_1   i u \sigma^+_1 \sigma^-_1 \sigma^-_2 i u^{-1} \sigma^+_3 \sigma^-_3 \sigma^+_4    + \sigma^-_0 \sigma^+_1 \sigma^-_1 i u \sigma^+_2 \sigma^-_2 i u^{-1} \sigma^+_3 \sigma^-_3 \sigma^+_4                  +  i u \sigma^+_0 \sigma^-_0      \sigma^-_1 \sigma^+_1 \sigma^-_2      i u^{-1} \sigma^+_3 \sigma^-_3 \sigma^+_4 \\ + i u \sigma^+_0 \sigma^-_0 \sigma^-_1 i u \sigma^+_1 \sigma^-_1 i u \sigma^+_2 \sigma^-_2 i u \sigma^+_3 \sigma^-_3 \sigma^+_4 + \sigma^-_0 i u \sigma^+_1 \sigma^-_1 \sigma^+_1 \sigma^-_2        i u^{-1} \sigma^+_3 \sigma^-_3 \sigma^+_4    \\ + \sigma^-_0 i u \sigma^+_1 \sigma^-_1 i  u \sigma^+_1 \sigma^-_1 i u \sigma^+_2 \sigma^-_2 i u^{-1} \sigma^+_3 \sigma^-_3 \sigma^+_4     + \sigma^-_0 i  u \sigma^+_1 \sigma^-_1 i u \sigma^+_1 \sigma^-_1 i  u \sigma^+_2 \sigma^-_2 i u^{-1} \sigma^+_3 \sigma^-_3 \sigma^+_4 \\ + i u \sigma^+_0 \sigma^-_0 i  u \sigma^+_1 \sigma^-_1 i u \sigma^+_1 \sigma^-_1 \sigma^+_2 \sigma^-_3 i u^{-1} \sigma^+_4 \sigma^-_4      + \sigma^-_0 \sigma^+_1 i u \sigma^+_1 \sigma^-_1 \sigma^+_2 \sigma^-_3 i u^{-1} \sigma^+_4 \sigma^-_4 \\ - i u \sigma^+_0 \sigma^-_0 i u \sigma^+_1 \sigma^-_1 \sigma^+_1 i u \sigma^+_2 \sigma^-_2 \sigma^-_3        + \sigma^-_0 \sigma^+_1 \sigma^+_1 i  u \sigma^+_2 \sigma^-_2 \sigma^-_3 i u^{-1} \sigma^+_4 \sigma^-_4 + i u \sigma^+_0 \sigma^-_0 \sigma^-_! i u \sigma^+_1 \sigma^-_1 i u \sigma^+_2 \sigma^-_2 \\ \times  i u \sigma^+_3 \sigma^-_3 \sigma^+_4    + \sigma^-_0  iu \sigma^+_1 \sigma^-_1 \sigma^+_1 \sigma^-_2 i u^{-1} \sigma^+_3 \sigma^-_3 \sigma^+_4     + \sigma^-_0 i  u \sigma^+_1 \sigma^-_1 i u \sigma^+_1 \sigma^-_1 i  u \sigma^+_2 \sigma^-_2 i u^{-1} \sigma^+_3 \sigma^-_3 \sigma^+_4  \\    + i u \sigma^+_0 \sigma^-_0 i u \sigma^+_1 \sigma^-_1 i u \sigma^+_1 \sigma^-_1 \sigma^+_2 \sigma^-_3 i u^{-1} \sigma^+_4 \sigma^-_4    + \sigma^-_0 \sigma^+_1 i  u \sigma^+_1 \sigma^-_1 i u \sigma^+_2 \sigma^-_2 i u^{-1} \sigma^+_3 \sigma^-_3 \sigma^+_4 \\ + i  u \sigma^+_0 \sigma^-_0 i u \sigma^+_1 \sigma^-_1 i u \sigma^+_1 \sigma^-_1 \sigma^+_2 \sigma^-_4 i u^{-1} \sigma^+_4 \sigma^-_4 + \sigma^-_0 \sigma^+_1         i u \sigma^+_1 \sigma^-_1 \sigma^+_2 \sigma^-_3 i u^{-1} \sigma^+_4 \sigma^-_4 \\ - i  u \sigma^+_0 \sigma^-_0 i u \sigma^+_1 \sigma^-_1  \sigma^+_1 i u \sigma^+_2 \sigma^-_2 \sigma^-_3 i u^{-1} \sigma^+_4 \sigma^-_4 + \sigma^-_0 \sigma^+_1 \sigma^+_1              i u \sigma^+_2 \sigma^-_2 \sigma^-_3 i u^{-1} \sigma^+_4 \sigma^-_4 \\ + i  u \sigma^+_0 \sigma^+_1 \sigma^+_1 \sigma^-_2 \sigma^-_3 i u^{-1} \sigma^+_4 \sigma^-_4      + i u \sigma^+_0 \sigma^+_1 i u \sigma^+_1 \sigma^-_1 i u \sigma^+_2 \sigma^-_2 \sigma^-_3 i u ^{-1} \sigma^+_4 \sigma^-_4 \\ + \sigma^+_0    i u \sigma^+_1 \sigma^-_1 \sigma^+_1 \sigma^-_1 \sigma^+_1 \sigma^-_2  \sigma^-_3 i u^{-1} \sigma^+_4 \sigma^-_4 + \sigma^+_0 i u \sigma^+_1 \sigma^-_1 i u \sigma^+_1 \sigma^-_1      i u \sigma^+_2 \sigma^-_2 \sigma^-_3 i u^{-1} \sigma^+_4 \sigma^-_4 \\  +       i u \sigma^+_0 \sigma^-_0 i u \sigma^+_1 \sigma^-_1 i u \sigma^+_1 \sigma^-_1 \sigma^-_23 i u^{-1} \sigma^+_3 \sigma^-_3 + i u \sigma^+_0 \sigma^-_0 i u \sigma^+_1 \sigma^-_1 \sigma^-_1 i u \sigma^+_2 \sigma^-_2 i u^{-1} \sigma^+_3 \sigma^-_3 i u^{-1} \sigma^+_4 \sigma^-_4 \\ +    \sigma^-_0  \sigma^+_1 i u \sigma^+_1 \sigma^-_1 \sigma^-_2 i u^{-1} \sigma^+_3 \sigma^-_3 i u^{-1} \sigma^+_4 \sigma^-_4    + \sigma^-_0 \sigma^+_1 \sigma^-_1 i u \sigma^+_2 \sigma^-_2 i u^{-1 } \sigma^+_3 \sigma^-_3           i u^{-1} \sigma^+_4 \sigma^-_4 \end{align*}

      \begin{align*}  + i  u \sigma^+_0 \sigma^-_0 \sigma^-_1 \sigma^+_1 \sigma^-_2  i u^{-1} \sigma^+_3 \sigma^-_3         i u^{-1} \sigma^+_4 \sigma^-_4   + i u \sigma^+_0 \sigma^-_0 \sigma^-_1 i u \sigma^+_1 \sigma^-_1    i u \sigma^+_2 i u^{-1} \sigma^+_3 \sigma^-_3 i u^{-1} \sigma^+_4 \sigma^-_4 \\ + \sigma^-_0 i u \sigma^+_1 \sigma^-_1 \sigma^+_2 i u^{-1} \sigma^+_3 \sigma^-_3 i u^{-1 } \sigma^+_4 \sigma^-_4 + \sigma^-_0              i u \sigma^+_1 \sigma^-_1 i u \sigma^+_1 \sigma^-_1 i u \sigma^+_2 \sigma^-_2 i u^{-1} \sigma^+_3 \sigma^-_3 i u^{-1} \sigma^+_4 \sigma^-_4                        \text{. }
\end{align*}

\noindent From each entry above, to obtain the desired entries of the approximate product representation, we group together terms which have prefactors $i u \sigma^+_0 \sigma^-_0$, $\sigma^-_0 \sigma^+_1$, or $\sigma^-_0$. For $\underline{\mathcal{I}^{\prime\prime\prime}_1}$, such a grouping of terms implies, approximately,

\begin{align*}
 i u \sigma^+_0 \sigma^-_0 \bigg[      i u \sigma^+_1 \sigma^-_1 i u \sigma^+_2 \sigma^-_2 i u \sigma^+_3 \sigma^-_3  i u \sigma^+_4 \sigma^-_4 + i u \sigma^+_1 \sigma^-_1 \sigma^-_1   \sigma^+_2 i u \sigma^+_3 \sigma^-_3 \sigma^+_4 \sigma^-_4        + i u \sigma^+_1 \sigma^-_1 \sigma^-_1 \sigma^+_2 i u \sigma^+_3 \sigma^-_3 \sigma^+_4 \sigma^-_4 \\  + i u \sigma^+_1 \sigma^-_1 \sigma^-_2 \sigma^+_2 \sigma^-_3 i u \sigma^+_4 \sigma^-_4 + i u \sigma^+_1 \sigma^-_1 i u \sigma^+_2 \sigma^-_2 \sigma^-_3 i u \sigma^+_4 \sigma^-_4                  + i u \sigma^+_1 \sigma^-_1 i u \sigma^+_2 \sigma^-_2 \sigma^+_3 \sigma^+_4   +       i u \sigma^+_1 \sigma^-_1 \sigma^+_2 \sigma^+_3 \sigma^+_4 \\ 
 + i u \sigma^+_1 \sigma^-_1 \sigma^-_2 i u^{-1} \sigma^+_3 \sigma^-_3 \sigma^+_4     + i u \sigma^+_1 \sigma^-_1 i u \sigma^+_2 \sigma^-_2 i u^{-1} \sigma^+_3 \sigma^-_3 \sigma^+_4 + i u \sigma^+_1 \sigma^-_1 i u \sigma^+_2 \sigma^-_2 i u^{-1} \sigma^+_3 \sigma^-_3 \sigma^+_4  \\ + \sigma^-_1 \sigma^+_1        \sigma^-_2 i u^{-1} \sigma^+_3 \sigma^-_3 \sigma^+_4 + \sigma^-_1 i u \sigma^+_1 \sigma^-_1 i u \sigma^+_2 \sigma^-_2 i u^{-1} \sigma^+_3 \sigma^-_3 \sigma^+_4                      \bigg]   \text{, } \\ \\   \sigma^-_0 \sigma^+_1   \bigg[        i u \sigma^+_1 \sigma^-_1 i u \sigma^+_2 \sigma^-_2 \sigma^+_3 \sigma^+_4 + \sigma^-_1 \sigma^+_2    \sigma^-_3 \sigma^+_4 + i u  \sigma^+_1 \sigma^-_1 i u \sigma^+_2 \sigma^-_2 i u^{-1} \sigma^+_3 \sigma^-_3 \sigma^+_4 + \sigma^-_1 i u \sigma^+_2 \sigma^-_2 i u^{-1} \sigma^+_3 \\  \times                \sigma^-_3 \sigma^+_4      \bigg] \text{, } \end{align*}

 \begin{align*}
 \sigma^-_0         \bigg[           i u \sigma^+_1 \sigma^-_1 \sigma^+_1 \sigma^-_2 i u^{-1} \sigma^+_3 \sigma^-_3 \sigma^+_4 + i u \sigma^+_1 \sigma^-_1 i u \sigma^+_2 \sigma^-_2   i u^{-1} \sigma^+_3 \sigma^-_3 \sigma^+_4  +    \sigma^+_1 i u \sigma^+_1 \sigma^-_1 i u \sigma^+_2 \sigma^-_2 i u \sigma^+_3 \sigma^-_3 i u \sigma^+_4 \sigma^-_4 \\ + i u \sigma^+_1 \sigma^-_1 \sigma^+_2 i u \sigma^+_3 \sigma^-_3 i u \sigma^+_4 \sigma^-_4   + \sigma^+_1 \sigma^-_1 \sigma^+_2 i u \sigma^+_3 \sigma^-_3 i u \sigma^+_4 \sigma^-_4 + \sigma^+_1 i u \sigma^+_1 \sigma^-_1 i u \sigma^+_2 \sigma^-_2 \sigma^-_3 i u \sigma^+_4 \sigma^-_4   \bigg] \text{. }
\end{align*}

\noindent For $\underline{\mathcal{I}^{\prime\prime\prime}_2}$, one has, approximately,

\begin{align*}  
        i u \sigma^+_0 \sigma^-_0      \bigg[       i u \sigma^+_1 \sigma^-_1 i u \sigma^+_2 \sigma^-_2 i u \sigma^+_3 \sigma^-_3  \sigma^-_4   + i u \sigma^+_1 \sigma^-_1 \sigma^+_2 i u \sigma^+_3 \sigma^-_3 \sigma^-_4 + i u \sigma^+_1 \sigma^-_1 \sigma^+_2 i u \sigma^+_3 \sigma^-_3 \sigma^-_4 + i u \sigma^+_1 \sigma^-_1 \sigma^-_2 \sigma^+_2 \sigma^-_3 \sigma^-_4 \\ 
        + \sigma^+_1 \sigma^-_2 \sigma^-_3 \sigma^-_4 + i u \sigma^+_1 \sigma^-_1 i u \sigma^+_2 \sigma^-_2 \sigma^-_3 \sigma^-_4    + i u \sigma^+_1 \sigma^-_1 i u \sigma^+_2 \sigma^-_2 \sigma^+_3 i u^{-1 } \sigma^+_4 \sigma^-_4  +    i u \sigma^+_1 \sigma^-_1 \sigma^-_2 i u^{-1} \sigma^+_3 \sigma^-_3 i u^{-1} \sigma^+_4 \sigma^-_4               \\         + i u \sigma^+_1 \sigma^-_1 i u \sigma^+_2 \sigma^-_2 i u^{-1} \sigma^+_3 \sigma^-_3     i u^{-1} \sigma^+_4 \sigma^-_4    + \sigma^-_1 \sigma^+_1 \sigma^-_2 i u^{-1} \sigma^+_3 \sigma^-_3 i u^{-1} \sigma^+_4 \sigma^-_4             + \sigma^-_1 \sigma^+_1 i u^{-1} \sigma^+_3    i u^{-1} \sigma^+_4 \sigma^-_4  \\     +     \sigma^-_1 i u \sigma^+_1 \sigma^-_1 i u \sigma^+_2 \sigma^-_2 i u^{-1} \sigma^+_3 \sigma^-_3      i u^{-1} \sigma^-_4 \sigma^+_4                            \bigg]   \text{, } \\ \\  \sigma^-_0 \sigma^+_1  \bigg[  i u \sigma^+_1 \sigma^-_1 i u \sigma^+_2 \sigma^-_2 i u \sigma^+_3 \sigma^-_3 \sigma^-_4 + i u \sigma^+_1 \sigma^-_1 i u \sigma^+_2 \sigma^-_2 \sigma^-_3 \sigma^-_4 + \sigma^+_1 i u \sigma^+_2 \sigma^-_2 \sigma^-_3 \sigma^-_4                         \bigg]    \text{, } \\ \\      \sigma^+_0   \bigg[                         i u \sigma^+_1 \sigma^-_1 \sigma^-_2 \sigma^-_3 \sigma^-_4 + i u \sigma^+_1 \sigma^-_1 i u \sigma^+_2 \sigma^-_2 \sigma^-_3 \sigma^-_4       \bigg] \text{, } \\ \\     \sigma^-_0     \bigg[              \sigma^+_1 \sigma^-_1 i u \sigma^+_2 \sigma^-_2 i u^{-1} \sigma^+_3 \sigma^-_3 i u^{-1} \sigma^+_4 \sigma^-_4  + i u \sigma^+_1 \sigma^-_1 \sigma^-_2 i u^{-1} \sigma^+_3 \sigma^-_3 i u^{-1} \sigma^+_4 \sigma^-_4  + i u \sigma^+_1 \sigma^-_1 i u \sigma^+_2 \sigma^-_2  \\ \times i u^{-1} \sigma^+_3 \sigma^-_3 i u^{-1} \sigma^+_4 \sigma^-_4     \bigg] \text{. }  
\end{align*}

\noindent For $\underline{\mathcal{I}^{\prime\prime\prime}_3}$, one has, approximately,

\begin{align*}
    i u \sigma^+_0 \sigma^-_0  \bigg[            i u \sigma^+_1 \sigma^-_1 i u \sigma^+_2 \sigma^-_2 i u \sigma^+_3 \sigma^-_3 \sigma^+_4     + i u \sigma^+_1 \sigma^-_1 \sigma^+_2 i u \sigma^+_3 \sigma^-_3 \sigma^+_4   + i u \sigma^+_1 \sigma^-_1 \sigma^-_2      \sigma^+_2 \sigma^-_3 \sigma^+_4   + \sigma^+_1 \sigma^-_2 \sigma^-_3 \sigma^+_4      +            i u \sigma^+_1 \sigma^-_1 \end{align*}

      \begin{align*}  
    \times \sigma^-_2 \sigma^-_3        i u^{-1} \sigma^+_4 \sigma^-_4           + i u \sigma^+_1 \sigma^-_1 \sigma^-_1 \sigma^+_2 \sigma^-_3 i u^{-1} \sigma^+_4 \sigma^-_4        + i u \sigma^+_1 \sigma^-_1 \sigma^-_1 \sigma^+_2 i u^{-1} \sigma^+_4 \sigma^-_3 + i u \sigma^+_1 \sigma^-_1 \sigma^-_2 i u \sigma^+_3 \sigma^-_3 i u^{-1} \sigma^+_4 \sigma^-_4  \\  + i u \sigma^+_1 \sigma^-_1 i u \sigma^+_2 \sigma^-_2 i u^{-1} \sigma^+_3 \sigma^-_3 i u^{-1} \sigma^+_4 \sigma^-_4  +     i u \sigma^+_1 \sigma^-_1 i u \sigma^+_2 \sigma^-_2 i u^{-1} \sigma^+_3 \sigma^-_3 i u^{-1} \sigma^+_4 \sigma^-_4 + \sigma^-_1 \sigma^+_1 \sigma^-_2 i u^{-1} \sigma^+_3 \sigma^-_3  i u^{-1} \sigma^+_4 \sigma^-_4 \\ +       \sigma^-_0 i u \sigma^+_1 \sigma^-_1 \sigma^-_1 i u \sigma^+_2 \sigma^-_2 i u^{-1} \sigma^+_3 \sigma^-_3 i u^{-1} \sigma^+_4 \sigma^-_4 + \sigma^-_0 \sigma^-_1 \sigma^+_1 \sigma^-_2 i u^{-1} \sigma^+_3 \sigma^-_3 i u^{-1} \sigma^+_4 \sigma^-_4   + \sigma^-_1 i u \sigma^+_1 \sigma^-_1 i u \sigma^+_2 \sigma^-_2 \\ \times  i u^{-1} \sigma^+_3 \sigma^-_3 i u^{-1} \sigma^+_4 \sigma^-_4  + \sigma^-_1 \sigma^-_2 i u^{-1} \sigma^+_3 \sigma^-_3 i u^{-1} \sigma^+_4 \sigma^-_4  + i u \sigma^+_1 \sigma^-_1 i u \sigma^+_2 \sigma^-_2 i u^{-1} \sigma^+_3 \sigma^-_3 i u^{-1} \sigma^+_4 \sigma^-_4 \\  + i u \sigma^+_1 \sigma^-_1 i u \sigma^+_2 \sigma^-_2 i u^{-1} \sigma^+_3 \sigma^-_3 i u^{-1} \sigma^+_4 \sigma^-_4           \bigg]  \text{, }\\ \\ 
    \sigma^-_0 \sigma^+_1      \bigg[     i u \sigma^+_1 \sigma^-_1 i u \sigma^+_2 \sigma^-_2 i u \sigma^+_3 \sigma^-_3 \sigma^+_4 + \sigma^-_1 \sigma^+_2 i u \sigma^+_3 \sigma^-_3 \sigma^+_4 +         i u \sigma^+_1 \sigma^-_1 i u \sigma^+_2 \sigma^-_2 \sigma^-_3 \sigma^+_4   +         i u \sigma^+_2 \sigma^-_2 \sigma^-_3 \sigma^+_4 \\  + \sigma^-_1 \sigma^+_2 \sigma^-_3 i u^{-1} \sigma^+_4 \sigma^-_4  + \sigma^-_1 \sigma^+_2 \sigma^-_3           i u^{-1} \sigma^+_4 \sigma^-_4  + i u \sigma^+_1 \sigma^-_1 \sigma^-_2 i u^{-1} \sigma^+_3 \sigma^-_3 \sigma^+_4 \sigma^-_4                 \bigg]  \text{, } \end{align*}

    \begin{align*}   \sigma^+_0          \bigg[         i u \sigma^+_1 \sigma^-_1 \sigma^-_2 \sigma^-_3 \sigma^+_4 + i u \sigma^+_1 \sigma^-_1 i u \sigma^+_2 \sigma^-_2 \sigma^-_3 \sigma^+_4      \bigg]  \text{, } \\ \\    \sigma^-_0          \bigg[   i u \sigma^+_1 \sigma^-_1 \sigma^-_2 i u \sigma^+_3 \sigma^-_3 i u^{-1} \sigma^+_4 \sigma^-_4   + i u \sigma^+_1 \sigma^-_1 i u \sigma^+_2 \sigma^-_2 i u \sigma^+_3 \sigma^-_3 i u^{-1} \sigma^+_4 \sigma^-_4                     \bigg]   \text{. } 
\end{align*}

\noindent For $\underline{\mathcal{I}^{\prime\prime\prime}_4}$, one has, approximately,

\begin{align*}
i u \sigma^+_0 \sigma^-_0   \bigg[     i u \sigma^+_1 \sigma^-_1 i u \sigma^+_2 \sigma^-_2 \sigma^+_3 \sigma^+_4 + i u \sigma^+_1 \sigma^-_1 \sigma^+_2 \sigma^+_3 \sigma^+_4     + i u \sigma^+_1 \sigma^-_1 \sigma^-_2 i u^{-1} \sigma^+_3 \sigma^-_3 \sigma^+_4 + i u      \sigma^+_1 \sigma^-_1 i u \sigma^+_2 \sigma^-_2  i u^{-1}      \sigma^+_3 \sigma^-_3 \sigma^+_4 \\  + \sigma^-_1     \sigma^+_1 \sigma^-_2 i u^{-1} \sigma^+_3 \sigma^-_3 \sigma^+_4 + i u \sigma^+_1 \sigma^-_1 i u \sigma^+_2 \sigma^-_2 i u \sigma^+_3 \sigma^-_3 \sigma^+_4            \bigg]  \text{, } \\ \\    \sigma^-_0 \sigma^+_1     \bigg[       i u \sigma^+_1 \sigma^-_1 i u \sigma^+_2 \sigma^-_2 \sigma^+_3 \sigma^+_4 + \sigma^-_1 \sigma^+_2 \sigma^-_3 \sigma^+_4                  + i u \sigma^+_1 \sigma^-_1 \sigma^-_2 i u^{-1} \sigma^+_3 \sigma^-_3 \sigma^+_4 + \sigma^-_1 i u \sigma^+_2 \sigma^-_2 i u^{-1} \sigma^+_3 \sigma^-_3 \sigma^+_4            \bigg]  \text{, } \end{align*}

\begin{align*}   \sigma^-_0 \bigg[          i u \sigma^+_1 \sigma^-_1 \sigma^-_2 i u^{-1} \sigma^+_3 \sigma^-_3        \sigma^+_4 + i u \sigma^+_1 \sigma^-_1 i u \sigma^+_2 \sigma^-_2 i u^{-1} \sigma^+_3 \sigma^-_3 \sigma^+_4 +   i u \sigma^+_1 \sigma^-_1 i u \sigma^+_2 \sigma^-_2 i u^{-1} \sigma^+_3 \sigma^-_3 \sigma^+_4  \\ + \sigma^+_1 i u \sigma^+_1 \sigma^-_1 \sigma^+_2 \sigma^-_3 i u^{-1} \sigma^+_4 \sigma^-_4 + \sigma^+_1 i u \sigma^+_2           \sigma^-_2 \sigma^-_3 i u^{-1} \sigma^+_4 \sigma^-_4 + i u \sigma^+_1 \sigma^-_! \sigma^-_2 i u^{-1} \sigma^+_3 \sigma^-_3 \sigma^+_4  \\
      + i u \sigma^+_1 \sigma^-_1 i u \sigma^+_2 \sigma^-_2 i u^{-1} \sigma^+_3 \sigma^-_3 \sigma^+_4 + \sigma^+_1 i u \sigma^+_1 \sigma^-_1 i u \sigma^+_2 \sigma^-_2 i u^{-1} \sigma^+_3 \sigma^-_3 \sigma^+_4        +       \sigma^+_1 i u \sigma^+_1 \sigma^-_1 \sigma^+_2 \sigma^-_3 i u^{-1} \sigma^+_4 \sigma^-_4 \\ + \sigma^+_1 i u \sigma^+_2 \sigma^-_2 \sigma^-_3 i u^{-1} \sigma^+_4 \sigma^-_4                    \bigg]  \text{, }  \\ \\  i u \sigma^+_0  \bigg[               \sigma^-_0 i u \sigma^+_1 \sigma^-_1 \sigma^+_2 \sigma^-_3 i u^{-1} \sigma^+_4 \sigma^-_4 - \sigma^-_0 i u \sigma^+_1 \sigma^-_1 i u \sigma^+_2 \sigma^-_2 \sigma^-_3 +     \sigma^-_0 \sigma^-_1 i u \sigma^+_1 \sigma^-_1 i u \sigma^+_2 \sigma^-_2 i u \sigma^+_2 \sigma^-_2 i u \sigma^+_3 \sigma^-_3       \sigma^+_4                  \\   +   \sigma^-_0 i u \sigma^+_1 \sigma^-_1 \sigma^+_2 \sigma^-_3 i u^{-1} \sigma^+_4 \sigma^-_4     + \sigma^-_0 i u \sigma^+_1 \sigma^-_1 \sigma^+_2 \sigma^-_3 i u^{-1} \sigma^+_4 \sigma^-_4     - \sigma^-_0 i u \sigma^+_1 \sigma^-_1       i u \sigma^+_2 \sigma^-_2 \sigma^-_3  i u^{-1} \sigma^+_4 \sigma^-_4 \\ + \sigma^+_1 \sigma^-_2 \sigma^-_3        i u^{-1} \sigma^+_4 \sigma^-_4 + \sigma^+_1 i u \sigma^+_1 \sigma^-_1 i u \sigma^+_2 \sigma^-_2 \sigma^-_3 i u^{-1} \sigma^+_4 \sigma^-_4 +  \sigma^-_0 i u \sigma^+_1 \sigma^-_1 \sigma^-_2 i u^{-1} \sigma^+_3 \sigma^-_3 \\ + \sigma^-_0 i u \sigma^+_1 \sigma^-_1 i u \sigma^+_2 \sigma^-_2 i u^{-1} \sigma^+_3 \sigma^-_3 i u^{-1} \sigma^+_4 \sigma^-_4 + \sigma^-_0 \sigma^-_1 \sigma^+_1 \sigma^-_2 i u^{-1} \sigma^+_3 \sigma^-_3 \sigma^+_4 \sigma^-_4  \\ + \sigma^-_0 \sigma^-_1 i u \sigma^+_1 \sigma^-_1 i u \sigma^+_2 \sigma^-_2 i u^{-1} \sigma^+_3 \sigma^-_3 i u^{-1} \sigma^+_4 \sigma^-_4          \bigg] \text{, } \\ \\    \sigma^-_0     \bigg[       i u \sigma^+_1 \sigma^-_1 \sigma^+_2 i u^{-1} \sigma^+_3 \sigma^-_3 i u^{-1} \sigma^+_4 \sigma^-_4 + i u \sigma^+_1 \sigma^-_1 i u \sigma^+_2 \sigma^-_2     i u^{-1} \sigma^+_3 \sigma^-_3 i u^{-1} \sigma^+_4 \sigma^-_4 + \sigma^+_1 i u \sigma^+_1 \sigma^-_1 \sigma^-_2  i u^{-1} \sigma^+_3 \sigma^-_3  i u^{-1} \sigma^+_4 \sigma^-_4 \\ + \sigma^+_1 \sigma^-_1 i u \sigma^+_2 \sigma^-_2 i u^{-1} \sigma^+_3 \sigma^-_3 i u^{-1} \sigma^+_4 \sigma^-_4             \bigg] \text{. } 
\end{align*}

\noindent With the series of four expressions for each entry of the product representation above, we group together terms as follows. For $\underline{\mathcal{I}^{\prime\prime\prime}_1}$, one has,

\begin{align*}
i u \sigma^+_0 \sigma^-_0    \bigg[   i u \sigma^+_1 \sigma^-_1 \bigg[ i u \sigma^+_2 \sigma^-_2  i u \sigma^+_3 \sigma^-_3 + \sigma^-_2 \sigma^+_2 \sigma^-_3         \bigg]     i u \sigma^+_4 \sigma^-_4 + i u \sigma^+_1 \sigma^-_1 \sigma^+_2 i u \sigma^+_3 \sigma^-_3    \sigma^+_4 \sigma^-_4 + i u \sigma^+_1 \sigma^-_1 i u \sigma^+_2 \sigma^-_2 \sigma^-_3 i u \sigma^+_4 \sigma^-_4 \\ + i u \sigma^+_1 \sigma^-_1 \bigg[  \sigma^-_2 + \sigma^+_2    \bigg]         \sigma^+_3 \sigma^+_4 +        i u \sigma^+_1 \sigma^-_1 \bigg[ \sigma^-_2 + i u \sigma^+_2 \sigma^-_2  \bigg]             i u^{-1} \sigma^+_3 \sigma^-_3 \sigma^+_4 
+      i u \sigma^+_1 \sigma^-_1 i u \sigma^+_2 \sigma^-_2 i u^{-1} \sigma^+_3 \sigma^-_3 \sigma^+_4 \\ + \sigma^-_1 \bigg[    \sigma^+_1 \sigma^-_2 +  i u \sigma^+_1 \sigma^-_1 i u \sigma^+_2 \sigma^-_2      \bigg]     i u^{-1} \sigma^+_3 \sigma^-_3 \sigma^+_4               \bigg] \text{, } \\ \\ 
\sigma^-_0 \sigma^+_1    \bigg[           i u \sigma^+_1 \sigma^-_1 i u \sigma^+_2 \sigma^-_2 \bigg[ \sigma^+_3 + i u^{-1} \sigma^+_3 \sigma^-_3 \bigg]      \sigma^+_4 + \sigma^-_1 \bigg[ \sigma^+_2 \sigma^-_3 + i u \sigma^+_2 \sigma^-_2 i u^{-1} \sigma^+_3 \sigma^-_3  \bigg]  \sigma^+_4                           \bigg] \text{, } \end{align*}

\begin{align*}    \sigma^-_0         \bigg[     i u \sigma^+_1 \sigma^-_1 \bigg[ \sigma^-_2 + i u \sigma^+_2 \sigma^-_2 \bigg]      i u^{-1} \sigma^+_3 \sigma^-_3 \sigma^+_4 + \sigma^+_1 \bigg[ i u \sigma^+_1 \sigma^-_1 i u \sigma^+_2 \sigma^-_2 + \sigma^-_1 \sigma^+_2 \bigg]   i u \sigma^+_3 \sigma^-_3  i u \sigma^+_4 \sigma^-_4           + \bigg[ \sigma^+_1 + i u \sigma^+_1 \bigg] \\ \times \bigg[ \sigma^+_2 i u \sigma^+_3 \sigma^-_3 + i u \sigma^+_1 \sigma^-_1  i u \sigma^+_2 \sigma^-_2 \sigma^-_3 \bigg]  i u \sigma^+_4 \sigma^-_4            \bigg] \text{, }  
\end{align*}

\noindent For $\underline{\mathcal{I}^{\prime\prime\prime}_2}$, one has,

\begin{align*}
 i u \sigma^+_0 \sigma^-_0  \bigg[  i u \sigma^+_1 \sigma^-_1 \bigg[ i u \sigma^+_2 \sigma^-_2 + \sigma^+_2 \bigg] i u \sigma^+_3 \sigma^-_3 \sigma^-_4 + i u \sigma^+_1 \sigma^-_1 \bigg[ \sigma^+_2 i u \sigma^+_3 \sigma^-_3  + \sigma^-_2 \sigma^+_2 \sigma^-_3 \bigg]     \sigma^-_4 + i u \sigma^+_1 \sigma^-_1 \bigg[ i u \sigma^+_2 \sigma^-_2 \bigg[ \sigma^-_3 \sigma^-_4 \\  + \sigma^+_3 i u^{-1} \sigma^+_4 \sigma^-_4  \bigg] \text{ } \bigg]    +     i u \sigma^+_1 \sigma^-_2 \sigma^+_3 \sigma^-_3 \sigma^-_4 + i u \sigma^+_1 \sigma^-_1 \bigg[ \sigma^-_2 + i u \sigma^+_2 \sigma^-_2 \bigg] i u^{-1} \sigma^+_3 \sigma^-_3 i u^{-1} \sigma^+_4 \sigma^-_4   \\ 
 + \sigma^-_1 \sigma^+_1 \bigg[ i u \sigma^+_2 \sigma^-_2 + \sigma^-_2 \bigg]    i u^{-1}   \sigma^+_3 \sigma^-_3 i u^{-1} \sigma^+_4 \sigma^-_4 + \sigma^-_1 i u \sigma^+_2 \sigma^-_2 i u^{-1} \sigma^+_3 \sigma^-_3 i u^{-1} \sigma^+_4 \sigma^-_4          \bigg]  \text{, } \\ \\    \sigma^-_0 \sigma^+_1   \bigg[             i u \sigma^+_1 \sigma^-_1 i u \sigma^+_2 \sigma^-_2 \bigg[ i u \sigma^+_3 \sigma^-_3 + \sigma^-_3 \bigg]   \sigma^-_4 + \sigma^+_1 i u \sigma^+_2 \sigma^-_2 \sigma^-_3 \sigma^-_4    \bigg]  \text{, }   \\ \\   \sigma^+_0   \bigg[       i u \sigma^+_1 \sigma^-_1 \bigg[ \sigma^-_2 + i u \sigma^+_2 \sigma^-_2 \bigg] \sigma^-_4       \bigg]  \text{, }   \\ \\ 
 \sigma^-_0  \bigg[        i u \sigma^+_1 \sigma^-_1 \bigg[     \sigma^-_2 + i u \sigma^+_2 \sigma^-_2    \bigg]  i u^{-1} \sigma^+_3 \sigma^-_3 i u^{-1} \sigma^+_4 \sigma^-_4 + \sigma^+_1 \sigma^-_1 i u \sigma^+_2 \sigma^-_2 i u^{-1} \sigma^+_3 \sigma^-_3 i u^{-1} \sigma^+_4 \sigma^-_4    \bigg] \text{. }
\end{align*}

\noindent For $\underline{\mathcal{I}^{\prime\prime\prime}_3}$, one has,

\begin{align*}
     i u \sigma^+_0 \sigma^-_0   \bigg[                 i u \sigma^+_1 \sigma^-_1 \bigg[          i u \sigma^+_2 \sigma^-_2 + \sigma^+_2 + \sigma^-_2            \bigg]     \sigma^-_3 \sigma^+_4 + i u \sigma^+_1 \sigma^-_1 \bigg[ \sigma^-_2 + \sigma^+_2 \bigg]  i u^{-1} \sigma^+_3 \sigma^-_3 \sigma^+_4 \sigma^-_4   + i u \sigma^+_1 \sigma^-_1 i u \sigma^+_2 \sigma^-_2 \\ \times \bigg[ i \big[ u + u^{-1} \big]  \sigma^+_3 \sigma^-_3 \bigg]     i u^{-1} \sigma^+_4    \sigma^-_4 + \sigma^+_1 \sigma^-_2 \sigma^-_3 \sigma^+_4 + \sigma^-_0 \bigg[ \text{ } \bigg[  i u \sigma^+_1 \sigma^-_1 i u \sigma^+_2 \sigma^-_2 + \sigma^-_1 \sigma^+_1 \sigma^-_2  \bigg]     i u^{-1} \sigma^+_3 \sigma^-_3 i u^{-1} \sigma^+_4 \sigma^-_4        \bigg]   \\ +              \sigma^-_1 \bigg[ \text{ } \bigg[        i u \sigma^+_2 \sigma^-_2 + \sigma^-_2            \bigg]   i u^{-1} \sigma^+_3 \sigma^-_3 i u^{-1} \sigma^+_4 \sigma^-_4             \bigg]   + i u \sigma^+_1 \sigma^-_1 i u \sigma^+_2 \sigma^-_2 i u^{-1} \sigma^+_3 \sigma^-_3 i u^{-1} \sigma^+_4 \sigma^-_4                                                       \bigg]  \text{, }    \\ \\  \sigma^-_0 \sigma^+_1 \bigg[        i u \sigma^+_1 \sigma^-_1 \bigg[ i u \sigma^+_2 \sigma^-_2 \bigg[ i u \sigma^+_3 \sigma^-_3 + \sigma^-_3 \bigg] \sigma^+_4 \bigg]  + i u \sigma^+_1 \sigma^-_1 \sigma^-_2 i u^{-1} \sigma^+_3 \sigma^-_3 \sigma^+_4  + i u \sigma^+_2 \sigma^-_2 \sigma^-_3 \sigma^+_4 + \sigma^-_1 \sigma^+_2 \bigg[          i u \sigma^+_3 \sigma^-_3 \\ + \sigma^-_3      \bigg]  i u^{-1} \sigma^+_4 \sigma^-_4 +     i u \sigma^+_3 \sigma^-_3 \sigma^+_4           \bigg]  \text{, } \\ \\                   \sigma^-_0              \bigg[    i u \sigma^+_1 \sigma^-_1 \bigg[      \sigma^-_2 i u \sigma^+_3 \sigma^-_3 i u^{-1} \sigma^+_4 \sigma^-_4 + i u \sigma^+_2 \sigma^-_2 i u \sigma^+_3 \sigma^-_3 i u^{-1} \sigma^+_4 \sigma^-_4    \bigg]                            \text{ }        \bigg]  \text{, } \end{align*}

      \begin{align*}  
     \sigma^+_0  \bigg[   i u \sigma^+_1 \sigma^-_1 \bigg[ \sigma^-_2 + i u \sigma^+_2 \sigma^-_2 \bigg] \sigma^-_3 \sigma^+_4        \bigg]  \text{. }  
\end{align*}

\noindent For $\underline{\mathcal{I}^{\prime\prime\prime}_4}$, one has,

\begin{align*}
         \sigma^-_0    \bigg[   i u \sigma^+_1 \sigma^-_1 \bigg[    \sigma^+_2 + i u \sigma^+_2 \sigma^-_2         \bigg]                   i u^{-1} \sigma^+_3 \sigma^-_3 i u^{-1} \sigma^+_4 \sigma^-_4 + \sigma^+_1 \bigg[ i u \sigma^+_1 \sigma^-_1 + \sigma^-_1 \bigg] i u^{-1} \sigma^+_3 \sigma^-_3 i u^{-1} \sigma^+_4 \sigma^-_4                 \bigg]  \text{, }  \\ \\ 
         i u \sigma^+_0    \bigg[   \sigma^-_0 \bigg[ i u \sigma^+_1 \sigma^-_1  \bigg[    \sigma^+_2 + i u \sigma^+_2 \sigma^-_2 \bigg] \sigma^-_3 i u^{-1} \sigma^+_4 \sigma^-_4 \bigg]  + \sigma^-_0         \bigg[      i u \sigma^+_1 \sigma^-_1     \bigg[ \sigma^-_2 + i u \sigma^+_2 \sigma^-_2  \bigg]           \sigma^-_3 i u^{-1} \sigma^+_4 \sigma^-_4                   \bigg]      +  \sigma^-_0 \sigma^-_1 \\ \times \bigg[         i u \sigma^+_1 \sigma^-_1 i u \sigma^+_2 \sigma^-_2 i u \sigma^+_3 \sigma^-_3 \sigma^+_4 + \sigma^-_2 i u^{-1} \sigma^+_3 \sigma^-_3 \sigma^+_4 \sigma^-_4            \bigg]           +    \sigma^-_0 \sigma^-_1 i u \sigma^+_1 \sigma^-_1 i u \sigma^+_2 \sigma^-_2 i u^{-1} \sigma^+_3 \sigma^-_3 i u^{-1} \sigma^+_4 \sigma^-_4           \bigg]     \text{, } \end{align*}

\begin{align*}   \sigma^-_0    \bigg[        i u \sigma^+_1 \sigma^-_1 \bigg[ \sigma^-_2 + i u \sigma^+_2 \sigma^-_2  \bigg]      i u^{-1} \sigma^+_3 \sigma^-_3 \sigma^+_4                  + i u \sigma^+_1 \sigma^-_1 \bigg[ \sigma^-_2 + i u \sigma^+_2 \sigma^-_2  \bigg]      i u^{-1} \sigma^+_3 \sigma^-_3   \sigma^+_4   + \sigma^+_1 \bigg[ i u \sigma^+_1 \sigma^-_1         \sigma^+_2   + \sigma^+_2          \bigg] \\ \times  \sigma^-_3 i u^{-1}     \sigma^+_4 \sigma^-_4    + i u \sigma^+_1 \sigma^-_1 \bigg[ i u \sigma^+_2 \sigma^-_2 + \sigma^-_2 \bigg]  \sigma^+_3 \sigma^-_3 \sigma^+_4     +      \sigma^+_1 i u \sigma^+_1 \sigma^-_1 i u \sigma^+_2 \sigma^-_2 i u^{-1} \sigma^+_3 \sigma^-_3 \sigma^+_4       \\ +     \sigma^+_1 \bigg[ i u \sigma^+_1 \sigma^-_1 \sigma^+_2 + i u \sigma^+_2 \sigma^-_2    \bigg]                  \sigma^-_3 i u^{-1} \sigma^+_4 \sigma^-_4        \bigg]                       \text{, }  \\ \\ 
         \sigma^-_0 \sigma^+_1 \bigg[       i u \sigma^+_1 \sigma^-_1 \bigg[ i u \sigma^+_2 \sigma^-_2    \sigma^+_3 + \sigma^-_2 \sigma^-_3       \bigg]         \sigma^+_4 + \sigma^-_1 i u \sigma^+_2 \sigma^-_2  i u^{-1} \sigma^+_3 \sigma^-_3 \sigma^+_4 \bigg]                            \text{ }                                       \bigg]     \text{, } \\ \\  i u \sigma^+_0 \sigma^-_0 \bigg[         i u \sigma^+_1 \sigma^+_1 \bigg[         i u \sigma^+_2 \sigma^-_2 + \sigma^+_2        \bigg]  \sigma^+_3 \sigma^+_4 +          i u 
\sigma^+_1 \sigma^-_1 \bigg[    \sigma^-_2 + i u \sigma^+_2 \sigma^-_2      \bigg]     i u^{-1} \sigma^+_3 \sigma^-_3 \sigma^+_4    + \sigma^-_1 \sigma^+_1 \sigma^-_2 i u^{-1} \sigma^+_3 \sigma^-_3 \sigma^+_4 + i u \sigma^+_1 \\ \times          \sigma^-_1 i u \sigma^+_2 \sigma^-_2 i u \sigma^+_3 \sigma^-_3 \sigma^+_4        \bigg]  \text{. } 
\end{align*}

\noindent From the previous set of expressions for $\underline{\mathcal{I}^{\prime\prime\prime}_1}, \underline{\mathcal{I}^{\prime\prime\prime}_2}, \underline{\mathcal{I}^{\prime\prime\prime}_3}$ and $\underline{\mathcal{I}^{\prime\prime\prime}_4}$, one arrives to an approximation of entries for the product representation through identifying the manner in which lower order terms generalize to higher order approximations of the operators for the transfer matrix. By making use of the support sets for each of the four operators for the transfer matrix representation, the first term, $\underline{\mathcal{I}_1}$, of the lower order expansion for the first operator, $A \big( u \big)$, takes the following form. The first component of $\underline{\mathcal{I}_1}$ takes the form,

\begin{align*}
 \mathscr{I}^1_1 \equiv  i u \sigma^+_0 \sigma^-_0 \bigg[  \text{ } \bigg[   \underset{j \text{ odd}}{\underset{1 \leq  j \leq N-1}{\prod}}       i u \sigma^+_j \sigma^-_j \textbf{1}_{\{ j^{\prime}> j \text{ } : \text{ } i u \sigma^+_{j^{\prime}} \sigma^-_{j^{\prime}} \in \mathrm{support} ( A ( u ) )\} }       \big[ i u \sigma^+_{j^{\prime}} \sigma^-_{j^{\prime}} i u \sigma^+_{j^{\prime}+1} \sigma^-_{j^{\prime}+1} + \sigma^-_{j^{\prime}} \sigma^+_{j^{\prime}} \sigma^-_{j^{\prime}+1} \big]        \bigg]   \bigg[    \underset{j \equiv N}{\prod}     i u \sigma^+_j \sigma^-_j \bigg]     \\    +   i u \sigma^+_1 \sigma^-_1 \sigma^+_2          \bigg[        \underset{i \text{ odd}}{\underset{1 \leq j \leq N}{\prod}}      i u \sigma^+_j \sigma^-_j             \bigg]     \bigg[ \underset{j \equiv N}{\prod} \sigma^+_j \sigma^-_j  \bigg]                 +  \bigg[   \underset{j \text{ even}}{\underset{1 \leq j \leq N-1}{\prod}}   i u \sigma^+_j \sigma^-_j  \bigg]  \bigg[       \underset{j \text{ odd}}{\underset{3 \leq j \leq N-1}{\prod}}      i u^{-1} \sigma^+_j \sigma^-_j              \bigg]   \bigg[ \underset{j \equiv N}{\prod} i u \sigma^+_j \sigma^-_j   \bigg] \end{align*}

 \noindent The second component of $\underline{\mathcal{I}_1}$ takes the form,
 
 \begin{align*} i u \sigma^+_1 \sigma^-_1 \bigg[ \underset{2 \leq j \leq N-2}{\prod} \big[ \sigma^-_j + \sigma^+_j \big]  \bigg]  \bigg[     \underset{3 \leq j < j+1 \leq N+1}{\prod}      \sigma^+_j \sigma^+_{j+1}     \bigg]    + i u \sigma^+_1 \sigma^-_1 \bigg[ \underset{j \text{ } \text{not such that } j \text{ } \mathrm{mod} 4 \equiv 0}{\underset{j \text{ } \text{mod} 2 \equiv 0}{\underset{2 \leq j \leq N-1}{\prod}}} \big[    \sigma^-_j + i u \sigma^+_j \sigma^-_j      \big]  \bigg]      \\ \times \bigg[ 
 \underset{j \text{ } \text{mod} 3 \equiv 0}{\underset{3 \leq j \leq N-1}{\prod}} i u^{-1} \sigma^+_j \sigma^-_j \bigg]   \bigg[ \underset{j \text{ } \text{mod} 4 \equiv 0}{\underset{j \equiv N}{\prod}} \sigma^+_j  \bigg]    \end{align*}
 
 \noindent The third component of $\underline{\mathcal{I}_1}$ takes the form,
 
 \begin{align*}
 \bigg[ \underset{1 \leq j < j+1 \leq N-1}{\prod}  \big[    i u \sigma^+_j \sigma^-_j i u \sigma^+_{j+1} \sigma^-_{j+1}   \big]  \bigg]  \bigg[ \underset{j \text{ } \text{mod} 3 \equiv 0}{\underset{3 \leq j \leq N-1}{\prod}}  i u^{-1} \sigma^+_j \sigma^-_j  \bigg]    \bigg[ \underset{j \equiv N}{\prod}  \sigma^+_j \bigg]  + \sigma^-_1 \bigg[      \underset{1 \leq j \leq N-1}{\prod}  \big[                       \sigma^+_j \sigma^-_{j+1}   + i u \sigma^+_j \sigma^-_j i u \sigma^+_{j+1} \sigma^-_{j+1}  \big]  \bigg] \\ \times   \bigg[ \underset{j \text{ } \mathrm{mod} 3 \equiv 0}{\underset{3 \leq j \leq N-1}{\prod}} i u^{-1} \sigma^+_j \sigma^-_j \bigg]    \bigg[ \underset{j \equiv N}{\prod}\sigma^+_j  \bigg]                           \text{ }   \bigg]  \text{. }
\end{align*}

\noindent The second term takes the form,

\begin{align*}
  \mathscr{I}^1_2 \equiv   \sigma^-_0 \sigma^+_1 \bigg[ \text{ }   \bigg[            \underset{1 \leq j \leq N-1}{\prod}   i u \sigma^+_j \sigma^-_j \textbf{1}_{\{j^{\prime}>j \text{ } :\text{ }  i u \sigma^+_{j^{\prime}} \sigma^-_{j^{\prime}} \in \mathrm{support}( A ( u )) \}} \big[  \sigma^+_{j^{\prime}}      +   i u^{-1} \sigma^+_{j^{\prime}} \sigma^-_{j^{\prime}}             \big]               \bigg]    \bigg[ \underset{j \equiv N}{\prod} \sigma^+_j  \bigg]       \text{ } \\  + \sigma^-_1     \bigg[    \underset{1 \leq j < N}{\prod}     \textbf{1}_{\{ j^{\prime}> j \text{ } : \text{ } i u \sigma^+_{j^{\prime}} \sigma^-_{j^{\prime}} \in \mathrm{support} ( A(u)) \}} \bigg[        \sigma^+_{j^{\prime}} \sigma^-_{j^{\prime}+1}  + i u \sigma^+_{j^{\prime}} \sigma^-_{j^{\prime}}  \bigg[ \underset{j \text{ } \mathrm{mod}3 \equiv 0}{\underset{3 \leq j \leq N-1}{\prod}}  i u^{-1} \sigma^+_j \sigma^-_j  \bigg] \text{ }    \bigg]       \text{ }          \bigg]                 \bigg[ \underset{j \equiv N}{\prod} \sigma^+_j  \bigg]   \text{ }   \bigg]     \text{. }
\end{align*}

\noindent The third term takes the form,

\begin{align*}
  \mathscr{I}^1_3 \equiv  \bigg[ \underset{0 \leq j < j+1 < N-1}{\prod}  \sigma^-_{j} i u \sigma^+_{j+1} \sigma^-_{j+1}      \textbf{1}_{\{j^{\prime}> j+1 \text{ } : \text{ } i u \sigma^-_{j^{\prime}} \sigma^+_{j^{\prime}} \in \mathrm{support} ( A(u))\}}  \big[ 
       \sigma^-_{j^{\prime}} +  i u \sigma^+_{j^{\prime}} \sigma^-_{j^{\prime}}      \big]  \bigg]         \bigg[       \underset{j  \text{ } \mathrm{mod}3 \equiv 0}{\underset{3 \leq j \leq N-1}{\prod}}      i u^{-1} \sigma^+_j \sigma^-_j              \bigg]  \bigg[ \underset{j \equiv N}{\prod}  \sigma^+_j \bigg]  \\ +     \bigg[ \underset{0 \leq j < j+ 1 < N-1}{\prod}    \sigma^-_j \sigma^+_{j+1}     \textbf{1}_{\{j^{\prime}> j+1 \text{ } : \text{ } i u \sigma^+_{j^{\prime}} \sigma^-_{j^{\prime}} \in \mathrm{support} ( A( u ))\}}  \big[      i u \sigma^+_{j^{\prime}} \sigma^-_{j^{\prime}} i u \sigma^+_{j^{\prime}+1} \sigma^-_{j^{\prime}+1}  + \sigma^-_{j^{\prime}}\sigma^+_{j^{\prime}+1}  \big]                                 \bigg]  \\ \times \bigg[      \underset{N-1 \leq j \leq N}{\prod}       i u \sigma^+_j \sigma^-_j     \bigg]  +      \bigg[           \underset{1 \leq j\leq N-1}{\prod}        \big[         \sigma^+_j   +  i u \sigma^+_j        \big] \textbf{1}_{\{j^{\prime} > j \text{ } : \text{ } i u \sigma^+_j \sigma^-_j \in \mathrm{support} ( A ( u ))\}} \big[      \sigma^+_{j^{\prime}} i u \sigma^+_{j^{\prime}+1} \sigma^-_{j^{\prime}+1} \\ + i u \sigma^+_j \sigma^-_j i u  \sigma^+_{j^{\prime}} \sigma^-_{j^{\prime}} \sigma^-_{j^{\prime}+1}                             \big]                   \bigg]        \bigg[ \underset{j \equiv N}{\prod}       i u \sigma^+_j \sigma^-_j     \bigg]              \text{. }
\end{align*}

\noindent The desired representation for the first entry of the transfer matrix takes the form,

\begin{align*}
\underline{\mathcal{I}^{\prime\cdots\prime}_1}  =  \mathscr{I}^1_1 + \mathscr{I}^1_2 + \mathscr{I}^1_3  \text{. }
\end{align*}

\noindent One obtains, for the first term of $\underline{\mathcal{I}_2}$, which is denoted with $ \mathscr{I}^2_1$, in the product representation for the second operator $B \big( u \big)$, the first component of which consists of terms,

\begin{align*}
     \underset{0 \leq j \leq N-1}{\prod}   i u \sigma^+_j \sigma^-_j i u \sigma^+_{j+1} \sigma^-_{j+1} \bigg[ \textbf{1}_{\{ j^{\prime} > j \text{ } : \text{ } i u \sigma^+_{j^{\prime}} \sigma^-_{j^{\prime}}    \in \mathrm{support} ( B ( u ))\}}         \bigg[        \frac{  i u \sigma^+_{j^{\prime}} \sigma^-_{j^{\prime}} + \sigma^+_{j^{\prime}}        }{i u \sigma^+_{j+1} \sigma^-_{j+1}}    \bigg]   \bigg[          \underset{j \equiv N-1}{\prod}  i u \sigma^+_j \sigma^-_j \sigma^-_{j+1}            \bigg]   \\    +           \textbf{1}_{\{ j^{\prime} > j \text{ } : \text{ } i u \sigma^+_{j^{\prime}} \sigma^-_{j^{\prime}} \in \mathrm{support} ( B ( u ))  \}}       \bigg[        \frac{\sigma^+_{j^{\prime}} i u \sigma^+_{j^{\prime}+1} \sigma^-_{j^{\prime}+1}            + \sigma^-_{j^{\prime}}   \sigma^+_{j^{\prime}} \sigma^-_{j^{\prime}+1}}{i u \sigma^+_{j+1} \sigma^-_{j+1}}          \bigg]       \bigg[       \underset{j \equiv N}{\prod}          \sigma^-_j                         \bigg] \text{ }  \bigg] \text{, } \end{align*}
     
     \noindent the second component of which consists of the terms, 
     
     \begin{align*}
      \underset{0 \leq j \leq N-1}{\prod}     i u \sigma^+_j       \sigma^-_j   i u \sigma^+_{j+1} \sigma^-_{j+1} \sigma^+_{j+2} \sigma^-_{j+2}        \bigg[        \textbf{1}_{ \{j^{\prime} > j + 2 \text{ } : \text{ } i u \sigma^+_{j^{\prime}} \sigma^-_{j^{\prime}}    \in \mathrm{support} ( B ( u ))        \}}          \bigg[        \frac{\sigma^-_{j^{\prime}} \sigma^-_{j^{\prime}+1} + \sigma^+_{j^{\prime}} i u^{-1} \sigma^+_{j^{\prime}} \sigma^-_{j^{\prime}}    }{i u \sigma^+_j \sigma^-_j i u \sigma^+_{j+1 } \sigma^-_{j+1}}            \bigg]      \text{ }             \bigg]          \\           + \bigg[ \underset{1 \leq j \leq N-1}{\prod}       i u \sigma^+_j \sigma^-_j     \bigg] \bigg[ \underset{j \equiv N}{\prod}         \sigma^+_{j-1} \sigma^-_{j-1} \sigma^-_j    \bigg]         +    \bigg[        \underset{j \text{ } \mathrm{mod}2 \equiv 0}{\underset{0 \leq j \leq N-1}{\prod}}         i u \sigma^+_{j} \sigma^-_j \sigma^+_{j+1} \sigma^-_{j+1}             \textbf{1}_{\{j^{\prime} > j \text{ } : \text{ } i u \sigma^+_{j^{\prime}} \sigma^-_{j^{\prime}} \in \mathrm{support} ( B( u ))\}} \big[       \sigma^-_{j^{\prime}} + i u \sigma^+_{j^{\prime}} \sigma^-_{j^{\prime}}          \big]               \bigg]  \\ \times  \bigg[ \text{ } \underset{j \text{ } \mathrm{mod}3 \equiv 0}{\underset{0 \leq j \leq N-1}{\prod}} \bigg[  i u^{-1} \sigma^+_j \sigma^-_j       i u^{-1} \sigma^+_{j+1} \sigma^-_{j+1}      \bigg] \text{ } \bigg] \text{, } \end{align*}
      
      \noindent and the third component of which consists of the terms,
      
      \begin{align*}
        \underset{0 \leq j \leq N-1}{\prod} i u \sigma^+_j \sigma^-_j \sigma^+_{j+1} \sigma^-_{j+1}  \bigg[ 
 \underset{j^{\prime} \text{ } \mathrm{mod}2 \equiv 0}{\underset{0 \leq j^{\prime} \leq N-1}{\prod}}   \textbf{1}_{\{ j^{\prime} > j \text{ } : \text{ } i u \sigma^+_{j^{\prime}} \sigma^-_{j^{\prime}} \in \mathrm{support} ( B ( u ) )  \}}  \big[  i u \sigma^+_{j^{\prime}} \sigma^-_{j^{\prime}} + \sigma^-_{j^{\prime}}               \big]       \bigg]  \\ \times   \bigg[ \text{ } \underset{j \text{ } \mathrm{mod}3 \equiv 0}{\underset{0 \leq j \leq N-1}{\prod}} \bigg[  \frac{i u^{-1} \sigma^+_j \sigma^-_j       i u^{-1} \sigma^+_{j+1} \sigma^-_{j+1}}{i u \sigma^+_j \sigma^-_j }      \bigg] \text{ } \bigg]    +         \bigg[  \underset{0 \leq j \leq N-1}{\prod}  i u \sigma^+_j \sigma^-_j \sigma^-_{j+1} i u \sigma^+_{j+2} \sigma^-_{j+2}       \bigg] \\
 \times \bigg[ 
 \underset{j \text{ } \mathrm{mod}3 \equiv 0}{\underset{0 \leq j \leq N-1}{\prod}} \bigg[  \frac{i u^{-1} \sigma^+_j \sigma^-_j       i u^{-1} \sigma^+_{j+1} \sigma^-_{j+1}}{i u \sigma^+_j \sigma^-_j } 
 \bigg]  \text{ } \bigg]  \text{. }   \end{align*}

\noindent The second term takes the form,

\begin{align*}
     \mathscr{I}^2_2 \equiv         \underset{0 \leq j \leq N-1}{\prod}    \sigma^-_j \sigma^+_j i u \sigma^-_{j+1} \sigma^+_{j+1} i u \sigma^+_{j+1} \sigma^-_{j+1}  \bigg[               \textbf{1}_{\{  j^{\prime} > j    \text{ } : j^{\prime} \text{ } \mathrm{mod} 3 \equiv 0, \text{ } i u \sigma^+_{j^{\prime}} \sigma^-_{j^{\prime}} \in \mathrm{support} ( B( u ))        \}}      \bigg[   \frac{i u \sigma^+_{j^{\prime}} \sigma^-_{j^{\prime}} + \sigma^-_{j^{\prime}}   }{\sigma^-_j \sigma^+_j iu\sigma^+_{j+1} \sigma^-_{j+1} } \bigg] \end{align*}

     \begin{align*}
     \times  \bigg[ \underset{j \equiv N}{\prod}  \sigma^-_j \bigg] \text{ }     \bigg]  +    \bigg[        \underset{0 \leq j \leq N-1}{\prod} \sigma^-_j \sigma^+_{j+1}   i u \sigma^+_{j+2} \sigma^-_{j+2} \bigg] \bigg[  \underset{j \equiv N}{\prod}   \sigma^-_{j-1} \sigma^-_j     \bigg]                        \text{. } 
\end{align*}

\noindent The third term takes the form,

\begin{align*}
   \mathscr{I}^2_3 \approx  \underset{0 \leq j \leq N-1}{\prod}  \sigma^+_{j} i u \sigma^+_{j+1} \sigma^-_{j+1} \bigg[ \textbf{1}_{\{ j^{\prime} > j  
            \text{ } : \text{ }      i u \sigma^+_{j^{\prime}} i u \sigma^-_{j^{\prime}} \in \mathrm{support} ( B( u ))    \}}      \bigg[ \frac{\sigma^-_{j^{\prime}} + i u \sigma^+_{j^{\prime}} \sigma^-_{j^{\prime}}}{i u \sigma^+_j \sigma^-_j } \bigg]              \bigg] \bigg[   \underset{j \equiv N}{\prod}  \sigma^-_j     \bigg]  \text{. } 
\end{align*}

\noindent The fourth term takes the form,

\begin{align*}
   \mathscr{I}^2_4 \approx              \underset{0 \leq j \leq N-1}{\prod}   \sigma^-_j i u \sigma^+_{j+1 } \sigma^-_{j+1} \bigg[ \textbf{1}_{\{ j^{\prime} > j \text{ } : \text{ }  i u \sigma^+_{j^{\prime}} \sigma^-_{j^{\prime}} \in \mathrm{support} ( B( u ))      \}}            \bigg[ \frac{\sigma^-_{j^{\prime}+1} + i u \sigma^+_{j^{\prime}+1}   \sigma^-_{j^{\prime}+1}}{i u \sigma^+_{j+1} \sigma^-_{j+1}}   \bigg]  \text{ }   \bigg] 
 \bigg[ \underset{j \equiv N}{\prod}  i u^{-1} \sigma^+_{j-1} \sigma^-_{j-1} i u^{-1} \sigma^+_{j} \sigma^-_j    \bigg]   \\ +               \bigg[   \underset{0 \leq j \leq N-1}{\prod}                  \sigma^-_j \sigma^+_{j+1}   i u \sigma^+_{j+2} \sigma^-_{j+2}   \bigg] \bigg[     \underset{j \equiv N}{\prod}   i u^{-1} \sigma^+_{j-1} \sigma^-_{j-1} i u^{-1} \sigma^+_j \sigma^-_j  \bigg]                         \text{. } 
\end{align*}

\noindent The desired representation for the second entry of the transfer matrix takes the form,

\begin{align*}
\underline{\mathcal{I}^{\prime\cdots\prime}_2}  =  \mathscr{I}^2_1 + \mathscr{I}^2_2 + \mathscr{I}^2_3 + \mathscr{I}^2_4   \text{. }
\end{align*}

\noindent One obtains, for the first term of $\underline{\mathcal{I}_3}$, which is denoted with  $\mathscr{I}^3_1$, in the product representation for the second operator $C \big( u \big)$, the first component of which consists of the terms,

\begin{align*}
      \bigg[ \underset{0 \leq j \leq N-1}{\prod}  i u \sigma^+_j \sigma^-_j i u \sigma^+_{j+1} \sigma^-_{j+1} \bigg[ \textbf{1}_{\{  j^{\prime} > j \text{ } : \text{ } i u \sigma^+_{j^{\prime}} \sigma^-_{j^{\prime}} \in \mathrm{support } ( C ( u ))       \}} \bigg[             \frac{ i u \sigma^+_{j^{\prime}} \sigma^-_{j^{\prime}} + \sigma^+_{j^{\prime}} + \sigma^-_{j^{\prime}}          }{i u \sigma^+_j \sigma^-_j }     \bigg] \text{ }          \bigg]   \bigg[ \underset{j \text{ } \mathrm{mod}3 \equiv 0}{\underset{0 \leq j \leq N-1}{\prod}}          \sigma^-_j \sigma^+_{j+1}     \bigg]   \\ +     \underset{0 \leq j \leq N-1}{\prod}  i u \sigma^+_j \sigma^-_j i u \sigma^+_{j+1} \sigma^-_{j+1} \bigg[ \textbf{1}_{\{  j^{\prime} > j \text{ } : \text{ } i u \sigma^+_{j^{\prime}} \sigma^-_{j^{\prime}} \in \mathrm{support } ( C ( u ))       \}}     \bigg[  \frac{\sigma^-_{j^{\prime}} + \sigma^+_{j^{\prime}}}{i u \sigma^+_j \sigma^-_j}     \bigg] \text{ } \bigg] \bigg[ \underset{j \text{ } \mathrm{mod} 3 \equiv 0}{\underset{0 \leq j \leq N-1}{\prod}} i u^{-1} \sigma^+_j \sigma^-_j  \bigg]             \text{ } \bigg]        \bigg[    \underset{j \text{ } \mathrm{mod}4 \equiv 0}{\underset{0 \leq j \leq N-1}{\prod}}  \sigma^+_j \sigma^-_j   \bigg] \text{, }  \end{align*}
      
      \noindent the second component of which consists of the terms,
      
      \begin{align*}
        \bigg[   \underset{1 \leq j \leq N-1}{\prod}   i u \sigma^+_j \sigma^-_j i u \sigma^+_{j+1} \sigma^-_{j+1}      \bigg]  \bigg[       \underset{j \text{ } \mathrm{mod}3 \equiv 0}{\underset{0 \leq j \leq N-1}{\prod}}  i \big[ u + u^{-1} \big] \sigma^+_j \sigma^-_j    \bigg]  +     \underset{0 \leq j \leq N-4}{\prod}   \sigma^+_j \sigma^-_{j+1} \sigma^-_{j+2} \sigma^+_{j+3}        \end{align*}

        \noindent and of,
        
        \begin{align*}
        i u \sigma^+_0   \sigma^-_0 \bigg[  \text{ } \bigg[    \underset{0 \leq j \leq N-1}{\prod}     i u \sigma^+_j \sigma^-_j i u \sigma^+_{j+1} \sigma^-_{j+1} \bigg] \bigg[  \underset{j \equiv N}{\prod} i u^{-1} \sigma^+_{j-1} \sigma^-_{j-1} i u^{-1} \sigma^+_j \sigma^-_j     \bigg]  + \bigg[  \underset{1 \leq j \leq N-1}{\prod} \sigma^-_j \sigma^+_j \sigma^-_{j+1} \bigg] \\ \times  \bigg[    \underset{j \equiv N}{\prod} i u^{-1} \sigma^+_{j-1} \sigma^-_{j-1} i u^{-1} \sigma^+_j \sigma^-_j       \bigg] \text{ } \bigg] \text{. } \end{align*} 
        
        \noindent The third component consists of the terms,
        
        \begin{align*}
        \bigg[ \underset{0 \leq j \leq N-1}{\prod}  i u \sigma^+_j \sigma^-_j \sigma^-_{j+1} \bigg[ \textbf{1}_{\{ j > j^{\prime} \text{ } : \text{ } i u \sigma^+_{j^{\prime}} \sigma^-_{j^{\prime}} \in \mathrm{support} ( C ( u ))  \}}           \bigg[ \frac{i u \sigma^+_{j^{\prime}} \sigma^-_{j^{\prime}} + \sigma^-_{j^{\prime}}}{ i u \sigma^+_j \sigma^-_j }         \bigg]  \text{ }                \bigg]  \text{ } \bigg]   \bigg[ \underset{j \equiv N}{\prod}  i u^{-1} \sigma^+_{j-1} \sigma^-_{j-1} i u^{-1} \sigma^+_j \sigma^-_j \bigg] \\  + \bigg[ \underset{0 \leq j \leq N-1}{\prod}  i u \sigma^+_j \sigma^-_j i u \sigma^+_{j+1} \sigma^-_{j+1}     \bigg]   \bigg[ \underset{j \equiv N}{\prod} iu^{-1} \sigma^+_j \sigma^-_{j-1} \bigg]     \text{. }
\end{align*}

\noindent The second term takes the form,

\begin{align*}
  \mathscr{I}^3_2 \approx   
  \sigma^-_0 \sigma^+_1 \bigg[ \text{ } \bigg[ \underset{0 \leq j \leq N-1}{\prod}  i u \sigma^+_j \sigma^-_j \bigg]  \bigg[ \underset{j \equiv N}{\prod}     \sigma^-_j \sigma^+_{j+1} \bigg] +   \bigg[ \underset{0 \leq j \leq N-2}{\prod}  i u \sigma^+_j \sigma^-_j  \bigg] \bigg[ \underset{j \text{ } \mathrm{mod} 3 \equiv 0}{\underset{0 \leq j \leq N-1}{\prod}}  \sigma^-_j \bigg]   \bigg[ \underset{j \equiv N}{\prod}     \sigma^-_j \sigma^+_{j+1} \bigg]  +             \bigg[ \underset{1 \leq j \leq N-1}{\prod}         i u \sigma^+_j \\ \times  \sigma^-_j      \bigg]    \bigg[     \underset{j \text{ } \mathrm{mod}2 \equiv 0}{\underset{0 \leq j \leq N-1}{\prod}}     \sigma^-_j    \bigg] \bigg[     \underset{j \text{ } \mathrm{mod}3 \equiv 0}{\underset{0 \leq j \leq N-1}{\prod}}  i u^{-1} \sigma^+_j \sigma^-_j        \bigg]  \bigg[ \underset{j \equiv N}{\prod} \sigma^+_j  \bigg]    +   \bigg[            \underset{2 \leq j \leq N-1}{\prod} i u \sigma^+_j \sigma^-_j     \bigg] \bigg[ \underset{j \text{ } \mathrm{mod} 3 \equiv 0}{\underset{ 0 \leq j \leq N-1}{\prod}}   \sigma^-_j   \bigg] \bigg[ \underset{j \equiv N}{\prod}      \sigma^+_j  \bigg]  \text{ } \bigg]         
\\ 
  +   \sigma^-_0 \sigma^+_1 \sigma^-_1 \sigma^+_2    \bigg[    \underset{j \text{ } \mathrm{mod}3 \equiv 0}{\underset{0 \leq j \leq N-1}{\prod}}     \bigg[               i u \sigma^+_j \sigma^-_j + \sigma^-_j               \bigg] \bigg[ \underset{j \text{ } \mathrm{mod} 4 \equiv 0}{\underset{0 \leq j \leq N-1}{\prod}}  i u^{-1} \sigma^+_j \sigma^-_j \bigg] + \bigg[  \underset{j \equiv N}{\prod}   i u \sigma^+_{j-1} \sigma^-_{j-1} \sigma^+_j  \bigg]  \text{ }  \bigg] \text{. }
\end{align*}

\noindent The third term takes the form,

\begin{align*}
 \mathscr{I}^3_3 \approx           \sigma^-_0 i u \sigma^+_1 \sigma^-_2 \sigma^-_2  \bigg[ \underset{j \equiv N}{\prod}       i u \sigma^+_{j-1} \sigma^-_{j-1} i u^{-1} \sigma^+_j \sigma^-_j       \bigg]  + \sigma^-_0 i u \sigma^+_1 \sigma^-_1 \bigg[     \underset{1 \leq j \leq N-1}{\prod}  i u \sigma^+_j \sigma^-_j     \bigg]  \bigg[ \underset{j \equiv N}{\prod}         i u^{-1} \sigma^+_j \sigma^-_j    \bigg]  \text{. }
\end{align*}

\noindent The fourth term takes the form,

\begin{align*}
 \mathscr{I}^3_4 \approx           \sigma^+_0 \bigg[   \underset{0 \leq j \leq N-1}{\prod}   i u \sigma^+_j \sigma^-_j  \bigg[ \text{ } \bigg[ \textbf{1}_{\{  j^{\prime} > j \text{ } : \text{ } i u \sigma^+_{j^{\prime}} \sigma^-_{j^{\prime}} \in \mathrm{support} ( C ( u ) )  \}}    \frac{  \sigma^-_{j^{\prime}} + i u \sigma^+_{j^{\prime}} \sigma^-_{j^{\prime}} }{i u \sigma^+_{j} \sigma^-_j }   \bigg] \bigg[ \underset{ j \equiv N}{\prod}   \sigma^-_{j-1} \sigma^+_j       \bigg]   \text{ } \bigg] \text{ } \bigg]     \text{. }
\end{align*}

\noindent The desired representation for the third entry of the transfer matrix takes the form,

\begin{align*}
\underline{\mathcal{I}^{\prime\cdots\prime}_3}  =  \mathscr{I}^3_1 + \mathscr{I}^3_2 + \mathscr{I}^3_3 + \mathscr{I}^3_4   \text{. }
\end{align*}

\noindent One obtains, for the first term of $\underline{\mathcal{I}_4}$, which is denoted with  $\mathscr{I}^4_1$, in the product representation for the second operator $D \big( u \big)$, the first component of which consists of the terms,

 \begin{align*}
   \mathscr{I}^4_1 \approx   \sigma^-_0 \bigg[   i u \sigma^+_1 \sigma^-_1            \bigg[ \underset{0 \leq j \leq N-1}{\prod}            \textbf{1}_{\{ j^{\prime} > j \text{ } : \text{ } i u \sigma^+_{j^{\prime}} \sigma^-_{j^{\prime}} \in \mathrm{support} ( D ( u ) ) \}}  \bigg[  \sigma^+_{j^{\prime}} + i u \sigma^+_{j^{\prime}} \sigma^-_{j^{\prime}} \bigg] \text{ }                   \bigg]  \bigg[ \underset{j \equiv N}{\prod} i u^{-1} \sigma^+_{j-1} \sigma^-_{j+1} i u^{-1} \sigma^+_j \sigma^-_j  \bigg]                          \\ +           \sigma^+_1 \bigg[  \underset{1 \leq j \leq N-1}{\prod}     \textbf{1}_{\{j^{\prime} > j \text{ } : \text{ }      i u \sigma^+_{j^{\prime}} \sigma^-_{j^{\prime}} \in \mathrm{support} ( D ( u ))         \}}    \bigg[         i u \sigma^+_{j^{\prime}} \sigma^-_{j^{\prime}} + \sigma^-_{j^{\prime}} \bigg]  \text{ }      \bigg]     \bigg[          \underset{j \equiv N}{\prod}        i u^{-1} \sigma^+_{j-1} \sigma^-_{j-1} i u^{-1} \sigma^+_j \sigma^-_j     \bigg] \text{ }        \bigg]    \text{. }
 \end{align*}

 \noindent The second term takes the form,

  \begin{align*}
   \mathscr{I}^4_2 \approx    i u \sigma^+_0 \bigg[        \underset{0 \leq j \leq N-1}{\prod}    \sigma^-_j \bigg[        \textbf{1}_{\{    j^{\prime} > j \text{ } : \text{ } i u \sigma^+_{j^{\prime} } \sigma^-_{j^{\prime}} \in \mathrm{support} ( D ( u ))    \}}   \bigg[ \frac{\sigma^+_{j^{\prime}} + i u \sigma^+_{j^{\prime}} \sigma^-_{j^{\prime}}}{\sigma^-_j } \bigg] \text{ }    \bigg]     + \underset{0 \leq. j\leq N-2}{\prod} \sigma^-_j i u \sigma^+_{j+1} \sigma^-_{j+1} \\ \times  \bigg[ \textbf{1}_{\{ j^{\prime} > j \text{ } : \text{ } i u \sigma^+_{j^{\prime}} \sigma^-_{j^{\prime}} \in \mathrm{suppoort } ( D ( u ) )  \}}  \bigg[  \sigma^-_{j^{\prime}} + i u \sigma^+_{j^{\prime} } \sigma^-_{j^{\prime}}          \bigg]  \bigg[  
   \underset{j \equiv N}{\prod}   \sigma^-_{j-1} i u^{-1} \sigma^+_j \sigma^-_j      \bigg] \text{ }               \bigg] +     \underset{0 \leq j \leq N-1}{\prod}  \sigma^-_j \sigma^-_{j+1} \\ \times \textbf{1}_{\{ j^{\prime} > j \text{ } : \text{ }   \sigma^+_{j^{\prime}} \sigma^-_{j^{\prime}} \in \mathrm{support} ( D ( u ))        \}}     \bigg[                i u \sigma^+_{j^{\prime}} \sigma^-_{j^{\prime}} i u \sigma^+_{j^{\prime}+1} \sigma^-_{j^{\prime}+1} i u \sigma^+_{j^{\prime}+2} \sigma^-_{j^{\prime}+2}  \sigma^+_{j^{\prime}+3} + \sigma^+_{j+1} i u^{-1} \sigma^+_{j+2} \sigma^-_{j+2} \sigma^+_{j+3} \sigma^-_{j+3}      \bigg] \\ +    \underset{0 \leq j \leq N-4}{\prod}  \sigma^+_j \sigma^-_{j+1}      i u \sigma^+_{j+1} \sigma^-_{j+1} i u \sigma^+_{j+2} \sigma^-_{j+2} i u^{-1} \sigma^+_{j+3} \sigma^-_{j+3} i u^{-1} \sigma^+_{j+4} \sigma^-_{j+4}            \bigg]                    \text{. }
 \end{align*}

 \noindent The third term takes the form,

  \begin{align*}
   \mathscr{I}^4_3 \approx \sigma^-_0      \bigg[            \underset{1 \leq j \leq N-1}{\prod}      i u \sigma^+_j \sigma^-_j           \bigg[ \textbf{1}_{\{j^{\prime} > j \text{ } : \text{ } i u \sigma^+_{j^{\prime}} \sigma^-_{j^{\prime}} \in \mathrm{support} ( D ( u )) \}}     \bigg[ \sigma^-_{j^{\prime}} + i u \sigma^+_{j^{\prime}} \sigma^-_{j^{\prime}}  \bigg]  \text{ }         \bigg]  \bigg[    \underset{j \equiv N-1}{\prod}    i u^{-1} \sigma^+_{j-1} \sigma^-_{j-1} \sigma^+_j                              \bigg]                 \\ +   \underset{1 \leq j \leq N-1}{\prod}      i u \sigma^+_j \sigma^-_j \bigg[ \textbf{1}_{\{ j^{\prime} > j \text{ } : \text{ } i u \sigma^+_{j^{\prime}} \sigma^-_{j^{\prime}} \in \mathrm{support} ( D(u )) \}}   \bigg[    \sigma^-_{j^{\prime}} + i u \sigma^+_{j^{\prime}} \sigma^-_{j^{\prime}}           \bigg] \text{ }              \bigg]   \bigg[ \underset{j \equiv N-1}{\prod}  i u^{-1} \sigma^+_j \sigma^-_j \sigma^+_{j+1} \bigg]              \end{align*}

   \begin{align*}
   +       \underset{1 \leq j \leq N-1}{\prod}   \sigma^+_j \bigg[     \textbf{1}_{\{ j^{\prime} > j \text{ } : \text{ } i u \sigma^+_{j^{\prime}} \sigma^-_{j^{\prime} \in \mathrm{support} (D ( u )) } \}}     \bigg[   i u \sigma^+_{j^{\prime}} \sigma^-_{j^{\prime}} \sigma^+_{j^{\prime}+1} + \sigma^+_{j^{\prime}+1}         \bigg]  \text{ } \bigg]             \bigg[ \underset{j \equiv N-1}{\prod}  \sigma^-_{j} i u^{-1} \sigma^+_{j+1} \sigma^-_{j+1} \bigg] \\ +         \underset{1 \leq j \leq N-1}{\prod}   i u \sigma^+_j \sigma^-_j \bigg[ \textbf{1}_{\{ j^{\prime} > j \text{ } : \text{ } i u \sigma^+_{j^{\prime}} \sigma^-_{j^{\prime}} \in \mathrm{support} ( D( u )) \}}    \bigg[    i u \sigma^+_{j^{\prime}} \sigma^-_{j^{\prime}} + \sigma^-_{j^{\prime}} \bigg] \text{ }    \bigg]        \bigg[  \underset{j \equiv N}{\prod} \sigma^+_j \sigma^-_j \sigma^+_{j+1} \bigg]    \\ +  \underset{1 \leq j \leq N-4}{\prod}          \sigma^+_j                   i u \sigma^+_j  \sigma^-_j i u \sigma^+_{j+1} \sigma^-_{j+1} \sigma^+_{j+2} \sigma^-_{j+2} \sigma^+_{j+3} +                   \underset{1 \leq j \leq N-1}{\prod}  \sigma^+_j     \bigg[            \textbf{1}_{\{ j^{\prime} > j \text{ } : \text{ } i u \sigma^+_{j^{\prime}} \sigma^-_{j^{\prime}} \in \mathrm{support} ( D ( u ))  \}}           \\ \times       \bigg[ i u \sigma^+_{j^{\prime}} \sigma^-_{j^{\prime}} \sigma^+_{j^{\prime}+1} + i u  \sigma^+_{j^{\prime}+1} \sigma^-_{j^{\prime}+1}\bigg]  \text{ }    \bigg] \bigg[        \underset{j \equiv N-1}{\prod}    \sigma^-_j i u^{-1} \sigma^+_{j+1} \sigma^-_{j+1}  \bigg] \text{ }                    \bigg]       \text{. }
 \end{align*}

  \noindent The fourth term takes the form,

  \begin{align*}
   \mathscr{I}^4_4 \approx          \sigma^-_0 \sigma^+_1 \bigg[   i u \sigma^+_1 \sigma^-_1 \bigg[   \underset{0 \leq j \leq N-1}{\prod}   \bigg[\textbf{1}_{\{ j^{\prime} > j \text{ } : \text{ } i u \sigma^+_{j^{\prime}} \sigma^-_{j^{\prime}} \in \mathrm{support} ( D( u )) \}}      \bigg[           i u \sigma^+_{j^{\prime}+1} \sigma^-_{j^{\prime}+1} + \sigma^-_{j^{\prime}+1} \sigma^-_{j^{\prime}+2 } \bigg] \text{ }            \bigg] \bigg[ \underset{j \equiv N}{\prod}    \sigma^+_j      \bigg]  \text{ } \bigg]   + \sigma^-_0 \sigma^+_1 \sigma^-_1  \\ \times    \bigg[   \underset{2 \leq j \leq N-2}{\prod}  i u \sigma^+_j \sigma^-_j i u \sigma^+_{j+1} \sigma^-_{j+1}     \bigg]    \bigg[                         \underset{j \equiv N}{\prod}   \sigma^+_j             \bigg]  \text{ }       \bigg]   \text{. }
 \end{align*}

  \noindent The fifth term takes the form,

  \begin{align*}
   \mathscr{I}^4_5 \approx   i u \sigma^+_0 \sigma^-_0        \bigg[  \underset{1 \leq j \leq N-2}{\prod} i u \sigma^+_j \sigma^-_j             \bigg[   \textbf{1}_{\{ j^{\prime} > j \text{ } : \text{ } i u \sigma^+_{j^{\prime}} \sigma^-_{j^{\prime}} \in \mathrm{support} ( D ( u )) \}}    \bigg[            i u \sigma^+_{j^{\prime}} \sigma^-_{j^{\prime}}  + \sigma^+_{j^{\prime}} \bigg]  \text{ }    \bigg] \text{ }          \bigg]         \bigg[ \underset{N-1 \leq j \leq N}{\prod}  \sigma^+_{j} \sigma^+_{j+1} \bigg]     \\ +      \underset{1 \leq j \leq N-2}{\prod}               i u \sigma^+_j \sigma^-_j \bigg[ \textbf{1}_{\{ j^{\prime} > j \text{ } : \text{ } i u \sigma^+_{j^{\prime}}  \sigma^-_{j^{\prime}} \in \mathrm{support} ( D ( u ))  \}}         \bigg[  \sigma^-_{j^{\prime}} + i u \sigma^+_{j^{\prime}} \sigma^-_{j^{\prime}} \bigg] \text{ }         \bigg] \bigg[ \underset{j \equiv N}{\prod}  i u^{-1} \sigma^+_j \sigma^-_j \sigma^+_{j+1} \bigg] \text{ }   \\ +  \underset{1 \leq j \leq N-3}{\prod}  \sigma^-_j \sigma^+_j i u^{-1} \sigma^+_{j+1} \sigma^-_{j+1} \sigma^+_{j+2} \sigma^-_{j+2} \sigma^+_{j+3}  + \underset{1 \leq j \leq N-3}{\prod} i u \sigma^+_j \sigma^-_j i u \sigma^+_{j+1} \sigma^-_{j+1} i u \sigma^+_{j+2} \sigma^-_{j+2} \sigma^+_{j+3} \bigg]        \text{. }
 \end{align*}

 \noindent The desired representation for the fourth entry of the transfer matrix takes the form,

\begin{align*}
\underline{\mathcal{I}^{\prime\cdots\prime}_4}  =  \mathscr{I}^4_1 + \mathscr{I}^4_2 + \mathscr{I}^4_3 + \mathscr{I}^4_4  + \mathscr{I}^4_5 \text{, }
\end{align*}

\noindent from which we conclude the argument. \boxed{}

\subsection{Theorem}

\noindent \textit{Proof of Theorem}. It suffices to check that the approximation for each Poisson bracket above holds by direct computation. From the first bracket appearing from the 4-vertex Poisson structure, 

\begin{align*}
         \big\{ \underline{\mathcal{I}^{\prime\cdots\prime}_1 \big( u , u^{-1} \big)}, \underline{\mathcal{I}^{\prime\cdots\prime}_1 \big( u^{\prime} , \big( u^{\prime} \big)^{-1} \big)}\big\} \text{,} \end{align*}

         \noindent in order, one has,
         
         \begin{align*}
         \bigg\{     i u \sigma^+_0 \sigma^-_0   \bigg[    \underset{j \text{ } \mathrm{odd}}{\underset{1 \leq j \leq N-1}{\prod}}     i u \sigma^+_j \sigma^-_j \textbf{1}_{\{j^{\prime}> j \text{ } : \text{ } i u\sigma^+_{j^{\prime}} \sigma^-_{j^{\prime} \in \mathrm{support} ( A ( u ) ) } \}} i u \sigma^+_{j^{\prime}}   \sigma^-_{j^{\prime}} i u \sigma^+_{j^{\prime}+1} \sigma^-_{j^{\prime}+1}  \bigg]               ,       i u^{\prime} \sigma^+_0 \sigma^-_0    \bigg[    \underset{j \text{ } \mathrm{odd}}{\underset{1 \leq j \leq N-1}{\prod}}   \\ \times   i u^{\prime} \sigma^+_j \sigma^-_j \textbf{1}_{\{j^{\prime}> j \text{ } : \text{ } i u^{\prime} \sigma^+_{j^{\prime}} \sigma^-_{j^{\prime} \in \mathrm{support} ( A ( u ) ) } \}} i u^{\prime} \sigma^+_{j^{\prime}} \sigma^-_{j^{\prime}} i u^{\prime} \\ \times \sigma^+_{j^{\prime}+1} \sigma^-_{j^{\prime}+1}  \bigg]                 \bigg\}  + \bigg\{   i u \sigma^+_0 \sigma^-_0   \bigg[    \underset{j \text{ } \mathrm{odd}}{\underset{1 \leq j \leq N-1}{\prod}}     i u \sigma^+_j \sigma^-_j \textbf{1}_{\{j^{\prime}> j \text{ } : \text{ } i u\sigma^+_{j^{\prime}} \sigma^-_{j^{\prime} \in \mathrm{support} ( A ( u ) ) } \}} i u \sigma^+_{j^{\prime}} \sigma^-_{j^{\prime}} \\ \times  i u \sigma^+_{j^{\prime}+1} \sigma^-_{j^{\prime}+1}  \bigg] ,  i u^{\prime} \sigma^+_1 \sigma^-_1 \sigma^+_2 \bigg[     \underset{j \text{ } \mathrm{odd}}{\underset{1 \leq j \leq N}{\prod}}          i u^{\prime} \sigma^+_j \sigma^-_j                                    \bigg]  \bigg[    \underset{j \equiv N}{\prod}   \sigma^+_j \sigma^-_j     \bigg]                     \bigg\} + \bigg\{       i u \sigma^+_0 \sigma^-_0   \bigg[    \underset{j \text{ } \mathrm{odd}}{\underset{1 \leq j \leq N-1}{\prod}}     i u \sigma^+_j \sigma^-_j \\  \times \textbf{1}_{\{j^{\prime}> j \text{ } : \text{ } i u\sigma^+_{j^{\prime}} \sigma^-_{j^{\prime} \in \mathrm{support} ( A ( u ) ) } \}} i u \sigma^+_{j^{\prime}} \sigma^-_{j^{\prime}}   i u \sigma^+_{j^{\prime}+1} \sigma^-_{j^{\prime}+1}  \bigg]          ,              \bigg[ \underset{j \text{ } \mathrm{even}}{\underset{1 \leq j \leq N-1}{\prod}}           i u^{\prime}  \sigma^+_j \sigma^-_j       \bigg]   \bigg[ \underset{j \text{ } \mathrm{odd}}{\underset{3 \leq j\leq N-1}{\prod}}           i \big( u^{\prime} \big)^{-1} \sigma^+_j \sigma^-_j            \bigg]        \\  \times          \bigg[        \underset{j \equiv N}{\prod}  i  u^{\prime}  \sigma^+_j \sigma^-_j  \bigg]                  \bigg\}  + \bigg\{    i u \sigma^+_1 \sigma^-_1 \sigma^+_2 \bigg[ \underset{j \text{ } \mathrm{odd}}{\underset{1 \leq j \leq N}{\prod}}          i u \sigma^+_j \sigma^-_j  \bigg]     \bigg[ \underset{j \equiv N}{\prod}           \sigma^+_j \sigma^-_j     \bigg]  ,      i u^{\prime} \sigma^+_0 \sigma^-_0   \bigg[    \underset{j \text{ } \mathrm{odd}}{\underset{1 \leq j \leq N-1}{\prod}}     i u^{\prime} \sigma^+_j \sigma^-_j  \\  \times \textbf{1}_{\{j^{\prime}> j \text{ } : \text{ } i u^{\prime} \sigma^+_{j^{\prime}} \sigma^-_{j^{\prime} \in \mathrm{support} ( A ( u ) ) } \}}   i u^{\prime} \sigma^+_{j^{\prime}} \sigma^-_{j^{\prime}} i u^{\prime} \sigma^+_{j^{\prime}+1} \sigma^-_{j^{\prime}+1}  \bigg]         \bigg\} + \bigg\{         i u \sigma^+_1 \sigma^-_1 \sigma^+_2 \bigg[ \underset{j \text{ } \mathrm{odd}}{\underset{1 \leq j \leq N}{\prod}}          i u \sigma^+_j \sigma^-_j  \bigg]     \bigg[ \underset{j \equiv N}{\prod}           \sigma^+_j \sigma^-_j     \bigg] 
     ,          i u^{\prime} \sigma^+_1 \sigma^-_1 \\  \times  \sigma^+_2 \bigg[ \underset{j \text{ } \mathrm{odd}}{\underset{1 \leq j \leq N}{\prod}}          i u^{\prime} \sigma^+_j \sigma^-_j  \bigg]     \bigg[ \underset{j \equiv N}{\prod}           \sigma^+_j \sigma^-_j     \bigg]      \bigg\}  + \bigg\{         i u \sigma^+_1 \sigma^-_1 \sigma^+_2 \bigg[ \underset{j \text{ } \mathrm{odd}}{\underset{1 \leq j \leq N}{\prod}}          i u \sigma^+_j \sigma^-_j  \bigg]     \bigg[ \underset{j \equiv N}{\prod}           \sigma^+_j \sigma^-_j     \bigg] 
     ,   \\  \bigg[ \underset{j \text{ } \mathrm{even}}{\underset{1 \leq j \leq N-1}{\prod}}           i u^{\prime}  \sigma^+_j \sigma^-_j       \bigg]   \bigg[ \underset{j \text{ } \mathrm{odd}}{\underset{3 \leq j\leq N-1}{\prod}}           i \big( u^{\prime} \big)^{-1} \sigma^+_j \sigma^-_j            \bigg]               \bigg[        \underset{j \equiv N}{\prod}  i  u^{\prime}  \sigma^+_j \sigma^-_j  \bigg]                      \bigg\}  \\   + \bigg\{  \bigg[ \underset{j \text{ } \mathrm {odd}}{\underset{1 \leq j \leq N-1}{\prod}} i u \sigma^+_j \sigma^-_j  \bigg]    \bigg[  \underset{j \text{ } \mathrm{odd}}{\underset{3 \leq j \leq N-1}{\prod}}       i u^{-1} \sigma^+_j \sigma^-_j      \bigg] \bigg[ \underset{j \equiv N}{\prod}  i u \sigma^+_j \sigma^-_j  \bigg]               
     ,        i u^{\prime} \sigma^+_0 \sigma^-_0   \bigg[    \underset{j \text{ } \mathrm{odd}}{\underset{1 \leq j \leq N-1}{\prod}}     i u^{\prime} \sigma^+_j \sigma^-_j \textbf{1}_{\{j^{\prime}> j \text{ } : \text{ } i u^{\prime} \sigma^+_{j^{\prime}} \sigma^-_{j^{\prime} \in \mathrm{support} ( A ( u ) ) } \}} \\ \times  i u^{\prime} \sigma^+_{j^{\prime}} \sigma^-_{j^{\prime}} i u^{\prime} \sigma^+_{j^{\prime}+1} \sigma^-_{j^{\prime}+1}  \bigg]              \bigg\} \\ + \bigg\{  \bigg[ \underset{j \text{ } \mathrm {odd}}{\underset{1 \leq j \leq N-1}{\prod}} i u \sigma^+_j \sigma^-_j  \bigg]    \bigg[  \underset{j \text{ } \mathrm{odd}}{\underset{3 \leq j \leq N-1}{\prod}}       i u^{-1} \sigma^+_j \sigma^-_j      \bigg] \bigg[ \underset{j \equiv N}{\prod}  i u \sigma^+_j \sigma^-_j  \bigg]               
     ,    i u^{\prime} \sigma^+_1 \sigma^-_1 \sigma^+_2 \bigg[ \underset{j \text{ } \mathrm{odd}}{\underset{1 \leq j\leq N}{\prod}}                 i u \sigma^+_j \sigma^-_j    \bigg] \bigg[ \underset{j \equiv N}{\prod} \sigma^+_j \sigma^-_j  \bigg]       \bigg\}  \\ + \bigg\{  \bigg[ \underset{j \text{ } \mathrm {odd}}{\underset{1 \leq j \leq N-1}{\prod}} i u \sigma^+_j \sigma^-_j  \bigg]    \bigg[  \underset{j \text{ } \mathrm{odd}}{\underset{3 \leq j \leq N-1}{\prod}}       i u^{-1} \sigma^+_j \sigma^-_j      \bigg] \bigg[ \underset{j \equiv N}{\prod}  i u \sigma^+_j \sigma^-_j  \bigg]               
     ,    \bigg[ \underset{j \text{ } \mathrm {odd}}{\underset{1 \leq j \leq N-1}{\prod}} i u^{\prime} \sigma^+_j \sigma^-_j  \bigg]    \bigg[  \underset{j \text{ } \mathrm{odd}}{\underset{3 \leq j \leq N-1}{\prod}}       i \big(  u^{\prime} \big)^{-1} \sigma^+_j \sigma^-_j      \bigg]  \\ \times           \bigg[ \underset{j \equiv N}{\prod}  i u^{\prime} \sigma^+_j \sigma^-_j  \bigg]         \bigg\}   \text{. }  \end{align*}

\noindent For first above,

\begin{align*}
         \bigg\{     i u \sigma^+_0 \sigma^-_0   \bigg[    \underset{j \text{ } \mathrm{odd}}{\underset{1 \leq j \leq N-1}{\prod}}     i u \sigma^+_j \sigma^-_j \textbf{1}_{\{j^{\prime}> j \text{ } : \text{ } i u\sigma^+_{j^{\prime}} \sigma^-_{j^{\prime} \in \mathrm{support} ( A ( u ) ) } \}} i u \sigma^+_{j^{\prime}} \sigma^-_{j^{\prime}} i u \sigma^+_{j^{\prime}+1} \sigma^-_{j^{\prime}+1}  \bigg]               ,       i u^{\prime} \sigma^+_0 \sigma^-_0   \bigg[    \underset{j \text{ } \mathrm{odd}}{\underset{1 \leq j \leq N-1}{\prod}}     i u^{\prime} \sigma^+_j \end{align*}

      \begin{align*}  \times  \sigma^-_j \textbf{1}_{\{j^{\prime}> j \text{ } : \text{ } i u^{\prime} \sigma^+_{j^{\prime}} \sigma^-_{j^{\prime} \in \mathrm{support} ( A ( u ) ) } \}} i u^{\prime} \sigma^+_{j^{\prime}} \sigma^-_{j^{\prime}} i u^{\prime}  \sigma^+_{j^{\prime}+1} \sigma^-_{j^{\prime}+1}  \bigg]                 \bigg\} \text{, } \end{align*}

\noindent corresponding to the first term, to which an application of (LR) yields the superposition,
         
         \begin{align*}
         \overset{(\mathrm{LR})}{=}     i u \sigma^+_0 \sigma^-_0      \bigg\{   \bigg[    \underset{j \text{ } \mathrm{odd}}{\underset{1 \leq j \leq N-1}{\prod}}     i u \sigma^+_j \sigma^-_j \textbf{1}_{\{j^{\prime}> j \text{ } : \text{ } i u\sigma^+_{j^{\prime}} \sigma^-_{j^{\prime} \in \mathrm{support} ( A ( u ) ) } \}} i u \sigma^+_{j^{\prime}} \sigma^-_{j^{\prime}} i u \sigma^+_{j^{\prime}+1} \sigma^-_{j^{\prime}+1}  \bigg]               ,       i u^{\prime} \sigma^+_0 \sigma^-_0   \bigg[    \underset{j \text{ } \mathrm{odd}}{\underset{1 \leq j \leq N-1}{\prod}}     i u^{\prime} \sigma^+_j \end{align*}

 \begin{align*}  \times  \sigma^-_j \textbf{1}_{\{j^{\prime}> j \text{ } : \text{ } i u^{\prime} \sigma^+_{j^{\prime}} \sigma^-_{j^{\prime} \in \mathrm{support} ( A ( u ) ) } \}} i u^{\prime} \sigma^+_{j^{\prime}} \sigma^-_{j^{\prime}} i u^{\prime}  \sigma^+_{j^{\prime}+1} \sigma^-_{j^{\prime}+1}  \bigg]                 \bigg\}  +  \bigg\{     i u \sigma^+_0 \sigma^-_0                 ,       i u^{\prime} \sigma^+_0 \sigma^-_0   \bigg[    \underset{j \text{ } \mathrm{odd}}{\underset{1 \leq j \leq N-1}{\prod}}     i u^{\prime} \sigma^+_j   \sigma^-_j  \\ \times \textbf{1}_{\{j^{\prime}> j \text{ } : \text{ } i u^{\prime} \sigma^+_{j^{\prime}} \sigma^-_{j^{\prime} \in \mathrm{support} ( A ( u ) ) } \}}    i u^{\prime} \sigma^+_{j^{\prime}} \sigma^-_{j^{\prime}} i u^{\prime}  \sigma^+_{j^{\prime}+1} \sigma^-_{j^{\prime}+1}  \bigg]                 \bigg\}  \bigg[    \underset{j \text{ } \mathrm{odd}}{\underset{1 \leq j \leq N-1}{\prod}}     i u  \sigma^+_j \sigma^-_j \textbf{1}_{\{j^{\prime}> j \text{ } : \text{ } i u\sigma^+_{j^{\prime}} \sigma^-_{j^{\prime} \in \mathrm{support} ( A ( u ) ) } \}} i u \sigma^+_{j^{\prime}} \\ \times  \sigma^-_{j^{\prime}} i u \sigma^+_{j^{\prime}+1} \sigma^-_{j^{\prime}+1}  \bigg]                   \text{. } \end{align*}

\noindent Applying (AC) to the superposition above yields,

 \begin{align*}
  -   i u \sigma^+_0 \sigma^-_0      \bigg\{          i u^{\prime} \sigma^+_0 \sigma^-_0   \bigg[    \underset{j \text{ } \mathrm{odd}}{\underset{1 \leq j \leq N-1}{\prod}}     i u^{\prime} \sigma^+_j  
         \sigma^-_j \textbf{1}_{\{j^{\prime}> j \text{ } : \text{ } i u^{\prime} \sigma^+_{j^{\prime}} \sigma^-_{j^{\prime} \in \mathrm{support} ( A ( u ) ) } \}} i u^{\prime} \sigma^+_{j^{\prime}} \sigma^-_{j^{\prime}} i   u^{\prime}  \sigma^+_{j^{\prime}+1} \sigma^-_{j^{\prime}+1}  \bigg]   ,  \bigg[    \underset{j \text{ } \mathrm{odd}}{\underset{1 \leq j \leq N-1}{\prod}}      i u^{\prime} \sigma^+_j \sigma^-_j \\    \times  \textbf{1}_{\{j^{\prime}> j \text{ } : \text{ } i u\sigma^+_{j^{\prime}} \sigma^-_{j^{\prime} \in \mathrm{support} ( A ( u^{\prime} ) ) } \}} i u^{\prime} \sigma^+_{j^{\prime}} \sigma^-_{j^{\prime}} i u^{\prime} \sigma^+_{j^{\prime}+1} \sigma^-_{j^{\prime}+1}  \bigg]                            \bigg\}  \text{, } \end{align*}
         
         \noindent corresponding to the first term,

         \begin{align*} -  \bigg\{        i u \sigma^+_0 \sigma^-_0   \bigg[    \underset{j \text{ } \mathrm{odd}}{\underset{1 \leq j \leq N-1}{\prod}}     i u^{\prime} \sigma^+_j   \sigma^-_j   \textbf{1}_{\{j^{\prime}> j \text{ } : \text{ } i u^{\prime} \sigma^+_{j^{\prime}} \sigma^-_{j^{\prime} \in \mathrm{support} ( A ( u ) ) } \}}  \\ \times   i u \sigma^+_{j^{\prime}} \sigma^-_{j^{\prime}} i u  \sigma^+_{j^{\prime}+1} \sigma^-_{j^{\prime}+1}  \bigg]  ,    i u^{\prime} \sigma^+_0 \sigma^-_0                                \bigg\}  \bigg[    \underset{j \text{ } \mathrm{odd}}{\underset{1 \leq j \leq N-1}{\prod}}     i u  \sigma^+_j \sigma^-_j \textbf{1}_{\{j^{\prime}> j \text{ } : \text{ } i u\sigma^+_{j^{\prime}} \sigma^-_{j^{\prime} \in \mathrm{support} ( A ( u ) ) } \}} i u \sigma^+_{j^{\prime}}   \sigma^-_{j^{\prime}} i u \sigma^+_{j^{\prime}+1} \sigma^-_{j^{\prime}+1}  \bigg] \text{, }  \end{align*}

         \noindent corresponding to the second term. Applying (AC) yields,

         \begin{align*}
         -   i u \sigma^+_0 \sigma^-_0      \bigg\{          i u^{\prime} \sigma^+_0 \sigma^-_0     ,  \bigg[    \underset{j \text{ } \mathrm{odd}}{\underset{1 \leq j \leq N-1}{\prod}}      i u^{\prime} \sigma^+_j \sigma^-_j \textbf{1}_{\{j^{\prime}> j \text{ } : \text{ } i u^{\prime}\sigma^+_{j^{\prime}} \sigma^-_{j^{\prime}} \in \mathrm{support} ( A ( u^{\prime} ) )  \}} i u \sigma^+_{j^{\prime}} \sigma^-_{j^{\prime}} i u^{\prime} \sigma^+_{j^{\prime}+1} \sigma^-_{j^{\prime}+1}  \bigg]                            \bigg\} \\ \times    \bigg[    \underset{j \text{ } \mathrm{odd}}{\underset{1 \leq j \leq N-1}{\prod}}     i u^{\prime} \sigma^+_j  
         \sigma^-_j \textbf{1}_{\{j^{\prime}> j \text{ } : \text{ } i u^{\prime} \sigma^+_{j^{\prime}} \sigma^-_{j^{\prime} \in \mathrm{support} ( A ( u ) ) } \}} i u^{\prime} \sigma^+_{j^{\prime}}  \sigma^-_{j^{\prime}} i   u^{\prime}  \sigma^+_{j^{\prime}+1} \sigma^-_{j^{\prime}+1}  \bigg]      -   i \big(  u  + u^{\prime} \big) \sigma^+_0 \sigma^-_0   \\   \times  \bigg\{            \bigg[    \underset{j \text{ } \mathrm{odd}}{\underset{1 \leq j \leq N-1}{\prod}}     i u^{\prime} \sigma^+_j  
         \sigma^-_j \textbf{1}_{\{j^{\prime}> j \text{ } : \text{ } i u^{\prime} \sigma^+_{j^{\prime}} \sigma^-_{j^{\prime}} \in \mathrm{support} ( A ( u ) )  \}} i u^{\prime} \sigma^+_{j^{\prime}} \sigma^-_{j^{\prime}} i   u^{\prime}  \sigma^+_{j^{\prime}+1} \sigma^-_{j^{\prime}+1}  \bigg]  \\     ,  \bigg[    \underset{j \text{ } \mathrm{odd}}{\underset{1 \leq j \leq N-1}{\prod}}      i u^{\prime} \sigma^+_j \sigma^-_j   \textbf{1}_{\{j^{\prime}> j \text{ } : \text{ } i u\sigma^+_{j^{\prime}} \sigma^-_{j^{\prime} \in \mathrm{support} ( A ( u ) ) } \}} i u \sigma^+_{j^{\prime}} \sigma^-_{j^{\prime}} i u \sigma^+_{j^{\prime}+1} \sigma^-_{j^{\prime}+1}  \bigg]                            \bigg\}              \\  -  i u^{\prime} \sigma^+_0 \sigma^-_0  \bigg\{           \bigg[    \underset{j \text{ } \mathrm{odd}}{\underset{1 \leq j \leq N-1}{\prod}}     i u^{\prime} \sigma^+_j   \sigma^-_j   \textbf{1}_{\{j^{\prime}> j \text{ } : \text{ } i u^{\prime} \sigma^+_{j^{\prime}} \sigma^-_{j^{\prime} \in \mathrm{support} ( A ( u ) ) } \}}   i u^{\prime} \sigma^+_{j^{\prime}} \sigma^-_{j^{\prime}} i u^{\prime}  \sigma^+_{j^{\prime}+1} \sigma^-_{j^{\prime}+1}  \bigg]  ,    i u \sigma^+_0 \sigma^-_0                                \bigg\} \end{align*}

      \begin{align*}  \times  \bigg[    \underset{j \text{ } \mathrm{odd}}{\underset{1 \leq j \leq N-1}{\prod}}     i u  \sigma^+_j \sigma^-_j \textbf{1}_{\{j^{\prime}> j \text{ } : \text{ } i u\sigma^+_{j^{\prime}} \sigma^-_{j^{\prime} \in \mathrm{support} ( A ( u ) ) } \}} i u \sigma^+_{j^{\prime}}   \sigma^-_{j^{\prime}} i u \sigma^+_{j^{\prime}+1} \sigma^-_{j^{\prime}+1}  \bigg] \\  -   \bigg\{        i u^{\prime} \sigma^+_0 \sigma^-_0    ,    i u \sigma^+_0 \sigma^-_0                                \bigg\}  \bigg[    \underset{j \text{ } \mathrm{odd}}{\underset{1 \leq j \leq N-1}{\prod}}     i u^{\prime} \sigma^+_j   \sigma^-_j   \textbf{1}_{\{j^{\prime}> j \text{ } : \text{ } i u^{\prime} \sigma^+_{j^{\prime}} \sigma^-_{j^{\prime} \in \mathrm{support} ( A ( u ) ) } \}}   i u^{\prime} \sigma^+_{j^{\prime}} \sigma^-_{j^{\prime}} i u^{\prime}  \sigma^+_{j^{\prime}+1} \sigma^-_{j^{\prime}+1}  \bigg]  \\ \times  \bigg[    \underset{j \text{ } \mathrm{odd}}{\underset{1 \leq j \leq N-1}{\prod}}     i u  \sigma^+_j \sigma^-_j \textbf{1}_{\{j^{\prime}> j \text{ } : \text{ } i u\sigma^+_{j^{\prime}} \sigma^-_{j^{\prime} \in \mathrm{support} ( A ( u ) ) } \}} i u \sigma^+_{j^{\prime}}   \sigma^-_{j^{\prime}} i u \sigma^+_{j^{\prime}+1} \sigma^-_{j^{\prime}+1}  \bigg]      \text{, }
\end{align*}

\noindent corresponding to the first term, which can be approximated with,

    \begin{align*}
           \big( \mathscr{C}^1_1 \big)_1   \text{, }
    \end{align*}

    \noindent For the next term, one has,

\begin{align*}
  \bigg\{   i u \sigma^+_0 \sigma^-_0   \bigg[    \underset{j \text{ } \mathrm{odd}}{\underset{1 \leq j \leq N-1}{\prod}}     i u \sigma^+_j \sigma^-_j \textbf{1}_{\{j^{\prime}> j \text{ } : \text{ } i u\sigma^+_{j^{\prime}} \sigma^-_{j^{\prime} \in \mathrm{support} ( A ( u ) ) } \}} i u \sigma^+_{j^{\prime}} \sigma^-_{j^{\prime}}  i u \sigma^+_{j^{\prime}+1} \sigma^-_{j^{\prime}+1}  \bigg] ,  i u^{\prime} \sigma^+_1 \sigma^-_1 \sigma^+_2 \bigg[     \underset{j \text{ } \mathrm{odd}}{\underset{1 \leq j \leq N}{\prod}}          i u^{\prime} \sigma^+_j \sigma^-_j                                    \bigg]  \\ \times \bigg[    \underset{j \equiv N}{\prod}   \sigma^+_j \sigma^-_j     \bigg]                     \bigg\}  \overset{(\mathrm{LR})}{=}              i u \sigma^+_0 \sigma^-_0       \bigg\{    \bigg[    \underset{j \text{ } \mathrm{odd}}{\underset{1 \leq j \leq N-1}{\prod}}     i u \sigma^+_j \sigma^-_j \textbf{1}_{\{j^{\prime}> j \text{ } : \text{ } i u\sigma^+_{j^{\prime}} \sigma^-_{j^{\prime} \in \mathrm{support} ( A ( u ) ) } \}} i u \sigma^+_{j^{\prime}} \sigma^-_{j^{\prime}}  i u \sigma^+_{j^{\prime}+1} \sigma^-_{j^{\prime}+1}  \bigg]  \\ ,  i u^{\prime} \sigma^+_1 \sigma^-_1 \sigma^+_2 \bigg[     \underset{j \text{ } \mathrm{odd}}{\underset{1 \leq j \leq N}{\prod}}          i u^{\prime} \sigma^+_j \sigma^-_j                                    \bigg]  \bigg[    \underset{j \equiv N}{\prod}   \sigma^+_j \sigma^-_j     \bigg]                     \bigg\}  +   \bigg\{   i u \sigma^+_0 \sigma^-_0     ,  i u^{\prime} \sigma^+_1 \sigma^-_1 \sigma^+_2 \bigg[     \underset{j \text{ } \mathrm{odd}}{\underset{1 \leq j \leq N}{\prod}}          i u^{\prime} \sigma^+_j \sigma^-_j                                    \bigg]   \bigg[    \underset{j \equiv N}{\prod}   \sigma^+_j \sigma^-_j     \bigg]                     \bigg\}  \\  \times    \bigg[    \underset{j \text{ } \mathrm{odd}}{\underset{1 \leq j \leq N-1}{\prod}}     i u \sigma^+_j \sigma^-_j \textbf{1}_{\{j^{\prime}> j \text{ } : \text{ } i u\sigma^+_{j^{\prime}} \sigma^-_{j^{\prime} \in \mathrm{support} ( A ( u ) ) } \}}    i u \sigma^+_{j^{\prime}} \sigma^-_{j^{\prime}}  i u \sigma^+_{j^{\prime}+1}  \sigma^-_{j^{\prime}+1}  \bigg]              \\ \\  \overset{(\mathrm{AC})}{=}   -       i u^{\prime} \sigma^+_1 \sigma^-_1 \sigma^+_2       \bigg\{   \bigg[     \underset{j \text{ } \mathrm{odd}}{\underset{1 \leq j \leq N}{\prod}}          i u^{\prime} \sigma^+_j \sigma^-_j                                    \bigg]  \bigg[    \underset{j \equiv N}{\prod}   \sigma^+_j \sigma^-_j     \bigg]                 \bigg[    \underset{j \text{ } \mathrm{odd}}{\underset{1 \leq j \leq N-1}{\prod}}    i u \sigma^+_j \sigma^-_j \textbf{1}_{\{j^{\prime}> j \text{ } : \text{ } i u\sigma^+_{j^{\prime}} \sigma^-_{j^{\prime} \in \mathrm{support} ( A ( u ) ) } \}} \\  \times  i u \sigma^+_{j^{\prime}} \sigma^-_{j^{\prime}}  i u \sigma^+_{j^{\prime}+1} \sigma^-_{j^{\prime}+1}  \bigg]   ,   i u \sigma^+_0 \sigma^-_0   \bigg\}   -   \bigg\{   i u^{\prime} \sigma^+_1 \sigma^-_1 \sigma^+_2 \bigg[     \underset{j \text{ } \mathrm{odd}}{\underset{1 \leq j \leq N}{\prod}}          i u^{\prime} \sigma^+_j \sigma^-_j                                    \bigg]     \\  \times    \bigg[    \underset{j \equiv N}{\prod}   \sigma^+_j \sigma^-_j     \bigg]    ,  i u \sigma^+_0 \sigma^-_0                     \bigg\}    \bigg[    \underset{j \text{ } \mathrm{odd}}{\underset{1 \leq j \leq N-1}{\prod}}     i u \sigma^+_j \sigma^-_j \textbf{1}_{\{j^{\prime}> j \text{ } : \text{ } i u\sigma^+_{j^{\prime}} \sigma^-_{j^{\prime} \in \mathrm{support} ( A ( u ) ) } \}}    i u \sigma^+_{j^{\prime}} \sigma^-_{j^{\prime}}  i u \sigma^+_{j^{\prime}+1}  \sigma^-_{j^{\prime}+1}  \bigg]           \\ \\ \overset{(\mathrm{LR})}{=}   -     i u^{\prime} \sigma^+_1 \sigma^-_1 \sigma^+_2     \bigg[     \underset{j \text{ } \mathrm{odd}}{\underset{1 \leq j \leq N}{\prod}}          i u^{\prime} \sigma^+_j \sigma^-_j                                    \bigg]  \bigg[    \underset{j \equiv N}{\prod}   \sigma^+_j \sigma^-_j     \bigg]     \bigg\{                  \bigg[    \underset{j \text{ } \mathrm{odd}}{\underset{1 \leq j \leq N-1}{\prod}}    i u \sigma^+_j \sigma^-_j \textbf{1}_{\{j^{\prime}> j \text{ } : \text{ } i u\sigma^+_{j^{\prime}} \sigma^-_{j^{\prime} \in \mathrm{support} ( A ( u ) ) } \}} \\  \times  i u \sigma^+_{j^{\prime}} \sigma^-_{j^{\prime}}  i u \sigma^+_{j^{\prime}+1} \sigma^-_{j^{\prime}+1}  \bigg]   ,   i u \sigma^+_0 \sigma^-_0   \bigg\}                   \\     -       i u^{\prime} \sigma^+_1 \sigma^-_1 \sigma^+_2       \bigg\{   \bigg[     \underset{j \text{ } \mathrm{odd}}{\underset{1 \leq j \leq N}{\prod}}          i u^{\prime} \sigma^+_j \sigma^-_j                                    \bigg]  \bigg[    \underset{j \equiv N}{\prod}   \sigma^+_j \sigma^-_j     \bigg]       ,   i u \sigma^+_0 \sigma^-_0   \bigg\}             \bigg[    \underset{j \text{ } \mathrm{odd}}{\underset{1 \leq j \leq N-1}{\prod}}    i u \sigma^+_j \sigma^-_j \textbf{1}_{\{j^{\prime}> j \text{ } : \text{ } i u\sigma^+_{j^{\prime}} \sigma^-_{j^{\prime} \in \mathrm{support} ( A ( u ) ) } \}} \\ \times  i u \sigma^+_{j^{\prime}} \sigma^-_{j^{\prime}}  i u \sigma^+_{j^{\prime}+1} \sigma^-_{j^{\prime}+1}  \bigg]     \end{align*}

      \begin{align*}   -    i u^{\prime} \sigma^+_1 \sigma^-_1 \sigma^+_2 \bigg\{   \bigg[     \underset{j \text{ } \mathrm{odd}}{\underset{1 \leq j \leq N}{\prod}}          i u^{\prime} \sigma^+_j \sigma^-_j                                    \bigg]    \bigg[    \underset{j \equiv N}{\prod}   \sigma^+_j \sigma^-_j     \bigg]    ,  i u \sigma^+_0 \sigma^-_0                     \bigg\}    \bigg[    \underset{j \text{ } \mathrm{odd}}{\underset{1 \leq j \leq N-1}{\prod}}  i u \sigma^+_j \sigma^-_j \textbf{1}_{\{j^{\prime}> j \text{ } : \text{ } i u\sigma^+_{j^{\prime}} \sigma^-_{j^{\prime} \in \mathrm{support} ( A ( u ) ) } \}}    \\ \times     i u \sigma^+_{j^{\prime}} \sigma^-_{j^{\prime}}  i u \sigma^+_{j^{\prime}+1}  \sigma^-_{j^{\prime}+1}  \bigg]    -   \bigg\{   i u^{\prime} \sigma^+_1 \sigma^-_1 \sigma^+_2   ,  i u \sigma^+_0 \sigma^-_0                     \bigg\}    \bigg[    \underset{j \text{ } \mathrm{odd}}{\underset{1 \leq j \leq N-1}{\prod}}  i u \sigma^+_j \sigma^-_j \textbf{1}_{\{j^{\prime}> j \text{ } : \text{ } i u\sigma^+_{j^{\prime}} \sigma^-_{j^{\prime} \in \mathrm{support} ( A ( u ) ) } \}}    \\ \times    i u \sigma^+_{j^{\prime}} \sigma^-_{j^{\prime}}  i u \sigma^+_{j^{\prime}+1}  \sigma^-_{j^{\prime}+1}  \bigg]   \bigg[     \underset{j \text{ } \mathrm{odd}}{\underset{1 \leq j \leq N}{\prod}}          i u^{\prime} \sigma^+_j \sigma^-_j                                    \bigg]        \bigg[    \underset{j \equiv N}{\prod}   \sigma^+_j \sigma^-_j     \bigg]   \text{, }   \end{align*}

  \noindent after an application of (AC), and (LR). Another application of (LR) yields,
  
  \begin{align*}
        -     i u^{\prime} \sigma^+_1 \sigma^-_1 \sigma^+_2     \bigg[     \underset{j \text{ } \mathrm{odd}}{\underset{1 \leq j \leq N}{\prod}}          i u^{\prime} \sigma^+_j \sigma^-_j                                    \bigg]  \bigg[    \underset{j \equiv N}{\prod}   \sigma^+_j \sigma^-_j     \bigg]     \bigg\{                  \bigg[    \underset{j \text{ } \mathrm{odd}}{\underset{1 \leq j \leq N-1}{\prod}}    i u \sigma^+_j \sigma^-_j \textbf{1}_{\{j^{\prime}> j \text{ } : \text{ } i u\sigma^+_{j^{\prime}} \sigma^-_{j^{\prime} \in \mathrm{support} ( A ( u ) ) } \}} \\ 
        \times  i u \sigma^+_{j^{\prime}} \sigma^-_{j^{\prime}}  i u \sigma^+_{j^{\prime}+1} \sigma^-_{j^{\prime}+1}  \bigg]   ,   i u \sigma^+_0 \sigma^-_0   \bigg\}  \text{, }     \end{align*}

 \noindent corresponding to the first term,
 
 \begin{align*}        -    i u^{\prime} \sigma^+_1 \sigma^-_1 \sigma^+_2  \bigg[     \underset{j \text{ } \mathrm{odd}}{\underset{1 \leq j \leq N}{\prod}}          i u^{\prime} \sigma^+_j \sigma^-_j                                    \bigg]   \bigg\{    \bigg[    \underset{j \equiv N}{\prod}   \sigma^+_j \sigma^-_j     \bigg]    ,  i u \sigma^+_0 \sigma^-_0                     \bigg\}    \bigg[    \underset{j \text{ } \mathrm{odd}}{\underset{1 \leq j \leq N-1}{\prod}}  i u \sigma^+_j \sigma^-_j \textbf{1}_{\{j^{\prime}> j \text{ } : \text{ } i u\sigma^+_{j^{\prime}} \sigma^-_{j^{\prime} \in \mathrm{support} ( A ( u ) ) } \}}    \\ \times     i u \sigma^+_{j^{\prime}} \sigma^-_{j^{\prime}}  i u \sigma^+_{j^{\prime}+1}  \sigma^-_{j^{\prime}+1}  \bigg] \end{align*}

\noindent corresponding to the second term,
 
 \begin{align*}
 -    i u^{\prime} \sigma^+_1 \sigma^-_1 \sigma^+_2    \bigg\{   \bigg[     \underset{j \text{ } \mathrm{odd}}{\underset{1 \leq j \leq N}{\prod}}          i u^{\prime} \sigma^+_j \sigma^-_j                                    \bigg]      ,  i u \sigma^+_0 \sigma^-_0                     \bigg\}    \bigg[    \underset{j \text{ } \mathrm{odd}}{\underset{1 \leq j \leq N-1}{\prod}}  i u \sigma^+_j \sigma^-_j \textbf{1}_{\{j^{\prime}> j \text{ } : \text{ } i u\sigma^+_{j^{\prime}} \sigma^-_{j^{\prime} \in \mathrm{support} ( A ( u ) ) } \}}                  \\  \times     i u \sigma^+_{j^{\prime}} \sigma^-_{j^{\prime}}  i u \sigma^+_{j^{\prime}+1}  \sigma^-_{j^{\prime}+1}  \bigg]  \bigg[    \underset{j \equiv N}{\prod}   \sigma^+_j \sigma^-_j     \bigg]   
             \end{align*}

  \noindent corresponding to the third term,
  
  \begin{align*}   -    i u^{\prime} \sigma^+_1 \sigma^-_1 \sigma^+_2 \bigg\{   \bigg[     \underset{j \text{ } \mathrm{odd}}{\underset{1 \leq j \leq N}{\prod}}          i u^{\prime} \sigma^+_j \sigma^-_j                                    \bigg]     ,  i u \sigma^+_0 \sigma^-_0                     \bigg\}    \bigg[    \underset{j \equiv N}{\prod}   \sigma^+_j \sigma^-_j     \bigg]    \bigg[    \underset{j \text{ } \mathrm{odd}}{\underset{1 \leq j \leq N-1}{\prod}}  i u \sigma^+_j \sigma^-_j \textbf{1}_{\{j^{\prime}> j \text{ } : \text{ } i u\sigma^+_{j^{\prime}} \sigma^-_{j^{\prime} \in \mathrm{support} ( A ( u ) ) } \}}    \\ \times     i u \sigma^+_{j^{\prime}} \sigma^-_{j^{\prime}}  i u \sigma^+_{j^{\prime}+1}  \sigma^-_{j^{\prime}+1}  \bigg]     \end{align*}

  \noindent corresponding to the fourth term,
  
  \begin{align*}
 -    i u^{\prime} \sigma^+_1 \sigma^-_1 \sigma^+_2 \bigg\{   \bigg[     \underset{j \text{ } \mathrm{odd}}{\underset{1 \leq j \leq N}{\prod}}          i u^{\prime} \sigma^+_j \sigma^-_j                                    \bigg]     ,  i u \sigma^+_0 \sigma^-_0                     \bigg\}    \bigg[    \underset{j \equiv N}{\prod}   \sigma^+_j \sigma^-_j     \bigg]    \bigg[    \underset{j \text{ } \mathrm{odd}}{\underset{1 \leq j \leq N-1}{\prod}}  i u \sigma^+_j \sigma^-_j \textbf{1}_{\{j^{\prime}> j \text{ } : \text{ } i u\sigma^+_{j^{\prime}} \sigma^-_{j^{\prime} \in \mathrm{support} ( A ( u ) ) } \}}  \\  \times     i u \sigma^+_{j^{\prime}} \sigma^-_{j^{\prime}}  i u \sigma^+_{j^{\prime}+1}  \sigma^-_{j^{\prime}+1}  \bigg]   \end{align*}
 
 \noindent corresponding to the fifth term,
 
 \begin{align*}
 -   \bigg\{   i u^{\prime} \sigma^+_1 \sigma^-_1 \sigma^+_2   ,  i u \sigma^+_0 \sigma^-_0                     \bigg\}    \bigg[    \underset{j \text{ } \mathrm{odd}}{\underset{1 \leq j \leq N-1}{\prod}}  i u \sigma^+_j \sigma^-_j \textbf{1}_{\{j^{\prime}> j \text{ } : \text{ } i u\sigma^+_{j^{\prime}} \sigma^-_{j^{\prime} \in \mathrm{support} ( A ( u ) ) } \}}       i u \sigma^+_{j^{\prime}} \sigma^-_{j^{\prime}}  i u \sigma^+_{j^{\prime}+1}  \sigma^-_{j^{\prime}+1}  \bigg]   \\ \times  \bigg[     \underset{j \text{ } \mathrm{odd}}{\underset{1 \leq j \leq N}{\prod}}          i u^{\prime} \sigma^+_j \sigma^-_j                                    \bigg]        \bigg[    \underset{j \equiv N}{\prod}   \sigma^+_j \sigma^-_j     \bigg]              \text{, }
\end{align*}

\noindent corresponding to the second term, which can be approximated with,

    \begin{align*}
    \big( \mathscr{C}^1_1 \big)_2   \text{, }
    \end{align*}

    \noindent For the next term, one has,

\begin{align*}
  \bigg\{       i u \sigma^+_0 \sigma^-_0   \bigg[    \underset{j \text{ } \mathrm{odd}}{\underset{1 \leq j \leq N-1}{\prod}}     i u \sigma^+_j \sigma^-_j  \times \textbf{1}_{\{j^{\prime}> j \text{ } : \text{ } i u\sigma^+_{j^{\prime}} \sigma^-_{j^{\prime} \in \mathrm{support} ( A ( u ) ) } \}} i u \sigma^+_{j^{\prime}} \sigma^-_{j^{\prime}}   i u \sigma^+_{j^{\prime}+1} \sigma^-_{j^{\prime}+1}  \bigg]          ,              \bigg[ \underset{j \text{ } \mathrm{even}}{\underset{1 \leq j \leq N-1}{\prod}}           i u^{\prime}  \sigma^+_j \sigma^-_j       \bigg] \\ \times    \bigg[ \underset{j \text{ } \mathrm{odd}}{\underset{3 \leq j\leq N-1}{\prod}}           i \big( u^{\prime} \big)^{-1} \sigma^+_j \sigma^-_j            \bigg]              \bigg[        \underset{j \equiv N}{\prod}  i  u^{\prime}  \sigma^+_j \sigma^-_j  \bigg]                  \bigg\} \overset{(\mathrm{LR})}{=}                 i u \sigma^+_0 \sigma^-_0     \bigg\{       \bigg[    \underset{j \text{ } \mathrm{odd}}{\underset{1 \leq j \leq N-1}{\prod}}     i u \sigma^+_j \sigma^-_j  \textbf{1}_{\{j^{\prime}> j \text{ } : \text{ } i u\sigma^+_{j^{\prime}} \sigma^-_{j^{\prime} \in \mathrm{support} ( A ( u ) ) } \}}  \\  \times i u \sigma^+_{j^{\prime}} \sigma^-_{j^{\prime}}    i u \sigma^+_{j^{\prime}+1} \sigma^-_{j^{\prime}+1}  \bigg]          ,              \bigg[ \underset{j \text{ } \mathrm{even}}{\underset{1 \leq j \leq N-1}{\prod}}           i u^{\prime}  \sigma^+_j \sigma^-_j       \bigg]    \bigg[ \underset{j \text{ } \mathrm{odd}}{\underset{3 \leq j\leq N-1}{\prod}}           i \big( u^{\prime} \big)^{-1} \sigma^+_j \sigma^-_j            \bigg]              \bigg[        \underset{j \equiv N}{\prod}  i  u^{\prime}  \sigma^+_j \sigma^-_j  \bigg]                  \bigg\}  +   \bigg\{       i u \sigma^+_0 \sigma^-_0        \\    ,              \bigg[ \underset{j \text{ } \mathrm{even}}{\underset{1 \leq j \leq N-1}{\prod}}           i u^{\prime}  \sigma^+_j \sigma^-_j       \bigg]    \bigg[ \underset{j \text{ } \mathrm{odd}}{\underset{3 \leq j\leq N-1}{\prod}}           i \big( u^{\prime} \big)^{-1} \sigma^+_j \sigma^-_j            \bigg]              \bigg[        \underset{j \equiv N}{\prod}  i  u^{\prime}  \sigma^+_j \sigma^-_j  \bigg]                  \bigg\}   \bigg[    \underset{j \text{ } \mathrm{odd}}{\underset{1 \leq j \leq N-1}{\prod}}     i u \sigma^+_j \sigma^-_j  \textbf{1}_{\{j^{\prime}> j \text{ } : \text{ } i u\sigma^+_{j^{\prime}} \sigma^-_{j^{\prime} \in \mathrm{support} ( A ( u ) ) } \}} \\ \times   i u \sigma^+_{j^{\prime}} \sigma^-_{j^{\prime}}    i u \sigma^+_{j^{\prime}+1} \sigma^-_{j^{\prime}+1}  \bigg]   \text{. } \end{align*}

  \noindent An application of (AC), which after two applications of (LR), yields,
  
  \begin{align*}
  -   i u \sigma^+_0 \sigma^-_0     \bigg\{                        \bigg[ \underset{j \text{ } \mathrm{even}}{\underset{1 \leq j \leq N-1}{\prod}}           i u^{\prime}  \sigma^+_j \sigma^-_j       \bigg]    \bigg[ \underset{j \text{ } \mathrm{odd}}{\underset{3 \leq j\leq N-1}{\prod}}           i \big( u^{\prime} \big)^{-1} \sigma^+_j \sigma^-_j            \bigg]              \bigg[        \underset{j \equiv N}{\prod}  i  u^{\prime}  \sigma^+_j \sigma^-_j  \bigg]  , \bigg[    \underset{j \text{ } \mathrm{odd}}{\underset{1 \leq j \leq N-1}{\prod}}     i u \sigma^+_j \sigma^-_j   \\ \times \textbf{1}_{\{j^{\prime}> j \text{ } : \text{ } i u\sigma^+_{j^{\prime}} \sigma^-_{j^{\prime} \in \mathrm{support} ( A ( u ) ) } \}}   i u \sigma^+_{j^{\prime}} \sigma^-_{j^{\prime}}    i u \sigma^+_{j^{\prime}+1} \sigma^-_{j^{\prime}+1}  \bigg]                     \bigg\}    -   \bigg\{                  \bigg[ \underset{j \text{ } \mathrm{even}}{\underset{1 \leq j \leq N-1}{\prod}}           i u^{\prime}  \sigma^+_j \sigma^-_j       \bigg]    \bigg[ \underset{j \text{ } \mathrm{odd}}{\underset{3 \leq j\leq N-1}{\prod}}           i \big( u^{\prime} \big)^{-1} \sigma^+_j \sigma^-_j            \bigg]    \\ \times          \bigg[        \underset{j \equiv N}{\prod}  i  u^{\prime}  \sigma^+_j \sigma^-_j  \bigg]               , i u \sigma^+_0 \sigma^-_0             \bigg\}   \bigg[    \underset{j \text{ } \mathrm{odd}}{\underset{1 \leq j \leq N-1}{\prod}}     i u \sigma^+_j \sigma^-_j  \textbf{1}_{\{j^{\prime}> j \text{ } : \text{ } i u\sigma^+_{j^{\prime}} \sigma^-_{j^{\prime} \in \mathrm{support} ( A ( u ) ) } \}}  i u \sigma^+_{j^{\prime}} \sigma^-_{j^{\prime}}    i u \sigma^+_{j^{\prime}+1} \sigma^-_{j^{\prime}+1}  \bigg]              \\ \\ \overset{(\mathrm{LR})}{=}                  -   i u \sigma^+_0 \sigma^-_0     \bigg[ \underset{j \text{ } \mathrm{even}}{\underset{1 \leq j \leq N-1}{\prod}}           i u^{\prime}  \sigma^+_j \sigma^-_j       \bigg]      \bigg\{                       \bigg[ \underset{j \text{ } \mathrm{odd}}{\underset{3 \leq j\leq N-1}{\prod}}           i \big( u^{\prime} \big)^{-1} \sigma^+_j \sigma^-_j            \bigg]              \bigg[        \underset{j \equiv N}{\prod}  i  u^{\prime}  \sigma^+_j \sigma^-_j  \bigg]  , \bigg[    \underset{j \text{ } \mathrm{odd}}{\underset{1 \leq j \leq N-1}{\prod}}     i u \sigma^+_j \sigma^-_j   \\  \times \textbf{1}_{\{j^{\prime}> j \text{ } : \text{ } i u\sigma^+_{j^{\prime}} \sigma^-_{j^{\prime} \in \mathrm{support} ( A ( u ) ) } \}}   i u \sigma^+_{j^{\prime}} \sigma^-_{j^{\prime}}    i u \sigma^+_{j^{\prime}+1} \sigma^-_{j^{\prime}+1}  \bigg]                     \bigg\}  \\    -   i u \sigma^+_0 \sigma^-_0     \bigg\{                        \bigg[ \underset{j \text{ } \mathrm{even}}{\underset{1 \leq j \leq N-1}{\prod}}           i u^{\prime}  \sigma^+_j \sigma^-_j       \bigg]     , \bigg[    \underset{j \text{ } \mathrm{odd}}{\underset{1 \leq j \leq N-1}{\prod}}     i u \sigma^+_j \sigma^-_j  \textbf{1}_{\{j^{\prime}> j \text{ } : \text{ } i u\sigma^+_{j^{\prime}} \sigma^-_{j^{\prime} \in \mathrm{support} ( A ( u ) ) } \}}    i u \sigma^+_{j^{\prime}} \sigma^-_{j^{\prime}}    i u \sigma^+_{j^{\prime}+1} \sigma^-_{j^{\prime}+1}  \bigg]                     \bigg\}        \end{align*}

      \begin{align*}  
\times   \bigg[ \underset{j \text{ } \mathrm{odd}}{\underset{3 \leq j\leq N-1}{\prod}}           i \big( u^{\prime} \big)^{-1} \sigma^+_j \sigma^-_j            \bigg]              \bigg[        \underset{j \equiv N}{\prod}  i  u^{\prime}  \sigma^+_j \sigma^-_j  \bigg]   - \bigg\{                  \bigg[ \underset{j \text{ } \mathrm{even}}{\underset{1 \leq j \leq N-1}{\prod}}           i u^{\prime}  \sigma^+_j \sigma^-_j       \bigg]    \bigg[ \underset{j \text{ } \mathrm{odd}}{\underset{3 \leq j\leq N-1}{\prod}}           i \big( u^{\prime} \big)^{-1} \sigma^+_j \sigma^-_j            \bigg]        \\ \times      \bigg[        \underset{j \equiv N}{\prod}  i  u^{\prime}  \sigma^+_j \sigma^-_j  \bigg]               , i u \sigma^+_0 \sigma^-_0             \bigg\}    \bigg[    \underset{j \text{ } \mathrm{odd}}{\underset{1 \leq j \leq N-1}{\prod}}     i u \sigma^+_j \sigma^-_j  \textbf{1}_{\{j^{\prime}> j \text{ } : \text{ } i u\sigma^+_{j^{\prime}} \sigma^-_{j^{\prime} \in \mathrm{support} ( A ( u ) ) } \}}    \\ \times i u \sigma^+_{j^{\prime}} \sigma^-_{j^{\prime}}    i u \sigma^+_{j^{\prime}+1} \sigma^-_{j^{\prime}+1}  \bigg]   - \bigg\{                  \bigg[ \underset{j \text{ } \mathrm{even}}{\underset{1 \leq j \leq N-1}{\prod}}           i u^{\prime}  \sigma^+_j \sigma^-_j       \bigg]    \bigg[ \underset{j \text{ } \mathrm{odd}}{\underset{3 \leq j\leq N-1}{\prod}}           i \big( u^{\prime} \big)^{-1} \sigma^+_j \sigma^-_j            \bigg]  \\  \times      \bigg[        \underset{j \equiv N}{\prod}  i  u^{\prime}  \sigma^+_j \sigma^-_j  \bigg]               , i u \sigma^+_0 \sigma^-_0             \bigg\}    \bigg[    \underset{j \text{ } \mathrm{odd}}{\underset{1 \leq j \leq N-1}{\prod}}     i u \sigma^+_j \sigma^-_j  \textbf{1}_{\{j^{\prime}> j \text{ } : \text{ } i u\sigma^+_{j^{\prime}} \sigma^-_{j^{\prime} \in \mathrm{support} ( A ( u ) ) } \}}       \\  \times i u \sigma^+_{j^{\prime}} \sigma^-_{j^{\prime}}    i u \sigma^+_{j^{\prime}+1} \sigma^-_{j^{\prime}+1}  \bigg]  \\   \overset{(\mathrm{LR})}{=}                 -   i u \sigma^+_0 \sigma^-_0     \bigg[ \underset{j \text{ } \mathrm{even}}{\underset{1 \leq j \leq N-1}{\prod}}           i u^{\prime}  \sigma^+_j \sigma^-_j       \bigg]     \bigg[ \underset{j \text{ } \mathrm{odd}}{\underset{3 \leq j\leq N-1}{\prod}}           i \big( u^{\prime} \big)^{-1} \sigma^+_j \sigma^-_j            \bigg]         \bigg\{                              \bigg[        \underset{j \equiv N}{\prod}  i  u^{\prime}  \sigma^+_j \sigma^-_j  \bigg]  , \bigg[    \underset{j \text{ } \mathrm{odd}}{\underset{1 \leq j \leq N-1}{\prod}}     i u \sigma^+_j \sigma^-_j  \\  \times \textbf{1}_{\{j^{\prime}> j \text{ } : \text{ } i u\sigma^+_{j^{\prime}} \sigma^-_{j^{\prime} \in \mathrm{support} ( A ( u ) ) } \}}   i u \sigma^+_{j^{\prime}} \sigma^-_{j^{\prime}}    i u \sigma^+_{j^{\prime}+1} \sigma^-_{j^{\prime}+1}  \bigg]                     \bigg\}    \\    -   i u \sigma^+_0 \sigma^-_0     \bigg[ \underset{j \text{ } \mathrm{even}}{\underset{1 \leq j \leq N-1}{\prod}}           i u^{\prime}  \sigma^+_j \sigma^-_j       \bigg]      \bigg\{                       \bigg[ \underset{j \text{ } \mathrm{odd}}{\underset{3 \leq j\leq N-1}{\prod}}           i \big( u^{\prime} \big)^{-1} \sigma^+_j \sigma^-_j            \bigg]              , \bigg[    \underset{j \text{ } \mathrm{odd}}{\underset{1 \leq j \leq N-1}{\prod}}     i u \sigma^+_j \sigma^-_j  \textbf{1}_{\{j^{\prime}> j \text{ } : \text{ } i u\sigma^+_{j^{\prime}} \sigma^-_{j^{\prime} \in \mathrm{support} ( A ( u ) ) } \}}   \\   \times    i u \sigma^+_{j^{\prime}} \sigma^-_{j^{\prime}}    i u \sigma^+_{j^{\prime}+1} \sigma^-_{j^{\prime}+1}  \bigg]                     \bigg\}    \bigg[        \underset{j \equiv N}{\prod}  i  u^{\prime}  \sigma^+_j \sigma^-_j  \bigg]    \\                         - \bigg[ \underset{j \text{ } \mathrm{even}}{\underset{1 \leq j \leq N-1}{\prod}}           i u^{\prime}  \sigma^+_j \sigma^-_j       \bigg]    \bigg[ \underset{j \text{ } \mathrm{odd}}{\underset{3 \leq j\leq N-1}{\prod}}           i \big( u^{\prime} \big)^{-1} \sigma^+_j \sigma^-_j            \bigg]    \bigg\{                          \bigg[        \underset{j \equiv N}{\prod}  i  u^{\prime}  \sigma^+_j \sigma^-_j  \bigg]               , i u \sigma^+_0 \sigma^-_0             \bigg\}    \bigg[    \underset{j \text{ } \mathrm{odd}}{\underset{1 \leq j \leq N-1}{\prod}}     i u \sigma^+_j \sigma^-_j  \\ \times \textbf{1}_{\{j^{\prime}> j \text{ } : \text{ } i u\sigma^+_{j^{\prime}} \sigma^-_{j^{\prime} \in \mathrm{support} ( A ( u ) ) } \}}    i u \sigma^+_{j^{\prime}} \sigma^-_{j^{\prime}}    i u \sigma^+_{j^{\prime}+1} \sigma^-_{j^{\prime}+1}  \bigg]   \\  - \bigg\{                  \bigg[ \underset{j \text{ } \mathrm{even}}{\underset{1 \leq j \leq N-1}{\prod}}           i u^{\prime}  \sigma^+_j \sigma^-_j       \bigg]    \bigg[ \underset{j \text{ } \mathrm{odd}}{\underset{3 \leq j\leq N-1}{\prod}}           i \big( u^{\prime} \big)^{-1} \sigma^+_j \sigma^-_j            \bigg]            , i u \sigma^+_0 \sigma^-_0             \bigg\}    \bigg[    \underset{j \text{ } \mathrm{odd}}{\underset{1 \leq j \leq N-1}{\prod}}     i u \sigma^+_j \sigma^-_j  \\  \times \textbf{1}_{\{j^{\prime}> j \text{ } : \text{ } i u\sigma^+_{j^{\prime}} \sigma^-_{j^{\prime} \in \mathrm{support} ( A ( u ) ) } \}}    i u \sigma^+_{j^{\prime}} \sigma^-_{j^{\prime}}    i u \sigma^+_{j^{\prime}+1} \sigma^-_{j^{\prime}+1}  \bigg]                     \bigg[        \underset{j \equiv N}{\prod}  i  u^{\prime}  \sigma^+_j \sigma^-_j  \bigg]  \\  -                       \bigg[ \underset{j \text{ } \mathrm{even}}{\underset{1 \leq j \leq N-1}{\prod}}           i u^{\prime}  \sigma^+_j \sigma^-_j       \bigg]    \bigg[ \underset{j \text{ } \mathrm{odd}}{\underset{3 \leq j\leq N-1}{\prod}}           i \big( u^{\prime} \big)^{-1} \sigma^+_j \sigma^-_j            \bigg]      \bigg\{                         \bigg[        \underset{j \equiv N}{\prod}  i  u^{\prime}  \sigma^+_j \sigma^-_j  \bigg]                 , i u \sigma^+_0 \sigma^-_0             \bigg\}    \bigg[    \underset{j \text{ } \mathrm{odd}}{\underset{1 \leq j \leq N-1}{\prod}}     i u \sigma^+_j \sigma^-_j   \\  \times \textbf{1}_{\{j^{\prime}> j \text{ } : \text{ } i u\sigma^+_{j^{\prime}} \sigma^-_{j^{\prime} \in \mathrm{support} ( A ( u ) ) } \}}    i u \sigma^+_{j^{\prime}} \sigma^-_{j^{\prime}}    i u \sigma^+_{j^{\prime}+1} \sigma^-_{j^{\prime}+1}  \bigg]  \\  -                 i u \sigma^+_0 \sigma^-_0     \bigg\{                        \bigg[ \underset{j \text{ } \mathrm{even}}{\underset{1 \leq j \leq N-1}{\prod}}           i u^{\prime}  \sigma^+_j \sigma^-_j       \bigg]     , \bigg[    \underset{j \text{ } \mathrm{odd}}{\underset{1 \leq j \leq N-1}{\prod}}     i u \sigma^+_j \sigma^-_j  \textbf{1}_{\{j^{\prime}> j \text{ } : \text{ } i u\sigma^+_{j^{\prime}} \sigma^-_{j^{\prime} \in \mathrm{support} ( A ( u ) ) } \}}    i u \sigma^+_{j^{\prime}} \sigma^-_{j^{\prime}}    i u \sigma^+_{j^{\prime}+1} \sigma^-_{j^{\prime}+1}  \bigg]                     \bigg\} \\ \times   \bigg[ \underset{j \text{ } \mathrm{odd}}{\underset{3 \leq j\leq N-1}{\prod}}           i \big( u^{\prime} \big)^{-1} \sigma^+_j \sigma^-_j            \bigg]              \bigg[        \underset{j \equiv N}{\prod}  i  u^{\prime}  \sigma^+_j \sigma^-_j  \bigg]   - \bigg\{                  \bigg[ \underset{j \text{ } \mathrm{even}}{\underset{1 \leq j \leq N-1}{\prod}}           i u^{\prime}  \sigma^+_j \sigma^-_j       \bigg]    \bigg[ \underset{j \text{ } \mathrm{odd}}{\underset{3 \leq j\leq N-1}{\prod}}           i \big( u^{\prime} \big)^{-1} \sigma^+_j \sigma^-_j            \bigg]                      \text{, }
\end{align*}

\noindent corresponding to the third term, which can be approximated with,

\begin{align*}
 \big( \mathscr{C}^1_1 \big)_3 \text{. }
\end{align*}

    \noindent For the next term, one has,

\begin{align*}
  \bigg\{    i u \sigma^+_1 \sigma^-_1 \sigma^+_2 \bigg[ \underset{j \text{ } \mathrm{odd}}{\underset{1 \leq j \leq N}{\prod}}          i u \sigma^+_j \sigma^-_j  \bigg]     \bigg[ \underset{j \equiv N}{\prod}           \sigma^+_j \sigma^-_j     \bigg]  ,      i u^{\prime} \sigma^+_0 \sigma^-_0   \bigg[    \underset{j \text{ } \mathrm{odd}}{\underset{1 \leq j \leq N-1}{\prod}}     i u^{\prime} \sigma^+_j \sigma^-_j           \textbf{1}_{\{j^{\prime}> j \text{ } : \text{ } i u^{\prime} \sigma^+_{j^{\prime}} \sigma^-_{j^{\prime} \in \mathrm{support} ( A ( u ) ) } \}}  \\ \times    i u^{\prime} \sigma^+_{j^{\prime}} \sigma^-_{j^{\prime}} i u^{\prime} \sigma^+_{j^{\prime}+1} \sigma^-_{j^{\prime}+1}  \bigg]         \bigg\}  \overset{(\mathrm{LR})}{=}    i u \sigma^+_1 \sigma^-_1 \sigma^+_2  \bigg\{   \bigg[ \underset{j \text{ } \mathrm{odd}}{\underset{1 \leq j \leq N}{\prod}}          i u \sigma^+_j \sigma^-_j  \bigg]     \bigg[ \underset{j \equiv N}{\prod}           \sigma^+_j \sigma^-_j     \bigg]  \\ ,      i u^{\prime} \sigma^+_0 \sigma^-_0   \bigg[    \underset{j \text{ } \mathrm{odd}}{\underset{1 \leq j \leq N-1}{\prod}}     i u^{\prime} \sigma^+_j \sigma^-_j            \textbf{1}_{\{j^{\prime}> j \text{ } : \text{ } i u^{\prime} \sigma^+_{j^{\prime}} \sigma^-_{j^{\prime} \in \mathrm{support} ( A ( u ) ) } \}}    i u^{\prime} \sigma^+_{j^{\prime}} \sigma^-_{j^{\prime}} i u^{\prime} \sigma^+_{j^{\prime}+1} \sigma^-_{j^{\prime}+1}  \bigg]         \bigg\} +  \bigg\{    i u \sigma^+_1 \sigma^-_1 \sigma^+_2   \\ 
 ,      i u^{\prime} \sigma^+_0 \sigma^-_0   \bigg[    \underset{j \text{ } \mathrm{odd}}{\underset{1 \leq j \leq N-1}{\prod}}     i u^{\prime} \sigma^+_j \sigma^-_j    \textbf{1}_{\{j^{\prime}> j \text{ } : \text{ } i u^{\prime} \sigma^+_{j^{\prime}} \sigma^-_{j^{\prime} \in \mathrm{support} ( A ( u ) ) } \}}    i u^{\prime} \sigma^+_{j^{\prime}} \sigma^-_{j^{\prime}}        i u^{\prime} \sigma^+_{j^{\prime}+1} \sigma^-_{j^{\prime}+1}  \bigg]         \bigg\}  \bigg[ \underset{j \text{ } \mathrm{odd}}{\underset{1 \leq j \leq N}{\prod}}          i u \sigma^+_j \sigma^-_j  \bigg]  \\ \times   \bigg[ \underset{j \equiv N}{\prod}           \sigma^+_j \sigma^-_j     \bigg]  \text{. }\end{align*}

\noindent Applying (LR), followed by (AC), to the superposition above yields,

 \begin{align*}
    i u \sigma^+_1 \sigma^-_1 \sigma^+_2  \bigg[ \underset{j \text{ } \mathrm{odd}}{\underset{1 \leq j \leq N}{\prod}}          i u \sigma^+_j \sigma^-_j  \bigg]    \bigg\{     \bigg[ \underset{j \equiv N}{\prod}           \sigma^+_j \sigma^-_j     \bigg]   ,      i u^{\prime} \sigma^+_0 \sigma^-_0   \bigg[    \underset{j \text{ } \mathrm{odd}}{\underset{1 \leq j \leq N-1}{\prod}}     i u^{\prime} \sigma^+_j \sigma^-_j            \textbf{1}_{\{j^{\prime}> j \text{ } : \text{ } i u^{\prime} \sigma^+_{j^{\prime}} \sigma^-_{j^{\prime} \in \mathrm{support} ( A ( u ) ) } \}} \\ \times    i u^{\prime} \sigma^+_{j^{\prime}} \sigma^-_{j^{\prime}} i u^{\prime} \sigma^+_{j^{\prime}+1} \sigma^-_{j^{\prime}+1}  \bigg]         \bigg\}   \\ + i u \sigma^+_1 \sigma^-_1 \sigma^+_2  \bigg\{   \bigg[ \underset{j \text{ } \mathrm{odd}}{\underset{1 \leq j \leq N}{\prod}}          i u \sigma^+_j \sigma^-_j  \bigg]       ,      i u^{\prime} \sigma^+_0 \sigma^-_0   \bigg[    \underset{j \text{ } \mathrm{odd}}{\underset{1 \leq j \leq N-1}{\prod}}     i u^{\prime} \sigma^+_j \sigma^-_j            \textbf{1}_{\{j^{\prime}> j \text{ } : \text{ } i u^{\prime} \sigma^+_{j^{\prime}} \sigma^-_{j^{\prime} \in \mathrm{support} ( A ( u ) ) } \}} \\ \times    i u^{\prime} \sigma^+_{j^{\prime}} \sigma^-_{j^{\prime}} i u^{\prime} \sigma^+_{j^{\prime}+1} \sigma^-_{j^{\prime}+1}  \bigg]         \bigg\}  \bigg[ \underset{j \equiv N}{\prod}           \sigma^+_j \sigma^-_j     \bigg]  -  \bigg\{        i u^{\prime} \sigma^+_0 \sigma^-_0   \bigg[    \underset{j \text{ } \mathrm{odd}}{\underset{1 \leq j \leq N-1}{\prod}}     i u^{\prime} \sigma^+_j \sigma^-_j  \\ \times     \textbf{1}_{\{j^{\prime}> j \text{ } : \text{ } i u^{\prime} \sigma^+_{j^{\prime}} \sigma^-_{j^{\prime} \in \mathrm{support} ( A ( u ) ) } \}}   i u^{\prime} \sigma^+_{j^{\prime}} \sigma^-_{j^{\prime}}       i u^{\prime} \sigma^+_{j^{\prime}+1} \sigma^-_{j^{\prime}+1}  \bigg] ,  i u \sigma^+_1 \sigma^-_1 \sigma^+_2           \bigg\} \\ \times  \bigg[ \underset{j \text{ } \mathrm{odd}}{\underset{1 \leq j \leq N}{\prod}}          i u \sigma^+_j \sigma^-_j  \bigg]     \bigg[ \underset{j \equiv N}{\prod}           \sigma^+_j \sigma^-_j     \bigg]  \\ \\ 
 \overset{(\mathrm{AC}),(\mathrm{LR})}{=}                   -   i u \sigma^+_1 \sigma^-_1 \sigma^+_2  \bigg[ \underset{j \text{ } \mathrm{odd}}{\underset{1 \leq j \leq N}{\prod}}          i u \sigma^+_j \sigma^-_j  \bigg]    \bigg\{        i u^{\prime} \sigma^+_0 \sigma^-_0   \bigg[    \underset{j \text{ } \mathrm{odd}}{\underset{1 \leq j \leq N-1}{\prod}}     i u^{\prime} \sigma^+_j \sigma^-_j            \textbf{1}_{\{j^{\prime}> j \text{ } : \text{ } i u^{\prime} \sigma^+_{j^{\prime}} \sigma^-_{j^{\prime} \in \mathrm{support} ( A ( u ) ) } \}} \\ \times    i u^{\prime} \sigma^+_{j^{\prime}} \sigma^-_{j^{\prime}} i u^{\prime} \sigma^+_{j^{\prime}+1} \sigma^-_{j^{\prime}+1}  \bigg]   ,   \bigg[ \underset{j \equiv N}{\prod}           \sigma^+_j \sigma^-_j     \bigg]         \bigg\}  \\ - i u \sigma^+_1 \sigma^-_1 \sigma^+_2  \bigg\{             i u^{\prime} \sigma^+_0 \sigma^-_0   \bigg[    \underset{j \text{ } \mathrm{odd}}{\underset{1 \leq j \leq N-1}{\prod}}     i u^{\prime} \sigma^+_j \sigma^-_j            \textbf{1}_{\{j^{\prime}> j \text{ } : \text{ } i u^{\prime} \sigma^+_{j^{\prime}} \sigma^-_{j^{\prime} \in \mathrm{support} ( A ( u ) ) } \}}    i u^{\prime} \sigma^+_{j^{\prime}} \sigma^-_{j^{\prime}} i u^{\prime} \sigma^+_{j^{\prime}+1} \sigma^-_{j^{\prime}+1}  \bigg] \\ , \bigg[ \underset{j \text{ } \mathrm{odd}}{\underset{1 \leq j \leq N}{\prod}}          i u \sigma^+_j \sigma^-_j  \bigg]          \bigg\}  \bigg[ \underset{j \equiv N}{\prod}           \sigma^+_j \sigma^-_j     \bigg]  -  \bigg\{        i u^{\prime} \sigma^+_0 \sigma^-_0    ,  i u \sigma^+_1 \sigma^-_1 \sigma^+_2           \bigg\} \\  \times  \bigg[ \underset{j \text{ } \mathrm{odd}}{\underset{1 \leq j \leq N}{\prod}}          i u \sigma^+_j \sigma^-_j  \bigg]     \bigg[ \underset{j \equiv N}{\prod}           \sigma^+_j \sigma^-_j     \bigg]   -     i u^{\prime} \sigma^+_0 \sigma^-_0     \bigg\{     \bigg[    \underset{j \text{ } \mathrm{odd}}{\underset{1 \leq j \leq N-1}{\prod}}     i u^{\prime} \sigma^+_j \sigma^-_j       \textbf{1}_{\{j^{\prime}> j \text{ } : \text{ } i u^{\prime} \sigma^+_{j^{\prime}} \sigma^-_{j^{\prime} \in \mathrm{support} ( A ( u ) ) } \}}  \end{align*}

      \begin{align*}   \times   i u^{\prime} \sigma^+_{j^{\prime}} \sigma^-_{j^{\prime}}       i u^{\prime} \sigma^+_{j^{\prime}+1} \sigma^-_{j^{\prime}+1}  \bigg]  ,  i u \sigma^+_1 \sigma^-_1 \sigma^+_2           \bigg\}  \bigg[    \underset{j \text{ } \mathrm{odd}}{\underset{1 \leq j \leq N-1}{\prod}}     i u^{\prime} \sigma^+_j \sigma^-_j       \textbf{1}_{\{j^{\prime}> j \text{ } : \text{ } i u^{\prime} \sigma^+_{j^{\prime}} \sigma^-_{j^{\prime} \in \mathrm{support} ( A ( u ) ) } \}}  \\  \times    i u^{\prime} \sigma^+_{j^{\prime}} \sigma^-_{j^{\prime}}       i u^{\prime} \sigma^+_{j^{\prime}+1} \sigma^-_{j^{\prime}+1}  \bigg]   \bigg[ \underset{j \text{ } \mathrm{odd}}{\underset{1 \leq j \leq N}{\prod}}          i u \sigma^+_j \sigma^-_j  \bigg]     \bigg[ \underset{j \equiv N}{\prod}           \sigma^+_j \sigma^-_j     \bigg] \text{. }  \end{align*}

 \noindent Another application of (LR) to the superpostion above yields,
 
 \begin{align*} 
   -   i u \sigma^+_1 \sigma^-_1 \sigma^+_2  \bigg[ \underset{j \text{ } \mathrm{odd}}{\underset{1 \leq j \leq N}{\prod}}          i u \sigma^+_j \sigma^-_j  \bigg]    \bigg\{        i u^{\prime} \sigma^+_0 \sigma^-_0     ,   \bigg[ \underset{j \equiv N}{\prod}           \sigma^+_j \sigma^-_j     \bigg]         \bigg\}  \bigg[    \underset{j \text{ } \mathrm{odd}}{\underset{1 \leq j \leq N-1}{\prod}}     i u^{\prime} \sigma^+_j \sigma^-_j            \textbf{1}_{\{j^{\prime}> j \text{ } : \text{ } i u^{\prime} \sigma^+_{j^{\prime}} \sigma^-_{j^{\prime} \in \mathrm{support} ( A ( u ) ) } \}}   \\ \times     i u^{\prime} \sigma^+_{j^{\prime}} \sigma^-_{j^{\prime}} i u^{\prime} \sigma^+_{j^{\prime}+1} \sigma^-_{j^{\prime}+1}  \bigg] \text{, }  \end{align*}
   
   \noindent corresponding to the first term,
   
   \begin{align*}
   -   i u \sigma^+_1 \sigma^-_1 \sigma^+_2  \bigg[ \underset{j \text{ } \mathrm{odd}}{\underset{1 \leq j \leq N}{\prod}}          i u \sigma^+_j \sigma^-_j  \bigg]    i u^{\prime} \sigma^+_0 \sigma^-_0     \bigg\{       \bigg[    \underset{j \text{ } \mathrm{odd}}{\underset{1 \leq j \leq N-1}{\prod}}     i u^{\prime} \sigma^+_j \sigma^-_j            \textbf{1}_{\{j^{\prime}> j \text{ } : \text{ } i u^{\prime} \sigma^+_{j^{\prime}} \sigma^-_{j^{\prime} \in \mathrm{support} ( A ( u ) ) } \}}   \\ 
 \times    i u^{\prime} \sigma^+_{j^{\prime}} \sigma^-_{j^{\prime}} i u^{\prime} \sigma^+_{j^{\prime}+1} \sigma^-_{j^{\prime}+1}  \bigg]   ,   \bigg[ \underset{j \equiv N}{\prod}           \sigma^+_j \sigma^-_j     \bigg]         \bigg\}    \text{, }        \end{align*}
   
   \noindent corresponding to the second term,
   
   \begin{align*}  - i u \sigma^+_1 \sigma^-_1 \sigma^+_2  \bigg\{             i u^{\prime} \sigma^+_0 \sigma^-_0   \bigg[    \underset{j \text{ } \mathrm{odd}}{\underset{1 \leq j \leq N-1}{\prod}}     i u^{\prime} \sigma^+_j \sigma^-_j            \textbf{1}_{\{j^{\prime}> j \text{ } : \text{ } i u^{\prime} \sigma^+_{j^{\prime}} \sigma^-_{j^{\prime} \in \mathrm{support} ( A ( u ) ) } \}}    i u^{\prime} \sigma^+_{j^{\prime}} \sigma^-_{j^{\prime}} i u^{\prime} \sigma^+_{j^{\prime}+1} \sigma^-_{j^{\prime}+1}  \bigg]  \\ , \bigg[ \underset{j \text{ } \mathrm{odd}}{\underset{1 \leq j \leq N}{\prod}}          i u \sigma^+_j \sigma^-_j  \bigg]          \bigg\}  \bigg[ \underset{j \equiv N}{\prod}           \sigma^+_j \sigma^-_j     \bigg]  \text{, } \end{align*}
   
   \noindent corresponding to the third term,
   
   \begin{align*} -  \bigg\{        i u^{\prime} \sigma^+_0 \sigma^-_0    ,  i u \sigma^+_1 \sigma^-_1 \sigma^+_2           \bigg\}   \bigg[ \underset{j \text{ } \mathrm{odd}}{\underset{1 \leq j \leq N}{\prod}}          i u \sigma^+_j \sigma^-_j  \bigg]     \bigg[ \underset{j \equiv N}{\prod}           \sigma^+_j \sigma^-_j     \bigg]   -     i u^{\prime} \sigma^+_0 \sigma^-_0     \bigg\{     \bigg[    \underset{j \text{ } \mathrm{odd}}{\underset{1 \leq j \leq N-1}{\prod}}     i u^{\prime} \sigma^+_j \sigma^-_j     \\  \times   \textbf{1}_{\{j^{\prime}> j \text{ } : \text{ } i u^{\prime} \sigma^+_{j^{\prime}} \sigma^-_{j^{\prime} \in \mathrm{support} ( A ( u ) ) } \}}  i u^{\prime} \sigma^+_{j^{\prime}} \sigma^-_{j^{\prime}}       i u^{\prime} \sigma^+_{j^{\prime}+1} \sigma^-_{j^{\prime}+1}  \bigg]  ,  i u \sigma^+_1 \sigma^-_1 \sigma^+_2           \bigg\}  \bigg[    \underset{j \text{ } \mathrm{odd}}{\underset{1 \leq j \leq N-1}{\prod}}     i u^{\prime} \sigma^+_j \sigma^-_j    \\ \times      \textbf{1}_{\{j^{\prime}> j \text{ } : \text{ } i u^{\prime} \sigma^+_{j^{\prime}} \sigma^-_{j^{\prime} \in \mathrm{support} ( A ( u ) ) } \}}    i u^{\prime} \sigma^+_{j^{\prime}} \sigma^-_{j^{\prime}}       i u^{\prime} \sigma^+_{j^{\prime}+1} \sigma^-_{j^{\prime}+1}  \bigg]   \bigg[ \underset{j \text{ } \mathrm{odd}}{\underset{1 \leq j \leq N}{\prod}}          i u \sigma^+_j \sigma^-_j  \bigg]     \bigg[ \underset{j \equiv N}{\prod}           \sigma^+_j \sigma^-_j     \bigg]   \text{, } \end{align*}
   
   \noindent corresponding to the fourth term. Below, a final application of (LR) yields,
   
   \begin{align*}  \overset{(\mathrm{LR})}{=}     -   i u \sigma^+_1 \sigma^-_1 \sigma^+_2  \bigg[ \underset{j \text{ } \mathrm{odd}}{\underset{1 \leq j \leq N}{\prod}}          i u \sigma^+_j \sigma^-_j  \bigg]    \bigg\{        i u^{\prime} \sigma^+_0 \sigma^-_0     ,   \bigg[ \underset{j \equiv N}{\prod}           \sigma^+_j \sigma^-_j     \bigg]         \bigg\}  \bigg[    \underset{j \text{ } \mathrm{odd}}{\underset{1 \leq j \leq N-1}{\prod}}     i u^{\prime} \sigma^+_j \sigma^-_j            \textbf{1}_{\{j^{\prime}> j \text{ } : \text{ } i u^{\prime} \sigma^+_{j^{\prime}} \sigma^-_{j^{\prime} \in \mathrm{support} ( A ( u ) ) } \}} \\    \times     i u^{\prime} \sigma^+_{j^{\prime}} \sigma^-_{j^{\prime}} i u^{\prime} \sigma^+_{j^{\prime}+1} \sigma^-_{j^{\prime}+1}  \bigg] \end{align*}

      \begin{align*}    -   i u \sigma^+_1 \sigma^-_1 \sigma^+_2  \bigg[ \underset{j \text{ } \mathrm{odd}}{\underset{1 \leq j \leq N}{\prod}}          i u \sigma^+_j \sigma^-_j  \bigg]    i u^{\prime} \sigma^+_0 \sigma^-_0     \bigg\{       \bigg[    \underset{j \text{ } \mathrm{odd}}{\underset{1 \leq j \leq N-1}{\prod}}     i u^{\prime} \sigma^+_j \sigma^-_j            \textbf{1}_{\{j^{\prime}> j \text{ } : \text{ } i u^{\prime} \sigma^+_{j^{\prime}} \sigma^-_{j^{\prime} \in \mathrm{support} ( A ( u ) ) } \}}       i u^{\prime} \sigma^+_{j^{\prime}} \sigma^-_{j^{\prime}} i u^{\prime} \sigma^+_{j^{\prime}+1} \sigma^-_{j^{\prime}+1}  \bigg]      \\ ,   \bigg[ \underset{j \equiv N}{\prod}           \sigma^+_j \sigma^-_j     \bigg]         \bigg\}        \\    - i u \sigma^+_1 \sigma^-_1 \sigma^+_2  
  i u^{\prime} \sigma^+_0 \sigma^-_0 \bigg\{              \bigg[    \underset{j \text{ } \mathrm{odd}}{\underset{1 \leq j \leq N-1}{\prod}}     i u^{\prime} \sigma^+_j \sigma^-_j            \textbf{1}_{\{j^{\prime}> j \text{ } : \text{ } i u^{\prime} \sigma^+_{j^{\prime}} \sigma^-_{j^{\prime} \in \mathrm{support} ( A ( u ) ) } \}}    i u^{\prime} \sigma^+_{j^{\prime}} \sigma^-_{j^{\prime}} i u^{\prime} \sigma^+_{j^{\prime}+1} \sigma^-_{j^{\prime}+1}  \bigg]   , \bigg[ \underset{j \text{ } \mathrm{odd}}{\underset{1 \leq j \leq N}{\prod}}          i u \sigma^+_j \sigma^-_j  \bigg]          \bigg\} \\ \times  \bigg[ \underset{j \equiv N}{\prod}           \sigma^+_j \sigma^-_j     \bigg]  \\  - i u \sigma^+_1 \sigma^-_1 \sigma^+_2  \bigg\{             i u^{\prime} \sigma^+_0 \sigma^-_0    , \bigg[ \underset{j \text{ } \mathrm{odd}}{\underset{1 \leq j \leq N}{\prod}}          i u \sigma^+_j \sigma^-_j  \bigg]          \bigg\}  \bigg[    \underset{j \text{ } \mathrm{odd}}{\underset{1 \leq j \leq N-1}{\prod}}     i u^{\prime} \sigma^+_j \sigma^-_j            \textbf{1}_{\{j^{\prime}> j \text{ } : \text{ } i u^{\prime} \sigma^+_{j^{\prime}} \sigma^-_{j^{\prime} \in \mathrm{support} ( A ( u ) ) } \}}    i u^{\prime} \sigma^+_{j^{\prime}} \sigma^-_{j^{\prime}} i u^{\prime} \sigma^+_{j^{\prime}+1} \sigma^-_{j^{\prime}+1}  \bigg]   \end{align*}

  \begin{align*}
  \times  \bigg[ \underset{j \equiv N}{\prod}           \sigma^+_j \sigma^-_j     \bigg]  -  \bigg\{        i u^{\prime} \sigma^+_0 \sigma^-_0    ,  i u \sigma^+_1 \sigma^-_1 \sigma^+_2           \bigg\}    \bigg[ \underset{j \text{ } \mathrm{odd}}{\underset{1 \leq j \leq N}{\prod}}          i u \sigma^+_j \sigma^-_j  \bigg]     \bigg[ \underset{j \equiv N}{\prod}           \sigma^+_j \sigma^-_j     \bigg]   -     i u^{\prime} \sigma^+_0 \sigma^-_0     \bigg\{     \bigg[    \underset{j \text{ } \mathrm{odd}}{\underset{1 \leq j \leq N-1}{\prod}}     i u^{\prime} \sigma^+_j \sigma^-_j    \\   \times    \textbf{1}_{\{j^{\prime}> j \text{ } : \text{ } i u^{\prime} \sigma^+_{j^{\prime}} \sigma^-_{j^{\prime} \in \mathrm{support} ( A ( u ) ) } \}} \\ \times   i u^{\prime} \sigma^+_{j^{\prime}} \sigma^-_{j^{\prime}}       i u^{\prime} \sigma^+_{j^{\prime}+1} \sigma^-_{j^{\prime}+1}  \bigg]  ,  i u \sigma^+_1 \sigma^-_1 \sigma^+_2           \bigg\}  \bigg[    \underset{j \text{ } \mathrm{odd}}{\underset{1 \leq j \leq N-1}{\prod}}     i u^{\prime} \sigma^+_j \sigma^-_j       \textbf{1}_{\{j^{\prime}> j \text{ } : \text{ } i u^{\prime} \sigma^+_{j^{\prime}} \sigma^-_{j^{\prime} \in \mathrm{support} ( A ( u ) ) } \}} \\ \times    i u^{\prime} \sigma^+_{j^{\prime}} \sigma^-_{j^{\prime}}       i u^{\prime} \sigma^+_{j^{\prime}+1} \sigma^-_{j^{\prime}+1}  \bigg]   \bigg[ \underset{j \text{ } \mathrm{odd}}{\underset{1 \leq j \leq N}{\prod}}          i u \sigma^+_j \sigma^-_j  \bigg]     \bigg[ \underset{j \equiv N}{\prod}           \sigma^+_j \sigma^-_j     \bigg]      \text{, }
\end{align*}

\noindent corresponding to the fourth term, which can be approximated with, 

\begin{align*}
\big( \mathscr{C}^1_1 \big)_4 \text{. }
\end{align*}

    \noindent For the next term, one has,

\begin{align*}
       \bigg\{         i u \sigma^+_1 \sigma^-_1 \sigma^+_2 \bigg[ \underset{j \text{ } \mathrm{odd}}{\underset{1 \leq j \leq N}{\prod}}          i u \sigma^+_j \sigma^-_j  \bigg]     \bigg[ \underset{j \equiv N}{\prod}           \sigma^+_j \sigma^-_j     \bigg] 
     ,          i u^{\prime} \sigma^+_1 \sigma^-_1  \sigma^+_2 \bigg[ \underset{j \text{ } \mathrm{odd}}{\underset{1 \leq j \leq N}{\prod}}          i u^{\prime} \sigma^+_j \sigma^-_j  \bigg]     \bigg[ \underset{j \equiv N}{\prod}           \sigma^+_j \sigma^-_j     \bigg]      \bigg\} \overset{(\mathrm{LR})}{=}         i u \sigma^+_1 \sigma^-_1 \sigma^+_2  \\
     \times \bigg\{  \bigg[ \underset{j \text{ } \mathrm{odd}}{\underset{1 \leq j \leq N}{\prod}}          i u \sigma^+_j \sigma^-_j  \bigg]     \bigg[ \underset{j \equiv N}{\prod}           \sigma^+_j \sigma^-_j     \bigg] 
     ,          i u^{\prime} \sigma^+_1 \sigma^-_1  \sigma^+_2  \bigg[ \underset{j \text{ } \mathrm{odd}}{\underset{1 \leq j \leq N}{\prod}}          i u^{\prime} \sigma^+_j \sigma^-_j  \bigg]     \bigg[ \underset{j \equiv N}{\prod}           \sigma^+_j \sigma^-_j     \bigg]      \bigg\}  + \bigg\{         i u \sigma^+_1 \sigma^-_1 \sigma^+_2 \\  ,          i u^{\prime} \sigma^+_1 \sigma^-_1  \sigma^+_2 \bigg[ \underset{j \text{ } \mathrm{odd}}{\underset{1 \leq j \leq N}{\prod}}          i u^{\prime} \sigma^+_j \sigma^-_j  \bigg]     \bigg[ \underset{j \equiv N}{\prod}           \sigma^+_j \sigma^-_j     \bigg]      \bigg\}     \bigg[ \underset{j \text{ } \mathrm{odd}}{\underset{1 \leq j \leq N}{\prod}}          i u \sigma^+_j \sigma^-_j  \bigg]     \bigg[ \underset{j \equiv N}{\prod}           \sigma^+_j \sigma^-_j     \bigg] 
    \\ \\ \overset{(\mathrm{LR})}{=}          i u \sigma^+_1 \sigma^-_1 \sigma^+_2  \bigg[ \underset{j \text{ } \mathrm{odd}}{\underset{1 \leq j \leq N}{\prod}}          i u \sigma^+_j \sigma^-_j  \bigg]     \bigg\{      \bigg[ \underset{j \equiv N}{\prod}           \sigma^+_j \sigma^-_j     \bigg] 
     ,          i u^{\prime} \sigma^+_1 \sigma^-_1  \sigma^+_2  \bigg[ \underset{j \text{ } \mathrm{odd}}{\underset{1 \leq j \leq N}{\prod}}          i u^{\prime} \sigma^+_j \sigma^-_j  \bigg]     \bigg[ \underset{j \equiv N}{\prod}           \sigma^+_j \sigma^-_j     \bigg]      \bigg\}   \\ +    i u \sigma^+_1 \sigma^-_1 \sigma^+_2     \bigg\{  \bigg[ \underset{j \text{ } \mathrm{odd}}{\underset{1 \leq j \leq N}{\prod}}          i u \sigma^+_j \sigma^-_j  \bigg]    
     ,          i u^{\prime} \sigma^+_1 \sigma^-_1  \sigma^+_2  \bigg[ \underset{j \text{ } \mathrm{odd}}{\underset{1 \leq j \leq N}{\prod}}          i u^{\prime} \sigma^+_j \sigma^-_j  \bigg]     \bigg[ \underset{j \equiv N}{\prod}           \sigma^+_j \sigma^-_j     \bigg]      \bigg\}     \bigg[ \underset{j \equiv N}{\prod}           \sigma^+_j \sigma^-_j     \bigg]   \end{align*}

      \begin{align*}  
     -  \bigg\{                  i u^{\prime} \sigma^+_1 \sigma^-_1  \sigma^+_2 ,  i u \sigma^+_1 \sigma^-_1 \sigma^+_2    \bigg\}    \bigg[ \underset{j \text{ } \mathrm{odd}}{\underset{1 \leq j \leq N}{\prod}}          i u^{\prime} \sigma^+_j \sigma^-_j  \bigg]     \bigg[ \underset{j \equiv N}{\prod}           \sigma^+_j \sigma^-_j     \bigg]     \bigg[ \underset{j \text{ } \mathrm{odd}}{\underset{1 \leq j \leq N}{\prod}}          i u \sigma^+_j \sigma^-_j  \bigg]     \bigg[ \underset{j \equiv N}{\prod}           \sigma^+_j \sigma^-_j     \bigg]  \\   -           i u^{\prime} \sigma^+_1 \sigma^-_1  \sigma^+_2    \bigg\{        \bigg[ \underset{j \text{ } \mathrm{odd}}{\underset{1 \leq j \leq N}{\prod}}          i u^{\prime} \sigma^+_j \sigma^-_j  \bigg]     \bigg[ \underset{j \equiv N}{\prod}           \sigma^+_j \sigma^-_j     \bigg]  ,  i u \sigma^+_1 \sigma^-_1 \sigma^+_2    \bigg\}     \bigg[ \underset{j \text{ } \mathrm{odd}}{\underset{1 \leq j \leq N}{\prod}}          i u \sigma^+_j \sigma^-_j  \bigg]     \bigg[ \underset{j \equiv N}{\prod}           \sigma^+_j \sigma^-_j     \bigg] \text{. }  \end{align*}
     
\noindent Applying (AC) to the last superposition above yields,

     \begin{align*}      -   i u \sigma^+_1 \sigma^-_1 \sigma^+_2  \bigg[ \underset{j \text{ } \mathrm{odd}}{\underset{1 \leq j \leq N}{\prod}}          i u \sigma^+_j \sigma^-_j  \bigg]     \bigg\{   
              i u^{\prime} \sigma^+_1 \sigma^-_1  \sigma^+_2  \bigg[ \underset{j \text{ } \mathrm{odd}}{\underset{1 \leq j \leq N}{\prod}}          i u^{\prime} \sigma^+_j \sigma^-_j  \bigg]     \bigg[ \underset{j \equiv N}{\prod}           \sigma^+_j \sigma^-_j     \bigg]    ,    \bigg[ \underset{j \equiv N}{\prod}           \sigma^+_j \sigma^-_j     \bigg]   \bigg\}   \\ -     i u \sigma^+_1 \sigma^-_1 \sigma^+_2     \bigg\{     
              i u^{\prime} \sigma^+_1 \sigma^-_1  \sigma^+_2  \bigg[ \underset{j \text{ } \mathrm{odd}}{\underset{1 \leq j \leq N}{\prod}}          i u^{\prime} \sigma^+_j \sigma^-_j  \bigg]     \bigg[ \underset{j \equiv N}{\prod}           \sigma^+_j \sigma^-_j     \bigg]  ,  \bigg[ \underset{j \text{ } \mathrm{odd}}{\underset{1 \leq j \leq N}{\prod}}          i u \sigma^+_j \sigma^-_j  \bigg]    \bigg\}     \bigg[ \underset{j \equiv N}{\prod}           \sigma^+_j \sigma^-_j     \bigg]     \\ -  \bigg\{                  i u^{\prime} \sigma^+_1 \sigma^-_1  \sigma^+_2 ,  i u \sigma^+_1 \sigma^-_1 \sigma^+_2    \bigg\}    \bigg[ \underset{j \text{ } \mathrm{odd}}{\underset{1 \leq j \leq N}{\prod}}          i u^{\prime} \sigma^+_j \sigma^-_j  \bigg]     \bigg[ \underset{j \equiv N}{\prod}           \sigma^+_j \sigma^-_j     \bigg]     \bigg[ \underset{j \text{ } \mathrm{odd}}{\underset{1 \leq j \leq N}{\prod}}          i u \sigma^+_j \sigma^-_j  \bigg]     \bigg[ \underset{j \equiv N}{\prod}           \sigma^+_j \sigma^-_j     \bigg]   \end{align*}

              \begin{align*}
     -           i u^{\prime} \sigma^+_1 \sigma^-_1  \sigma^+_2    \bigg\{        \bigg[ \underset{j \text{ } \mathrm{odd}}{\underset{1 \leq j \leq N}{\prod}}          i u^{\prime} \sigma^+_j \sigma^-_j  \bigg]     \bigg[ \underset{j \equiv N}{\prod}           \sigma^+_j \sigma^-_j     \bigg]  ,  i u \sigma^+_1 \sigma^-_1 \sigma^+_2    \bigg\}     \bigg[ \underset{j \text{ } \mathrm{odd}}{\underset{1 \leq j \leq N}{\prod}}          i u \sigma^+_j \sigma^-_j  \bigg]     \bigg[ \underset{j \equiv N}{\prod}           \sigma^+_j \sigma^-_j     \bigg]        \text{, }    \end{align*}

\noindent to which two applications of (LR) yield,

     \begin{align*}
     \overset{(\mathrm{LR})}{=}      -   i u \sigma^+_1 \sigma^-_1 \sigma^+_2  \bigg[ \underset{j \text{ } \mathrm{odd}}{\underset{1 \leq j \leq N}{\prod}}          i u \sigma^+_j \sigma^-_j  \bigg]     \bigg\{   
              i u^{\prime} \sigma^+_1 \sigma^-_1  \sigma^+_2     ,    \bigg[ \underset{j \equiv N}{\prod}           \sigma^+_j \sigma^-_j     \bigg]   \bigg\}  \bigg[ \underset{j \text{ } \mathrm{odd}}{\underset{1 \leq j \leq N}{\prod}}          i u^{\prime} \sigma^+_j \sigma^-_j  \bigg]     \bigg[ \underset{j \equiv N}{\prod}           \sigma^+_j \sigma^-_j     \bigg]    \\  -   i u \sigma^+_1 \sigma^-_1 \sigma^+_2  \bigg[ \underset{j \text{ } \mathrm{odd}}{\underset{1 \leq j \leq N}{\prod}}          i u \sigma^+_j \sigma^-_j  \bigg]      i u^{\prime} \sigma^+_1 \sigma^-_1  \sigma^+_2  \bigg\{   
              \bigg[ \underset{j \text{ } \mathrm{odd}}{\underset{1 \leq j \leq N}{\prod}}          i u^{\prime} \sigma^+_j \sigma^-_j  \bigg]     \bigg[ \underset{j \equiv N}{\prod}           \sigma^+_j \sigma^-_j     \bigg]    ,    \bigg[ \underset{j \equiv N}{\prod}           \sigma^+_j \sigma^-_j     \bigg]   \bigg\}    \\    -  \big[    i u \sigma^+_1 \sigma^-_1 \sigma^+_2  \big]^2     \bigg\{     
              \bigg[ \underset{j \text{ } \mathrm{odd}}{\underset{1 \leq j \leq N}{\prod}}          i u^{\prime} \sigma^+_j \sigma^-_j  \bigg]     \bigg[ \underset{j \equiv N}{\prod}           \sigma^+_j \sigma^-_j     \bigg]  ,  \bigg[ \underset{j \text{ } \mathrm{odd}}{\underset{1 \leq j \leq N}{\prod}}          i u \sigma^+_j \sigma^-_j  \bigg]    \bigg\}     \bigg[ \underset{j \equiv N}{\prod}           \sigma^+_j \sigma^-_j     \bigg]   \\  -    i u \sigma^+_1 \sigma^-_1 \sigma^+_2     \bigg\{     
              i u^{\prime} \sigma^+_1 \sigma^-_1  \sigma^+_2  ,  \bigg[ \underset{j \text{ } \mathrm{odd}}{\underset{1 \leq j \leq N}{\prod}}          i u \sigma^+_j \sigma^-_j  \bigg]    \bigg\}   \bigg[ \underset{j \text{ } \mathrm{odd}}{\underset{1 \leq j \leq N}{\prod}}          i u^{\prime} \sigma^+_j \sigma^-_j  \bigg]     \bigg[ \underset{j \equiv N}{\prod}           \sigma^+_j \sigma^-_j     \bigg]    \bigg[ \underset{j \equiv N}{\prod}           \sigma^+_j \sigma^-_j     \bigg]   \\  -  \bigg\{                  i u^{\prime} \sigma^+_1 \sigma^-_1  \sigma^+_2 ,  i u \sigma^+_1 \sigma^-_1 \sigma^+_2    \bigg\}    \bigg[ \underset{j \text{ } \mathrm{odd}}{\underset{1 \leq j \leq N}{\prod}}          i u^{\prime} \sigma^+_j \sigma^-_j  \bigg]     \bigg[ \underset{j \equiv N}{\prod}           \sigma^+_j \sigma^-_j     \bigg]     \bigg[ \underset{j \text{ } \mathrm{odd}}{\underset{1 \leq j \leq N}{\prod}}          i u \sigma^+_j \sigma^-_j  \bigg]     \bigg[ \underset{j \equiv N}{\prod}           \sigma^+_j \sigma^-_j     \bigg]   \\ 
     -           i u^{\prime} \sigma^+_1 \sigma^-_1  \sigma^+_2  \bigg[ \underset{j \text{ } \mathrm{odd}}{\underset{1 \leq j \leq N}{\prod}}          i u^{\prime} \sigma^+_j \sigma^-_j  \bigg]   \bigg\{             \bigg[ \underset{j \equiv N}{\prod}           \sigma^+_j \sigma^-_j     \bigg]  ,  i u \sigma^+_1 \sigma^-_1 \sigma^+_2    \bigg\}     \bigg[ \underset{j \text{ } \mathrm{odd}}{\underset{1 \leq j \leq N}{\prod}}          i u \sigma^+_j \sigma^-_j  \bigg]     \bigg[ \underset{j \equiv N}{\prod}           \sigma^+_j \sigma^-_j     \bigg]             \\    -           i u^{\prime} \sigma^+_1 \sigma^-_1  \sigma^+_2    \bigg\{        \bigg[ \underset{j \text{ } \mathrm{odd}}{\underset{1 \leq j \leq N}{\prod}}          i u^{\prime} \sigma^+_j \sigma^-_j  \bigg]     ,  i u \sigma^+_1 \sigma^-_1 \sigma^+_2    \bigg\}    \bigg[ \underset{j \equiv N}{\prod}           \sigma^+_j \sigma^-_j     \bigg]     \bigg[ \underset{j \text{ } \mathrm{odd}}{\underset{1 \leq j \leq N}{\prod}}          i u \sigma^+_j \sigma^-_j  \bigg]     \bigg[ \underset{j \equiv N}{\prod}           \sigma^+_j \sigma^-_j     \bigg]   \\ \\ \overset{(\mathrm{LR})}{=}      -   i u \sigma^+_1 \sigma^-_1 \sigma^+_2  \bigg[ \underset{j \text{ } \mathrm{odd}}{\underset{1 \leq j \leq N}{\prod}}          i u \sigma^+_j \sigma^-_j  \bigg]     \bigg\{   
              i u^{\prime} \sigma^+_1 \sigma^-_1  \sigma^+_2     ,    \bigg[ \underset{j \equiv N}{\prod}           \sigma^+_j \sigma^-_j     \bigg]   \bigg\}  \bigg[ \underset{j \text{ } \mathrm{odd}}{\underset{1 \leq j \leq N}{\prod}}          i u^{\prime} \sigma^+_j \sigma^-_j  \bigg]     \bigg[ \underset{j \equiv N}{\prod}           \sigma^+_j \sigma^-_j     \bigg]  \\   -   i u \sigma^+_1 \sigma^-_1 \sigma^+_2  \bigg[ \underset{j \text{ } \mathrm{odd}}{\underset{1 \leq j \leq N}{\prod}}          i u \sigma^+_j \sigma^-_j  \bigg]      i u^{\prime} \sigma^+_1 \sigma^-_1  \sigma^+_2  \bigg\{   
              \bigg[ \underset{j \text{ } \mathrm{odd}}{\underset{1 \leq j \leq N}{\prod}}          i u^{\prime} \sigma^+_j \sigma^-_j  \bigg]      ,    \bigg[ \underset{j \equiv N}{\prod}           \sigma^+_j \sigma^-_j     \bigg]   \bigg\}   \bigg[ \underset{j \equiv N}{\prod}           \sigma^+_j \sigma^-_j     \bigg]  \end{align*}

      \begin{align*}   -   i u \sigma^+_1 \sigma^-_1 \sigma^+_2  \bigg[ \underset{j \text{ } \mathrm{odd}}{\underset{1 \leq j \leq N}{\prod}}          i u \sigma^+_j \sigma^-_j  \bigg]      i u^{\prime} \sigma^+_1 \sigma^-_1  \sigma^+_2   \bigg[ \underset{j \text{ } \mathrm{odd}}{\underset{1 \leq j \leq N}{\prod}}          i u^{\prime} \sigma^+_j \sigma^-_j  \bigg]   \bigg\{   
                 \bigg[ \underset{j \equiv N}{\prod}           \sigma^+_j \sigma^-_j     \bigg]    ,    \bigg[ \underset{j \equiv N}{\prod}           \sigma^+_j \sigma^-_j     \bigg]   \bigg\}    \\ -  \big[    i u \sigma^+_1 \sigma^-_1 \sigma^+_2  \big]^2     \bigg\{     
              \bigg[ \underset{j \text{ } \mathrm{odd}}{\underset{1 \leq j \leq N}{\prod}}          i u^{\prime} \sigma^+_j \sigma^-_j  \bigg]    ,  \bigg[ \underset{j \text{ } \mathrm{odd}}{\underset{1 \leq j \leq N}{\prod}}          i u \sigma^+_j \sigma^-_j  \bigg]    \bigg\}   \bigg[ \underset{j \equiv N}{\prod}           \sigma^+_j \sigma^-_j     \bigg]      \bigg[ \underset{j \equiv N}{\prod}           \sigma^+_j \sigma^-_j     \bigg] \\    -  \big[    i u \sigma^+_1 \sigma^-_1 \sigma^+_2  \big]^2  
              \bigg[ \underset{j \text{ } \mathrm{odd}}{\underset{1 \leq j \leq N}{\prod}}          i u^{\prime} \sigma^+_j \sigma^-_j  \bigg]     \bigg\{         \bigg[ \underset{j \equiv N}{\prod}           \sigma^+_j \sigma^-_j     \bigg]  ,  \bigg[ \underset{j \text{ } \mathrm{odd}}{\underset{1 \leq j \leq N}{\prod}}          i u \sigma^+_j \sigma^-_j  \bigg]    \bigg\}     \bigg[ \underset{j \equiv N}{\prod}           \sigma^+_j \sigma^-_j     \bigg]   \\ -     i u \sigma^+_1 \sigma^-_1 \sigma^+_2     \bigg\{     
              i u^{\prime} \sigma^+_1 \sigma^-_1  \sigma^+_2  ,  \bigg[ \underset{j \text{ } \mathrm{odd}}{\underset{1 \leq j \leq N}{\prod}}          i u \sigma^+_j \sigma^-_j  \bigg]    \bigg\}   \bigg[ \underset{j \text{ } \mathrm{odd}}{\underset{1 \leq j \leq N}{\prod}}          i u^{\prime} \sigma^+_j \sigma^-_j  \bigg]     \bigg[ \underset{j \equiv N}{\prod}           \sigma^+_j \sigma^-_j     \bigg]    \bigg[ \underset{j \equiv N}{\prod}           \sigma^+_j \sigma^-_j     \bigg]      \\ -  \bigg\{                  i u^{\prime} \sigma^+_1 \sigma^-_1  \sigma^+_2 ,  i u \sigma^+_1 \sigma^-_1 \sigma^+_2    \bigg\}    \bigg[ \underset{j \text{ } \mathrm{odd}}{\underset{1 \leq j \leq N}{\prod}}          i u^{\prime} \sigma^+_j \sigma^-_j  \bigg]     \bigg[ \underset{j \equiv N}{\prod}           \sigma^+_j \sigma^-_j     \bigg]     \bigg[ \underset{j \text{ } \mathrm{odd}}{\underset{1 \leq j \leq N}{\prod}}          i u \sigma^+_j \sigma^-_j  \bigg]     \bigg[ \underset{j \equiv N}{\prod}           \sigma^+_j \sigma^-_j     \bigg] \\  
     -           i u^{\prime} \sigma^+_1 \sigma^-_1  \sigma^+_2  \bigg[ \underset{j \text{ } \mathrm{odd}}{\underset{1 \leq j \leq N}{\prod}}          i u^{\prime} \sigma^+_j \sigma^-_j  \bigg]   \bigg\{             \bigg[ \underset{j \equiv N}{\prod}           \sigma^+_j \sigma^-_j     \bigg]  ,  i u \sigma^+_1 \sigma^-_1 \sigma^+_2    \bigg\}     \bigg[ \underset{j \text{ } \mathrm{odd}}{\underset{1 \leq j \leq N}{\prod}}          i u \sigma^+_j \sigma^-_j  \bigg]     \bigg[ \underset{j \equiv N}{\prod}           \sigma^+_j \sigma^-_j     \bigg]             \\  -           i u^{\prime} \sigma^+_1 \sigma^-_1  \sigma^+_2    \bigg\{        \bigg[ \underset{j \text{ } \mathrm{odd}}{\underset{1 \leq j \leq N}{\prod}}          i u^{\prime} \sigma^+_j \sigma^-_j  \bigg]     ,  i u \sigma^+_1 \sigma^-_1 \sigma^+_2    \bigg\}    \bigg[ \underset{j \equiv N}{\prod}           \sigma^+_j \sigma^-_j     \bigg]     \bigg[ \underset{j \text{ } \mathrm{odd}}{\underset{1 \leq j \leq N}{\prod}}          i u \sigma^+_j \sigma^-_j  \bigg]     \bigg[ \underset{j \equiv N}{\prod}           \sigma^+_j \sigma^-_j     \bigg]        \text{, }
    \end{align*}

    \noindent corresponding to the fifth term, which can be approximated with,

    \begin{align*}
 \big( \mathscr{C}^1_1 \big)_5    \text{. }
    \end{align*}

    \noindent For the next term, one has,

\begin{align*}
         \bigg\{         i u \sigma^+_1 \sigma^-_1 \sigma^+_2 \bigg[ \underset{j \text{ } \mathrm{odd}}{\underset{1 \leq j \leq N}{\prod}}          i u \sigma^+_j \sigma^-_j  \bigg]     \bigg[ \underset{j \equiv N}{\prod}           \sigma^+_j \sigma^-_j     \bigg] 
     ,          i u^{\prime} \sigma^+_1 \sigma^-_1  \sigma^+_2 \bigg[ \underset{j \text{ } \mathrm{odd}}{\underset{1 \leq j \leq N}{\prod}}          i u^{\prime} \sigma^+_j \sigma^-_j  \bigg]     \bigg[ \underset{j \equiv N}{\prod}           \sigma^+_j \sigma^-_j     \bigg]      \bigg\} \overset{(\mathrm{LR})}{=}          i u \sigma^+_1 \sigma^-_1 \sigma^+_2  \\ 
     \times    \bigg\{      \bigg[ \underset{j \text{ } \mathrm{odd}}{\underset{1 \leq j \leq N}{\prod}}          i u \sigma^+_j \sigma^-_j  \bigg]     \bigg[ \underset{j \equiv N}{\prod}           \sigma^+_j \sigma^-_j     \bigg] 
     ,          i u^{\prime} \sigma^+_1 \sigma^-_1  \sigma^+_2 \bigg[ \underset{j \text{ } \mathrm{odd}}{\underset{1 \leq j \leq N}{\prod}}          i u^{\prime} \sigma^+_j \sigma^-_j  \bigg]     \bigg[ \underset{j \equiv N}{\prod}           \sigma^+_j \sigma^-_j     \bigg]      \bigg\}             +     \bigg\{         i u \sigma^+_1 \sigma^-_1 \sigma^+_2  ,          i u^{\prime} \sigma^+_1 \sigma^-_1  \sigma^+_2 \\ 
     \times \bigg[ \underset{j \text{ } \mathrm{odd}}{\underset{1 \leq j \leq N}{\prod}}          i u^{\prime} \sigma^+_j \sigma^-_j  \bigg]     \bigg[ \underset{j \equiv N}{\prod}           \sigma^+_j \sigma^-_j     \bigg]      \bigg\} \bigg[ \underset{j \text{ } \mathrm{odd}}{\underset{1 \leq j \leq N}{\prod}}          i u \sigma^+_j \sigma^-_j  \bigg]     \bigg[ \underset{j \equiv N}{\prod}           \sigma^+_j \sigma^-_j     \bigg] 
       \\ \\ \overset{(\mathrm{AC}),(\mathrm{LR})}{=}                     i u \sigma^+_1 \sigma^-_1 \sigma^+_2     \bigg[ \underset{j \text{ } \mathrm{odd}}{\underset{1 \leq j \leq N}{\prod}}          i u \sigma^+_j \sigma^-_j  \bigg]      \bigg\{      \bigg[ \underset{j \equiv N}{\prod}           \sigma^+_j \sigma^-_j     \bigg] 
     ,          i u^{\prime} \sigma^+_1 \sigma^-_1  \sigma^+_2 \bigg[ \underset{j \text{ } \mathrm{odd}}{\underset{1 \leq j \leq N}{\prod}}          i u^{\prime} \sigma^+_j \sigma^-_j  \bigg]       \bigg[ \underset{j \equiv N}{\prod}           \sigma^+_j \sigma^-_j     \bigg]      \bigg\}   \\   +      i u \sigma^+_1 \sigma^-_1 \sigma^+_2    \bigg\{      \bigg[ \underset{j \text{ } \mathrm{odd}}{\underset{1 \leq j \leq N}{\prod}}          i u \sigma^+_j \sigma^-_j  \bigg]   
     ,          i u^{\prime} \sigma^+_1 \sigma^-_1  \sigma^+_2 \bigg[ \underset{j \text{ } \mathrm{odd}}{\underset{1 \leq j \leq N}{\prod}}          i u^{\prime} \sigma^+_j \sigma^-_j  \bigg]       \bigg[ \underset{j \equiv N}{\prod}           \sigma^+_j \sigma^-_j     \bigg]      \bigg\}   \bigg[ \underset{j \equiv N}{\prod}           \sigma^+_j \sigma^-_j     \bigg]    \\     -     \bigg\{           i u^{\prime} \sigma^+_1 \sigma^-_1  \sigma^+_2  \bigg[ \underset{j \text{ } \mathrm{odd}}{\underset{1 \leq j \leq N}{\prod}}          i u^{\prime} \sigma^+_j \sigma^-_j  \bigg]     \bigg[ \underset{j \equiv N}{\prod}           \sigma^+_j \sigma^-_j     \bigg]  ,        i u \sigma^+_1 \sigma^-_1 \sigma^+_2       \bigg\} \bigg[ \underset{j \text{ } \mathrm{odd}}{\underset{1 \leq j \leq N}{\prod}}          i u \sigma^+_j \sigma^-_j  \bigg]     \bigg[ \underset{j \equiv N}{\prod}           \sigma^+_j \sigma^-_j     \bigg] \\ \\ \overset{(\mathrm{LR})}{=}                i u \sigma^+_1 \sigma^-_1 \sigma^+_2     \bigg[ \underset{j \text{ } \mathrm{odd}}{\underset{1 \leq j \leq N}{\prod}}          i u \sigma^+_j \sigma^-_j  \bigg]      \bigg\{      \bigg[ \underset{j \equiv N}{\prod}           \sigma^+_j \sigma^-_j     \bigg] 
     ,          i u^{\prime} \sigma^+_1 \sigma^-_1  \sigma^+_2 \bigg[ \underset{j \text{ } \mathrm{odd}}{\underset{1 \leq j \leq N}{\prod}}          i u^{\prime} \sigma^+_j \sigma^-_j  \bigg]       \bigg[ \underset{j \equiv N}{\prod}           \sigma^+_j \sigma^-_j     \bigg]      \bigg\}       \end{align*}

      \begin{align*}   +      i u \sigma^+_1 \sigma^-_1 \sigma^+_2    \bigg\{      \bigg[ \underset{j \text{ } \mathrm{odd}}{\underset{1 \leq j \leq N}{\prod}}          i u \sigma^+_j \sigma^-_j  \bigg]   
     ,          i u^{\prime} \sigma^+_1 \sigma^-_1  \sigma^+_2 \bigg[ \underset{j \text{ } \mathrm{odd}}{\underset{1 \leq j \leq N}{\prod}}          i u^{\prime} \sigma^+_j \sigma^-_j  \bigg]       \bigg[ \underset{j \equiv N}{\prod}           \sigma^+_j \sigma^-_j     \bigg]      \bigg\}   \bigg[ \underset{j \equiv N}{\prod}           \sigma^+_j \sigma^-_j     \bigg]    \\    -     \bigg\{           i u^{\prime} \sigma^+_1 \sigma^-_1  \sigma^+_2      ,        i u \sigma^+_1 \sigma^-_1 \sigma^+_2       \bigg\}   \bigg[ \underset{j \text{ } \mathrm{odd}}{\underset{1 \leq j \leq N}{\prod}}          i u^{\prime} \sigma^+_j \sigma^-_j  \bigg]   \bigg[ \underset{j \equiv N}{\prod}           \sigma^+_j \sigma^-_j     \bigg]  \bigg[ \underset{j \text{ } \mathrm{odd}}{\underset{1 \leq j \leq N}{\prod}}          i u \sigma^+_j \sigma^-_j  \bigg]     \bigg[ \underset{j \equiv N}{\prod}           \sigma^+_j \sigma^-_j     \bigg]  \\  -        i u^{\prime} \sigma^+_1 \sigma^-_1  \sigma^+_2  \bigg\{         \bigg[ \underset{j \text{ } \mathrm{odd}}{\underset{1 \leq j \leq N}{\prod}}          i u^{\prime} \sigma^+_j \sigma^-_j  \bigg]     \bigg[ \underset{j \equiv N}{\prod}           \sigma^+_j \sigma^-_j     \bigg]  ,        i u \sigma^+_1 \sigma^-_1 \sigma^+_2       \bigg\} \bigg[ \underset{j \text{ } \mathrm{odd}}{\underset{1 \leq j \leq N}{\prod}}          i u \sigma^+_j \sigma^-_j  \bigg]     \bigg[ \underset{j \equiv N}{\prod}           \sigma^+_j \sigma^-_j     \bigg]       \\ \\      \overset{(\mathrm{AC}),(\mathrm{LR})}{=}   i u \sigma^+_1 \sigma^-_1 \sigma^+_2     \bigg[ \underset{j \text{ } \mathrm{odd}}{\underset{1 \leq j \leq N}{\prod}}          i u \sigma^+_j \sigma^-_j  \bigg]      \bigg\{        i u^{\prime} \sigma^+_1 \sigma^-_1  \sigma^+_2 \bigg[ \underset{j \text{ } \mathrm{odd}}{\underset{1 \leq j \leq N}{\prod}}          i u^{\prime} \sigma^+_j \sigma^-_j  \bigg]       \bigg[ \underset{j \equiv N}{\prod}           \sigma^+_j \sigma^-_j     \bigg]  ,     \bigg[ \underset{j \equiv N}{\prod}           \sigma^+_j \sigma^-_j     \bigg] 
        \bigg\}             \end{align*}

     \begin{align*}  -      i u \sigma^+_1 \sigma^-_1 \sigma^+_2 i u^{\prime} \sigma^+_1 \sigma^-_1  \sigma^+_2     \bigg\{         
             \bigg[ \underset{j \text{ } \mathrm{odd}}{\underset{1 \leq j \leq N}{\prod}}          i u^{\prime} \sigma^+_j \sigma^-_j  \bigg]       \bigg[ \underset{j \equiv N}{\prod}           \sigma^+_j \sigma^-_j     \bigg]  , \bigg[ \underset{j \text{ } \mathrm{odd}}{\underset{1 \leq j \leq N}{\prod}}          i u \sigma^+_j \sigma^-_j  \bigg]     \bigg\}   \bigg[ \underset{j \equiv N}{\prod}           \sigma^+_j \sigma^-_j     \bigg]   \\  -      i u \sigma^+_1 \sigma^-_1 \sigma^+_2    \bigg\{         
             i u^{\prime} \sigma^+_1 \sigma^-_1  \sigma^+_2   , \bigg[ \underset{j \text{ } \mathrm{odd}}{\underset{1 \leq j \leq N}{\prod}}          i u \sigma^+_j \sigma^-_j  \bigg]     \bigg\}  \bigg[ \underset{j \text{ } \mathrm{odd}}{\underset{1 \leq j \leq N}{\prod}}          i u^{\prime} \sigma^+_j \sigma^-_j  \bigg]       \bigg[ \underset{j \equiv N}{\prod}           \sigma^+_j \sigma^-_j     \bigg]  \bigg[ \underset{j \equiv N}{\prod}           \sigma^+_j \sigma^-_j     \bigg]    \\ 
     -     \bigg\{           i u^{\prime} \sigma^+_1 \sigma^-_1  \sigma^+_2      ,        i u \sigma^+_1 \sigma^-_1 \sigma^+_2       \bigg\}   \bigg[ \underset{j \text{ } \mathrm{odd}}{\underset{1 \leq j \leq N}{\prod}}          i u^{\prime} \sigma^+_j \sigma^-_j  \bigg]   \bigg[ \underset{j \equiv N}{\prod}           \sigma^+_j \sigma^-_j     \bigg]  \bigg[ \underset{j \text{ } \mathrm{odd}}{\underset{1 \leq j \leq N}{\prod}}          i u \sigma^+_j \sigma^-_j  \bigg]     \bigg[ \underset{j \equiv N}{\prod}           \sigma^+_j \sigma^-_j     \bigg]  \\ -        i u^{\prime} \sigma^+_1 \sigma^-_1  \sigma^+_2  \bigg\{         \bigg[ \underset{j \text{ } \mathrm{odd}}{\underset{1 \leq j \leq N}{\prod}}          i u^{\prime} \sigma^+_j \sigma^-_j  \bigg]      ,        i u \sigma^+_1 \sigma^-_1 \sigma^+_2       \bigg\}  \bigg[ \underset{j \equiv N}{\prod}           \sigma^+_j \sigma^-_j     \bigg]  \bigg[ \underset{j \text{ } \mathrm{odd}}{\underset{1 \leq j \leq N}{\prod}}          i u \sigma^+_j \sigma^-_j  \bigg]     \bigg[ \underset{j \equiv N}{\prod}           \sigma^+_j \sigma^-_j     \bigg]  \\  -        i u^{\prime} \sigma^+_1 \sigma^-_1  \sigma^+_2    \bigg[ \underset{j \text{ } \mathrm{odd}}{\underset{1 \leq j \leq N}{\prod}}          i u^{\prime} \sigma^+_j \sigma^-_j  \bigg]  \bigg\{            \bigg[ \underset{j \equiv N}{\prod}           \sigma^+_j \sigma^-_j     \bigg]  ,        i u \sigma^+_1 \sigma^-_1 \sigma^+_2       \bigg\} \bigg[ \underset{j \text{ } \mathrm{odd}}{\underset{1 \leq j \leq N}{\prod}}          i u \sigma^+_j \sigma^-_j  \bigg]     \bigg[ \underset{j \equiv N}{\prod}           \sigma^+_j \sigma^-_j     \bigg]                     \text{, }
\end{align*}

\noindent corresponding to the sixth term, which can be approximated with, 

\begin{align*}
    \big( \mathscr{C}^1_1 \big)_6 \text{. }
    \end{align*}

    \noindent For the next term, one has,

\begin{align*}
  \bigg\{  \bigg[ \underset{j \text{ } \mathrm {odd}}{\underset{1 \leq j \leq N-1}{\prod}} i u \sigma^+_j \sigma^-_j  \bigg]    \bigg[  \underset{j \text{ } \mathrm{odd}}{\underset{3 \leq j \leq N-1}{\prod}}       i u^{-1} \sigma^+_j \sigma^-_j      \bigg] \bigg[ \underset{j \equiv N}{\prod}  i u \sigma^+_j \sigma^-_j  \bigg]               
     ,        i u^{\prime} \sigma^+_0 \sigma^-_0   \bigg[    \underset{j \text{ } \mathrm{odd}}{\underset{1 \leq j \leq N-1}{\prod}}     i u^{\prime} \sigma^+_j \sigma^-_j \textbf{1}_{{\{j^{\prime}> j \text{ } : \text{ } i u^{\prime} \sigma^+_{j^{\prime}} \sigma^-_{j^{\prime}} \in \mathrm{support} ( A ( u ) ) } \}} \\  \times   i u^{\prime} \sigma^+_{j^{\prime}} \sigma^-_{j^{\prime}} i u^{\prime} \sigma^+_{j^{\prime}+1} \sigma^-_{j^{\prime}+1}  \bigg]              \bigg\} \overset{(\mathrm{LR})}{=}    \bigg[ \underset{j \text{ } \mathrm {odd}}{\underset{1 \leq j \leq N-1}{\prod}} i u \sigma^+_j \sigma^-_j  \bigg]    \bigg[  \underset{j \text{ } \mathrm{odd}}{\underset{3 \leq j \leq N-1}{\prod}}       i u^{-1} \sigma^+_j \sigma^-_j      \bigg] 
  \bigg\{    \bigg[ \underset{j \equiv N}{\prod}  i u \sigma^+_j \sigma^-_j  \bigg]               
     ,        i u^{\prime} \sigma^+_0 \sigma^-_0  \\  \times   \bigg[    \underset{j \text{ } \mathrm{odd}}{\underset{1 \leq j \leq N-1}{\prod}}     i u^{\prime} \sigma^+_j \sigma^-_j \textbf{1}_{{\{j^{\prime}> j \text{ } : \text{ } i u^{\prime} \sigma^+_{j^{\prime}} \sigma^-_{j^{\prime}} \in \mathrm{support} ( A ( u ) ) } \}}  \bigg] \bigg\}  +  \bigg\{  \bigg[ \underset{j \text{ } \mathrm {odd}}{\underset{1 \leq j \leq N-1}{\prod}} i u \sigma^+_j \sigma^-_j  \bigg]   \bigg[  \underset{j \text{ } \mathrm{odd}}{\underset{3 \leq j \leq N-1}{\prod}}       i u^{-1} \sigma^+_j \sigma^-_j      \bigg] 
                \\     ,        i u^{\prime} \sigma^+_0 \sigma^-_0    \bigg[    \underset{j \text{ } \mathrm{odd}}{\underset{1 \leq j \leq N-1}{\prod}}     i u^{\prime} \sigma^+_j \sigma^-_j \textbf{1}_{{\{j^{\prime}> j \text{ } : \text{ } i u^{\prime} \sigma^+_{j^{\prime}} \sigma^-_{j^{\prime}} \in \mathrm{support} ( A ( u ) ) } \}} \bigg]  \bigg\}  \bigg[ \underset{j \equiv N}{\prod}  i u            \sigma^+_j \sigma^-_j  \bigg]      \end{align*}

      \begin{align*}    \overset{(\mathrm{AC}),(\mathrm{LR})}{=}        -     \bigg[ \underset{j \text{ } \mathrm {odd}}{\underset{1 \leq j \leq N-1}{\prod}} i u \sigma^+_j \sigma^-_j  \bigg]    \bigg[  \underset{j \text{ } \mathrm{odd}}{\underset{3 \leq j \leq N-1}{\prod}}       i u^{-1} \sigma^+_j \sigma^-_j      \bigg] 
                    i u^{\prime} \sigma^+_0 \sigma^-_0 \bigg\{   \bigg[    \underset{j \text{ } \mathrm{odd}}{\underset{1 \leq j \leq N-1}{\prod}}     i u^{\prime} \sigma^+_j \sigma^-_j \textbf{1}_{{\{j^{\prime}> j \text{ } : \text{ } i u^{\prime} \sigma^+_{j^{\prime}} \sigma^-_{j^{\prime}} \in \mathrm{support} ( A ( u ) ) } \}}  \bigg]  \\  , \bigg[ \underset{j \equiv N}{\prod}  i u \sigma^+_j \sigma^-_j \bigg]          \bigg\}     -      \bigg[ \underset{j \text{ } \mathrm {odd}}{\underset{1 \leq j \leq N-1}{\prod}} i u \sigma^+_j \sigma^-_j  \bigg]    \bigg[  \underset{j \text{ } \mathrm{odd}}{\underset{3 \leq j \leq N-1}{\prod}}       i u^{-1} \sigma^+_j \sigma^-_j      \bigg]  \bigg\{  i u^{\prime} \sigma^+_0 \sigma^-_0   , \bigg[ \underset{j \equiv N}{\prod}  i u \sigma^+_j \sigma^-_j \bigg]          \bigg\} \\ \times  \bigg[    \underset{j \text{ } \mathrm{odd}}{\underset{1 \leq j \leq N-1}{\prod}}     i u^{\prime} \sigma^+_j \sigma^-_j \textbf{1}_{{\{j^{\prime}> j \text{ } : \text{ } i u^{\prime} \sigma^+_{j^{\prime}} \sigma^-_{j^{\prime}} \in \mathrm{support} ( A ( u ) ) } \}}  \bigg]  + \bigg[ \underset{j \text{ } \mathrm {odd}}{\underset{1 \leq j \leq N-1}{\prod}} i u \sigma^+_j \sigma^-_j  \bigg]    \bigg\{    \bigg[  \underset{j \text{ } \mathrm{odd}}{\underset{3 \leq j \leq N-1}{\prod}}       i u^{-1} \sigma^+_j \sigma^-_j      \bigg] 
                \\       ,        i u^{\prime} \sigma^+_0 \sigma^-_0    \bigg[    \underset{j \text{ } \mathrm{odd}}{\underset{1 \leq j \leq N-1}{\prod}}     i u^{\prime} \sigma^+_j \sigma^-_j \textbf{1}_{{\{j^{\prime}> j \text{ } : \text{ } i u^{\prime} \sigma^+_{j^{\prime}} \sigma^-_{j^{\prime}} \in \mathrm{support} ( A ( u ) ) } \}} \bigg]  \bigg\}  \bigg[ \underset{j \equiv N}{\prod}  i u            \sigma^+_j \sigma^-_j  \bigg]    \\ + 
                 \bigg\{  \bigg[ \underset{j \text{ } \mathrm {odd}}{\underset{1 \leq j \leq N-1}{\prod}} i u \sigma^+_j \sigma^-_j  \bigg]   ,        i u^{\prime} \sigma^+_0 \sigma^-_0    \bigg[    \underset{j \text{ } \mathrm{odd}}{\underset{1 \leq j \leq N-1}{\prod}}     i u^{\prime} \sigma^+_j \sigma^-_j \textbf{1}_{{\{j^{\prime}> j \text{ } : \text{ } i u^{\prime} \sigma^+_{j^{\prime}} \sigma^-_{j^{\prime}} \in \mathrm{support} ( A ( u ) ) } \}} \bigg]  \bigg\}   \bigg[  \underset{j \text{ } \mathrm{odd}}{\underset{3 \leq j \leq N-1}{\prod}}       i u^{-1} \sigma^+_j \sigma^-_j      \bigg] 
\end{align*}

                \begin{align*}  \times  \bigg[ \underset{j \equiv N}{\prod}  i u \sigma^+_j \sigma^-_j \bigg] \\  \overset{(\mathrm{AC}),(\mathrm{LR})}{=}         -     \bigg[ \underset{j \text{ } \mathrm {odd}}{\underset{1 \leq j \leq N-1}{\prod}} i u \sigma^+_j \sigma^-_j  \bigg]    \bigg[  \underset{j \text{ } \mathrm{odd}}{\underset{3 \leq j \leq N-1}{\prod}}       i u^{-1} \sigma^+_j \sigma^-_j      \bigg] 
                    i u^{\prime} \sigma^+_0 \sigma^-_0 \bigg\{   \bigg[    \underset{j \text{ } \mathrm{odd}}{\underset{1 \leq j \leq N-1}{\prod}}     i u^{\prime} \sigma^+_j \sigma^-_j \textbf{1}_{{\{j^{\prime}> j \text{ } : \text{ } i u^{\prime} \sigma^+_{j^{\prime}} \sigma^-_{j^{\prime}} \in \mathrm{support} ( A ( u ) ) } \}}  \bigg]  \\ , \bigg[ \underset{j \equiv N}{\prod}  i u \sigma^+_j \sigma^-_j \bigg]          \bigg\}     -      \bigg[ \underset{j \text{ } \mathrm {odd}}{\underset{1 \leq j \leq N-1}{\prod}} i u \sigma^+_j \sigma^-_j  \bigg]    \bigg[  \underset{j \text{ } \mathrm{odd}}{\underset{3 \leq j \leq N-1}{\prod}}       i u^{-1} \sigma^+_j \sigma^-_j      \bigg]  \bigg\{  i u^{\prime} \sigma^+_0 \sigma^-_0   , \bigg[ \underset{j \equiv N}{\prod}  i u \sigma^+_j \sigma^-_j \bigg]          \bigg\} \\  \times  \bigg[    \underset{j \text{ } \mathrm{odd}}{\underset{1 \leq j \leq N-1}{\prod}}     i u^{\prime} \sigma^+_j \sigma^-_j \textbf{1}_{{\{j^{\prime}> j \text{ } : \text{ } i u^{\prime} \sigma^+_{j^{\prime}} \sigma^-_{j^{\prime}} \in \mathrm{support} ( A ( u ) ) } \}}  \bigg] - \bigg[ \underset{j \text{ } \mathrm {odd}}{\underset{1 \leq j \leq N-1}{\prod}} i u \sigma^+_j \sigma^-_j  \bigg]       i u^{\prime} \sigma^+_0 \sigma^-_0    \\   \times   \bigg\{        \bigg[    \underset{j \text{ } \mathrm{odd}}{\underset{1 \leq j \leq N-1}{\prod}}     i u^{\prime} \sigma^+_j \sigma^-_j \textbf{1}_{{\{j^{\prime}> j \text{ } : \text{ } i u^{\prime} \sigma^+_{j^{\prime}} \sigma^-_{j^{\prime}} \in \mathrm{support} ( A ( u ) ) } \}} \bigg] ,  \bigg[  \underset{j \text{ } \mathrm{odd}}{\underset{3 \leq j \leq N-1}{\prod}}       i u^{-1} \sigma^+_j \sigma^-_j      \bigg] 
                 \bigg\}  \bigg[ \underset{j \equiv N}{\prod}  i u            \sigma^+_j \sigma^-_j  \bigg]     \\   -    \bigg[ \underset{j \text{ } \mathrm {odd}}{\underset{1 \leq j \leq N-1}{\prod}} i u \sigma^+_j \sigma^-_j  \bigg]    \bigg\{         i u^{\prime} \sigma^+_0 \sigma^-_0  ,  \bigg[  \underset{j \text{ } \mathrm{odd}}{\underset{3 \leq j \leq N-1}{\prod}}       i u^{-1} \sigma^+_j \sigma^-_j      \bigg] 
                 \bigg\}  \end{align*}

              \begin{align*}  \times  \bigg[    \underset{j \text{ } \mathrm{odd}}{\underset{1 \leq j \leq N-1}{\prod}}     i u^{\prime} \sigma^+_j \sigma^-_j \textbf{1}_{{\{j^{\prime}> j \text{ } : \text{ } i u^{\prime} \sigma^+_{j^{\prime}} \sigma^-_{j^{\prime}} \in \mathrm{support} ( A ( u ) ) } \}} \bigg] \bigg[ \underset{j \equiv N}{\prod}  i u            \sigma^+_j \sigma^-_j  \bigg]     \\  - i u^{\prime} \sigma^+_0 \sigma^-_0   \bigg\{      \bigg[    \underset{j \text{ } \mathrm{odd}}{\underset{1 \leq j \leq N-1}{\prod}}     i u^{\prime} \sigma^+_j \sigma^-_j \textbf{1}_{{\{j^{\prime}> j \text{ } : \text{ } i u^{\prime} \sigma^+_{j^{\prime}} \sigma^-_{j^{\prime}} \in \mathrm{support} ( A ( u ) ) } \}} \bigg] 
  ,  \bigg[ \underset{j \text{ } \mathrm {odd}}{\underset{1 \leq j \leq N-1}{\prod}} i u \sigma^+_j \sigma^-_j  \bigg]    \bigg\}   \bigg[  \underset{j \text{ } \mathrm{odd}}{\underset{3 \leq j \leq N-1}{\prod}}       i u^{-1} \sigma^+_j \sigma^-_j      \bigg]   \\  \times  \bigg[ \underset{j \equiv N}{\prod}  i u \sigma^+_j \sigma^-_j \bigg]  -  \bigg\{       i u^{\prime} \sigma^+_0 \sigma^-_0    
  ,  \bigg[ \underset{j \text{ } \mathrm {odd}}{\underset{1 \leq j \leq N-1}{\prod}} i u \sigma^+_j \sigma^-_j  \bigg]    \bigg\} \\  \times   \bigg[    \underset{j \text{ } \mathrm{odd}}{\underset{1 \leq j \leq N-1}{\prod}}     i u^{\prime} \sigma^+_j \sigma^-_j \textbf{1}_{{\{j^{\prime}> j \text{ } : \text{ } i u^{\prime} \sigma^+_{j^{\prime}} \sigma^-_{j^{\prime}} \in \mathrm{support} ( A ( u ) ) } \}} \bigg]   \bigg[  \underset{j \text{ } \mathrm{odd}}{\underset{3 \leq j \leq N-1}{\prod}}       i u^{-1} \sigma^+_j \sigma^-_j      \bigg] 
               \bigg[ \underset{j \equiv N}{\prod}  i u \sigma^+_j \sigma^-_j \bigg]                               \text{, }
\end{align*}

\noindent corresponding to the seventh term, which can be approximated with, 

\begin{align*}
  \big( \mathscr{C}^1_1 \big)_7  \text{. }
    \end{align*}

    \noindent For the next term, after one application of (LR) one has,

\begin{align*}
     \bigg[ \underset{j \text{ } \mathrm {odd}}{\underset{1 \leq j \leq N-1}{\prod}} i u \sigma^+_j \sigma^-_j  \bigg]    \bigg[  \underset{j \text{ } \mathrm{odd}}{\underset{3 \leq j \leq N-1}{\prod}}       i u^{-1} \sigma^+_j \sigma^-_j      \bigg]   \bigg\{       \bigg[ \underset{j \equiv N}{\prod}  i u \sigma^+_j \sigma^-_j  \bigg]                     ,  i u^{\prime} \sigma^+_1 \sigma^-_1 \sigma^+_2 \\ \times \bigg[ \underset{j \text{ } \mathrm{odd}}{\underset{1 \leq j\leq N}{\prod}}                 i u \sigma^+_j \sigma^-_j    \bigg] \bigg[ \underset{j \equiv N}{\prod} \sigma^+_j \sigma^-_j  \bigg]  \bigg\}     \\ + \bigg\{      \bigg[ \underset{j \text{ } \mathrm {odd}}{\underset{1 \leq j \leq N-1}{\prod}} i u \sigma^+_j \sigma^-_j  \bigg]    \bigg[  \underset{j \text{ } \mathrm{odd}}{\underset{3 \leq j \leq N-1}{\prod}}       i u^{-1} \sigma^+_j \sigma^-_j      \bigg]  ,   i u^{\prime} \sigma^+_1 \sigma^-_1 \sigma^+_2  \\ \times \bigg[ \underset{j \text{ } \mathrm{odd}}{\underset{1 \leq j\leq N}{\prod}}                 i u \sigma^+_j \sigma^-_j    \bigg] \bigg[ \underset{j \equiv N}{\prod} \sigma^+_j \sigma^-_j  \bigg]  \bigg\}    \bigg[ \underset{j \equiv N}{\prod}  i u \sigma^+_j \sigma^-_j  \bigg]         \end{align*}

\begin{align*} \overset{(\mathrm{AC}),(\mathrm{LR})}{=}                       -  \bigg[ \underset{j \text{ } \mathrm {odd}}{\underset{1 \leq j \leq N-1}{\prod}} i u \sigma^+_j \sigma^-_j  \bigg]    \bigg[  \underset{j \text{ } \mathrm{odd}}{\underset{3 \leq j \leq N-1}{\prod}}       i u^{-1} \sigma^+_j \sigma^-_j      \bigg] 
                    i u^{\prime} \sigma^+_1 \sigma^-_1 \sigma^+_2 \bigg[ \underset{j \text{ } \mathrm{odd}}{\underset{1 \leq j\leq N}{\prod}}                 i u \sigma^+_j \sigma^-_j    \bigg]        \bigg\{ \bigg[       \underset{j \equiv N}{\prod}   \sigma^+_j \sigma^-_j    \bigg] , \bigg[   \underset{j \equiv N}{\prod}    i u \sigma^+_j \sigma^-_j            \bigg]  \bigg\}         \\ 
                - \bigg[ \underset{j \text{ } \mathrm {odd}}{\underset{1 \leq j \leq N-1}{\prod}} i u \sigma^+_j \sigma^-_j  \bigg]    \bigg[  \underset{j \text{ } \mathrm{odd}}{\underset{3 \leq j \leq N-1}{\prod}}       i u^{-1} \sigma^+_j \sigma^-_j      \bigg]         \bigg\{   i u^{\prime} \sigma^+_1 \sigma^-_1 \sigma^+_2 \bigg[ \underset{j \text{ } \mathrm{odd}}{\underset{1 \leq j\leq N}{\prod}}                 i u \sigma^+_j \sigma^-_j    \bigg]    , \bigg[   \underset{j \equiv N}{\prod}    i u \sigma^+_j \sigma^-_j            \bigg]  \bigg\}       \bigg[       \underset{j \equiv N}{\prod}   \sigma^+_j \sigma^-_j    \bigg] 
   \\   +  \bigg[ \underset{j \text{ } \mathrm {odd}}{\underset{1 \leq j \leq N-1}{\prod}} i u \sigma^+_j \sigma^-_j  \bigg]   \bigg\{         \bigg[  \underset{j \text{ } \mathrm{odd}}{\underset{3 \leq j \leq N-1}{\prod}}       i u^{-1} \sigma^+_j \sigma^-_j      \bigg]  ,   i u^{\prime} \sigma^+_1 \sigma^-_1 \sigma^+_2 \bigg[ \underset{j \text{ } \mathrm{odd}}{\underset{1 \leq j\leq N}{\prod}}                 i u \sigma^+_j \sigma^-_j    \bigg] \bigg[ \underset{j \equiv N}{\prod} \sigma^+_j \sigma^-_j  \bigg]  \bigg\}    \bigg[ \underset{j \equiv N}{\prod}  i u \sigma^+_j \sigma^-_j  \bigg] \\ + \bigg\{      \bigg[ \underset{j \text{ } \mathrm {odd}}{\underset{1 \leq j \leq N-1}{\prod}} i u \sigma^+_j \sigma^-_j  \bigg]    \bigg[  \underset{j \text{ } \mathrm{odd}}{\underset{3 \leq j \leq N-1}{\prod}}       i u^{-1} \sigma^+_j \sigma^-_j      \bigg]  ,   i u^{\prime} \sigma^+_1 \sigma^-_1 \sigma^+_2 \bigg[ \underset{j \text{ } \mathrm{odd}}{\underset{1 \leq j\leq N}{\prod}}                 i u \sigma^+_j \sigma^-_j    \bigg] \bigg[ \underset{j \equiv N}{\prod} \sigma^+_j \sigma^-_j  \bigg]  \bigg\}    \bigg[ \underset{j \equiv N}{\prod}  i u \sigma^+_j \sigma^-_j  \bigg]          \end{align*}

                    \begin{align*} \overset{(\mathrm{AC}),(\mathrm{LR})}{=}      -  \bigg[ \underset{j \text{ } \mathrm {odd}}{\underset{1 \leq j \leq N-1}{\prod}} i u \sigma^+_j \sigma^-_j  \bigg]    \bigg[  \underset{j \text{ } \mathrm{odd}}{\underset{3 \leq j \leq N-1}{\prod}}       i u^{-1} \sigma^+_j \sigma^-_j      \bigg] 
                    i u^{\prime} \sigma^+_1 \sigma^-_1 \sigma^+_2 \bigg[ \underset{j \text{ } \mathrm{odd}}{\underset{1 \leq j\leq N}{\prod}}                 i u \sigma^+_j \sigma^-_j    \bigg]        \bigg\{ \bigg[       \underset{j \equiv N}{\prod}   \sigma^+_j \sigma^-_j    \bigg] , \bigg[   \underset{j \equiv N}{\prod}    i u \sigma^+_j \sigma^-_j            \bigg]  \bigg\}           \\ - \bigg[ \underset{j \text{ } \mathrm {odd}}{\underset{1 \leq j \leq N-1}{\prod}} i u \sigma^+_j \sigma^-_j  \bigg]    \bigg[  \underset{j \text{ } \mathrm{odd}}{\underset{3 \leq j \leq N-1}{\prod}}       i u^{-1} \sigma^+_j \sigma^-_j      \bigg]         \bigg\{   i u^{\prime} \sigma^+_1 \sigma^-_1 \sigma^+_2 \bigg[ \underset{j \text{ } \mathrm{odd}}{\underset{1 \leq j\leq N}{\prod}}                 i u \sigma^+_j \sigma^-_j    \bigg]    , \bigg[   \underset{j \equiv N}{\prod}    i u \sigma^+_j \sigma^-_j            \bigg]  \bigg\}       \bigg[       \underset{j \equiv N}{\prod}   \sigma^+_j \sigma^-_j    \bigg]                        \text{, }
\end{align*}

\noindent \noindent corresponding to the eighth term, , which can be approximated with, 

\begin{align*}
  \big( \mathscr{C}^1_1 \big)_8   \text{, }
    \end{align*}

    \noindent and,

\begin{align*}
\bigg\{  \bigg[ \underset{j \text{ } \mathrm {odd}}{\underset{1 \leq j \leq N-1}{\prod}} i u \sigma^+_j \sigma^-_j  \bigg]    \bigg[  \underset{j \text{ } \mathrm{odd}}{\underset{3 \leq j \leq N-1}{\prod}}       i u^{-1} \sigma^+_j \sigma^-_j      \bigg] \bigg[ \underset{j \equiv N}{\prod}  i u \sigma^+_j \sigma^-_j  \bigg]               
     ,    \bigg[ \underset{j \text{ } \mathrm {odd}}{\underset{1 \leq j \leq N-1}{\prod}} i u^{\prime} \sigma^+_j \sigma^-_j  \bigg]    \bigg[  \underset{j \text{ } \mathrm{odd}}{\underset{3 \leq j \leq N-1}{\prod}}       i \big(  u^{\prime} \big)^{-1} \sigma^+_j \sigma^-_j      \bigg]  \\ \times           \bigg[ \underset{j \equiv N}{\prod}  i u^{\prime} \sigma^+_j \sigma^-_j  \bigg]         \bigg\} \overset{(\mathrm{LR})}{=}      \bigg[ \underset{j \text{ } \mathrm {odd}}{\underset{1 \leq j \leq N-1}{\prod}} i u \sigma^+_j \sigma^-_j  \bigg]    \bigg[  \underset{j \text{ } \mathrm{odd}}{\underset{3 \leq j \leq N-1}{\prod}}       i u^{-1} \sigma^+_j \sigma^-_j      \bigg]    \bigg\{  \bigg[ \underset{j \equiv N}{\prod} i u \sigma^+_j \sigma^-_j   \bigg]                ,  \bigg[ \underset{j \text{ } \mathrm {odd}}{\underset{1 \leq j \leq N-1}{\prod}} i u^{\prime} \sigma^+_j \sigma^-_j  \bigg]   \\ \times          \bigg[  \underset{j \text{ } \mathrm{odd}}{\underset{3 \leq j \leq N-1}{\prod}}       i \big(  u^{\prime} \big)^{-1} \sigma^+_j \sigma^-_j      \bigg]     \bigg[ \underset{j \equiv N}{\prod}  i u^{\prime} \sigma^+_j \sigma^-_j  \bigg]         \bigg\}   +     \bigg\{   \bigg[ \underset{j \text{ } \mathrm {odd}}{\underset{1 \leq j \leq N-1}{\prod}} i u \sigma^+_j \sigma^-_j  \bigg]    \bigg[  \underset{j \text{ } \mathrm{odd}}{\underset{3 \leq j \leq N-1}{\prod}}       i u^{-1} \sigma^+_j \sigma^-_j      \bigg] ,  \bigg[ \underset{j \text{ } \mathrm {odd}}{\underset{1 \leq j \leq N-1}{\prod}} i u^{\prime} \sigma^+_j \sigma^-_j  \bigg] \\  \times          \bigg[  \underset{j \text{ } \mathrm{odd}}{\underset{3 \leq j \leq N-1}{\prod}}       i \big(  u^{\prime} \big)^{-1} \sigma^+_j \sigma^-_j      \bigg]     \bigg[ \underset{j \equiv N}{\prod}  i u^{\prime} \sigma^+_j \sigma^-_j  \bigg]         \bigg\}  \bigg[ \underset{j \equiv N}{\prod} i u \sigma^+_j \sigma^-_j   \bigg]       \end{align*}

      \begin{align*}     \overset{(\mathrm{AC}),(\mathrm{LR})}{=}       -    \bigg[ \underset{j \text{ } \mathrm {odd}}{\underset{1 \leq j \leq N-1}{\prod}} i u \sigma^+_j \sigma^-_j  \bigg]    \bigg[  \underset{j \text{ } \mathrm{odd}}{\underset{3 \leq j \leq N-1}{\prod}}       i u^{-1} \sigma^+_j \sigma^-_j      \bigg]   \bigg[ \underset{j \text{ } \mathrm {odd}}{\underset{1 \leq j \leq N-1}{\prod}} i u^{\prime} \sigma^+_j \sigma^-_j  \bigg]          \bigg[  \underset{j \text{ } \mathrm{odd}}{\underset{3 \leq j \leq N-1}{\prod}}       i \big(  u^{\prime} \big)^{-1} \sigma^+_j \sigma^-_j      \bigg]    \bigg\{     \bigg[ \underset{j \equiv N}{\prod}  i u^{\prime} \sigma^+_j \sigma^-_j  \bigg]    \\ ,  \bigg[ \underset{j \equiv N}{\prod} i u \sigma^+_j \sigma^-_j   \bigg]                     \bigg\}   -    \bigg[ \underset{j \text{ } \mathrm {odd}}{\underset{1 \leq j \leq N-1}{\prod}} i u \sigma^+_j \sigma^-_j  \bigg]  \bigg[  \underset{j \text{ } \mathrm{odd}}{\underset{3 \leq j \leq N-1}{\prod}}       i u^{-1} \sigma^+_j \sigma^-_j      \bigg]     \bigg\{                \bigg[  \underset{j \text{ } \mathrm{odd}}{\underset{3 \leq j \leq N-1}{\prod}}       i   u^{\prime}\sigma^+_j \sigma^-_j      \bigg]       \\ 
     ,  \bigg[ \underset{j \equiv N}{\prod} i u \sigma^+_j \sigma^-_j   \bigg]                     \bigg\}      \bigg[ \underset{j \equiv N}{\prod}  i u^{\prime} \sigma^+_j \sigma^-_j  \bigg]   +  \bigg[ \underset{j \text{ } \mathrm {odd}}{\underset{1 \leq j \leq N-1}{\prod}} i u \sigma^+_j \sigma^-_j  \bigg]  \bigg\{   \bigg[  \underset{j \text{ } \mathrm{odd}}{\underset{3 \leq j \leq N-1}{\prod}}       i u^{-1} \sigma^+_j \sigma^-_j      \bigg] ,  \bigg[ \underset{j \text{ } \mathrm {odd}}{\underset{1 \leq j \leq N-1}{\prod}} i u^{\prime} \sigma^+_j \sigma^-_j  \bigg]   \\ \times          \bigg[  \underset{j \text{ } \mathrm{odd}}{\underset{3 \leq j \leq N-1}{\prod}}       i \big(  u^{\prime} \big)^{-1} \sigma^+_j \sigma^-_j      \bigg]     \bigg[ \underset{j \equiv N}{\prod}  i u^{\prime} \sigma^+_j \sigma^-_j  \bigg]         \bigg\}  \bigg[ \underset{j \equiv N}{\prod} i u \sigma^+_j \sigma^-_j   \bigg]   +    \bigg\{  \bigg[ \underset{j \text{ } \mathrm {odd}}{\underset{1 \leq j \leq N-1}{\prod}} i u \sigma^+_j \sigma^-_j  \bigg] \\  ,  \bigg[ \underset{j \text{ } \mathrm {odd}}{\underset{1 \leq j \leq N-1}{\prod}} i u^{\prime} \sigma^+_j \sigma^-_j  \bigg]           \bigg[  \underset{j \text{ } \mathrm{odd}}{\underset{3 \leq j \leq N-1}{\prod}}       i \big(  u^{\prime} \big)^{-1} \sigma^+_j \sigma^-_j      \bigg]     \bigg[ \underset{j \equiv N}{\prod}  i u^{\prime} \sigma^+_j \sigma^-_j  \bigg]         \bigg\}  \bigg[ \underset{j \equiv N}{\prod} i u \sigma^+_j \sigma^-_j   \bigg]    \bigg[  \underset{j \text{ } \mathrm{odd}}{\underset{3 \leq j \leq N-1}{\prod}}       i u^{-1} \sigma^+_j \sigma^-_j      \bigg] \\ \\ \overset{(\mathrm{AC}),(\mathrm{LR})}{=}                      -    \bigg[ \underset{j \text{ } \mathrm {odd}}{\underset{1 \leq j \leq N-1}{\prod}} i u \sigma^+_j \sigma^-_j  \bigg]    \bigg[  \underset{j \text{ } \mathrm{odd}}{\underset{3 \leq j \leq N-1}{\prod}}       i u^{-1} \sigma^+_j \sigma^-_j      \bigg]   \bigg[ \underset{j \text{ } \mathrm {odd}}{\underset{1 \leq j \leq N-1}{\prod}} i u^{\prime} \sigma^+_j \sigma^-_j  \bigg]          \bigg[  \underset{j \text{ } \mathrm{odd}}{\underset{3 \leq j \leq N-1}{\prod}}       i \big(  u^{\prime} \big)^{-1} \sigma^+_j \sigma^-_j      \bigg]    \bigg\{     \bigg[ \underset{j \equiv N}{\prod}  i u^{\prime} \sigma^+_j \sigma^-_j  \bigg]    \\ ,  \bigg[ \underset{j \equiv N}{\prod} i u \sigma^+_j \sigma^-_j   \bigg]                     \bigg\}   -    \bigg[ \underset{j \text{ } \mathrm {odd}}{\underset{1 \leq j \leq N-1}{\prod}} i u \sigma^+_j \sigma^-_j  \bigg]  \bigg[  \underset{j \text{ } \mathrm{odd}}{\underset{3 \leq j \leq N-1}{\prod}}       i u^{-1} \sigma^+_j \sigma^-_j      \bigg]     \bigg\{                \bigg[  \underset{j \text{ } \mathrm{odd}}{\underset{3 \leq j \leq N-1}{\prod}}       i   u^{\prime}\sigma^+_j \sigma^-_j      \bigg]        \\ ,  \bigg[ \underset{j \equiv N}{\prod} i u \sigma^+_j \sigma^-_j   \bigg]                     \bigg\}      \bigg[ \underset{j \equiv N}{\prod}  i u^{\prime} \sigma^+_j \sigma^-_j  \bigg]  -    \bigg[ \underset{j \text{ } \mathrm {odd}}{\underset{1 \leq j \leq N-1}{\prod}} i u \sigma^+_j \sigma^-_j  \bigg]  \bigg\{   \bigg[  \underset{j \text{ } \mathrm{odd}}{\underset{3 \leq j \leq N-1}{\prod}}       i u^{-1} \sigma^+_j \sigma^-_j      \bigg] ,  \bigg[ \underset{j \text{ } \mathrm {odd}}{\underset{1 \leq j \leq N-1}{\prod}} i u^{\prime} \sigma^+_j \sigma^-_j  \bigg]     \\   \times          \bigg[  \underset{j \text{ } \mathrm{odd}}{\underset{3 \leq j \leq N-1}{\prod}}       i \big(  u^{\prime} \big)^{-1} \sigma^+_j \sigma^-_j      \bigg]     \bigg[ \underset{j \equiv N}{\prod}  i u^{\prime} \sigma^+_j \sigma^-_j  \bigg]         \bigg\}  \bigg[ \underset{j \equiv N}{\prod} i u \sigma^+_j \sigma^-_j   \bigg]   -       \bigg[ \underset{j \text{ } \mathrm {odd}}{\underset{1 \leq j \leq N-1}{\prod}} i u^{\prime} \sigma^+_j \sigma^-_j  \bigg]   \\
   \times           \bigg[ \underset{j \equiv N}{\prod}  i u^{\prime} \sigma^+_j \sigma^-_j  \bigg]   , \bigg[  \underset{j \text{ } \mathrm{odd}}{\underset{3 \leq j \leq N-1}{\prod}}       i u^{-1} \sigma^+_j \sigma^-_j      \bigg]       \bigg\}  \bigg[ \underset{j \equiv N}{\prod} i u \sigma^+_j \sigma^-_j   \bigg]   -       \bigg[ \underset{j \text{ } \mathrm {odd}}{\underset{1 \leq j \leq N-1}{\prod}} i u^{\prime} \sigma^+_j \sigma^-_j  \bigg]     \\  \times  \bigg[  \text{ }    \bigg[  \underset{j \text{ } \mathrm{odd}}{\underset{3 \leq j \leq N-1}{\prod}}       i \big(  u^{\prime} \big)^{-1} \sigma^+_j \sigma^-_j      \bigg]     \bigg\{      \bigg[ \underset{j \equiv N}{\prod}  i u^{\prime} \sigma^+_j \sigma^-_j  \bigg] ,  \bigg[ \underset{j \text{ } \mathrm {odd}}{\underset{1 \leq j \leq N-1}{\prod}} i u^{-1} \sigma^+_j \sigma^-_j  \bigg]        \bigg\} \\ +    \bigg\{      \bigg[  \underset{j \text{ } \mathrm{odd}}{\underset{3 \leq j \leq N-1}{\prod}}       i \big(  u^{\prime} \big)^{-1} \sigma^+_j \sigma^-_j      \bigg]   ,  \bigg[ \underset{j \text{ } \mathrm {odd}}{\underset{1 \leq j \leq N-1}{\prod}} i u^{-1} \sigma^+_j \sigma^-_j  \bigg]        \bigg\}    \bigg[ \underset{j \equiv N}{\prod}  i u^{\prime} \sigma^+_j \sigma^-_j  \bigg] \text{ }  \bigg]  \bigg[ \underset{j \equiv N}{\prod} i u \sigma^+_j \sigma^-_j   \bigg]    \\ -     \bigg\{    \bigg[ \underset{j \text{ } \mathrm {odd}}{\underset{1 \leq j \leq N-1}{\prod}} i u^{\prime} \sigma^+_j \sigma^-_j  \bigg]             ,  \bigg[ \underset{j \text{ } \mathrm {odd}}{\underset{1 \leq j \leq N-1}{\prod}} i u \sigma^+_j \sigma^-_j  \bigg]     \bigg\}   \\ \times \bigg[  \underset{j \text{ } \mathrm{odd}}{\underset{3 \leq j \leq N-1}{\prod}}       i \big(  u^{\prime} \big)^{-1} \sigma^+_j \sigma^-_j      \bigg]     \bigg[ \underset{j \equiv N}{\prod}  i u^{\prime} \sigma^+_j \sigma^-_j  \bigg]   \bigg[ \underset{j \equiv N}{\prod} i u \sigma^+_j \sigma^-_j   \bigg]    \bigg[  \underset{j \text{ } \mathrm{odd}}{\underset{3 \leq j \leq N-1}{\prod}}       i u^{-1} \sigma^+_j \sigma^-_j      \bigg]    \\ \\ \overset{(\mathrm{LR})}{=}             -    \bigg[ \underset{j \text{ } \mathrm {odd}}{\underset{1 \leq j \leq N-1}{\prod}} i u \sigma^+_j \sigma^-_j  \bigg]    \bigg[  \underset{j \text{ } \mathrm{odd}}{\underset{3 \leq j \leq N-1}{\prod}}       i u^{-1} \sigma^+_j \sigma^-_j      \bigg]   \bigg[ \underset{j \text{ } \mathrm {odd}}{\underset{1 \leq j \leq N-1}{\prod}} i u^{\prime} \sigma^+_j \sigma^-_j  \bigg]          \bigg[  \underset{j \text{ } \mathrm{odd}}{\underset{3 \leq j \leq N-1}{\prod}}       i \big(  u^{\prime} \big)^{-1} \sigma^+_j \sigma^-_j      \bigg]    \bigg\{     \bigg[ \underset{j \equiv N}{\prod}  i u^{\prime} \sigma^+_j \sigma^-_j  \bigg]     \\  ,  \bigg[ \underset{j \equiv N}{\prod} i u \sigma^+_j \sigma^-_j   \bigg]                     \bigg\}   -    \bigg[ \underset{j \text{ } \mathrm {odd}}{\underset{1 \leq j \leq N-1}{\prod}} i u \sigma^+_j \sigma^-_j  \bigg]  \bigg[  \underset{j \text{ } \mathrm{odd}}{\underset{3 \leq j \leq N-1}{\prod}}       i u^{-1} \sigma^+_j \sigma^-_j      \bigg]     \bigg\{                \bigg[  \underset{j \text{ } \mathrm{odd}}{\underset{3 \leq j \leq N-1}{\prod}}       i   u^{\prime}\sigma^+_j \sigma^-_j      \bigg]        \end{align*}

     \begin{align*} ,  \bigg[ \underset{j \equiv N}{\prod} i u \sigma^+_j \sigma^-_j   \bigg]                     \bigg\}      \bigg[ \underset{j \equiv N}{\prod}  i u^{\prime} \sigma^+_j \sigma^-_j  \bigg]  +    \bigg[ \underset{j \text{ } \mathrm {odd}}{\underset{1 \leq j \leq N-1}{\prod}} i u \sigma^+_j \sigma^-_j  \bigg]  \bigg[  \text{ }   \bigg[ \underset{j \text{ } \mathrm {odd}}{\underset{1 \leq j \leq N-1}{\prod}} i u^{\prime} \sigma^+_j \sigma^-_j  \bigg]  \bigg\{      \bigg[  \underset{j \text{ } \mathrm{odd}}{\underset{3 \leq j \leq N-1}{\prod}}       i \big(  u^{\prime} \big)^{-1} \sigma^+_j \sigma^-_j      \bigg]      \\   \times           \bigg[ \underset{j \equiv N}{\prod}  i u^{\prime} \sigma^+_j \sigma^-_j  \bigg]   , \bigg[  \underset{j \text{ } \mathrm{odd}}{\underset{3 \leq j \leq N-1}{\prod}}       i u^{-1} \sigma^+_j \sigma^-_j      \bigg]       \bigg\}  +           \bigg\{     \bigg[ \underset{j \text{ } \mathrm {odd}}{\underset{1 \leq j \leq N-1}{\prod}} i u^{\prime} \sigma^+_j \sigma^-_j  \bigg]    \bigg[  \underset{j \text{ } \mathrm{odd}}{\underset{3 \leq j \leq N-1}{\prod}}       i \big(  u^{\prime} \big)^{-1} \sigma^+_j \sigma^-_j      \bigg]  \\   , \bigg[  \underset{j \text{ } \mathrm{odd}}{\underset{3 \leq j \leq N-1}{\prod}}       i u^{-1} \sigma^+_j \sigma^-_j      \bigg]       \bigg\}              \bigg[ \underset{j \equiv N}{\prod}  i u^{\prime} \sigma^+_j \sigma^-_j  \bigg] \text{ }           \bigg]   \bigg[ \underset{j \equiv N}{\prod} i u \sigma^+_j \sigma^-_j   \bigg]   -       \bigg[ \underset{j \text{ } \mathrm {odd}}{\underset{1 \leq j \leq N-1}{\prod}} i u^{\prime} \sigma^+_j \sigma^-_j  \bigg]    \\   \times  \bigg[  \text{ }    \bigg[  \underset{j \text{ } \mathrm{odd}}{\underset{3 \leq j \leq N-1}{\prod}}       i \big(  u^{\prime} \big)^{-1} \sigma^+_j \sigma^-_j      \bigg]     \bigg\{      \bigg[ \underset{j \equiv N}{\prod}  i u^{\prime} \sigma^+_j \sigma^-_j  \bigg] ,  \bigg[ \underset{j \text{ } \mathrm {odd}}{\underset{1 \leq j \leq N-1}{\prod}} i u^{-1} \sigma^+_j \sigma^-_j  \bigg]        \bigg\}       \\    +    \bigg\{      \bigg[  \underset{j \text{ } \mathrm{odd}}{\underset{3 \leq j \leq N-1}{\prod}}       i \big(  u^{\prime} \big)^{-1} \sigma^+_j \sigma^-_j      \bigg]   ,  \bigg[ \underset{j \text{ } \mathrm {odd}}{\underset{1 \leq j \leq N-1}{\prod}} i u^{-1} \sigma^+_j \sigma^-_j  \bigg]        \bigg\}    \bigg[ \underset{j \equiv N}{\prod}  i u^{\prime} \sigma^+_j \sigma^-_j  \bigg] \text{ }  \bigg]  \bigg[ \underset{j \equiv N}{\prod} i u \sigma^+_j \sigma^-_j   \bigg]    \\ -     \bigg\{    \bigg[ \underset{j \text{ } \mathrm {odd}}{\underset{1 \leq j \leq N-1}{\prod}} i u^{\prime} \sigma^+_j \sigma^-_j  \bigg]             ,  \bigg[ \underset{j \text{ } \mathrm {odd}}{\underset{1 \leq j \leq N-1}{\prod}} i u \sigma^+_j \sigma^-_j  \bigg]     \bigg\}   \\ \times \bigg[  \underset{j \text{ } \mathrm{odd}}{\underset{3 \leq j \leq N-1}{\prod}}       i \big(  u^{\prime} \big)^{-1} \sigma^+_j \sigma^-_j      \bigg]     \bigg[ \underset{j \equiv N}{\prod}  i u^{\prime} \sigma^+_j \sigma^-_j  \bigg]   \bigg[ \underset{j \equiv N}{\prod} i u \sigma^+_j \sigma^-_j   \bigg]    \bigg[  \underset{j \text{ } \mathrm{odd}}{\underset{3 \leq j \leq N-1}{\prod}}       i u^{-1} \sigma^+_j \sigma^-_j      \bigg]  \\ \\ \overset{(\mathrm{LR})}{=}               -    \bigg[ \underset{j \text{ } \mathrm {odd}}{\underset{1 \leq j \leq N-1}{\prod}} i u \sigma^+_j \sigma^-_j  \bigg]    \bigg[  \underset{j \text{ } \mathrm{odd}}{\underset{3 \leq j \leq N-1}{\prod}}       i u^{-1} \sigma^+_j \sigma^-_j      \bigg]   \bigg[ \underset{j \text{ } \mathrm {odd}}{\underset{1 \leq j \leq N-1}{\prod}} i u^{\prime} \sigma^+_j \sigma^-_j  \bigg]          \bigg[  \underset{j \text{ } \mathrm{odd}}{\underset{3 \leq j \leq N-1}{\prod}}       i \big(  u^{\prime} \big)^{-1} \sigma^+_j \sigma^-_j      \bigg]    \bigg\{     \bigg[ \underset{j \equiv N}{\prod}  i u^{\prime} \sigma^+_j \sigma^-_j  \bigg]   \\
   ,  \bigg[ \underset{j \equiv N}{\prod} i u \sigma^+_j \sigma^-_j   \bigg]                     \bigg\}   -    \bigg[ \underset{j \text{ } \mathrm {odd}}{\underset{1 \leq j \leq N-1}{\prod}} i u \sigma^+_j \sigma^-_j  \bigg]  \bigg[  \underset{j \text{ } \mathrm{odd}}{\underset{3 \leq j \leq N-1}{\prod}}       i u^{-1} \sigma^+_j \sigma^-_j      \bigg]     \bigg\{                \bigg[  \underset{j \text{ } \mathrm{odd}}{\underset{3 \leq j \leq N-1}{\prod}}       i   u^{\prime}\sigma^+_j \sigma^-_j      \bigg]          \\  ,  \bigg[ \underset{j \equiv N}{\prod} i u \sigma^+_j \sigma^-_j   \bigg]                     \bigg\}      \bigg[ \underset{j \equiv N}{\prod}  i u^{\prime} \sigma^+_j \sigma^-_j  \bigg]  +    \bigg[ \underset{j \text{ } \mathrm {odd}}{\underset{1 \leq j \leq N-1}{\prod}} i u \sigma^+_j \sigma^-_j  \bigg]  \bigg[  \text{ }   \bigg[ \underset{j \text{ } \mathrm {odd}}{\underset{1 \leq j \leq N-1}{\prod}} i u^{\prime} \sigma^+_j \sigma^-_j  \bigg]  \bigg[ \text{ }      \bigg[  \underset{j \text{ } \mathrm{odd}}{\underset{3 \leq j \leq N-1}{\prod}}       i \big(  u^{\prime} \big)^{-1} \sigma^+_j \sigma^-_j      \bigg]   \\    \times \bigg\{      \bigg[ \underset{j \equiv N}{\prod}  i u^{\prime} \sigma^+_j \sigma^-_j  \bigg]   , \bigg[  \underset{j \text{ } \mathrm{odd}}{\underset{3 \leq j \leq N-1}{\prod}}       i u^{-1} \sigma^+_j \sigma^-_j      \bigg]       \bigg\}  +                  \bigg\{      \bigg[  \underset{j \text{ } \mathrm{odd}}{\underset{3 \leq j \leq N-1}{\prod}}       i \big(  u^{\prime} \big)^{-1} \sigma^+_j \sigma^-_j      \bigg]           , \bigg[  \underset{j \text{ } \mathrm{odd}}{\underset{3 \leq j \leq N-1}{\prod}}       i u^{-1} \sigma^+_j \sigma^-_j      \bigg]       \bigg\}  \\ \times   \bigg[ \underset{j \equiv N}{\prod}  i u^{\prime} \sigma^+_j \sigma^-_j  \bigg]         \text{ }           \bigg]    +       \bigg[ \text{ } \bigg[ \underset{j \text{ } \mathrm {odd}}{\underset{1 \leq j \leq N-1}{\prod}} i u^{\prime} \sigma^+_j \sigma^-_j  \bigg]      \bigg\{       \bigg[  \underset{j \text{ } \mathrm{odd}}{\underset{3 \leq j \leq N-1}{\prod}}       i \big(  u^{\prime} \big)^{-1} \sigma^+_j \sigma^-_j      \bigg]     , \bigg[  \underset{j \text{ } \mathrm{odd}}{\underset{3 \leq j \leq N-1}{\prod}}       i u^{-1} \sigma^+_j \sigma^-_j      \bigg]       \bigg\}  \\  +     \bigg\{   \bigg[ \underset{j \text{ } \mathrm {odd}}{\underset{1 \leq j \leq N-1}{\prod}} i u^{\prime} \sigma^+_j \sigma^-_j  \bigg]        ,    \bigg[  \underset{j \text{ } \mathrm{odd}}{\underset{3 \leq j \leq N-1}{\prod}}       i u^{-1} \sigma^+_j \sigma^-_j      \bigg] \bigg\}  \bigg[ \underset{j \text{ } \mathrm{odd}}{\underset{3 \leq j \leq N-1}{\prod}}                 i \big( u^{\prime} \big)^{-1} \sigma^+_j \sigma^-_j        \bigg]  \text{ }   \bigg]  \bigg[ \underset{j \equiv N}{\prod}      i  \big[ u^{\prime} + u \big]  \sigma^+_j \sigma^-_j           \bigg]                \\  \times    \bigg[ \underset{j \equiv N}{\prod}  i u^{\prime} \sigma^+_j \sigma^-_j  \bigg]   \bigg[ \underset{j \equiv N}{\prod} i u \sigma^+_j \sigma^-_j   \bigg]   -       \bigg[ \underset{j \text{ } \mathrm {odd}}{\underset{1 \leq j \leq N-1}{\prod}} i u^{\prime} \sigma^+_j \sigma^-_j  \bigg]      \bigg[  \text{ }    \bigg[  \underset{j \text{ } \mathrm{odd}}{\underset{3 \leq j \leq N-1}{\prod}}       i \big(  u^{\prime} \big)^{-1} \sigma^+_j \sigma^-_j      \bigg]     \bigg\{      \bigg[ \underset{j \equiv N}{\prod}  i u^{\prime} \sigma^+_j \sigma^-_j  \bigg] \\ ,  \bigg[ \underset{j \text{ } \mathrm {odd}}{\underset{1 \leq j \leq N-1}{\prod}} i u^{-1} \sigma^+_j \sigma^-_j  \bigg]        \bigg\}  +    \bigg\{      \bigg[  \underset{j \text{ } \mathrm{odd}}{\underset{3 \leq j \leq N-1}{\prod}}       i \big(  u^{\prime} \big)^{-1} \sigma^+_j \sigma^-_j      \bigg]   ,  \bigg[ \underset{j \text{ } \mathrm {odd}}{\underset{1 \leq j \leq N-1}{\prod}} i \big( u^{\prime}\big)^{-1} \sigma^+_j \sigma^-_j  \bigg]        \bigg\}    \bigg[ \underset{j \equiv N}{\prod}  i u^{\prime} \sigma^+_j \sigma^-_j  \bigg] \text{ }  \bigg] \\ \times   \bigg[ \underset{j \equiv N}{\prod} i u \sigma^+_j \sigma^-_j   \bigg]    -     \bigg\{    \bigg[ \underset{j \text{ } \mathrm {odd}}{\underset{1 \leq j \leq N-1}{\prod}} i u^{\prime} \sigma^+_j \sigma^-_j  \bigg]             ,  \bigg[ \underset{j \text{ } \mathrm {odd}}{\underset{1 \leq j \leq N-1}{\prod}} i u \sigma^+_j \sigma^-_j  \bigg]     \bigg\}   \end{align*}

              \begin{align*}   \times \bigg[  \underset{j \text{ } \mathrm{odd}}{\underset{3 \leq j \leq N-1}{\prod}}       i \big(  u^{\prime} \big)^{-1} \sigma^+_j \sigma^-_j      \bigg]     \bigg[ \underset{j \equiv N}{\prod}  i u^{\prime} \sigma^+_j \sigma^-_j  \bigg]   \bigg[ \underset{j \equiv N}{\prod} i u \sigma^+_j \sigma^-_j   \bigg]    \bigg[  \underset{j \text{ } \mathrm{odd}}{\underset{3 \leq j \leq N-1}{\prod}}       i u^{-1} \sigma^+_j \sigma^-_j      \bigg]                                         \text{, }
\end{align*}

\noindent corresponding to the ninth term, which can be approximated with, 

\begin{align*}
   \big( \mathscr{C}^1_1 \big)_9  \text{. }
    \end{align*}

\noindent From computations with the Poisson bracket above, additional terms can be approximated with very similar applications of (BL), (AC), and (LR). Such brackets take the form,

\begin{align*}
    \big\{ \mathscr{I}^1_1 \big( \underline{u} \big) , \mathscr{I}^1_3 \big( \underline{u^{\prime}} \big)   \big\}    \text{, } \\  \big\{ \mathscr{I}^1_1 \big( \underline{u} \big) , \mathscr{I}^1_3 \big( \underline{u^{\prime}} \big)   \big\}  \text{, } \\  \big\{ \mathscr{I}^1_1 \big( \underline{u} \big) , \mathscr{I}^1_4 \big( \underline{u^{\prime}} \big)   \big\} \text{. }
\end{align*}

\noindent For contributions from $\mathscr{I}^1_2$ with itself, 

\begin{align*}
 \big\{ \mathscr{I}^1_2 \big( \underline{u} \big) , \mathscr{I}^1_2 \big( \underline{u^{\prime}} \big) \big\}    \text{, }
\end{align*}

\noindent decomposes into a collection of Poisson brackets that one must approximate, which take the form,

\begin{align*}
            \bigg\{ \sigma^-_0 \sigma^+_1   \bigg[       \underset{1 \leq j \leq N-1}{\prod}   i u \sigma^+_j \sigma^-_j \textbf{1}_{\{ j^{\prime} > j \text{ } : \text{ } i u \sigma^+_{j^{\prime}} \sigma^-_{j^{\prime}} \in \mathrm{support} ( A ( u ))  \}}    \big[  \sigma^+_{j^{\prime}} + i u^{-1} \sigma^+_{j^{\prime}} \sigma^-_{j^{\prime}}     \big]    \bigg] \bigg[ \underset{j \equiv N}{\prod} \sigma^+_j  \bigg] , \sigma^-_0 \sigma^+_1 \bigg[      \text{ }  \bigg[       \underset{1 \leq j \leq N-1}{\prod}   i u^{\prime} \sigma^+_j \sigma^-_j  \\ \times 
        \textbf{1}_{\{ j^{\prime} > j \text{ } : \text{ } i u^{\prime} \sigma^+_{j^{\prime}} \sigma^-_{j^{\prime}} \in \mathrm{support} ( A ( u^{\prime} ))  \}}    \big[  \sigma^+_{j^{\prime}} + i \big( u^{\prime} \big)^{-1} \sigma^+_{j^{\prime}} \sigma^-_{j^{\prime}}     \big]    \bigg] \bigg[ \underset{j \equiv N}{\prod} \sigma^+_j  \bigg] \bigg\}     \overset{(\mathrm{LR})}{=} \sigma^-_0   \bigg\{  \sigma^+_1  \bigg[      \text{ }  \bigg[       \underset{1 \leq j \leq N-1}{\prod}   i u \sigma^+_j \sigma^-_j \\
        \times \textbf{1}_{\{ j^{\prime} > j \text{ } : \text{ } i u \sigma^+_{j^{\prime}} \sigma^-_{j^{\prime}} \in \mathrm{support} ( A ( u ))  \}}    \big[  \sigma^+_{j^{\prime}} + i u^{-1} \sigma^+_{j^{\prime}} \sigma^-_{j^{\prime}}     \big]    \bigg] \bigg[ \underset{j \equiv N}{\prod} \sigma^+_j  \bigg] , \sigma^-_0 \sigma^+_1 \bigg[      \text{ }  \bigg[       \underset{1 \leq j \leq N-1}{\prod}   i u^{\prime} \sigma^+_j \sigma^-_j  \\ \times 
        \textbf{1}_{\{ j^{\prime} > j \text{ } : \text{ } i u^{\prime} \sigma^+_{j^{\prime}} \sigma^-_{j^{\prime}} \in \mathrm{support} ( A ( u^{\prime} ))  \}}    \big[  \sigma^+_{j^{\prime}} + i \big( u^{\prime} \big)^{-1} \sigma^+_{j^{\prime}} \sigma^-_{j^{\prime}}     \big]    \bigg] \bigg[ \underset{j \equiv N}{\prod} \sigma^+_j  \bigg] \bigg\} \\ +   \bigg\{ \sigma^-_0   , \sigma^-_0 \sigma^+_1 \bigg[           \underset{1 \leq j \leq N-1}{\prod}   i u^{\prime} \sigma^+_j \sigma^-_j         \textbf{1}_{\{ j^{\prime} > j \text{ } : \text{ } i u^{\prime} \sigma^+_{j^{\prime}} \sigma^-_{j^{\prime}} \in \mathrm{support} ( A ( u^{\prime} ))  \}}   \big[  \sigma^+_{j^{\prime}} + i \big( u^{\prime} \big)^{-1} \sigma^+_{j^{\prime}} \sigma^-_{j^{\prime}}     \big]    \bigg] \bigg[ \underset{j \equiv N}{\prod} \sigma^+_j  \bigg] \bigg\} \\ \times  \sigma^+_1  \bigg[      \text{ }  \bigg[       \underset{1 \leq j \leq N-1}{\prod}   i u \sigma^+_j \sigma^-_j \textbf{1}_{\{ j^{\prime} > j \text{ } : \text{ } i u \sigma^+_{j^{\prime}} \sigma^-_{j^{\prime}} \in \mathrm{support} ( A ( u ))  \}}    \big[  \sigma^+_{j^{\prime}} + i u^{-1} \sigma^+_{j^{\prime}} \sigma^-_{j^{\prime}}     \big]    \bigg] \bigg[ \underset{j \equiv N}{\prod} \sigma^+_j  \bigg] \\ \\ \overset{(\mathrm{LR})}{=}     \sigma^-_0  \sigma^+_1    \bigg\{    \bigg[       \underset{1 \leq j \leq N-1}{\prod}   i u \sigma^+_j \sigma^-_j  \textbf{1}_{\{ j^{\prime} > j \text{ } : \text{ } i u \sigma^+_{j^{\prime}} \sigma^-_{j^{\prime}} \in \mathrm{support} ( A ( u ))  \}}    \big[  \sigma^+_{j^{\prime}} + i u^{-1} \sigma^+_{j^{\prime}} \sigma^-_{j^{\prime}}     \big]    \bigg] \bigg[ \underset{j \equiv N}{\prod} \sigma^+_j  \bigg] , \sigma^-_0 \sigma^+_1 \\ 
        \times  \bigg[      \text{ }  \bigg[       \underset{1 \leq j \leq N-1}{\prod}   i u^{\prime} \sigma^+_j \sigma^-_j  
        \textbf{1}_{\{ j^{\prime} > j \text{ } : \text{ } i u^{\prime} \sigma^+_{j^{\prime}} \sigma^-_{j^{\prime}} \in \mathrm{support} ( A ( u^{\prime} ))  \}}    \big[  \sigma^+_{j^{\prime}} + i \big( u^{\prime} \big)^{-1} \sigma^+_{j^{\prime}} \sigma^-_{j^{\prime}}     \big]    \bigg] \bigg[ \underset{j \equiv N}{\prod} \sigma^+_j  \bigg] \bigg\}  \\  +            \sigma^-_0   \bigg\{  \sigma^+_1  , \sigma^-_0 \sigma^+_1   \bigg[      \text{ }  \bigg[       \underset{1 \leq j \leq N-1}{\prod}   i u^{\prime} \sigma^+_j \sigma^-_j  
        \textbf{1}_{\{ j^{\prime} > j \text{ } : \text{ } i u^{\prime} \sigma^+_{j^{\prime}} \sigma^-_{j^{\prime}} \in \mathrm{support} ( A ( u^{\prime} ))  \}}    \big[  \sigma^+_{j^{\prime}} + i \big( u^{\prime} \big)^{-1} \sigma^+_{j^{\prime}} \sigma^-_{j^{\prime}}     \big]    \bigg] \bigg[ \underset{j \equiv N}{\prod} \sigma^+_j  \bigg] \bigg\}  \\ \times   \bigg[       \underset{1 \leq j \leq N-1}{\prod}   i u \sigma^+_j \sigma^-_j  \textbf{1}_{\{ j^{\prime} > j \text{ } : \text{ } i u \sigma^+_{j^{\prime}} \sigma^-_{j^{\prime}} \in \mathrm{support} ( A ( u ))  \}}    \big[  \sigma^+_{j^{\prime}} + i u^{-1} \sigma^+_{j^{\prime}} \sigma^-_{j^{\prime}}     \big]    \bigg] \bigg[ \underset{j \equiv N}{\prod} \sigma^+_j  \bigg]               \text{. }       \end{align*}

       \noindent An application of (AC), followed by (LR), yields,

        \begin{align*}
        \sigma^-_0  \sigma^+_1  \bigg[       \underset{1 \leq j \leq N-1}{\prod}   i u \sigma^+_j \sigma^-_j  \textbf{1}_{\{ j^{\prime} > j \text{ } : \text{ } i u \sigma^+_{j^{\prime}} \sigma^-_{j^{\prime}} \in \mathrm{support} ( A ( u ))  \}}    \big[  \sigma^+_{j^{\prime}} + i u^{-1} \sigma^+_{j^{\prime}} \sigma^-_{j^{\prime}}     \big]    \bigg]  \bigg\{      \bigg[ \underset{j \equiv N}{\prod} \sigma^+_j  \bigg]   , \sigma^-_0 \sigma^+_1 \\ \times  \bigg[      \text{ }  \bigg[       \underset{1 \leq j \leq N-1}{\prod}   i u^{\prime} \sigma^+_j \sigma^-_j  
        \textbf{1}_{\{ j^{\prime} > j \text{ } : \text{ } i u^{\prime} \sigma^+_{j^{\prime}} \sigma^-_{j^{\prime}} \in \mathrm{support} ( A ( u^{\prime} ))  \}}    \big[  \sigma^+_{j^{\prime}} + i \big( u^{\prime} \big)^{-1} \sigma^+_{j^{\prime}} \sigma^-_{j^{\prime}}     \big]    \bigg] \bigg[ \underset{j \equiv N}{\prod} \sigma^+_j  \bigg]  \text{ } \bigg] 
    \bigg\} \text{, } \end{align*}
    
    \noindent corresponding to the first term,
    
    \begin{align*}
   \sigma^-_0  \sigma^+_1    \bigg\{     \bigg[       \underset{1 \leq j \leq N-1}{\prod}   i u \sigma^+_j \sigma^-_j  \textbf{1}_{\{ j^{\prime} > j \text{ } : \text{ } i u \sigma^+_{j^{\prime}} \sigma^-_{j^{\prime}} \in \mathrm{support} ( A ( u ))  \}}    \big[  \sigma^+_{j^{\prime}} + i u^{-1} \sigma^+_{j^{\prime}} \sigma^-_{j^{\prime}}     \big]    \bigg]   , \sigma^-_0 \sigma^+_1 \\ \times  \bigg[      \text{ }  \bigg[       \underset{1 \leq j \leq N-1}{\prod}   i u^{\prime} \sigma^+_j \sigma^-_j  
        \textbf{1}_{\{ j^{\prime} > j \text{ } : \text{ } i u^{\prime} \sigma^+_{j^{\prime}} \sigma^-_{j^{\prime}} \in \mathrm{support} ( A ( u^{\prime} ))  \}}    \big[  \sigma^+_{j^{\prime}} + i \big( u^{\prime} \big)^{-1} \sigma^+_{j^{\prime}} \sigma^-_{j^{\prime}}     \big]    \bigg] \bigg[ \underset{j \equiv N}{\prod} \sigma^+_j  \bigg] \text{ } \bigg]   \bigg\}     \bigg[ \underset{j \equiv N}{\prod} \sigma^+_j  \bigg]    \\ \\                \overset{(\mathrm{AC}),(\mathrm{LR})}{=}                \big[   \sigma^-_0  \sigma^+_1  \big]^2  \bigg[       \underset{1 \leq j \leq N-1}{\prod}   i u \sigma^+_j \sigma^-_j  \textbf{1}_{\{ j^{\prime} > j \text{ } : \text{ } i u \sigma^+_{j^{\prime}} \sigma^-_{j^{\prime}} \in \mathrm{support} ( A ( u ))  \}}    \big[  \sigma^+_{j^{\prime}} + i u^{-1} \sigma^+_{j^{\prime}} \sigma^-_{j^{\prime}}     \big]    \bigg]  \\ 
\times \bigg\{    \bigg[       \underset{1 \leq j \leq N-1}{\prod}   i u^{\prime} \sigma^+_j \sigma^-_j  
        \textbf{1}_{\{ j^{\prime} > j \text{ } : \text{ } i u^{\prime} \sigma^+_{j^{\prime}} \sigma^-_{j^{\prime}} \in \mathrm{support} ( A ( u^{\prime} ))  \}}    \big[  \sigma^+_{j^{\prime}} + i \big( u^{\prime} \big)^{-1} \sigma^+_{j^{\prime}} \sigma^-_{j^{\prime}}     \big]    \bigg] \bigg[ \underset{j \equiv N}{\prod} \sigma^+_j  \bigg]  \text{ } \bigg] ,  \bigg[ \underset{j \equiv N}{\prod} \sigma^+_j  \bigg]
    \bigg\}      \end{align*}

              \begin{align*}   -    \sigma^-_0  \sigma^+_1  \bigg[       \underset{1 \leq j \leq N-1}{\prod}   i u \sigma^+_j \sigma^-_j  \textbf{1}_{\{ j^{\prime} > j \text{ } : \text{ } i u \sigma^+_{j^{\prime}} \sigma^-_{j^{\prime}} \in \mathrm{support} ( A ( u ))  \}}    \big[  \sigma^+_{j^{\prime}} + i u^{-1} \sigma^+_{j^{\prime}} \sigma^-_{j^{\prime}}     \big]    \bigg]  \bigg\{       \sigma^-_0 \sigma^+_1    , \bigg[ \underset{j \equiv N}{\prod} \sigma^+_j  \bigg]  
    \bigg\} \\ \times   \bigg[     \underset{1 \leq j \leq N-1}{\prod}   i u  \sigma^+_j \sigma^-_j  
        \textbf{1}_{\{ j^{\prime} > j \text{ } : \text{ } i u \sigma^+_{j^{\prime}} \sigma^-_{j^{\prime}} \in \mathrm{support} ( A ( u ))  \}}    \big[  \sigma^+_{j^{\prime}} + i \big( u^{\prime} \big)^{-1} \sigma^+_{j^{\prime}} \sigma^-_{j^{\prime}}     \big]    \bigg] \bigg[ \underset{j \equiv N}{\prod} \sigma^+_j  \bigg]  \text{ } \bigg]   \text{, } \\ \\ 
        \bigg\{        \sigma^-_0 \sigma^+_1  \bigg[      \underset{1 \leq j \leq N-1}{\prod}   i u \sigma^+_j \sigma^-_j \textbf{1}_{\{ j^{\prime} > j \text{ } : \text{ } i u \sigma^+_{j^{\prime}} \sigma^-_{j^{\prime}} \in \mathrm{support} ( A ( u ))  \}}    \big[  \sigma^+_{j^{\prime}} + i u^{-1} \sigma^+_{j^{\prime}} \sigma^-_{j^{\prime}}     \big]    \bigg] \bigg[ \underset{j \equiv N}{\prod} \sigma^+_j  \bigg]     ,                \sigma^-_1 \\ 
            \times \bigg[           \underset{1 \leq j \leq N}{\prod}    \textbf{1}_{\{j^{\prime} > j \text{ } : \text{ } i u^{\prime} \sigma^+_{j^{\prime}} \sigma^-_{j^{\prime}} \in \mathrm{support}(A(u^{\prime}))\}} 
\bigg[ \sigma^+_{j^{\prime}}    \sigma^-_{j^{\prime}+1} + i u^{\prime} \sigma^+_{j^{\prime}} \sigma^-_{j^{\prime}} \bigg[      \underset{j \mathrm{mod}3 \equiv 0}{\underset{3 \leq j \leq N-1}{\prod}}   i \big( u^{\prime} \big)^{-1} \sigma^+_j \sigma^-_j   \bigg]  \text{ } \bigg]  \bigg[   \underset{ j \equiv N}{\prod}  \sigma^+_j   \bigg] \text{ }  \bigg]                            \bigg\}  \\ \overset{(\mathrm{LR})}{=}       \sigma^-_0 \sigma^+_1     \bigg\{           \bigg[       \underset{1 \leq j \leq N-1}{\prod}   i u \sigma^+_j \sigma^-_j \textbf{1}_{\{ j^{\prime} > j \text{ } : \text{ } i u \sigma^+_{j^{\prime}} \sigma^-_{j^{\prime}} \in \mathrm{support} ( A ( u ))  \}}    \big[  \sigma^+_{j^{\prime}} + i u^{-1} \sigma^+_{j^{\prime}} \sigma^-_{j^{\prime}}     \big]    \bigg] \bigg[ \underset{j \equiv N}{\prod} \sigma^+_j  \bigg]     ,                \sigma^-_1 \\ 
            \times \bigg[           \underset{1 \leq j \leq N}{\prod}    \textbf{1}_{\{j^{\prime} > j \text{ } : \text{ } i u^{\prime}  \sigma^+_{j^{\prime}} \sigma^-_{j^{\prime}} \in \mathrm{support}(A(u^{\prime}))\}} 
\bigg[ \sigma^+_{j^{\prime}}    \sigma^-_{j^{\prime}+1} + i u^{\prime} \sigma^+_{j^{\prime}} \sigma^-_{j^{\prime}} \bigg[      \underset{j \mathrm{mod}3 \equiv 0}{\underset{3 \leq j \leq N-1}{\prod}}   i \big( u^{\prime} \big)^{-1} \sigma^+_j \sigma^-_j   \bigg]  \text{ } \bigg]  \bigg[   \underset{ j \equiv N}{\prod}  \sigma^+_j   \bigg] \text{ }  \bigg]                            \bigg\} \\ +   \bigg\{        \sigma^-_0 \sigma^+_1       ,                \sigma^-_1  
            \bigg[           \underset{1 \leq j \leq N}{\prod}    \textbf{1}_{\{j^{\prime} > j \text{ } : \text{ } i u^{\prime} \sigma^+_{j^{\prime}} \sigma^-_{j^{\prime}} \in \mathrm{support}(A(u^{\prime}))\}} 
\bigg[ \sigma^+_{j^{\prime}}    \sigma^-_{j^{\prime}+1} + i u^{\prime}  \sigma^+_{j^{\prime}} \sigma^-_{j^{\prime}} \bigg[      \underset{j \mathrm{mod}3 \equiv 0}{\underset{3 \leq j \leq N-1}{\prod}}   i \big( u^{\prime} \big)^{-1} \sigma^+_j \sigma^-_j   \bigg]  \text{ } \bigg]  \bigg[   \underset{ j \equiv N}{\prod}  \sigma^+_j   \bigg] \text{ }  \bigg]                            \bigg\}        \\ \times \bigg[      \text{ }  \bigg[       \underset{1 \leq j \leq N-1}{\prod}   i u \sigma^+_j \sigma^-_j \textbf{1}_{\{ j^{\prime} > j \text{ } : \text{ } i u \sigma^+_{j^{\prime}} \sigma^-_{j^{\prime}} \in \mathrm{support} ( A ( u ))  \}}    \big[  \sigma^+_{j^{\prime}} + i u^{-1} \sigma^+_{j^{\prime}} \sigma^-_{j^{\prime}}     \big]    \bigg] \bigg[ \underset{j \equiv N}{\prod} \sigma^+_j  \bigg] \text{ } \bigg]                          \text{. } \end{align*}

\noindent Applying (LR) to the superposition of brackets above is composed of several contributions, the first of which is, 

\begin{align*}
 \sigma^-_0 \sigma^+_1     \bigg\{         \bigg[      \text{ }  \bigg[       \underset{1 \leq j \leq N-1}{\prod}   i u \sigma^+_j \sigma^-_j \textbf{1}_{\{ j^{\prime} > j \text{ } : \text{ } i u \sigma^+_{j^{\prime}} \sigma^-_{j^{\prime}} \in \mathrm{support} ( A ( u ))  \}}    \big[  \sigma^+_{j^{\prime}} + i u^{-1} \sigma^+_{j^{\prime}} \sigma^-_{j^{\prime}}     \big]    \bigg]   ,                \sigma^-_1 \\ 
            \times \bigg[           \underset{1 \leq j \leq N}{\prod}    \textbf{1}_{\{j^{\prime} > j \text{ } : \text{ } i u^{\prime} \sigma^+_{j^{\prime}} \sigma^-_{j^{\prime}} \in \mathrm{support}(A(u^{\prime}))\}} 
\bigg[ \sigma^+_{j^{\prime}}    \sigma^-_{j^{\prime}+1} + i u^{\prime}  \sigma^+_{j^{\prime}} \sigma^-_{j^{\prime}} \bigg[      \underset{j \mathrm{mod}3 \equiv 0}{\underset{3 \leq j \leq N-1}{\prod}}   i \big( u^{\prime} \big)^{-1} \sigma^+_j \sigma^-_j   \bigg]  \text{ } \bigg]  \bigg[   \underset{ j \equiv N}{\prod}  \sigma^+_j   \bigg] \text{ }  \bigg]                            \bigg\} 
\\ \times   \bigg[ \underset{j \equiv N}{\prod} \sigma^+_j  \bigg]  
  +  \sigma^-_0 \sigma^+_1 
 \bigg[       \underset{1 \leq j \leq N-1}{\prod}   i u \sigma^+_j \sigma^-_j \textbf{1}_{\{ j^{\prime} > j \text{ } : \text{ } i u \sigma^+_{j^{\prime}} \sigma^-_{j^{\prime}} \in \mathrm{support} ( A ( u ))  \}}    \big[  \sigma^+_{j^{\prime}} + i u^{-1} \sigma^+_{j^{\prime}} \sigma^-_{j^{\prime}}     \big]    \bigg]  \bigg\{      \bigg[ \underset{j \equiv N}{\prod} \sigma^+_j  \bigg]     ,                \sigma^-_1 \\ 
            \times \bigg[           \underset{1 \leq j \leq N}{\prod}    \textbf{1}_{\{j^{\prime} > j \text{ } : \text{ } i u^{\prime}  \sigma^+_{j^{\prime}} \sigma^-_{j^{\prime}} \in \mathrm{support}(A(u^{\prime}))\}} 
\bigg[ \sigma^+_{j^{\prime}}    \sigma^-_{j^{\prime}+1} + i u^{\prime} \sigma^+_{j^{\prime}} \sigma^-_{j^{\prime}} \bigg[      \underset{j \mathrm{mod}3 \equiv 0}{\underset{3 \leq j \leq N-1}{\prod}}   i \big( u^{\prime} \big)^{-1} \sigma^+_j \sigma^-_j   \bigg]  \text{ } \bigg]  \bigg[   \underset{ j \equiv N}{\prod}  \sigma^+_j   \bigg] \text{ }  \bigg]                            \bigg\}   \text{, } \end{align*}

\noindent and the second of which is,

\begin{align*}
\bigg\{        \sigma^-_0 \sigma^+_1       ,                \sigma^-_1  
            \bigg[           \underset{1 \leq j \leq N}{\prod}    \textbf{1}_{\{j^{\prime} > j \text{ } : \text{ } i u^{\prime} \sigma^+_{j^{\prime}} \sigma^-_{j^{\prime}} \in \mathrm{support}(A(u^{\prime}))\}} 
\bigg[ \sigma^+_{j^{\prime}}    \sigma^-_{j^{\prime}+1} + i u^{\prime} \sigma^+_{j^{\prime}} \sigma^-_{j^{\prime}} \bigg[      \underset{j \mathrm{mod}3 \equiv 0}{\underset{3 \leq j \leq N-1}{\prod}}   i \big( u^{\prime} \big)^{-1} \sigma^+_j \sigma^-_j   \bigg]  \text{ } \bigg]  \bigg[   \underset{ j \equiv N}{\prod}  \sigma^+_j   \bigg] \text{ }  \bigg]                            \bigg\}        \\ \times \bigg[      \text{ }  \bigg[       \underset{1 \leq j \leq N-1}{\prod}   i u \sigma^+_j \sigma^-_j \textbf{1}_{\{ j^{\prime} > j \text{ } : \text{ } i u \sigma^+_{j^{\prime}} \sigma^-_{j^{\prime}} \in \mathrm{support} ( A ( u ))  \}}    \big[  \sigma^+_{j^{\prime}} + i u^{-1} \sigma^+_{j^{\prime}} \sigma^-_{j^{\prime}}     \big]    \bigg] \bigg[ \underset{j \equiv N}{\prod} \sigma^+_j  \bigg] \text{ } \bigg]  \text{. }    \end{align*}

\noindent Next, applying (AC) to the two terms above yields,

\begin{align*}
    -  \sigma^-_0 \sigma^+_1     \bigg\{                         \sigma^-_1
            \bigg[           \underset{1 \leq j \leq N}{\prod}    \textbf{1}_{\{j^{\prime} > j \text{ } : \text{ } i u \sigma^+_{j^{\prime}} \sigma^-_{j^{\prime}} \in \mathrm{support}(A(u))\}} 
\bigg[ \sigma^+_{j^{\prime}}    \sigma^-_{j^{\prime}+1} + i u \sigma^+_{j^{\prime}} \sigma^-_{j^{\prime}} \bigg[      \underset{j \mathrm{mod}3 \equiv 0}{\underset{3 \leq j \leq N-1}{\prod}}   i \big( u^{\prime} \big)^{-1} \sigma^+_j \sigma^-_j   \bigg]  \text{ } \bigg]  \bigg[   \underset{ j \equiv N}{\prod}  \sigma^+_j   \bigg] \text{ }  \bigg]              \end{align*}

\begin{align*}
, \bigg[      \text{ }  \bigg[       \underset{1 \leq j \leq N-1}{\prod}   i u^{\prime} \sigma^+_j \sigma^-_j \textbf{1}_{\{ j^{\prime} > j \text{ } : \text{ } i u^{\prime} \sigma^+_{j^{\prime}} \sigma^-_{j^{\prime}} \in \mathrm{support} ( A ( u^{\prime} ))  \}}    \big[  \sigma^+_{j^{\prime}} + i \big( u^{\prime}\big)^{-1} \sigma^+_{j^{\prime}} \sigma^-_{j^{\prime}}     \big]    \bigg]             \bigg\}    \bigg[ \underset{j \equiv N}{\prod} \sigma^+_j  \bigg]  \end{align*}

\noindent corresponding to the first term,

\begin{align*}
-  \sigma^-_0 \sigma^+_1 
 \bigg[       \underset{1 \leq j \leq N-1}{\prod}   i u \sigma^+_j \sigma^-_j \textbf{1}_{\{ j^{\prime} > j \text{ } : \text{ } i u \sigma^+_{j^{\prime}} \sigma^-_{j^{\prime}} \in \mathrm{support} ( A ( u ))  \}}    \big[  \sigma^+_{j^{\prime}} + i u^{-1} \sigma^+_{j^{\prime}} \sigma^-_{j^{\prime}}     \big]    \bigg]  \bigg\{               \sigma^-_1 \\    
            \times \bigg[           \underset{1 \leq j \leq N}{\prod}    \textbf{1}_{\{j^{\prime} > j \text{ } : \text{ } i u \sigma^+_{j^{\prime}} \sigma^-_{j^{\prime}} \in \mathrm{support}(A(u))\}} 
\bigg[ \sigma^+_{j^{\prime}}    \sigma^-_{j^{\prime}+1} + i u \sigma^+_{j^{\prime}} \sigma^-_{j^{\prime}} \bigg[      \underset{j \mathrm{mod}3 \equiv 0}{\underset{3 \leq j \leq N-1}{\prod}}   i \big( u^{\prime} \big)^{-1} \sigma^+_j \sigma^-_j   \bigg]  \text{ } \bigg]  \bigg[   \underset{ j \equiv N}{\prod}  \sigma^+_j   \bigg] \text{ }  \bigg]            ,  \bigg[ \underset{j \equiv N}{\prod} \sigma^+_j  \bigg]                     \bigg\}  \text{, } \end{align*}

    \noindent corresponding to the third term,

\begin{align*}
   \bigg\{        \sigma^-_0 \sigma^+_1       ,                \sigma^-_1  
            \bigg[           \underset{1 \leq j \leq N}{\prod}    \textbf{1}_{\{j^{\prime} > j \text{ } : \text{ } i u^{\prime} \sigma^+_{j^{\prime}} \sigma^-_{j^{\prime}} \in \mathrm{support}(A(u^{\prime}))\}} 
\bigg[ \sigma^+_{j^{\prime}}    \sigma^-_{j^{\prime}+1} + i u^{\prime} \sigma^+_{j^{\prime}} \sigma^-_{j^{\prime}} \bigg[      \underset{j \mathrm{mod}3 \equiv 0}{\underset{3 \leq j \leq N-1}{\prod}}   i \big( u^{\prime} \big)^{-1} \sigma^+_j \sigma^-_j   \bigg]  \text{ } \bigg]  \bigg[   \underset{ j \equiv N}{\prod}  \sigma^+_j   \bigg] \text{ }  \bigg]                            \bigg\}      \\ \times \bigg[      \text{ }  \bigg[       \underset{1 \leq j \leq N-1}{\prod}   i u \sigma^+_j \sigma^-_j \textbf{1}_{\{ j^{\prime} > j \text{ } : \text{ } i u \sigma^+_{j^{\prime}} \sigma^-_{j^{\prime}} \in \mathrm{support} ( A ( u ))  \}}    \big[  \sigma^+_{j^{\prime}} + i u^{-1} \sigma^+_{j^{\prime}} \sigma^-_{j^{\prime}}     \big]    \bigg] \bigg[ \underset{j \equiv N}{\prod} \sigma^+_j  \bigg] \text{ } \bigg] \text{, }   \end{align*}

\noindent corresponding to the fourth term. Applying (LR) to these terms yields,

\begin{align*}
-  \sigma^-_0 \sigma^+_1  \sigma^-_1     \bigg\{                         
            \bigg[           \underset{1 \leq j \leq N}{\prod}    \textbf{1}_{\{j^{\prime} > j \text{ } : \text{ } i u \sigma^+_{j^{\prime}} \sigma^-_{j^{\prime}} \in \mathrm{support}(A(u))\}} 
\bigg[ \sigma^+_{j^{\prime}}    \sigma^-_{j^{\prime}+1} + i u \sigma^+_{j^{\prime}} \sigma^-_{j^{\prime}} \bigg[      \underset{j \mathrm{mod}3 \equiv 0}{\underset{3 \leq j \leq N-1}{\prod}}   i \big( u^{\prime} \big)^{-1} \sigma^+_j \sigma^-_j   \bigg]  \text{ } \bigg]  \bigg[   \underset{ j \equiv N}{\prod}  \sigma^+_j   \bigg] \text{ }  \bigg]                  \\ , \bigg[      \text{ }  \bigg[       \underset{1 \leq j \leq N-1}{\prod}   i u^{\prime} \sigma^+_j \sigma^-_j \textbf{1}_{\{ j^{\prime} > j \text{ } : \text{ } i u^{\prime} \sigma^+_{j^{\prime}} \sigma^-_{j^{\prime}} \in \mathrm{support} ( A (  u^{\prime}  ))  \}}    \big[  \sigma^+_{j^{\prime}} + i \big( u^{\prime}\big)^{-1} \sigma^+_{j^{\prime}} \sigma^-_{j^{\prime}}     \big]    \bigg]             \bigg\} 
  \bigg[ \underset{j \equiv N}{\prod} \sigma^+_j  \bigg]  
    \end{align*}
    
  \noindent corresponding to the first term,  
    
    \begin{align*}
    -  \sigma^-_0 \sigma^+_1     \bigg\{                         \sigma^-_1
     , \bigg[      \text{ }  \bigg[       \underset{1 \leq j \leq N-1}{\prod}   i u^{\prime}  \sigma^+_j \sigma^-_j \textbf{1}_{\{ j^{\prime} > j \text{ } : \text{ } i u^{\prime} \sigma^+_{j^{\prime}} \sigma^-_{j^{\prime}} \in \mathrm{support} ( A ( u^{\prime} ))  \}}    \big[  \sigma^+_{j^{\prime}} + i \big( u^{\prime}\big)^{-1} \sigma^+_{j^{\prime}} \sigma^-_{j^{\prime}}     \big]    \bigg]             \bigg\} \\ \times    \bigg[           \underset{1 \leq j \leq N}{\prod}    \textbf{1}_{\{j^{\prime} > j \text{ } : \text{ } i u \sigma^+_{j^{\prime}} \sigma^-_{j^{\prime}} \in \mathrm{support}(A(u))\}} 
\bigg[ \sigma^+_{j^{\prime}}    \sigma^-_{j^{\prime}+1} + i u \sigma^+_{j^{\prime}} \sigma^-_{j^{\prime}} \bigg[      \underset{j \mathrm{mod}3 \equiv 0}{\underset{3 \leq j \leq N-1}{\prod}}   i \big( u^{\prime} \big)^{-1} \sigma^+_j \sigma^-_j   \bigg]  \text{ } \bigg]  \bigg[   \underset{ j \equiv N}{\prod}  \sigma^+_j   \bigg] \text{ }  \bigg]    \bigg[ \underset{j \equiv N}{\prod} \sigma^+_j  \bigg]  
    \end{align*}

\noindent corresponding to the second term,

    \begin{align*}
    - \bigg[  \sigma^-_0 \sigma^+_1 
     \underset{1 \leq j \leq N-1}{\prod}   i u \sigma^+_j \sigma^-_j \textbf{1}_{\{ j^{\prime} > j \text{ } : \text{ } i u \sigma^+_{j^{\prime}} \sigma^-_{j^{\prime}} \in \mathrm{support} ( A ( u ))  \}}    \big[  \sigma^+_{j^{\prime}} + i u^{-1} \sigma^+_{j^{\prime}} \sigma^-_{j^{\prime}}     \big]    \bigg]       \sigma^-_1    \end{align*}

              \begin{align*}     \times   \bigg\{                  \bigg[           \underset{1 \leq j \leq N}{\prod}    \textbf{1}_{\{j^{\prime} > j \text{ } : \text{ } i u \sigma^+_{j^{\prime}} \sigma^-_{j^{\prime}} \in \mathrm{support}(A(u))\}} 
\bigg[ \sigma^+_{j^{\prime}}    \sigma^-_{j^{\prime}+1} + i u \sigma^+_{j^{\prime}} \sigma^-_{j^{\prime}} \bigg[      \underset{j \mathrm{mod}3 \equiv 0}{\underset{3 \leq j \leq N-1}{\prod}}   i \big( u^{\prime} \big)^{-1} \sigma^+_j \sigma^-_j   \bigg]  \text{ } \bigg]  \bigg[   \underset{ j \equiv N}{\prod}  \sigma^+_j   \bigg] \text{ }  \bigg]            ,  \bigg[ \underset{j \equiv N}{\prod} \sigma^+_j  \bigg]                     \bigg\} \text{, }\end{align*} 

\noindent corresponding to the third term,

\begin{align*}
-  \sigma^-_0 \sigma^+_1 
 \bigg[       \underset{1 \leq j \leq N-1}{\prod}   i u \sigma^+_j \sigma^-_j \textbf{1}_{\{ j^{\prime} > j \text{ } : \text{ } i u \sigma^+_{j^{\prime}} \sigma^-_{j^{\prime}} \in \mathrm{support} ( A ( u ))  \}}    \big[  \sigma^+_{j^{\prime}} + i u^{-1} \sigma^+_{j^{\prime}} \sigma^-_{j^{\prime}}     \big]    \bigg]  \bigg\{               \sigma^-_1           \\    ,  \bigg[ \underset{j \equiv N}{\prod} \sigma^+_j  \bigg]                     \bigg\}  \bigg[           \underset{1 \leq j \leq N}{\prod}    \textbf{1}_{\{j^{\prime} > j \text{ } : \text{ } i u \sigma^+_{j^{\prime}} \sigma^-_{j^{\prime}} \in \mathrm{support}(A(u))\}} 
\bigg[ \sigma^+_{j^{\prime}}    \sigma^-_{j^{\prime}+1} + i u \sigma^+_{j^{\prime}} \sigma^-_{j^{\prime}} \bigg[      \underset{j \mathrm{mod}3 \equiv 0}{\underset{3 \leq j \leq N-1}{\prod}}   i \big( u^{\prime} \big)^{-1} \sigma^+_j \sigma^-_j   \bigg]  \text{ } \bigg]  \bigg[   \underset{ j \equiv N}{\prod}  \sigma^+_j   \bigg] \text{ }  \bigg]     \text{, }   \end{align*} 
\noindent corresponding to the fourth term,

\begin{align*}    \bigg\{        \sigma^-_0      ,                \sigma^-_1  
            \bigg[           \underset{1 \leq j \leq N}{\prod}    \textbf{1}_{\{j^{\prime} > j \text{ } : \text{ } i u^{\prime} \sigma^+_{j^{\prime}} \sigma^-_{j^{\prime}} \in \mathrm{support}(A(u^{\prime}))\}} 
\bigg[ \sigma^+_{j^{\prime}}    \sigma^-_{j^{\prime}+1} + i u^{\prime} \sigma^+_{j^{\prime}} \sigma^-_{j^{\prime}} \bigg[      \underset{j \mathrm{mod}3 \equiv 0}{\underset{3 \leq j \leq N-1}{\prod}}   i \big( u^{\prime} \big)^{-1} \sigma^+_j \sigma^-_j   \bigg]  \text{ } \bigg]  \bigg[   \underset{ j \equiv N}{\prod}  \sigma^+_j   \bigg] \text{ }  \bigg]                            \bigg\}       \sigma^+_1    \\ \times \bigg[      \text{ }  \bigg[       \underset{1 \leq j \leq N-1}{\prod}   i u \sigma^+_j \sigma^-_j \textbf{1}_{\{ j^{\prime} > j \text{ } : \text{ } i u \sigma^+_{j^{\prime}} \sigma^-_{j^{\prime}} \in \mathrm{support} ( A ( u ))  \}}    \big[  \sigma^+_{j^{\prime}} + i u^{-1} \sigma^+_{j^{\prime}} \sigma^-_{j^{\prime}}     \big]    \bigg] \bigg[ \underset{j \equiv N}{\prod} \sigma^+_j  \bigg] \text{ } \bigg]   \\ + 
\sigma^-_0  \bigg\{     \sigma^+_1       ,                \sigma^-_1  
            \bigg[           \underset{1 \leq j \leq N}{\prod}    \textbf{1}_{\{j^{\prime} > j \text{ } : \text{ } i u^{\prime} \sigma^+_{j^{\prime}} \sigma^-_{j^{\prime}} \in \mathrm{support}(A(u^{\prime}))\}} 
\bigg[ \sigma^+_{j^{\prime}}    \sigma^-_{j^{\prime}+1} + i u^{\prime} \sigma^+_{j^{\prime}} \sigma^-_{j^{\prime}} \bigg[      \underset{j \mathrm{mod}3 \equiv 0}{\underset{3 \leq j \leq N-1}{\prod}}   i \big( u^{\prime} \big)^{-1} \sigma^+_j \sigma^-_j   \bigg]  \text{ } \bigg]  \bigg[   \underset{ j \equiv N}{\prod}  \sigma^+_j   \bigg] \text{ }  \bigg]                            \bigg\}        \\ \times \bigg[      \text{ }  \bigg[       \underset{1 \leq j \leq N-1}{\prod}   i u \sigma^+_j \sigma^-_j \textbf{1}_{\{ j^{\prime} > j \text{ } : \text{ } i u \sigma^+_{j^{\prime}} \sigma^-_{j^{\prime}} \in \mathrm{support} ( A ( u ))  \}}    \big[  \sigma^+_{j^{\prime}} + i u^{-1} \sigma^+_{j^{\prime}} \sigma^-_{j^{\prime}}     \big]    \bigg] \bigg[ \underset{j \equiv N}{\prod} \sigma^+_j  \bigg] \text{ } \bigg]   \text{, } \end{align*}

\noindent corresponding to the fifth term. Applying a combination of (AC), followed by (LR), yields,

\begin{align*}
    -  \sigma^-_0 \sigma^+_1  \sigma^-_1     \bigg\{                         
            \bigg[           \underset{1 \leq j \leq N}{\prod}    \textbf{1}_{\{j^{\prime} > j \text{ } : \text{ } i u \sigma^+_{j^{\prime}} \sigma^-_{j^{\prime}} \in \mathrm{support}(A(u))\}} 
\bigg[ \sigma^+_{j^{\prime}}    \sigma^-_{j^{\prime}+1} + i u \sigma^+_{j^{\prime}} \sigma^-_{j^{\prime}} \bigg[      \underset{j \mathrm{mod}3 \equiv 0}{\underset{3 \leq j \leq N-1}{\prod}}   i \big( u^{\prime} \big)^{-1} \sigma^+_j \sigma^-_j   \bigg]  \text{ } \bigg]  \bigg[   \underset{ j \equiv N}{\prod}  \sigma^+_j   \bigg] \text{ }  \bigg]                  \\ , \bigg[      \text{ }  \bigg[       \underset{1 \leq j \leq N-1}{\prod}   i u^{\prime}  \sigma^+_j \sigma^-_j \textbf{1}_{\{ j^{\prime} > j \text{ } : \text{ } i u^{\prime} \sigma^+_{j^{\prime}} \sigma^-_{j^{\prime}} \in \mathrm{support} ( A ( u^{\prime} ))  \}}    \big[  \sigma^+_{j^{\prime}} + i \big( u^{\prime}\big)^{-1} \sigma^+_{j^{\prime}} \sigma^-_{j^{\prime}}     \big]    \bigg]             \bigg\} 
  \bigg[ \underset{j \equiv N}{\prod} \sigma^+_j  \bigg]  
     \end{align*} 
     
     \noindent corresponding to the first term,

     \begin{align*} -  \sigma^-_0 \sigma^+_1     \bigg\{                         \sigma^-_1
     , \bigg[          \underset{1 \leq j \leq N-1}{\prod}   i u^{\prime}  \sigma^+_j \sigma^-_j \textbf{1}_{\{ j^{\prime} > j \text{ } : \text{ } i u^{\prime} \sigma^+_{j^{\prime}} \sigma^-_{j^{\prime}} \in \mathrm{support} ( A ( u^{\prime} ))  \}}    \big[  \sigma^+_{j^{\prime}} + i \big( u^{\prime}\big)^{-1} \sigma^+_{j^{\prime}} \sigma^-_{j^{\prime}}     \big]    \bigg]             \bigg\} \\
     \times    \bigg[           \underset{1 \leq j \leq N}{\prod}    \textbf{1}_{\{j^{\prime} > j \text{ } : \text{ } i u \sigma^+_{j^{\prime}} \sigma^-_{j^{\prime}} \in \mathrm{support}(A(u))\}} 
\bigg[ \sigma^+_{j^{\prime}}    \sigma^-_{j^{\prime}+1} + i u \sigma^+_{j^{\prime}} \sigma^-_{j^{\prime}} \bigg[      \underset{j \mathrm{mod}3 \equiv 0}{\underset{3 \leq j \leq N-1}{\prod}}   i \big( u^{\prime} \big)^{-1} \sigma^+_j \sigma^-_j   \bigg]  \text{ } \bigg]  \bigg[   \underset{ j \equiv N}{\prod}  \sigma^+_j   \bigg] \text{ }  \bigg]    \bigg[ \underset{j \equiv N}{\prod} \sigma^+_j  \bigg]  
    \end{align*}

    \noindent corresponding to the first term,
    
    \begin{align*}
    -  \sigma^-_0 \sigma^+_1 
 \bigg[       \underset{1 \leq j \leq N-1}{\prod}   i u \sigma^+_j \sigma^-_j \textbf{1}_{\{ j^{\prime} > j \text{ } : \text{ } i u \sigma^+_{j^{\prime}} \sigma^-_{j^{\prime}} \in \mathrm{support} ( A ( u ))  \}}    \big[  \sigma^+_{j^{\prime}} + i u^{-1} \sigma^+_{j^{\prime}} \sigma^-_{j^{\prime}}     \big]    \bigg]       \sigma^-_1   \\   \times   \bigg\{                  \bigg[           \underset{1 \leq j \leq N}{\prod}    \textbf{1}_{\{j^{\prime} > j \text{ } : \text{ } i u \sigma^+_{j^{\prime}} \sigma^-_{j^{\prime}} \in \mathrm{support}(A(u))\}} 
\bigg[ \sigma^+_{j^{\prime}}    \sigma^-_{j^{\prime}+1} + i u \sigma^+_{j^{\prime}} \sigma^-_{j^{\prime}} \bigg[      \underset{j \mathrm{mod}3 \equiv 0}{\underset{3 \leq j \leq N-1}{\prod}}   i \big( u^{\prime} \big)^{-1} \sigma^+_j \sigma^-_j   \bigg]  \text{ } \bigg]  \bigg[   \underset{ j \equiv N}{\prod}  \sigma^+_j   \bigg] \text{ }  \bigg]            ,  \bigg[ \underset{j \equiv N}{\prod} \sigma^+_j  \bigg]                     \bigg\} \text{, } \end{align*} 

\noindent corresponding to the second term,

 \begin{align*}      -  \sigma^-_0 \sigma^+_1 
 \bigg[       \underset{1 \leq j \leq N-1}{\prod}   i u \sigma^+_j \sigma^-_j \textbf{1}_{\{ j^{\prime} > j \text{ } : \text{ } i u \sigma^+_{j^{\prime}} \sigma^-_{j^{\prime}} \in \mathrm{support} ( A ( u ))  \}}    \big[  \sigma^+_{j^{\prime}} + i u^{-1} \sigma^+_{j^{\prime}} \sigma^-_{j^{\prime}}     \big]    \bigg]  \bigg\{               \sigma^-_1               ,  \bigg[ \underset{j \equiv N}{\prod} \sigma^+_j  \bigg]                     \bigg\} \\ \times  \bigg[           \underset{1 \leq j \leq N}{\prod}    \textbf{1}_{\{j^{\prime} > j \text{ } : \text{ } i u \sigma^+_{j^{\prime}} \sigma^-_{j^{\prime}} \in \mathrm{support}(A(u))\}} 
\bigg[ \sigma^+_{j^{\prime}}    \sigma^-_{j^{\prime}+1} + i u \sigma^+_{j^{\prime}} \sigma^-_{j^{\prime}} \bigg[      \underset{j \mathrm{mod}3 \equiv 0}{\underset{3 \leq j \leq N-1}{\prod}}   i \big( u^{\prime} \big)^{-1} \sigma^+_j \sigma^-_j   \bigg]  \text{ } \bigg]  \bigg[   \underset{ j \equiv N}{\prod}  \sigma^+_j   \bigg] \text{ }  \bigg]    \text{, }      \end{align*}

\noindent corresponding to the third term,

\begin{align*}
-   \bigg\{            \sigma^-_1  
            ,      \sigma^-_0                                \bigg\}   \bigg[           \underset{1 \leq j \leq N}{\prod}    \textbf{1}_{\{j^{\prime} > j \text{ } : \text{ } i u \sigma^+_{j^{\prime}} \sigma^-_{j^{\prime}} \in \mathrm{support}(A(u))\}} 
\bigg[ \sigma^+_{j^{\prime}}    \sigma^-_{j^{\prime}+1} + i u \sigma^+_{j^{\prime}} \sigma^-_{j^{\prime}} \bigg[      \underset{j \mathrm{mod}3 \equiv 0}{\underset{3 \leq j \leq N-1}{\prod}}   i \big( u^{\prime} \big)^{-1} \sigma^+_j \sigma^-_j   \bigg]  \text{ } \bigg]  \bigg[   \underset{ j \equiv N}{\prod}  \sigma^+_j   \bigg] \text{ }  \bigg]       \sigma^+_1    \\ \times \bigg[      \text{ }  \bigg[       \underset{1 \leq j \leq N-1}{\prod}   i u \sigma^+_j \sigma^-_j \textbf{1}_{\{ j^{\prime} > j \text{ } : \text{ } i u \sigma^+_{j^{\prime}} \sigma^-_{j^{\prime}} \in \mathrm{support} ( A ( u ))  \}}    \big[  \sigma^+_{j^{\prime}} + i u^{-1} \sigma^+_{j^{\prime}} \sigma^-_{j^{\prime}}     \big]    \bigg] \bigg[ \underset{j \equiv N}{\prod} \sigma^+_j  \bigg] \text{ } \bigg] \end{align*}

\noindent corresponding to the fourth term,

\begin{align*}
-    \sigma^-_1  
            \bigg\{                       \bigg[           \underset{1 \leq j \leq N}{\prod}    \textbf{1}_{\{j^{\prime} > j \text{ } : \text{ } i u \sigma^+_{j^{\prime}} \sigma^-_{j^{\prime}} \in \mathrm{support}(A(u))\}} 
\bigg[ \sigma^+_{j^{\prime}}    \sigma^-_{j^{\prime}+1} + i u \sigma^+_{j^{\prime}} \sigma^-_{j^{\prime}} \bigg[      \underset{j \mathrm{mod}3 \equiv 0}{\underset{3 \leq j \leq N-1}{\prod}}   i \big( u^{\prime} \big)^{-1} \sigma^+_j \sigma^-_j   \bigg]  \text{ } \bigg]  \bigg[   \underset{ j \equiv N}{\prod}  \sigma^+_j   \bigg] \text{ }  \bigg]                          , \sigma^-_0       \bigg\}       \sigma^+_1    \\ \times \bigg[      \text{ }  \bigg[       \underset{1 \leq j \leq N-1}{\prod}   i u \sigma^+_j \sigma^-_j \textbf{1}_{\{ j^{\prime} > j \text{ } : \text{ } i u \sigma^+_{j^{\prime}} \sigma^-_{j^{\prime}} \in \mathrm{support} ( A ( u ))  \}}    \big[  \sigma^+_{j^{\prime}} + i u^{-1} \sigma^+_{j^{\prime}} \sigma^-_{j^{\prime}}     \big]    \bigg] \bigg[ \underset{j \equiv N}{\prod} \sigma^+_j  \bigg] \text{ } \bigg]  \end{align*}

\noindent corresponding to the fifth term,

\begin{align*}
-      \sigma^-_0  \bigg\{                  \sigma^-_1  
            ,    \sigma^+_1                               \bigg\}      \bigg[           \underset{1 \leq j \leq N}{\prod}    \textbf{1}_{\{j^{\prime} > j \text{ } : \text{ } i u \sigma^+_{j^{\prime}} \sigma^-_{j^{\prime}} \in \mathrm{support}(A(u))\}} 
\bigg[ \sigma^+_{j^{\prime}}    \sigma^-_{j^{\prime}+1} + i u \sigma^+_{j^{\prime}} \sigma^-_{j^{\prime}}  \bigg[      \underset{j \mathrm{mod}3 \equiv 0}{\underset{3 \leq j \leq N-1}{\prod}}   i \big( u^{\prime} \big)^{-1}  \\ \times  \sigma^+_j \sigma^-_j   \bigg]  \text{ } \bigg]  \bigg[   \underset{ j \equiv N}{\prod}  \sigma^+_j   \bigg] \text{ }  \bigg]     \bigg[      \text{ }  \bigg[       \underset{1 \leq j \leq N-1}{\prod}   i u \sigma^+_j \sigma^-_j \textbf{1}_{\{ j^{\prime} > j \text{ } : \text{ } i u \sigma^+_{j^{\prime}} \sigma^-_{j^{\prime}} \in \mathrm{support} ( A ( u ))  \}}    \big[  \sigma^+_{j^{\prime}} + i u^{-1} \sigma^+_{j^{\prime}} \sigma^-_{j^{\prime}}     \big]    \bigg] \bigg[ \underset{j \equiv N}{\prod} \sigma^+_j  \bigg] \text{ } \bigg]  \end{align*}

\noindent corresponding to the sixth term,

\begin{align*}
-       \sigma^-_0   \sigma^-_1   \bigg\{        
            \bigg[           \underset{1 \leq j \leq N}{\prod}    \textbf{1}_{\{j^{\prime} > j \text{ } : \text{ } i u \sigma^+_{j^{\prime}} \sigma^-_{j^{\prime}} \in \mathrm{support}(A(u))\}} 
\bigg[ \sigma^+_{j^{\prime}}    \sigma^-_{j^{\prime}+1} + i u \sigma^+_{j^{\prime}} \sigma^-_{j^{\prime}} \bigg[      \underset{j \mathrm{mod}3 \equiv 0}{\underset{3 \leq j \leq N-1}{\prod}}   i \big( u^{\prime} \big)^{-1} \sigma^+_j \sigma^-_j   \bigg]  \text{ } \bigg]  \bigg[   \underset{ j \equiv N}{\prod}  \sigma^+_j   \bigg] \text{ }  \bigg]    ,   \sigma^+_1                               \bigg\}        \\ \times \bigg[      \text{ }  \bigg[       \underset{1 \leq j \leq N-1}{\prod}   i u \sigma^+_j \sigma^-_j \textbf{1}_{\{ j^{\prime} > j \text{ } : \text{ } i u \sigma^+_{j^{\prime}} \sigma^-_{j^{\prime}} \in \mathrm{support} ( A ( u ))  \}}    \big[  \sigma^+_{j^{\prime}} + i u^{-1} \sigma^+_{j^{\prime}} \sigma^-_{j^{\prime}}     \big]    \bigg] \bigg[ \underset{j \equiv N}{\prod} \sigma^+_j  \bigg] \text{ } \bigg]     \text{, } \end{align*}

\noindent corresponding to the seventh term, 
\begin{align*}
   \bigg\{ \sigma^-_1  \bigg[           \underset{1 \leq j \leq N}{\prod}    \textbf{1}_{j^{\prime} > j \text{ } : \text{ } i u \sigma^+_{j^{\prime}} \sigma^-_{j^{\prime}} \in \mathrm{support}(A(u))} 
\bigg[ \sigma^+_{j^{\prime}}    \sigma^-_{j^{\prime}+1} + i u \sigma^+_{j^{\prime}} \sigma^-_{j^{\prime}} \bigg[      \underset{j \mathrm{mod}3 \equiv 0}{\underset{3 \leq j \leq N-1}{\prod}}   i \big( u^{\prime} \big)^{-1} \sigma^+_j \sigma^-_j   \bigg]  \text{ } \bigg]  \bigg[   \underset{ j \equiv N}{\prod}  \sigma^+_j   \bigg] \text{ }  \bigg]    ,  \sigma^-_1   \\  \times  \bigg[           \underset{1 \leq j \leq N}{\prod}    \textbf{1}_{\{j^{\prime} > j \text{ } : \text{ } i u^{\prime}  \sigma^+_{j^{\prime}} \sigma^-_{j^{\prime}} \in \mathrm{support}(A(u^{\prime}))\}} 
\bigg[ \sigma^+_{j^{\prime}}    \sigma^-_{j^{\prime}+1} + i u^{\prime} \sigma^+_{j^{\prime}} \sigma^-_{j^{\prime}} \bigg[      \underset{j \mathrm{mod}3 \equiv 0}{\underset{3 \leq j \leq N-1}{\prod}}   i \big( u^{\prime} \big)^{-1} \sigma^+_j \sigma^-_j   \bigg]  \text{ } \bigg]  \bigg[   \underset{ j \equiv N}{\prod}  \sigma^+_j   \bigg] \text{ }  \bigg]      \bigg\} \text{. } \end{align*}

\noindent corresponding to the ninth term. Applying (LR) to the collection of nine terms above yields,

\begin{align*}
\sigma^-_1    \bigg\{  \bigg[           \underset{1 \leq j \leq N}{\prod}    \textbf{1}_{j^{\prime} > j \text{ } : \text{ } i u \sigma^+_{j^{\prime}} \sigma^-_{j^{\prime}} \in \mathrm{support}(A(u))} 
\bigg[ \sigma^+_{j^{\prime}}    \sigma^-_{j^{\prime}+1} + i u \sigma^+_{j^{\prime}} \sigma^-_{j^{\prime}} \bigg[      \underset{j \mathrm{mod}3 \equiv 0}{\underset{3 \leq j \leq N-1}{\prod}}   i \big( u^{\prime} \big)^{-1} \sigma^+_j \sigma^-_j   \bigg]  \text{ } \bigg]  \bigg[   \underset{ j \equiv N}{\prod}  \sigma^+_j   \bigg] \text{ }  \bigg]    ,  \sigma^-_1    \\  \times  \bigg[           \underset{1 \leq j \leq N}{\prod}    \textbf{1}_{\{j^{\prime} > j \text{ } : \text{ } i u^{\prime}  \sigma^+_{j^{\prime}} \sigma^-_{j^{\prime}} \in \mathrm{support}(A(u^{\prime}))\}} 
\bigg[ \sigma^+_{j^{\prime}}    \sigma^-_{j^{\prime}+1} + i u^{\prime} \sigma^+_{j^{\prime}} \sigma^-_{j^{\prime}} \bigg[      \underset{j \mathrm{mod}3 \equiv 0}{\underset{3 \leq j \leq N-1}{\prod}}   i \big( u^{\prime} \big)^{-1} \sigma^+_j \sigma^-_j   \bigg]  \text{ } \bigg]  \bigg[   \underset{ j \equiv N}{\prod}  \sigma^+_j   \bigg] \text{ }  \bigg]      \bigg\}      \text{, } \end{align*}

\noindent corresponding to the first term,

\begin{align*}
\bigg\{ \sigma^-_1      ,  \sigma^-_1   \bigg[           \underset{1 \leq j \leq N}{\prod}    \textbf{1}_{\{j^{\prime} > j \text{ } : \text{ } i u^{\prime}  \sigma^+_{j^{\prime}} \sigma^-_{j^{\prime}} \in \mathrm{support}(A(u^{\prime}))\}} 
\bigg[ \sigma^+_{j^{\prime}}    \sigma^-_{j^{\prime}+1} + i u^{\prime} \sigma^+_{j^{\prime}} \sigma^-_{j^{\prime}}  \bigg[      \underset{j \mathrm{mod}3 \equiv 0}{\underset{3 \leq j \leq N-1}{\prod}}   i \big( u^{\prime} \big)^{-1} \sigma^+_j \sigma^-_j   \bigg]  \text{ } \bigg]  \bigg[   \underset{ j \equiv N}{\prod}  \sigma^+_j   \bigg] \text{ }  \bigg]      \bigg\} \\  \times  \bigg[           \underset{1 \leq j \leq N}{\prod}    \textbf{1}_{j^{\prime} > j \text{ } : \text{ } i u \sigma^+_{j^{\prime}} \sigma^-_{j^{\prime}} \in \mathrm{support}(A(u))} 
 \bigg[ \sigma^+_{j^{\prime}}    \sigma^-_{j^{\prime}+1} + i u \sigma^+_{j^{\prime}} \sigma^-_{j^{\prime}}  \bigg[      \underset{j \mathrm{mod}3 \equiv 0}{\underset{3 \leq j \leq N-1}{\prod}}   i \big( u^{\prime} \big)^{-1} \sigma^+_j \sigma^-_j   \bigg]  \text{ } \bigg]  \bigg[   \underset{ j \equiv N}{\prod}  \sigma^+_j   \bigg] \text{ }  \bigg] \text{, } \end{align*}

 \noindent corresponding to the second term. Applying (AC) yields,

\begin{align*}        - \sigma^-_1    \bigg\{      \sigma^-_1   \bigg[           \underset{1 \leq j \leq N}{\prod}    \textbf{1}_{\{j^{\prime} > j \text{ } : \text{ } i u \sigma^+_{j^{\prime}} \sigma^-_{j^{\prime}} \in \mathrm{support}(A(u^{\prime}))\}} 
\bigg[ \sigma^+_{j^{\prime}}    \sigma^-_{j^{\prime}+1} + i u^{\prime} \sigma^+_{j^{\prime}} \sigma^-_{j^{\prime}} \bigg[      \underset{j \mathrm{mod}3 \equiv 0}{\underset{3 \leq j \leq N-1}{\prod}}   i \big( u^{\prime} \big)^{-1} \sigma^+_j \sigma^-_j   \bigg]  \text{ } \bigg]  \bigg[   \underset{ j \equiv N}{\prod}  \sigma^+_j   \bigg] \text{ }  \bigg]    \\  , \bigg[           \underset{1 \leq j \leq N}{\prod}    \textbf{1}_{j^{\prime} > j \text{ } : \text{ } i u^{\prime} \sigma^+_{j^{\prime}} \sigma^-_{j^{\prime}} \in \mathrm{support}(A(u^{\prime}))} 
\bigg[ \sigma^+_{j^{\prime}}    \sigma^-_{j^{\prime}+1} + i u^{\prime} \sigma^+_{j^{\prime}} \sigma^-_{j^{\prime}} \bigg[      \underset{j \mathrm{mod}3 \equiv 0}{\underset{3 \leq j \leq N-1}{\prod}}   i \big( u^{\prime} \big)^{-1} \sigma^+_j \sigma^-_j   \bigg]  \text{ } \bigg]  \bigg[   \underset{ j \equiv N}{\prod}  \sigma^+_j   \bigg] \text{ }  \bigg]        \bigg\} \end{align*}

\noindent corresponding to the first term, 

\begin{align*}
- \bigg\{  \sigma^-_1   \bigg[           \underset{1 \leq j \leq N}{\prod}    \textbf{1}_{\{j^{\prime} > j \text{ } : \text{ } i u \sigma^+_{j^{\prime}} \sigma^-_{j^{\prime}} \in \mathrm{support}(A(u^{\prime}))\}} 
\bigg[ \sigma^+_{j^{\prime}}    \sigma^-_{j^{\prime}+1} + i u^{\prime} \sigma^+_{j^{\prime}} \sigma^-_{j^{\prime}}  \bigg[      \underset{j \mathrm{mod}3 \equiv 0}{\underset{3 \leq j \leq N-1}{\prod}}   i \big( u^{\prime} \big)^{-1} \sigma^+_j \sigma^-_j   \bigg]  \text{ } \bigg]  \bigg[   \underset{ j \equiv N}{\prod}  \sigma^+_j   \bigg] \text{ }  \bigg]    , \sigma^-_1       \bigg\} \\ \times  \bigg[           \underset{1 \leq j \leq N}{\prod}    \textbf{1}_{j^{\prime} > j \text{ } : \text{ } i u \sigma^+_{j^{\prime}} \sigma^-_{j^{\prime}} \in \mathrm{support}(A(u))} 
 \bigg[ \sigma^+_{j^{\prime}}    \sigma^-_{j^{\prime}+1} + i u \sigma^+_{j^{\prime}} \sigma^-_{j^{\prime}}  \bigg[      \underset{j \mathrm{mod}3 \equiv 0}{\underset{3 \leq j \leq N-1}{\prod}}   i \big( u^{\prime} \big)^{-1} \sigma^+_j \sigma^-_j   \bigg]  \text{ } \bigg]  \bigg[   \underset{ j \equiv N}{\prod}  \sigma^+_j   \bigg] \text{ }  \bigg]   \end{align*}
 
 \noindent corresponding to the second term. Applying (LR) yields,
 
 \begin{align*}
 - \big[ \sigma^-_1  \big]^2   \bigg\{        \bigg[           \underset{1 \leq j \leq N}{\prod}    \textbf{1}_{\{j^{\prime} > j \text{ } : \text{ } i u \sigma^+_{j^{\prime}} \sigma^-_{j^{\prime}} \in \mathrm{support}(A(u^{\prime}))\}} 
\bigg[ \sigma^+_{j^{\prime}}    \sigma^-_{j^{\prime}+1} + i u^{\prime} \sigma^+_{j^{\prime}} \sigma^-_{j^{\prime}} \bigg[      \underset{j \mathrm{mod}3 \equiv 0}{\underset{3 \leq j \leq N-1}{\prod}}   i \big( u^{\prime} \big)^{-1} \sigma^+_j \sigma^-_j   \bigg]  \text{ } \bigg]  \bigg[   \underset{ j \equiv N}{\prod}  \sigma^+_j   \bigg] \text{ }  \bigg] \\ , \bigg[           \underset{1 \leq j \leq N}{\prod}    \textbf{1}_{j^{\prime} > j \text{ } : \text{ } i u^{\prime} \sigma^+_{j^{\prime}} \sigma^-_{j^{\prime}} \in \mathrm{support}(A(u^{\prime}))} 
\bigg[ \sigma^+_{j^{\prime}}    \sigma^-_{j^{\prime}+1} + i u^{\prime}  \sigma^+_{j^{\prime}} \sigma^-_{j^{\prime}} \bigg[      \underset{j \mathrm{mod}3 \equiv 0}{\underset{3 \leq j \leq N-1}{\prod}}   i \big( u^{\prime} \big)^{-1} \sigma^+_j \sigma^-_j   \bigg]  \text{ } \bigg]  \bigg[   \underset{ j \equiv N}{\prod}  \sigma^+_j   \bigg] \text{ }  \bigg]        \bigg\} \text{, } \end{align*}

\noindent corresponding to the first term,

\begin{align*}
- \sigma^-_1    \bigg\{      \sigma^-_1    , \bigg[           \underset{1 \leq j \leq N}{\prod}    \textbf{1}_{j^{\prime} > j \text{ } : \text{ } i u^{\prime} \sigma^+_{j^{\prime}} \sigma^-_{j^{\prime}} \in \mathrm{support}(A(u^{\prime}))} 
\bigg[ \sigma^+_{j^{\prime}}    \sigma^-_{j^{\prime}+1} + i u^{\prime} \sigma^+_{j^{\prime}} \sigma^-_{j^{\prime}} \bigg[      \underset{j \mathrm{mod}3 \equiv 0}{\underset{3 \leq j \leq N-1}{\prod}}   i \big( u^{\prime} \big)^{-1} \sigma^+_j \sigma^-_j   \bigg]  \text{ } \bigg]  \bigg[   \underset{ j \equiv N}{\prod}  \sigma^+_j   \bigg] \text{ }  \bigg]        \bigg\} \\ \times \bigg[           \underset{1 \leq j \leq N}{\prod}    \textbf{1}_{\{j^{\prime} > j \text{ } : \text{ } i u \sigma^+_{j^{\prime}} \sigma^-_{j^{\prime}} \in \mathrm{support}(A(u^{\prime}))\}} 
\bigg[ \sigma^+_{j^{\prime}}    \sigma^-_{j^{\prime}+1} + i u^{\prime} \sigma^+_{j^{\prime}} \sigma^-_{j^{\prime}} \bigg[      \underset{j \mathrm{mod}3 \equiv 0}{\underset{3 \leq j \leq N-1}{\prod}}   i \big( u^{\prime} \big)^{-1} \sigma^+_j \sigma^-_j   \bigg]  \text{ } \bigg]  \bigg[   \underset{ j \equiv N}{\prod}  \sigma^+_j   \bigg] \text{ }  \bigg] \text{, }  \end{align*}

\noindent corresponding to the second term,

\begin{align*}
- \bigg\{  \sigma^-_1    , \sigma^-_1       \bigg\}   \bigg[           \underset{1 \leq j \leq N}{\prod}    \textbf{1}_{\{j^{\prime} > j \text{ } : \text{ } i u \sigma^+_{j^{\prime}} \sigma^-_{j^{\prime}} \in \mathrm{support}(A(u^{\prime}))\}} 
\bigg[ \sigma^+_{j^{\prime}}    \sigma^-_{j^{\prime}+1} + i u^{\prime} \sigma^+_{j^{\prime}} \sigma^-_{j^{\prime}}  \bigg[      \underset{j \mathrm{mod}3 \equiv 0}{\underset{3 \leq j \leq N-1}{\prod}}   i \big( u^{\prime} \big)^{-1} \sigma^+_j \sigma^-_j   \bigg]  \text{ } \bigg]\\ 
\times   \bigg[   \underset{ j \equiv N}{\prod}  \sigma^+_j   \bigg] \text{ }  \bigg]   \bigg[           \underset{1 \leq j \leq N}{\prod}    \textbf{1}_{j^{\prime} > j \text{ } : \text{ } i u \sigma^+_{j^{\prime}} \sigma^-_{j^{\prime}} \in \mathrm{support}(A(u))} 
 \bigg[ \sigma^+_{j^{\prime}}    \sigma^-_{j^{\prime}+1} + i u \sigma^+_{j^{\prime}} \sigma^-_{j^{\prime}}  \bigg[      \underset{j \mathrm{mod}3 \equiv 0}{\underset{3 \leq j \leq N-1}{\prod}}   i \big( u^{\prime} \big)^{-1} \sigma^+_j \sigma^-_j   \bigg]  \text{ } \bigg]   \\  \times   \bigg[   \underset{ j \equiv N}{\prod}  \sigma^+_j   \bigg] \text{ }  \bigg]                   \end{align*} 

\noindent corresponding to the third term,

\begin{align*}  -  \sigma^-_1  \bigg\{    \bigg[           \underset{1 \leq j \leq N}{\prod}    \textbf{1}_{\{j^{\prime} > j \text{ } : \text{ } i u \sigma^+_{j^{\prime}} \sigma^-_{j^{\prime}} \in \mathrm{support}(A(u^{\prime}))\}} 
\bigg[ \sigma^+_{j^{\prime}}    \sigma^-_{j^{\prime}+1} + i u^{\prime} \sigma^+_{j^{\prime}} \sigma^-_{j^{\prime}}  \bigg[      \underset{j \mathrm{mod}3 \equiv 0}{\underset{3 \leq j \leq N-1}{\prod}}   i \big( u^{\prime} \big)^{-1} \sigma^+_j \sigma^-_j   \bigg]  \text{ } \bigg]  \bigg[   \underset{ j \equiv N}{\prod}  \sigma^+_j   \bigg] \text{ }  \bigg]    , \sigma^-_1       \bigg\} \\ \times  \bigg[           \underset{1 \leq j \leq N}{\prod}    \textbf{1}_{j^{\prime} > j \text{ } : \text{ } i u \sigma^+_{j^{\prime}} \sigma^-_{j^{\prime}} \in \mathrm{support}(A(u))} 
 \bigg[ \sigma^+_{j^{\prime}}    \sigma^-_{j^{\prime}+1} + i u \sigma^+_{j^{\prime}} \sigma^-_{j^{\prime}}  \bigg[      \underset{j \mathrm{mod}3 \equiv 0}{\underset{3 \leq j \leq N-1}{\prod}}   i \big( u^{\prime} \big)^{-1} \sigma^+_j \sigma^-_j   \bigg]  \text{ } \bigg]  \bigg[   \underset{ j \equiv N}{\prod}  \sigma^+_j   \bigg] \text{ }  \bigg]  \text{,} 
\end{align*} 

\noindent which can be approximated with,

\begin{align*}
    \big( \mathscr{C}^1_1 \big)_{10}    \text{. }
\end{align*}

\noindent For contributions of $\mathscr{I}^1_3$, the last term of the decomposition of $\mathcal{I}^1$, one has,

\begin{align*}
 \bigg\{    \bigg[ \underset{0 \leq j < j+1 < N-1}{\prod}  \sigma^-_{j} i u \sigma^+_{j+1} \sigma^-_{j+1}      \textbf{1}_{\{j^{\prime}> j+1 \text{ } : \text{ } i u \sigma^-_{j^{\prime}} \sigma^+_{j^{\prime}} \in \mathrm{support} ( A(u))\}}  \big[ 
       \sigma^-_{j^{\prime}} +  i u \sigma^+_{j^{\prime}} \sigma^-_{j^{\prime}}      \big]  \bigg]         \bigg[       \underset{j  \text{ } \mathrm{mod}3 \equiv 0}{\underset{3 \leq j \leq N-1}{\prod}}      i u^{-1} \sigma^+_j \sigma^-_j              \bigg]   \\ 
       \times \bigg[ \underset{j \equiv N}{\prod}  \sigma^+_j \bigg]    , \bigg[ \underset{0 \leq j < j+1 < N-1}{\prod}  \sigma^-_{j} i u^{\prime}  \sigma^+_{j+1} \sigma^-_{j+1}      \textbf{1}_{\{j^{\prime}> j+1 \text{ } : \text{ } i u^{\prime} \sigma^-_{j^{\prime}} \sigma^+_{j^{\prime}} \in \mathrm{support} ( A(u^{\prime}))\}}  \big[ 
       \sigma^-_{j^{\prime}} +  i u^{\prime} \sigma^+_{j^{\prime}} \sigma^-_{j^{\prime}}      \big]  \bigg]         \bigg[       \underset{j  \text{ } \mathrm{mod}3 \equiv 0}{\underset{3 \leq j \leq N-1}{\prod}}      i \big( u^{\prime}\big)^{-1} \sigma^+_j \sigma^-_j              \bigg]  \\ \times \bigg[ \underset{j \equiv N}{\prod}  \sigma^+_j \bigg]        \bigg\} \\ \\   
       \overset{(\mathrm{LR}),(\mathrm{AC})}{=}                   \bigg[ \underset{0 \leq j < j+1 < N-1}{\prod}  \sigma^-_{j} i u \sigma^+_{j+1} \sigma^-_{j+1}      \textbf{1}_{\{j^{\prime}> j+1 \text{ } : \text{ } i u \sigma^-_{j^{\prime}} \sigma^+_{j^{\prime}} \in \mathrm{support} ( A(u))\}}  \big[ 
       \sigma^-_{j^{\prime}} +  i u \sigma^+_{j^{\prime}} \sigma^-_{j^{\prime}}      \big]  \bigg]      \bigg\{       \bigg[       \underset{j  \text{ } \mathrm{mod}3 \equiv 0}{\underset{3 \leq j \leq N-1}{\prod}}      i u^{-1} \sigma^+_j \sigma^-_j              \bigg] \\ \times   \bigg[ \underset{j \equiv N}{\prod}  \sigma^+_j \bigg]   \\  , \bigg[ \underset{0 \leq j < j+1 < N-1}{\prod}  \sigma^-_{j} i u^{\prime} \sigma^+_{j+1} \sigma^-_{j+1}      \textbf{1}_{\{j^{\prime}> j+1 \text{ } : \text{ } i u^{\prime} \sigma^-_{j^{\prime}} \sigma^+_{j^{\prime}} \in \mathrm{support} ( A(u^{\prime}))\}}  \big[ 
       \sigma^-_{j^{\prime}} +  i u^{\prime} \sigma^+_{j^{\prime}} \sigma^-_{j^{\prime}}      \big]  \bigg]         \bigg[       \underset{j  \text{ } \mathrm{mod}3 \equiv 0}{\underset{3 \leq j \leq N-1}{\prod}}      i \big( u^{\prime}\big)^{-1} \sigma^+_j \sigma^-_j              \bigg]  \\ \times  \bigg[ \underset{j \equiv N}{\prod}  \sigma^+_j \bigg]        \bigg\}                \\ - \bigg[       \underset{j  \text{ } \mathrm{mod}3 \equiv 0}{\underset{3 \leq j \leq N-1}{\prod}}      i \big( u^{\prime}\big)^{-1} \sigma^+_j \sigma^-_j              \bigg]  \bigg[ \underset{j \equiv N}{\prod}  \sigma^+_j \bigg] \bigg\{         \bigg[ \underset{0 \leq j < j+1 < N-1}{\prod}  \sigma^-_{j} i u^{\prime} \sigma^+_{j+1} \sigma^-_{j+1}      \textbf{1}_{\{j^{\prime}> j+1 \text{ } : \text{ } i u^{\prime} \sigma^-_{j^{\prime}} \sigma^+_{j^{\prime}} \in \mathrm{support} ( A(u^{\prime}))\}}  \\ \times \big[ 
       \sigma^-_{j^{\prime}} +  i u^{\prime} \sigma^+_{j^{\prime}} \sigma^-_{j^{\prime}}      \big]  \bigg]  \\ 
       \times      \bigg[       \underset{j  \text{ } \mathrm{mod}3 \equiv 0}{\underset{3 \leq j \leq N-1}{\prod}}      i \big( u^{\prime}\big)^{-1} \sigma^+_j \sigma^-_j              \bigg]  \bigg[ \underset{j \equiv N}{\prod}  \sigma^+_j \bigg]   ,   \bigg[ \underset{0 \leq j < j+1 < N-1}{\prod}  \sigma^-_{j} i u^{\prime} \sigma^+_{j+1} \sigma^-_{j+1}      \textbf{1}_{\{j^{\prime}> j+1 \text{ } : \text{ } i u^{\prime} \sigma^-_{j^{\prime}} \sigma^+_{j^{\prime}} \in \mathrm{support} ( A(u^{\prime}))\}} \\ \times  \big[ 
       \sigma^-_{j^{\prime}} +  i u^{\prime} \sigma^+_{j^{\prime}} \sigma^-_{j^{\prime}}      \big]  \bigg]          \bigg\} \text{, } \end{align*}
       
       \noindent to which an application of (LR) yields,
       
       \begin{align*}
       \bigg[ \underset{0 \leq j < j+1 < N-1}{\prod}  \sigma^-_{j} i u \sigma^+_{j+1} \sigma^-_{j+1}      \textbf{1}_{\{j^{\prime}> j+1 \text{ } : \text{ } i u \sigma^-_{j^{\prime}} \sigma^+_{j^{\prime}} \in \mathrm{support} ( A(u))\}}  \big[ 
       \sigma^-_{j^{\prime}} +  i u \sigma^+_{j^{\prime}} \sigma^-_{j^{\prime}}      \big]  \bigg]           \bigg[       \underset{j  \text{ } \mathrm{mod}3 \equiv 0}{\underset{3 \leq j \leq N-1}{\prod}}      i u^{-1} \sigma^+_j \sigma^-_j              \bigg] \\ \times    \bigg\{  \bigg[ \underset{j \equiv N}{\prod}  \sigma^+_j \bigg] \end{align*}

              \begin{align*}  
       , \bigg[ \underset{0 \leq j < j+1 < N-1}{\prod}  \sigma^-_{j} i u^{\prime} \sigma^+_{j+1} \sigma^-_{j+1}      \textbf{1}_{\{j^{\prime}> j+1 \text{ } : \text{ } i u^{\prime} \sigma^-_{j^{\prime}} \sigma^+_{j^{\prime}} \in \mathrm{support} ( A(u^{\prime}))\}}  \big[ 
       \sigma^-_{j^{\prime}} +  i u^{\prime} \sigma^+_{j^{\prime}} \sigma^-_{j^{\prime}}      \big]  \bigg]         \bigg[       \underset{j  \text{ } \mathrm{mod}3 \equiv 0}{\underset{3 \leq j \leq N-1}{\prod}}      i \big( u^{\prime}\big)^{-1} \sigma^+_j \sigma^-_j              \bigg]  \\ \times  \bigg[ \underset{j \equiv N}{\prod}  \sigma^+_j \bigg]        \bigg\}    \text{, }  \end{align*} 
       
       \noindent corresponding to the first term,

       \begin{align*}
       \bigg[ \underset{0 \leq j < j+1 < N-1}{\prod}  \sigma^-_{j} i u \sigma^+_{j+1} \sigma^-_{j+1}      \textbf{1}_{\{j^{\prime}> j+1 \text{ } : \text{ } i u \sigma^-_{j^{\prime}} \sigma^+_{j^{\prime}} \in \mathrm{support} ( A(u))\}}  \big[ 
       \sigma^-_{j^{\prime}} +  i u \sigma^+_{j^{\prime}} \sigma^-_{j^{\prime}}      \big]  \bigg]      \bigg\{       \bigg[       \underset{j  \text{ } \mathrm{mod}3 \equiv 0}{\underset{3 \leq j \leq N-1}{\prod}}      i u^{-1} \sigma^+_j \sigma^-_j              \bigg] \\  , \bigg[ \underset{0 \leq j < j+1 < N-1}{\prod}  \sigma^-_{j} i u^{\prime} \sigma^+_{j+1} \sigma^-_{j+1}      \textbf{1}_{\{j^{\prime}> j+1 \text{ } : \text{ } i u^{\prime} \sigma^-_{j^{\prime}} \sigma^+_{j^{\prime}} \in \mathrm{support} ( A(u^{\prime}))\}}  \big[ 
       \sigma^-_{j^{\prime}} +  i u^{\prime} \sigma^+_{j^{\prime}} \sigma^-_{j^{\prime}}      \big]  \bigg]         \bigg[       \underset{j  \text{ } \mathrm{mod}3 \equiv 0}{\underset{3 \leq j \leq N-1}{\prod}}      i \big( u^{\prime} \big)^{-1} \sigma^+_j \sigma^-_j              \bigg] \\   \times  \bigg[ \underset{j \equiv N}{\prod}  \sigma^+_j \bigg]        \bigg\} \bigg[ \underset{j \equiv N}{\prod}  \sigma^+_j \bigg] \end{align*} 
       
       \noindent corresponding to the second term, 
       
       \begin{align*}
       - \bigg[       \underset{j  \text{ } \mathrm{mod}3 \equiv 0}{\underset{3 \leq j \leq N-1}{\prod}}      i u^{-1} \sigma^+_j \sigma^-_j              \bigg]  \bigg[ \underset{j \equiv N}{\prod}  \sigma^+_j \bigg]       \bigg[ \underset{0 \leq j < j+1 < N-1}{\prod}  \sigma^-_{j} i u \sigma^+_{j+1} \sigma^-_{j+1}      \textbf{1}_{\{j^{\prime}> j+1 \text{ } : \text{ } i u \sigma^-_{j^{\prime}} \sigma^+_{j^{\prime}} \in \mathrm{support} ( A(u))\}}   \\  \times \big[ 
       \sigma^-_{j^{\prime}} +  i u \sigma^+_{j^{\prime}} \sigma^-_{j^{\prime}}      \big]  \bigg] \\ 
       \times   \bigg\{       \bigg[       \underset{j  \text{ } \mathrm{mod}3 \equiv 0}{\underset{3 \leq j \leq N-1}{\prod}}      i u^{-1} \sigma^+_j \sigma^-_j              \bigg]  \bigg[ \underset{j \equiv N}{\prod}  \sigma^+_j \bigg]   ,   \bigg[ \underset{0 \leq j < j+1 < N-1}{\prod}  \sigma^-_{j} i u^{\prime} \sigma^+_{j+1} \sigma^-_{j+1}      \textbf{1}_{\{j^{\prime}> j+1 \text{ } : \text{ } i u^{\prime} \sigma^-_{j^{\prime}} \sigma^+_{j^{\prime}} \in \mathrm{support} ( A(u^{\prime}))\}} \\ \times  \big[ 
       \sigma^-_{j^{\prime}} +  i u^{\prime} \sigma^+_{j^{\prime}} \sigma^-_{j^{\prime}}      \big]  \bigg]          \bigg\}   \end{align*}
       
     \noindent corresponding to the third term,   
       
       \begin{align*}
       - \bigg[       \underset{j  \text{ } \mathrm{mod}3 \equiv 0}{\underset{3 \leq j \leq N-1}{\prod}}      i u^{-1} \sigma^+_j \sigma^-_j              \bigg]  \bigg[ \underset{j \equiv N}{\prod}  \sigma^+_j \bigg] \bigg\{         \bigg[ \underset{0 \leq j < j+1 < N-1}{\prod}  \sigma^-_{j} i u \sigma^+_{j+1} \sigma^-_{j+1}      \textbf{1}_{\{j^{\prime}> j+1 \text{ } : \text{ } i u \sigma^-_{j^{\prime}} \sigma^+_{j^{\prime}} \in \mathrm{support} ( A(u))\}}  \\ \times \big[ 
       \sigma^-_{j^{\prime}} +  i u \sigma^+_{j^{\prime}} \sigma^-_{j^{\prime}}      \big]  \bigg]   ,   \bigg[ \underset{0 \leq j < j+1 < N-1}{\prod}  \sigma^-_{j} i u^{\prime} \sigma^+_{j+1} \sigma^-_{j+1}      \textbf{1}_{\{j^{\prime}> j+1 \text{ } : \text{ } i u^{\prime} \sigma^-_{j^{\prime}} \sigma^+_{j^{\prime}} \in \mathrm{support} ( A(u^{\prime}))\}}     \big[ 
       \sigma^-_{j^{\prime}} +  i u^{\prime} \sigma^+_{j^{\prime}} \sigma^-_{j^{\prime}}      \big]  \bigg]          \bigg\}  \\ 
       \times     \bigg[       \underset{j  \text{ } \mathrm{mod}3 \equiv 0}{\underset{3 \leq j \leq N-1}{\prod}}      i u^{-1} \sigma^+_j \sigma^-_j              \bigg]  \bigg[ \underset{j \equiv N}{\prod}  \sigma^+_j \bigg] \text{, }     \end{align*}

       \noindent corresponding to the fourth term. Applying (AC), followed by (LR), to the superposition above yields,
       
       \begin{align*}
       -   \bigg[ \underset{0 \leq j < j+1 < N-1}{\prod}  \sigma^-_{j} i u \sigma^+_{j+1} \sigma^-_{j+1}      \textbf{1}_{\{j^{\prime}> j+1 \text{ } : \text{ } i u \sigma^-_{j^{\prime}} \sigma^+_{j^{\prime}} \in \mathrm{support} ( A(u))\}}  \big[ 
       \sigma^-_{j^{\prime}} +  i u \sigma^+_{j^{\prime}} \sigma^-_{j^{\prime}}      \big]  \bigg]           \bigg[       \underset{j  \text{ } \mathrm{mod}3 \equiv 0}{\underset{3 \leq j \leq N-1}{\prod}}      i u^{-1} \sigma^+_j \sigma^-_j              \bigg] \\ \times     \bigg[ \underset{0 \leq j < j+1 < N-1}{\prod}  \sigma^-_{j} i u \sigma^+_{j+1} \sigma^-_{j+1}      \textbf{1}_{\{j^{\prime}> j+1 \text{ } : \text{ } i u \sigma^-_{j^{\prime}} \sigma^+_{j^{\prime}} \in \mathrm{support} ( A(u))\}}  \big[ 
       \sigma^-_{j^{\prime}} +  i u \sigma^+_{j^{\prime}} \sigma^-_{j^{\prime}}      \big]  \bigg]    \bigg\{         \bigg[       \underset{j  \text{ } \mathrm{mod}3 \equiv 0}{\underset{3 \leq j \leq N-1}{\prod}}      i u^{-1} \sigma^+_j \sigma^-_j              \bigg]  \\ \times  \bigg[ \underset{j \equiv N}{\prod}  \sigma^+_j \bigg] , \bigg[ \underset{j \equiv N}{\prod}  \sigma^+_j \bigg]         \bigg\}   \end{align*}

       \noindent corresponding to the first term, 
       \begin{align*}
       -   \bigg[ \underset{0 \leq j < j+1 < N-1}{\prod}  \sigma^-_{j} i u \sigma^+_{j+1} \sigma^-_{j+1}      \textbf{1}_{\{j^{\prime}> j+1 \text{ } : \text{ } i u \sigma^-_{j^{\prime}} \sigma^+_{j^{\prime}} \in \mathrm{support} ( A(u))\}}  \big[ 
       \sigma^-_{j^{\prime}} +  i u \sigma^+_{j^{\prime}} \sigma^-_{j^{\prime}}      \big]  \bigg]           \bigg[       \underset{j  \text{ } \mathrm{mod}3 \equiv 0}{\underset{3 \leq j \leq N-1}{\prod}}      i u^{-1} \sigma^+_j \sigma^-_j              \bigg] \\ \times    \bigg\{   \bigg[ \underset{0 \leq j < j+1 < N-1}{\prod}  \sigma^-_{j} i u \sigma^+_{j+1} \sigma^-_{j+1}      \textbf{1}_{\{j^{\prime}> j+1 \text{ } : \text{ } i u \sigma^-_{j^{\prime}} \sigma^+_{j^{\prime}} \in \mathrm{support} ( A(u))\}}  \big[ 
       \sigma^-_{j^{\prime}} +  i u \sigma^+_{j^{\prime}} \sigma^-_{j^{\prime}}      \big]  \bigg]         , \bigg[ \underset{j \equiv N}{\prod}  \sigma^+_j \bigg]         \bigg\} \end{align*}

       \begin{align*}
       \times \bigg[       \underset{j  \text{ } \mathrm{mod}3 \equiv 0}{\underset{3 \leq j \leq N-1}{\prod}}      i u^{-1} \sigma^+_j \sigma^-_j              \bigg]   \bigg[ \underset{j \equiv N}{\prod}  \sigma^+_j \bigg]   \text{, }   \end{align*}
       
       \noindent corresponding to the second term,
       
       \begin{align*}
       -     \bigg[ \underset{0 \leq j < j+1 < N-1}{\prod}  \sigma^-_{j} i u \sigma^+_{j+1} \sigma^-_{j+1}      \textbf{1}_{\{j^{\prime}> j+1 \text{ } : \text{ } i u \sigma^-_{j^{\prime}} \sigma^+_{j^{\prime}} \in \mathrm{support} ( A(u))\}}  \big[ 
       \sigma^-_{j^{\prime}} +  i u \sigma^+_{j^{\prime}} \sigma^-_{j^{\prime}}      \big]  \bigg]    \bigg[ \underset{0 \leq j < j+1 < N-1}{\prod}  \sigma^-_{j} i u \sigma^+_{j+1} \sigma^-_{j+1}   \\ \times    \textbf{1}_{\{j^{\prime}> j+1 \text{ } : \text{ } i u \sigma^-_{j^{\prime}} \sigma^+_{j^{\prime}} \in \mathrm{support} ( A(u))\}}  \big[ 
       \sigma^-_{j^{\prime}} +  i u \sigma^+_{j^{\prime}} \sigma^-_{j^{\prime}}      \big]  \bigg]     \bigg\{       \bigg[       \underset{j  \text{ } \mathrm{mod}3 \equiv 0}{\underset{3 \leq j \leq N-1}{\prod}}      i u^{-1} \sigma^+_j \sigma^-_j              \bigg]  \\ \times   \bigg[ \underset{j \equiv N}{\prod}  \sigma^+_j \bigg]  ,  \bigg[       \underset{j  \text{ } \mathrm{mod}3 \equiv 0}{\underset{3 \leq j \leq N-1}{\prod}}      i \big( u^{\prime} \big)^{-1} \sigma^+_j \sigma^-_j              \bigg]        \bigg\}  \bigg[ \underset{j \equiv N}{\prod}  \sigma^+_j \bigg]  \text{, }   \end{align*}
       
       \noindent corresponding to the third term,
       
       \begin{align*}    -     \bigg[ \underset{0 \leq j < j+1 < N-1}{\prod}  \sigma^-_{j} i u \sigma^+_{j+1} \sigma^-_{j+1}      \textbf{1}_{\{j^{\prime}> j+1 \text{ } : \text{ } i u \sigma^-_{j^{\prime}} \sigma^+_{j^{\prime}} \in \mathrm{support} ( A(u))\}}  \big[ 
       \sigma^-_{j^{\prime}} +  i u \sigma^+_{j^{\prime}} \sigma^-_{j^{\prime}}      \big]  \bigg]      \bigg\{      \bigg[ \underset{0 \leq j < j+1 < N-1}{\prod}  \sigma^-_{j} i u \sigma^+_{j+1} \sigma^-_{j+1}  \\   \times    \textbf{1}_{\{j^{\prime}> j+1 \text{ } : \text{ } i u \sigma^-_{j^{\prime}} \sigma^+_{j^{\prime}} \in \mathrm{support} ( A(u))\}}  \big[ 
       \sigma^-_{j^{\prime}} +  i u \sigma^+_{j^{\prime}} \sigma^-_{j^{\prime}}      \big]  \bigg] \\   ,  \bigg[       \underset{j  \text{ } \mathrm{mod}3 \equiv 0}{\underset{3 \leq j \leq N-1}{\prod}}      i u^{-1} \sigma^+_j \sigma^-_j              \bigg]        \bigg\}   \bigg[       \underset{j  \text{ } \mathrm{mod}3 \equiv 0}{\underset{3 \leq j \leq N-1}{\prod}}      i \big( u^{\prime} \big)^{-1} \sigma^+_j \sigma^-_j              \bigg]    \bigg[ \underset{j \equiv N}{\prod}  \sigma^+_j \bigg]^2  \text{, }   \end{align*}
       
       \noindent corresponding to the fourth term,
       
       \begin{align*}   
                     - \bigg[       \underset{j  \text{ } \mathrm{mod}3 \equiv 0}{\underset{3 \leq j \leq N-1}{\prod}}      i u^{-1} \sigma^+_j \sigma^-_j              \bigg]  \bigg[ \underset{j \equiv N}{\prod}  \sigma^+_j \bigg]       \bigg[ \underset{0 \leq j < j+1 < N-1}{\prod}  \sigma^-_{j} i u \sigma^+_{j+1} \sigma^-_{j+1}      \textbf{1}_{\{j^{\prime}> j+1 \text{ } : \text{ } i u \sigma^-_{j^{\prime}} \sigma^+_{j^{\prime}} \in \mathrm{support} ( A(u))\}}  \\  \times \big[ 
       \sigma^-_{j^{\prime}} +  i u \sigma^+_{j^{\prime}} \sigma^-_{j^{\prime}}      \big]  \bigg]       \\     \times  \bigg[       \underset{j  \text{ } \mathrm{mod}3 \equiv 0}{\underset{3 \leq j \leq N-1}{\prod}}      i u^{-1} \sigma^+_j \sigma^-_j              \bigg]    \bigg\{       \bigg[ \underset{j \equiv N}{\prod}  \sigma^+_j \bigg]   ,   \bigg[ \underset{0 \leq j < j+1 < N-1}{\prod}  \sigma^-_{j} i u^{\prime} \sigma^+_{j+1} \sigma^-_{j+1}      \textbf{1}_{\{j^{\prime}> j+1 \text{ } : \text{ } i u^{\prime} \sigma^-_{j^{\prime}} \sigma^+_{j^{\prime}} \in \mathrm{support} ( A(u^{\prime}))\}}  \\  \times  \big[ 
       \sigma^-_{j^{\prime}} +  i u^{\prime} \sigma^+_{j^{\prime}} \sigma^-_{j^{\prime}}      \big]  \bigg]          \bigg\}     \text{, }   \end{align*}
       
       \noindent corresponding to the fifth term,
       
       \begin{align*}    - \bigg[       \underset{j  \text{ } \mathrm{mod}3 \equiv 0}{\underset{3 \leq j \leq N-1}{\prod}}      i u^{-1} \sigma^+_j \sigma^-_j              \bigg]  \bigg[ \underset{j \equiv N}{\prod}  \sigma^+_j \bigg]       \bigg[ \underset{0 \leq j < j+1 < N-1}{\prod}  \sigma^-_{j} i u \sigma^+_{j+1} \sigma^-_{j+1}      \textbf{1}_{\{j^{\prime}> j+1 \text{ } : \text{ } i u \sigma^-_{j^{\prime}} \sigma^+_{j^{\prime}} \in \mathrm{support} ( A(u))\}}  \\ \times \big[ 
       \sigma^-_{j^{\prime}} +  i u \sigma^+_{j^{\prime}} \sigma^-_{j^{\prime}}      \big]  \bigg]      \end{align*}

              \begin{align*}       \times   \bigg\{       \bigg[       \underset{j  \text{ } \mathrm{mod}3 \equiv 0}{\underset{3 \leq j \leq N-1}{\prod}}      i u^{-1} \sigma^+_j \sigma^-_j              \bigg]   ,   \bigg[ \underset{0 \leq j < j+1 < N-1}{\prod}  \sigma^-_{j} i u^{\prime} \sigma^+_{j+1} \sigma^-_{j+1}      \textbf{1}_{\{j^{\prime}> j+1 \text{ } : \text{ } i u^{\prime} \sigma^-_{j^{\prime}} \sigma^+_{j^{\prime}} \in \mathrm{support} ( A(u^{\prime}))\}}       \\   \times  \big[ 
       \sigma^-_{j^{\prime}} +  i u^{\prime} \sigma^+_{j^{\prime}} \sigma^-_{j^{\prime}}      \big]  \bigg]          \bigg\}  \bigg[ \underset{j \equiv N}{\prod}  \sigma^+_j \bigg] \text{, }   \end{align*}
       
       \noindent corresponding to the sixth term,
       
       \begin{align*}    - \bigg[       \underset{j  \text{ } \mathrm{mod}3 \equiv 0}{\underset{3 \leq j \leq N-1}{\prod}}      i u^{-1} \sigma^+_j \sigma^-_j              \bigg]  \bigg[ \underset{j \equiv N}{\prod}  \sigma^+_j \bigg] \bigg\{         \bigg[ \underset{0 \leq j < j+1 < N-1}{\prod}  \sigma^-_{j} i u \sigma^+_{j+1} \sigma^-_{j+1}      \textbf{1}_{\{j^{\prime}> j+1 \text{ } : \text{ } i u \sigma^-_{j^{\prime}} \sigma^+_{j^{\prime}} \in \mathrm{support} ( A(u))\}}  \\ \times \big[ 
       \sigma^-_{j^{\prime}} +  i u \sigma^+_{j^{\prime}} \sigma^-_{j^{\prime}}      \big]  \bigg]   ,   \bigg[ \underset{0 \leq j < j+1 < N-1}{\prod}  \sigma^-_{j} i u^{\prime} \sigma^+_{j+1} \sigma^-_{j+1}      \textbf{1}_{\{j^{\prime}> j+1 \text{ } : \text{ } i u^{\prime} \sigma^-_{j^{\prime}} \sigma^+_{j^{\prime}} \in \mathrm{support} ( A(u^{\prime}))\}}     \big[ 
       \sigma^-_{j^{\prime}} +  i u^{\prime} \sigma^+_{j^{\prime}} \sigma^-_{j^{\prime}}      \big]  \bigg]          \bigg\}  \\ 
       \times     \bigg[       \underset{j  \text{ } \mathrm{mod}3 \equiv 0}{\underset{3 \leq j \leq N-1}{\prod}}      i u^{-1} \sigma^+_j \sigma^-_j              \bigg]  \bigg[ \underset{j \equiv N}{\prod}  \sigma^+_j \bigg]  \text{, }   \end{align*}
       
       \noindent corresponding to the seventh term. An application of (LR) yields, 
       
       \begin{align*}             -   \bigg[ \underset{0 \leq j < j+1 < N-1}{\prod}  \sigma^-_{j} i u \sigma^+_{j+1} \sigma^-_{j+1}      \textbf{1}_{\{j^{\prime}> j+1 \text{ } : \text{ } i u \sigma^-_{j^{\prime}} \sigma^+_{j^{\prime}} \in \mathrm{support} ( A(u))\}}  \big[ 
       \sigma^-_{j^{\prime}} +  i u \sigma^+_{j^{\prime}} \sigma^-_{j^{\prime}}      \big]  \bigg]           \bigg[       \underset{j  \text{ } \mathrm{mod}3 \equiv 0}{\underset{3 \leq j \leq N-1}{\prod}}      i u^{-1} \sigma^+_j \sigma^-_j              \bigg] \\ \times     \bigg[ \underset{0 \leq j < j+1 < N-1}{\prod}  \sigma^-_{j} i u \sigma^+_{j+1} \sigma^-_{j+1}      \textbf{1}_{\{j^{\prime}> j+1 \text{ } : \text{ } i u \sigma^-_{j^{\prime}} \sigma^+_{j^{\prime}} \in \mathrm{support} ( A(u))\}}  \big[ 
       \sigma^-_{j^{\prime}} +  i u \sigma^+_{j^{\prime}} \sigma^-_{j^{\prime}}      \big]  \bigg]           \bigg[       \underset{j  \text{ } \mathrm{mod}3 \equiv 0}{\underset{3 \leq j \leq N-1}{\prod}}      i u^{-1} \sigma^+_j \sigma^-_j              \bigg]  \\ \times  \bigg\{   \bigg[ \underset{j \equiv N}{\prod}  \sigma^+_j \bigg] , \bigg[ \underset{j \equiv N}{\prod}  \sigma^+_j \bigg]         \bigg\} \end{align*}
       
       \noindent corresponding to the first term,

       \begin{align*} -   \bigg[ \underset{0 \leq j < j+1 < N-1}{\prod}  \sigma^-_{j} i u \sigma^+_{j+1} \sigma^-_{j+1}      \textbf{1}_{\{j^{\prime}> j+1 \text{ } : \text{ } i u \sigma^-_{j^{\prime}} \sigma^+_{j^{\prime}} \in \mathrm{support} ( A(u))\}}  \big[ 
       \sigma^-_{j^{\prime}} +  i u \sigma^+_{j^{\prime}} \sigma^-_{j^{\prime}}      \big]  \bigg]           \bigg[       \underset{j  \text{ } \mathrm{mod}3 \equiv 0}{\underset{3 \leq j \leq N-1}{\prod}}      i u^{-1} \sigma^+_j \sigma^-_j              \bigg] \\   \times     \bigg[ \underset{0 \leq j < j+1 < N-1}{\prod}  \sigma^-_{j} i u \sigma^+_{j+1} \sigma^-_{j+1}      \textbf{1}_{\{j^{\prime}> j+1 \text{ } : \text{ } i u \sigma^-_{j^{\prime}} \sigma^+_{j^{\prime}} \in \mathrm{support} ( A(u))\}}  \big[ 
       \sigma^-_{j^{\prime}} +  i u \sigma^+_{j^{\prime}} \sigma^-_{j^{\prime}}      \big]  \bigg]    \bigg\{         \bigg[       \underset{j  \text{ } \mathrm{mod}3 \equiv 0}{\underset{3 \leq j \leq N-1}{\prod}}      i u^{-1} \sigma^+_j \sigma^-_j              \bigg]    \\   , \bigg[ \underset{j \equiv N}{\prod}  \sigma^+_j \bigg]         \bigg\}  \bigg[ \underset{j \equiv N}{\prod}  \sigma^+_j \bigg]   \text{, }  \end{align*}

       \noindent corresponding to the second term,

        \begin{align*}      -   \bigg[ \underset{0 \leq j < j+1 < N-1}{\prod}  \sigma^-_{j} i u \sigma^+_{j+1} \sigma^-_{j+1}      \textbf{1}_{\{j^{\prime}> j+1 \text{ } : \text{ } i u \sigma^-_{j^{\prime}} \sigma^+_{j^{\prime}} \in \mathrm{support} ( A(u))\}}  \big[ 
       \sigma^-_{j^{\prime}} +  i u \sigma^+_{j^{\prime}} \sigma^-_{j^{\prime}}      \big]  \bigg]           \bigg[       \underset{j  \text{ } \mathrm{mod}3 \equiv 0}{\underset{3 \leq j \leq N-1}{\prod}}      i u^{-1} \sigma^+_j \sigma^-_j              \bigg] \\ \times    \bigg\{   \bigg[ \underset{0 \leq j < j+1 < N-1}{\prod}  \sigma^-_{j} i u \sigma^+_{j+1} \sigma^-_{j+1}      \textbf{1}_{\{j^{\prime}> j+1 \text{ } : \text{ } i u \sigma^-_{j^{\prime}} \sigma^+_{j^{\prime}} \in \mathrm{support} ( A(u))\}}  \big[ 
       \sigma^-_{j^{\prime}} +  i u \sigma^+_{j^{\prime}} \sigma^-_{j^{\prime}}      \big]  \bigg]         , \bigg[ \underset{j \equiv N}{\prod}  \sigma^+_j \bigg]         \bigg\}      \\  \times \bigg[       \underset{j  \text{ } \mathrm{mod}3 \equiv 0}{\underset{3 \leq j \leq N-1}{\prod}}      i u^{-1} \sigma^+_j \sigma^-_j              \bigg]   \bigg[ \underset{j \equiv N}{\prod}  \sigma^+_j \bigg]  \text{, }  \end{align*}

       \noindent corresponding to the third term,

        \begin{align*}     -     \bigg[ \underset{0 \leq j < j+1 < N-1}{\prod}  \sigma^-_{j} i u \sigma^+_{j+1} \sigma^-_{j+1}      \textbf{1}_{\{j^{\prime}> j+1 \text{ } : \text{ } i u \sigma^-_{j^{\prime}} \sigma^+_{j^{\prime}} \in \mathrm{support} ( A(u))\}}  \big[ 
       \sigma^-_{j^{\prime}} +  i u \sigma^+_{j^{\prime}} \sigma^-_{j^{\prime}}      \big]  \bigg]    \bigg[ \underset{0 \leq j < j+1 < N-1}{\prod}  \sigma^-_{j} i u \sigma^+_{j+1} \sigma^-_{j+1}  \\  \times    \textbf{1}_{\{j^{\prime}> j+1 \text{ } : \text{ } i u \sigma^-_{j^{\prime}} \sigma^+_{j^{\prime}} \in \mathrm{support} ( A(u))\}}  \big[ 
       \sigma^-_{j^{\prime}} +  i u \sigma^+_{j^{\prime}} \sigma^-_{j^{\prime}}      \big]  \bigg]    \bigg[       \underset{j  \text{ } \mathrm{mod}3 \equiv 0}{\underset{3 \leq j \leq N-1}{\prod}}      i u^{-1} \sigma^+_j \sigma^-_j              \bigg]         \end{align*}

       \begin{align*}
       \times  \bigg\{   \bigg[ \underset{j \equiv N}{\prod}  \sigma^+_j \bigg]  ,  \bigg[       \underset{j  \text{ } \mathrm{mod}3 \equiv 0}{\underset{3 \leq j \leq N-1}{\prod}}      i \big( u^{\prime}\big)^{-1} \sigma^+_j \sigma^-_j              \bigg]        \bigg\}  \bigg[ \underset{j \equiv N}{\prod}  \sigma^+_j \bigg]\\ -     \bigg[ \underset{0 \leq j < j+1 < N-1}{\prod}  \sigma^-_{j} i u \sigma^+_{j+1} \sigma^-_{j+1}      \textbf{1}_{\{j^{\prime}> j+1 \text{ } : \text{ } i u \sigma^-_{j^{\prime}} \sigma^+_{j^{\prime}} \in \mathrm{support} ( A(u))\}}  \big[ 
       \sigma^-_{j^{\prime}} +  i u \sigma^+_{j^{\prime}} \sigma^-_{j^{\prime}}      \big]  \bigg]    \bigg[ \underset{0 \leq j < j+1 < N-1}{\prod}  \sigma^-_{j} i u \sigma^+_{j+1} \sigma^-_{j+1}   \\  \times    \textbf{1}_{\{j^{\prime}> j+1 \text{ } : \text{ } i u \sigma^-_{j^{\prime}} \sigma^+_{j^{\prime}} \in \mathrm{support} ( A(u))\}}  \big[ 
       \sigma^-_{j^{\prime}} +  i u \sigma^+_{j^{\prime}} \sigma^-_{j^{\prime}}      \big]  \bigg]     \bigg\{       \bigg[       \underset{j  \text{ } \mathrm{mod}3 \equiv 0}{\underset{3 \leq j \leq N-1}{\prod}}      i u^{-1} \sigma^+_j \sigma^-_j              \bigg]   \\  ,  \bigg[       \underset{j  \text{ } \mathrm{mod}3 \equiv 0}{\underset{3 \leq j \leq N-1}{\prod}}      i \big( u^{\prime} \big)^{-1} \sigma^+_j \sigma^-_j              \bigg]        \bigg\}  
   \bigg[ \underset{j \equiv N}{\prod}  \sigma^+_j \bigg]^2   \text{, }  \end{align*}

       \noindent corresponding to the fourth term,

        \begin{align*}     -     \bigg[ \underset{0 \leq j < j+1 < N-1}{\prod}  \sigma^-_{j} i u \sigma^+_{j+1} \sigma^-_{j+1}      \textbf{1}_{\{j^{\prime}> j+1 \text{ } : \text{ } i u \sigma^-_{j^{\prime}} \sigma^+_{j^{\prime}} \in \mathrm{support} ( A(u))\}}  \big[ 
       \sigma^-_{j^{\prime}} +  i u \sigma^+_{j^{\prime}} \sigma^-_{j^{\prime}}      \big]  \bigg]      \bigg\{      \bigg[ \underset{0 \leq j < j+1 < N-1}{\prod}  \sigma^-_{j} i u \sigma^+_{j+1} \sigma^-_{j+1}   \\ \times    \textbf{1}_{\{j^{\prime}> j+1 \text{ } : \text{ } i u \sigma^-_{j^{\prime}} \sigma^+_{j^{\prime}} \in \mathrm{support} ( A(u))\}}  \big[ 
       \sigma^-_{j^{\prime}} +  i u \sigma^+_{j^{\prime}} \sigma^-_{j^{\prime}}      \big]  \bigg] \\   ,  \bigg[       \underset{j  \text{ } \mathrm{mod}3 \equiv 0}{\underset{3 \leq j \leq N-1}{\prod}}      i \big( u^{\prime} \big)^{-1} \sigma^+_j \sigma^-_j              \bigg]        \bigg\}   \bigg[       \underset{j  \text{ } \mathrm{mod}3 \equiv 0}{\underset{3 \leq j \leq N-1}{\prod}}      i u^{-1} \sigma^+_j \sigma^-_j              \bigg]    \bigg[ \underset{j \equiv N}{\prod}  \sigma^+_j \bigg]^2          \text{, }       \end{align*}
       
       \noindent corresponding to the fifth term,
       
       \begin{align*} - \bigg[       \underset{j  \text{ } \mathrm{mod}3 \equiv 0}{\underset{3 \leq j \leq N-1}{\prod}}      i u^{-1} \sigma^+_j \sigma^-_j              \bigg]  \bigg[ \underset{j \equiv N}{\prod}  \sigma^+_j \bigg]       \bigg[ \underset{0 \leq j < j+1 < N-1}{\prod}  \sigma^-_{j} i u \sigma^+_{j+1} \sigma^-_{j+1}      \textbf{1}_{\{j^{\prime}> j+1 \text{ } : \text{ } i u \sigma^-_{j^{\prime}} \sigma^+_{j^{\prime}} \in \mathrm{support} ( A(u))\}}  \\ \times \big[ 
       \sigma^-_{j^{\prime}} +  i u \sigma^+_{j^{\prime}} \sigma^-_{j^{\prime}}      \big]  \bigg]       \\     \times  \bigg[       \underset{j  \text{ } \mathrm{mod}3 \equiv 0}{\underset{3 \leq j \leq N-1}{\prod}}      i u^{-1} \sigma^+_j \sigma^-_j              \bigg]    \bigg\{       \bigg[ \underset{j \equiv N}{\prod}  \sigma^+_j \bigg]   ,   \bigg[ \underset{0 \leq j < j+1 < N-1}{\prod}  \sigma^-_{j} i u^{\prime} \sigma^+_{j+1} \sigma^-_{j+1}      \textbf{1}_{\{j^{\prime}> j+1 \text{ } : \text{ } i u^{\prime} \sigma^-_{j^{\prime}} \sigma^+_{j^{\prime}} \in \mathrm{support} ( A(u^{\prime}))\}} \\ \times  \big[ 
       \sigma^-_{j^{\prime}} +  i u^{\prime} \sigma^+_{j^{\prime}} \sigma^-_{j^{\prime}}      \big]  \bigg]          \bigg\} \text{, }  \end{align*}

\noindent corresponding to the sixth term,

       \begin{align*}
       - \bigg[       \underset{j  \text{ } \mathrm{mod}3 \equiv 0}{\underset{3 \leq j \leq N-1}{\prod}}      i u^{-1} \sigma^+_j \sigma^-_j              \bigg]  \bigg[ \underset{j \equiv N}{\prod}  \sigma^+_j \bigg]       \bigg[ \underset{0 \leq j < j+1 < N-1}{\prod}  \sigma^-_{j} i u \sigma^+_{j+1} \sigma^-_{j+1}      \textbf{1}_{\{j^{\prime}> j+1 \text{ } : \text{ } i u \sigma^-_{j^{\prime}} \sigma^+_{j^{\prime}} \in \mathrm{support} ( A(u))\}}   \\ \times \big[ 
       \sigma^-_{j^{\prime}} +  i u \sigma^+_{j^{\prime}} \sigma^-_{j^{\prime}}      \big]  \bigg]   \\ \times   \bigg\{       \bigg[       \underset{j  \text{ } \mathrm{mod}3 \equiv 0}{\underset{3 \leq j \leq N-1}{\prod}}      i u^{-1} \sigma^+_j \sigma^-_j              \bigg]   ,   \bigg[ \underset{0 \leq j < j+1 < N-1}{\prod}  \sigma^-_{j} i u^{\prime} \sigma^+_{j+1} \sigma^-_{j+1}      \textbf{1}_{\{j^{\prime}> j+1 \text{ } : \text{ } i u^{\prime} \sigma^-_{j^{\prime}} \sigma^+_{j^{\prime}} \in \mathrm{support} ( A(u^{\prime}))\}}      \big[ 
       \sigma^-_{j^{\prime}} +  i u^{\prime} \sigma^+_{j^{\prime}} \sigma^-_{j^{\prime}}      \big]  \bigg]          \bigg\}  \bigg[ \underset{j \equiv N}{\prod}  \sigma^+_j \bigg] \text{, }\end{align*}

\noindent corresponding to the seventh term,

       \begin{align*}
       - \bigg[       \underset{j  \text{ } \mathrm{mod}3 \equiv 0}{\underset{3 \leq j \leq N-1}{\prod}}      i u^{-1} \sigma^+_j \sigma^-_j              \bigg]  \bigg[ \underset{j \equiv N}{\prod}  \sigma^+_j \bigg] \bigg\{         \bigg[ \underset{0 \leq j < j+1 < N-1}{\prod}  \sigma^-_{j} i u \sigma^+_{j+1} \sigma^-_{j+1}      \textbf{1}_{\{j^{\prime}> j+1 \text{ } : \text{ } i u \sigma^-_{j^{\prime}} \sigma^+_{j^{\prime}} \in \mathrm{support} ( A(u))\}} \end{align*}

       \begin{align*}
       \times \big[ 
       \sigma^-_{j^{\prime}} +  i u \sigma^+_{j^{\prime}} \sigma^-_{j^{\prime}}      \big]  \bigg]   ,   \bigg[ \underset{0 \leq j < j+1 < N-1}{\prod}  \sigma^-_{j} i u^{\prime} \sigma^+_{j+1} \sigma^-_{j+1}      \textbf{1}_{\{j^{\prime}> j+1 \text{ } : \text{ } i u^{\prime} \sigma^-_{j^{\prime}} \sigma^+_{j^{\prime}} \in \mathrm{support} ( A(u^{\prime}))\}}     \big[ 
       \sigma^-_{j^{\prime}} +  i u^{\prime} \sigma^+_{j^{\prime}} \sigma^-_{j^{\prime}}      \big]  \bigg]          \bigg\}   \\    
       \times     \bigg[       \underset{j  \text{ } \mathrm{mod}3 \equiv 0}{\underset{3 \leq j \leq N-1}{\prod}}      i u^{-1} \sigma^+_j \sigma^-_j              \bigg]  \bigg[ \underset{j \equiv N}{\prod}  \sigma^+_j \bigg]        \text{. } \end{align*}

\noindent The next set of terms include,

       \begin{align*}
       \bigg\{  \bigg[ \underset{0 \leq j < j+1 < N-1}{\prod}  \sigma^-_{j} i u \sigma^+_{j+1} \sigma^-_{j+1}      \textbf{1}_{\{j^{\prime}> j+1 \text{ } : \text{ } i u \sigma^-_{j^{\prime}} \sigma^+_{j^{\prime}} \in \mathrm{support} ( A(u))\}}  \big[ 
       \sigma^-_{j^{\prime}} +  i u \sigma^+_{j^{\prime}} \sigma^-_{j^{\prime}}      \big]  \bigg]         \bigg[       \underset{j  \text{ } \mathrm{mod}3 \equiv 0}{\underset{3 \leq j \leq N-1}{\prod}}      i u^{-1} \sigma^+_j \sigma^-_j              \bigg]  \bigg[ \underset{j \equiv N}{\prod}  \sigma^+_j \bigg]     \\  ,  \bigg[ \underset{0 \leq j < j+1 < N-1}{\prod}     \sigma^-_j \sigma^+_j \textbf{1}_{\{ j^{\prime} > j+1 \text{ } : \text{ } i u^{\prime} \sigma^+_{j^{\prime}} \sigma^-_{j^{\prime}} \in \mathrm{support} ( A ( u^{\prime} )) \}} \big[ i u^{\prime} \sigma^+_{j^{\prime}} \sigma^-_{j^{\prime}} i u^{\prime}       \sigma^+_{j^{\prime}+1} \sigma^-_{j^{\prime}+1} + \sigma^-_{j^{\prime}} \sigma^+_{j^{\prime}+1}        \big]    \bigg]  \\ \times        \bigg[ \underset{N-1 \leq j \leq N}{\prod} i u^{\prime} \sigma^+_j \sigma^-_j  \bigg]    \bigg\}  \\ \\ \overset{(\mathrm{LR})}{=}   \bigg[ \underset{0 \leq j < j+1 < N-1}{\prod}  \sigma^-_{j} i u \sigma^+_{j+1} \sigma^-_{j+1}      \textbf{1}_{\{j^{\prime}> j+1 \text{ } : \text{ } i u \sigma^-_{j^{\prime}} \sigma^+_{j^{\prime}} \in \mathrm{support} ( A(u))\}}  \big[ 
       \sigma^-_{j^{\prime}} +  i u \sigma^+_{j^{\prime}} \sigma^-_{j^{\prime}}      \big]  \bigg]      \bigg\{      \bigg[       \underset{j  \text{ } \mathrm{mod}3 \equiv 0}{\underset{3 \leq j \leq N-1}{\prod}}      i u^{-1} \sigma^+_j \sigma^-_j              \bigg]  \bigg[ \underset{j \equiv N}{\prod}  \sigma^+_j \bigg]     \\  ,  \bigg[ \underset{0 \leq j < j+1 < N-1}{\prod}     \sigma^-_j \sigma^+_j \textbf{1}_{\{ j^{\prime} > j+1 \text{ } : \text{ } i u^{\prime} \sigma^+_{j^{\prime}} \sigma^-_{j^{\prime}} \in \mathrm{support} ( A ( u^{\prime} )) \}} \big[ i u^{\prime} \sigma^+_{j^{\prime}} \sigma^-_{j^{\prime}} i u^{\prime}       \sigma^+_{j^{\prime}+1} \sigma^-_{j^{\prime}+1} + \sigma^-_{j^{\prime}} \sigma^+_{j^{\prime}+1}        \big]    \bigg]  \\ \times        \bigg[ \underset{N-1 \leq j \leq N}{\prod} i u^{\prime} \sigma^+_j \sigma^-_j  \bigg]    \bigg\} \\ +  \bigg\{  \bigg[ \underset{0 \leq j < j+1 < N-1}{\prod}  \sigma^-_{j} i u \sigma^+_{j+1} \sigma^-_{j+1}      \textbf{1}_{\{j^{\prime}> j+1 \text{ } : \text{ } i u \sigma^-_{j^{\prime}} \sigma^+_{j^{\prime}} \in \mathrm{support} ( A(u))\}}  \big[ 
       \sigma^-_{j^{\prime}} +  i u \sigma^+_{j^{\prime}} \sigma^-_{j^{\prime}}      \big]  \bigg]           ,  \bigg[ \underset{0 \leq j < j+1 < N-1}{\prod}     \sigma^-_j \sigma^+_j \\ \times  \textbf{1}_{\{ j^{\prime} > j+1 \text{ } : \text{ } i u^{\prime} \sigma^+_{j^{\prime}} \sigma^-_{j^{\prime}} \in \mathrm{support} ( A ( u^{\prime} )) \}}  \big[ i u^{\prime} \sigma^+_{j^{\prime}} \sigma^-_{j^{\prime}} i u^{\prime}       \sigma^+_{j^{\prime}+1} \sigma^-_{j^{\prime}+1} + \sigma^-_{j^{\prime}} \sigma^+_{j^{\prime}+1}        \big]    \bigg]         \bigg[ \underset{N-1 \leq j \leq N}{\prod} i u^{\prime} \sigma^+_j \sigma^-_j  \bigg]    \bigg\}   \\ \times       \bigg[       \underset{j  \text{ } \mathrm{mod}3 \equiv 0}{\underset{3 \leq j \leq N-1}{\prod}}      i u^{-1} \sigma^+_j \sigma^-_j              \bigg]  \bigg[ \underset{j \equiv N}{\prod}  \sigma^+_j \bigg] \end{align*}
       
       \noindent An application of (AC), followed by (LR), yields,

       \begin{align*}
       \bigg[ \underset{0 \leq j < j+1 < N-1}{\prod}  \sigma^-_{j} i u \sigma^+_{j+1} \sigma^-_{j+1}      \textbf{1}_{\{j^{\prime}> j+1 \text{ } : \text{ } i u \sigma^-_{j^{\prime}} \sigma^+_{j^{\prime}} \in \mathrm{support} ( A(u))\}}  \big[ 
       \sigma^-_{j^{\prime}} +  i u \sigma^+_{j^{\prime}} \sigma^-_{j^{\prime}}      \big]  \bigg]   \\ \times   \bigg[       \underset{j  \text{ } \mathrm{mod}3 \equiv 0}{\underset{3 \leq j \leq N-1}{\prod}}      i u^{-1} \sigma^+_j \sigma^-_j              \bigg]  \bigg\{       \bigg[ \underset{j \equiv N}{\prod}  \sigma^+_j \bigg]    ,  \bigg[ \underset{0 \leq j < j+1 < N-1}{\prod}     \sigma^-_j \sigma^+_j \textbf{1}_{\{ j^{\prime} > j+1 \text{ } : \text{ } i u^{\prime} \sigma^+_{j^{\prime}} \sigma^-_{j^{\prime}} \in \mathrm{support} ( A ( u^{\prime} )) \}} \big[ i u^{\prime} \sigma^+_{j^{\prime}} \sigma^-_{j^{\prime}} \\ \times  i u^{\prime}       \sigma^+_{j^{\prime}+1} \sigma^-_{j^{\prime}+1}  + \sigma^-_{j^{\prime}} \sigma^+_{j^{\prime}+1}        \big]    \bigg]          \bigg[ \underset{N-1 \leq j \leq N}{\prod} i u^{\prime} \sigma^+_j \sigma^-_j  \bigg]    \bigg\}  \end{align*}

       \noindent corresponding to the first term,
       
       \begin{align*}
        \bigg[ \underset{0 \leq j < j+1 < N-1}{\prod}  \sigma^-_{j} i u \sigma^+_{j+1} \sigma^-_{j+1}      \textbf{1}_{\{j^{\prime}> j+1 \text{ } : \text{ } i u \sigma^-_{j^{\prime}} \sigma^+_{j^{\prime}} \in \mathrm{support} ( A(u))\}}  \big[ 
       \sigma^-_{j^{\prime}} +  i u \sigma^+_{j^{\prime}} \sigma^-_{j^{\prime}}      \big]  \bigg]  \\  \times    \bigg\{      \bigg[       \underset{j  \text{ } \mathrm{mod}3 \equiv 0}{\underset{3 \leq j \leq N-1}{\prod}}      i u^{-1} \sigma^+_j \sigma^-_j              \bigg]   ,  \bigg[ \underset{0 \leq j < j+1 < N-1}{\prod}     \sigma^-_j \sigma^+_j \textbf{1}_{\{ j^{\prime} > j+1 \text{ } : \text{ } i u^{\prime} \sigma^+_{j^{\prime}} \sigma^-_{j^{\prime}} \in \mathrm{support} ( A ( u^{\prime} )) \}} \big[ i u^{\prime} \sigma^+_{j^{\prime}} \sigma^-_{j^{\prime}}   \end{align*}

       \begin{align*}
       \times  i u^{\prime}       \sigma^+_{j^{\prime}+1} \sigma^-_{j^{\prime}+1}  + \sigma^-_{j^{\prime}} \sigma^+_{j^{\prime}+1}        \big]    \bigg]          \bigg[ \underset{N-1 \leq j \leq N}{\prod} i u^{\prime} \sigma^+_j \sigma^-_j  \bigg] \bigg\} \bigg[ \underset{j \equiv N}{\prod}  \sigma^+_j \bigg] \text{, }  \end{align*} 
       
       \noindent corresponding to the second term,
       
       \begin{align*} -   \bigg[ \underset{0 \leq j < j+1 < N-1}{\prod}     \sigma^-_j \sigma^+_j  \textbf{1}_{\{ j^{\prime} > j+1 \text{ } : \text{ } i u \sigma^+_{j^{\prime}} \sigma^-_{j^{\prime}} \in \mathrm{support} ( A ( u )) \}}  \big[ i u^{\prime} \sigma^+_{j^{\prime}} \sigma^-_{j^{\prime}} i u^{\prime}       \sigma^+_{j^{\prime}+1} \sigma^-_{j^{\prime}+1} + \sigma^-_{j^{\prime}} \sigma^+_{j^{\prime}+1}        \big]    \bigg]  \bigg\{        
            \bigg[ \underset{N-1 \leq j \leq N}{\prod} i u \sigma^+_j \sigma^-_j  \bigg]  \\  , \bigg[ \underset{0 \leq j < j+1 < N-1}{\prod}  \sigma^-_{j} i u^{\prime} \sigma^+_{j+1} \sigma^-_{j+1}      \textbf{1}_{\{j^{\prime}> j+1 \text{ } : \text{ } i u^{\prime} \sigma^-_{j^{\prime}} \sigma^+_{j^{\prime}} \in \mathrm{support} ( A(u^{\prime}))\}}  \big[ 
       \sigma^-_{j^{\prime}} +  i u^{\prime} \sigma^+_{j^{\prime}} \sigma^-_{j^{\prime}}      \big]  \bigg]      \bigg\}    \bigg[       \underset{j  \text{ } \mathrm{mod}3 \equiv 0}{\underset{3 \leq j \leq N-1}{\prod}}      i u^{-1} \sigma^+_j \sigma^-_j              \bigg]  \\
    \times      \bigg[ \underset{j \equiv N}{\prod}  \sigma^+_j \bigg]            \text{, }  \end{align*} 
       
       \noindent corresponding to the third term term,
       
       \begin{align*}
  -  \bigg\{        
     \bigg[ \underset{0 \leq j < j+1 < N-1}{\prod}     \sigma^-_j \sigma^+_j  \textbf{1}_{\{ j^{\prime} > j+1 \text{ } : \text{ } i u \sigma^+_{j^{\prime}} \sigma^-_{j^{\prime}} \in \mathrm{support} ( A ( u )) \}}  \big[ i u^{\prime} \sigma^+_{j^{\prime}} \sigma^-_{j^{\prime}} i u^{\prime}       \sigma^+_{j^{\prime}+1} \sigma^-_{j^{\prime}+1} + \sigma^-_{j^{\prime}} \sigma^+_{j^{\prime}+1}        \big]    \bigg]     \\ , \bigg[ \underset{0 \leq j < j+1 < N-1}{\prod}  \sigma^-_{j} i u^{\prime} \sigma^+_{j+1} \sigma^-_{j+1}      \textbf{1}_{\{j^{\prime}> j+1 \text{ } : \text{ } i u^{\prime} \sigma^-_{j^{\prime}} \sigma^+_{j^{\prime}} \in \mathrm{support} ( A(u^{\prime}))\}}  \big[ 
       \sigma^-_{j^{\prime}} +  i u^{\prime} \sigma^+_{j^{\prime}} \sigma^-_{j^{\prime}}      \big]  \bigg]      \bigg\}      \bigg[ \underset{N-1 \leq j \leq N}{\prod} i u \sigma^+_j \sigma^-_j  \bigg]  \\  \times       \bigg[       \underset{j  \text{ } \mathrm{mod}3 \equiv 0}{\underset{3 \leq j \leq N-1}{\prod}}      i u^{-1} \sigma^+_j \sigma^-_j              \bigg]  \bigg[ \underset{j \equiv N}{\prod}  \sigma^+_j \bigg]         \text{, } \end{align*}

       \noindent corresponding to the second term. The superposition above can be approximated with an application of $(\mathrm{BL})$, followed by $(\mathrm{LR})$, to
       
       \begin{align*}
        -   \bigg[ \underset{0 \leq j < j+1 < N-1}{\prod}     \sigma^-_j \sigma^+_j  \textbf{1}_{\{ j^{\prime} > j+1 \text{ } : \text{ } i u \sigma^+_{j^{\prime}} \sigma^-_{j^{\prime}} \in \mathrm{support} ( A ( u )) \}}  \big[ i u^{\prime} \sigma^+_{j^{\prime}} \sigma^-_{j^{\prime}} i u^{\prime}       \sigma^+_{j^{\prime}+1} \sigma^-_{j^{\prime}+1} + \sigma^-_{j^{\prime}} \sigma^+_{j^{\prime}+1}        \big]    \bigg]  \bigg\{        
            \bigg[ \underset{N-1 \leq j \leq N}{\prod} i u \sigma^+_j \sigma^-_j  \bigg]  \\ , \bigg[ \underset{0 \leq j < j+1 < N-1}{\prod}  \sigma^-_{j} i u^{\prime} \sigma^+_{j+1} \sigma^-_{j+1}      \textbf{1}_{\{j^{\prime}> j+1 \text{ } : \text{ } i u^{\prime} \sigma^-_{j^{\prime}} \sigma^+_{j^{\prime}} \in \mathrm{support} ( A(u^{\prime}))\}}  \big[ 
       \sigma^-_{j^{\prime}} +  i u^{\prime} \sigma^+_{j^{\prime}} \sigma^-_{j^{\prime}}      \big]  \bigg]      \bigg\}    \bigg[       \underset{j  \text{ } \mathrm{mod}3 \equiv 0}{\underset{3 \leq j \leq N-1}{\prod}}      i u^{-1} \sigma^+_j \sigma^-_j              \bigg]  \\
    \times      \bigg[ \underset{j \equiv N}{\prod}  \sigma^+_j \bigg]              \\ -  \bigg\{        
     \bigg[ \underset{0 \leq j < j+1 < N-1}{\prod}     \sigma^-_j \sigma^+_j  \textbf{1}_{\{ j^{\prime} > j+1 \text{ } : \text{ } i u \sigma^+_{j^{\prime}} \sigma^-_{j^{\prime}} \in \mathrm{support} ( A ( u )) \}}  \big[ i u^{\prime} \sigma^+_{j^{\prime}} \sigma^-_{j^{\prime}} i u^{\prime}       \sigma^+_{j^{\prime}+1} \sigma^-_{j^{\prime}+1} + \sigma^-_{j^{\prime}} \sigma^+_{j^{\prime}+1}        \big]    \bigg]     \\ , \bigg[ \underset{0 \leq j < j+1 < N-1}{\prod}  \sigma^-_{j} i u^{\prime} \sigma^+_{j+1} \sigma^-_{j+1}      \textbf{1}_{\{j^{\prime}> j+1 \text{ } : \text{ } i u^{\prime} \sigma^-_{j^{\prime}} \sigma^+_{j^{\prime}} \in \mathrm{support} ( A(u^{\prime}))\}}  \big[ 
       \sigma^-_{j^{\prime}} +  i u^{\prime} \sigma^+_{j^{\prime}} \sigma^-_{j^{\prime}}      \big]  \bigg]      \bigg\}      \bigg[ \underset{N-1 \leq j \leq N}{\prod} i u \sigma^+_j \sigma^-_j  \bigg]   \\
    \times       \bigg[       \underset{j  \text{ } \mathrm{mod}3 \equiv 0}{\underset{3 \leq j \leq N-1}{\prod}}      i u^{-1} \sigma^+_j \sigma^-_j              \bigg]  \bigg[ \underset{j \equiv N}{\prod}  \sigma^+_j \bigg]             \text{, } \end{align*}

\noindent which can be used to obtain the final approximation,

\begin{align*}
      \big( \mathscr{C}^1_1 \big)_{11}    \text{. }
\end{align*}

\noindent The remaining Poisson bracket for $\underline{\mathcal{I}^1}$ takes the form,

    \begin{align*}
\bigg\{  \bigg[ \underset{0 \leq j < j+1 < N-1}{\prod}  \sigma^-_{j} i u \sigma^+_{j+1} \sigma^-_{j+1}      \textbf{1}_{\{j^{\prime}> j+1 \text{ } : \text{ } i u \sigma^-_{j^{\prime}} \sigma^+_{j^{\prime}} \in \mathrm{support} ( A(u))\}}  \big[ 
       \sigma^-_{j^{\prime}} +  i u \sigma^+_{j^{\prime}} \sigma^-_{j^{\prime}}      \big]  \bigg]         \bigg[       \underset{j  \text{ } \mathrm{mod}3 \equiv 0}{\underset{3 \leq j \leq N-1}{\prod}}      i u^{-1} \sigma^+_j \sigma^-_j              \bigg]  \bigg[ \underset{j \equiv N}{\prod}  \sigma^+_j \bigg]    \end{align*}

       \begin{align*}    ,   \bigg[ \underset{1 \leq j \leq N-1}{\prod} \big[ \sigma^+_j + i u^{\prime} \sigma^+_j \big] \textbf{1}_{\{ j^{\prime} >j \text{ } : \text{ } i u^{\prime} \sigma^+_{j^{\prime}} \sigma^-_{j^{\prime}} \in \mathrm{support} ( A ( u^{\prime})) \}}  \big[ \sigma^+_{j^{\prime}} i u^{\prime} \sigma^+_{j^{\prime}+1} \sigma^-_{j^{\prime}+1}        + i u^{\prime}     \sigma^+_{j^{\prime}} \sigma^-_{j^{\prime}} \sigma^-_{j^{\prime}+1}        \big]    \bigg]  \bigg[ \underset{j \equiv N}{\prod} i u^{\prime} \sigma^+_j \sigma^-_j  \bigg]   \bigg\} \\ \\ 
       \overset{(\mathrm{LR})}{=}       \bigg[ \underset{0 \leq j < j+1 < N-1}{\prod}  \sigma^-_{j} i u \sigma^+_{j+1} \sigma^-_{j+1}      \textbf{1}_{\{j^{\prime}> j+1 \text{ } : \text{ } i u \sigma^-_{j^{\prime}} \sigma^+_{j^{\prime}} \in \mathrm{support} ( A(u))\}}  \big[ 
       \sigma^-_{j^{\prime}} +  i u \sigma^+_{j^{\prime}} \sigma^-_{j^{\prime}}      \big]  \bigg]          \bigg\{     \bigg[       \underset{j  \text{ } \mathrm{mod}3 \equiv 0}{\underset{3 \leq j \leq N-1}{\prod}}      i u^{-1} \sigma^+_j \sigma^-_j              \bigg]  \bigg[ \underset{j \equiv N}{\prod}  \sigma^+_j \bigg]    \\  ,   \bigg[ \underset{1 \leq j \leq N-1}{\prod} \big[ \sigma^+_j + i u^{\prime} \sigma^+_j \big] \textbf{1}_{\{ j^{\prime} >j \text{ } : \text{ } i u^{\prime} \sigma^+_{j^{\prime}} \sigma^-_{j^{\prime}} \in \mathrm{support} ( A ( u^{\prime})) \}}  \big[ \sigma^+_{j^{\prime}} i u^{\prime} \sigma^+_{j^{\prime}+1} \sigma^-_{j^{\prime}+1}        + i u^{\prime}     \sigma^+_{j^{\prime}} \sigma^-_{j^{\prime}} \sigma^-_{j^{\prime}+1}        \big]    \bigg]  \bigg[ \underset{j \equiv N}{\prod} i u^{\prime} \sigma^+_j \sigma^-_j  \bigg]   \bigg\} \\  +  \bigg\{  \bigg[ \underset{0 \leq j < j+1 < N-1}{\prod}  \sigma^-_{j} i u \sigma^+_{j+1} \sigma^-_{j+1}      \textbf{1}_{\{j^{\prime}> j+1 \text{ } : \text{ } i u \sigma^-_{j^{\prime}} \sigma^+_{j^{\prime}} \in \mathrm{support} ( A(u))\}}  \big[ 
       \sigma^-_{j^{\prime}} +  i u \sigma^+_{j^{\prime}} \sigma^-_{j^{\prime}}      \big]  \bigg]       ,   \bigg[ \underset{1 \leq j \leq N-1}{\prod} \big[ \sigma^+_j + i u^{\prime} \sigma^+_j \big] \\ \times    \textbf{1}_{\{ j^{\prime} >j \text{ } : \text{ } i u \sigma^+_{j^{\prime}} \sigma^-_{j^{\prime}} \in \mathrm{support} ( A ( u^{\prime})) \}}  \big[ \sigma^+_{j^{\prime}}  i u^{\prime} \sigma^+_{j^{\prime}+1} \sigma^-_{j^{\prime}+1}        + i u^{\prime}     \sigma^+_{j^{\prime}} \sigma^-_{j^{\prime}} \sigma^-_{j^{\prime}+1}        \big]    \bigg]  \bigg[ \underset{j \equiv N}{\prod} i u^{\prime} \sigma^+_j \sigma^-_j  \bigg]   \bigg\}      \bigg[       \underset{j  \text{ } \mathrm{mod}3 \equiv 0}{\underset{3 \leq j \leq N-1}{\prod}}      i u^{-1} \sigma^+_j \sigma^-_j              \bigg]  \bigg[ \underset{j \equiv N}{\prod}  \sigma^+_j \bigg]  \text{. } \end{align*}

       \noindent An application of (AC), followed by (LR), yields,
       
       \begin{align*}
       \bigg[ \underset{0 \leq j < j+1 < N-1}{\prod}  \sigma^-_{j} i u \sigma^+_{j+1} \sigma^-_{j+1}      \textbf{1}_{\{j^{\prime}> j+1 \text{ } : \text{ } i u \sigma^-_{j^{\prime}} \sigma^+_{j^{\prime}} \in \mathrm{support} ( A(u))\}}  \big[ 
       \sigma^-_{j^{\prime}} +  i u \sigma^+_{j^{\prime}} \sigma^-_{j^{\prime}}      \big]  \bigg]          \bigg\{     \bigg[       \underset{j  \text{ } \mathrm{mod}3 \equiv 0}{\underset{3 \leq j \leq N-1}{\prod}}      i u^{-1} \sigma^+_j \sigma^-_j              \bigg]    \\    ,   \bigg[ \underset{1 \leq j \leq N-1}{\prod} \big[ \sigma^+_j + i u^{\prime} \sigma^+_j \big] \textbf{1}_{\{ j^{\prime} >j \text{ } : \text{ } i u^{\prime} \sigma^+_{j^{\prime}} \sigma^-_{j^{\prime}} \in \mathrm{support} ( A ( u^{\prime})) \}}  \big[ \sigma^+_{j^{\prime}} i u^{\prime} \sigma^+_{j^{\prime}+1} \sigma^-_{j^{\prime}+1}        + i u^{\prime}     \sigma^+_{j^{\prime}} \sigma^-_{j^{\prime}} \sigma^-_{j^{\prime}+1}        \big]    \bigg]  \bigg[ \underset{j \equiv N}{\prod} i u^{\prime} \sigma^+_j \sigma^-_j  \bigg]   \bigg\} \bigg[ \underset{j \equiv N}{\prod}  \sigma^+_j \bigg] \text{, }  \end{align*}

       \noindent corresponding to the first term,

       \begin{align*}
           \bigg[ \underset{0 \leq j < j+1 < N-1}{\prod}  \sigma^-_{j} i u \sigma^+_{j+1} \sigma^-_{j+1}      \textbf{1}_{\{j^{\prime}> j+1 \text{ } : \text{ } i u \sigma^-_{j^{\prime}} \sigma^+_{j^{\prime}} \in \mathrm{support} ( A(u))\}}  \big[ 
       \sigma^-_{j^{\prime}} +  i u \sigma^+_{j^{\prime}} \sigma^-_{j^{\prime}}      \big]  \bigg]   \bigg[       \underset{j  \text{ } \mathrm{mod}3 \equiv 0}{\underset{3 \leq j \leq N-1}{\prod}}      i u^{-1} \sigma^+_j \sigma^-_j              \bigg]                \\  \times  \bigg\{  \bigg[ \underset{j \equiv N}{\prod}  \sigma^+_j \bigg]   ,   \bigg[ \underset{1 \leq j \leq N-1}{\prod} \big[ \sigma^+_j + i u^{\prime} \sigma^+_j \big] \textbf{1}_{\{ j^{\prime} >j \text{ } : \text{ } i u^{\prime} \sigma^+_{j^{\prime}} \sigma^-_{j^{\prime}} \in \mathrm{support} ( A ( u^{\prime})) \}}  \big[ \sigma^+_{j^{\prime}} i u^{\prime} \sigma^+_{j^{\prime}+1} \sigma^-_{j^{\prime}+1}        + i u^{\prime}     \sigma^+_{j^{\prime}} \sigma^-_{j^{\prime}} \sigma^-_{j^{\prime}+1}        \big]    \bigg] \\ \times   \bigg[ \underset{j \equiv N}{\prod} i u^{\prime} \sigma^+_j \sigma^-_j  \bigg]   \bigg\}  \text{, }  \end{align*}

       \noindent corresponding to the second term,

       \begin{align*} -  \bigg\{       \bigg[ \underset{1 \leq j \leq N-1}{\prod} \big[ \sigma^+_j + i u \sigma^+_j \big]   \textbf{1}_{\{ j^{\prime} >j \text{ } : \text{ } i u \sigma^+_{j^{\prime}} \sigma^-_{j^{\prime}} \in \mathrm{support} ( A ( u^{\prime})) \}}  \big[ \sigma^+_{j^{\prime}}  i u^{\prime} \sigma^+_{j^{\prime}+1} \sigma^-_{j^{\prime}+1}         + i u^{\prime}     \sigma^+_{j^{\prime}} \sigma^-_{j^{\prime}} \sigma^-_{j^{\prime}+1}        \big]    \bigg]  \bigg[ \underset{j \equiv N}{\prod} i u^{\prime} \sigma^+_j \sigma^-_j  \bigg]  \\ ,  \bigg[ \underset{0 \leq j < j+1 < N-1}{\prod}  \sigma^-_{j} i u^{\prime} \sigma^+_{j+1} \sigma^-_{j+1}      \textbf{1}_{\{j^{\prime}> j+1 \text{ } : \text{ } i u^{\prime} \sigma^-_{j^{\prime}} \sigma^+_{j^{\prime}} \in \mathrm{support} ( A(u^{\prime}))\}}  \big[ 
       \sigma^-_{j^{\prime}} +  i u^{\prime} \sigma^+_{j^{\prime}} \sigma^-_{j^{\prime}}      \big]  \bigg]     \bigg\}      \bigg[       \underset{j  \text{ } \mathrm{mod}3 \equiv 0}{\underset{3 \leq j \leq N-1}{\prod}}      i u^{-1} \sigma^+_j \sigma^-_j              \bigg]  \bigg[ \underset{j \equiv N}{\prod}  \sigma^+_j \bigg] \text{. } \end{align*}

       \noindent Proceeding, another application of (AC), followed by (LR), yields,

       \begin{align*}
         -       \bigg[ \underset{0 \leq j < j+1 < N-1}{\prod}  \sigma^-_{j} i u \sigma^+_{j+1} \sigma^-_{j+1}      \textbf{1}_{\{j^{\prime}> j+1 \text{ } : \text{ } i u \sigma^-_{j^{\prime}} \sigma^+_{j^{\prime}} \in \mathrm{support} ( A(u))\}}  \big[ 
       \sigma^-_{j^{\prime}} +  i u \sigma^+_{j^{\prime}} \sigma^-_{j^{\prime}}      \big]  \bigg]          \bigg\{      \bigg[ \underset{1 \leq j \leq N-1}{\prod} \big[ \sigma^+_j + i u \sigma^+_j \big] \\ \times \textbf{1}_{\{ j^{\prime} >j \text{ } : \text{ } i u \sigma^+_{j^{\prime}} \sigma^-_{j^{\prime}} \in \mathrm{support} ( A ( u^{\prime})) \}}  \big[ \sigma^+_{j^{\prime}} i u^{\prime} \sigma^+_{j^{\prime}+1} \sigma^-_{j^{\prime}+1}        + i u^{\prime}     \sigma^+_{j^{\prime}} \sigma^-_{j^{\prime}} \sigma^-_{j^{\prime}+1}        \big]    \bigg]  \bigg[ \underset{j \equiv N}{\prod} i u^{\prime} \sigma^+_j \sigma^-_j  \bigg]  , \bigg[       \underset{j  \text{ } \mathrm{mod}3 \equiv 0}{\underset{3 \leq j \leq N-1}{\prod}}      i \big( u^{\prime} \big)^{-1} \sigma^+_j \sigma^-_j              \bigg]      \bigg\} \\ \times  \bigg[ \underset{j \equiv N}{\prod}  \sigma^+_j \bigg]  \text{, }  \end{align*}

       \noindent corresponding to the first term,

       \begin{align*} -     \bigg[ \underset{0 \leq j < j+1 < N-1}{\prod}  \sigma^-_{j} i u \sigma^+_{j+1} \sigma^-_{j+1}      \textbf{1}_{\{j^{\prime}> j+1 \text{ } : \text{ } i u \sigma^-_{j^{\prime}} \sigma^+_{j^{\prime}} \in \mathrm{support} ( A(u))\}}  \big[ 
       \sigma^-_{j^{\prime}} +  i u \sigma^+_{j^{\prime}} \sigma^-_{j^{\prime}}      \big]  \bigg]   \bigg[       \underset{j  \text{ } \mathrm{mod}3 \equiv 0}{\underset{3 \leq j \leq N-1}{\prod}}      i u^{-1} \sigma^+_j \sigma^-_j              \bigg]                \\  \times  \bigg\{     \bigg[ \underset{1 \leq j \leq N-1}{\prod} \big[ \sigma^+_j + i u \sigma^+_j \big] \textbf{1}_{\{ j^{\prime} >j \text{ } : \text{ } i u \sigma^+_{j^{\prime}} \sigma^-_{j^{\prime}} \in \mathrm{support} ( A ( u^{\prime})) \}}  \big[ \sigma^+_{j^{\prime}} i u^{\prime} \sigma^+_{j^{\prime}+1} \sigma^-_{j^{\prime}+1}        + i u^{\prime}     \sigma^+_{j^{\prime}} \sigma^-_{j^{\prime}} \sigma^-_{j^{\prime}+1}        \big]    \bigg]  \\  \times   \bigg[ \underset{j \equiv N}{\prod} i u^{\prime} \sigma^+_j \sigma^-_j  \bigg] ,  \bigg[ \underset{j \equiv N}{\prod}  \sigma^+_j \bigg]     \bigg\}  \text{, }  \end{align*}

       \noindent corresponding to the second term,

       \begin{align*} -    \bigg[ \underset{1 \leq j \leq N-1}{\prod} \big[ \sigma^+_j + i u \sigma^+_j \big]   \textbf{1}_{\{ j^{\prime} >j \text{ } : \text{ } i u \sigma^+_{j^{\prime}} \sigma^-_{j^{\prime}} \in \mathrm{support} ( A ( u^{\prime})) \}}  \big[ \sigma^+_{j^{\prime}}  i u^{\prime} \sigma^+_{j^{\prime}+1} \sigma^-_{j^{\prime}+1}         + i u^{\prime}     \sigma^+_{j^{\prime}} \sigma^-_{j^{\prime}} \sigma^-_{j^{\prime}+1}        \big]    \bigg]   \bigg\{     \bigg[ \underset{j \equiv N}{\prod} i u^{\prime} \sigma^+_j \sigma^-_j  \bigg]   \\  ,  \bigg[ \underset{0 \leq j < j+1 < N-1}{\prod}  \sigma^-_{j} i u^{\prime} \sigma^+_{j+1} \sigma^-_{j+1}      \textbf{1}_{\{j^{\prime}> j+1 \text{ } : \text{ } i u^{\prime} \sigma^-_{j^{\prime}} \sigma^+_{j^{\prime}} \in \mathrm{support} ( A(u^{\prime}))\}}  \big[ 
       \sigma^-_{j^{\prime}} +  i u^{\prime} \sigma^+_{j^{\prime}} \sigma^-_{j^{\prime}}      \big]  \bigg]     \bigg\}      \bigg[       \underset{j  \text{ } \mathrm{mod}3 \equiv 0}{\underset{3 \leq j \leq N-1}{\prod}}      i u^{-1} \sigma^+_j \sigma^-_j              \bigg]  \bigg[ \underset{j \equiv N}{\prod}  \sigma^+_j \bigg]                         \text{, }  \end{align*}

       \noindent corresponding to the third term,

       \begin{align*}
        -  \bigg\{       \bigg[ \underset{1 \leq j \leq N-1}{\prod} \big[ \sigma^+_j + i u \sigma^+_j \big]   \textbf{1}_{\{ j^{\prime} >j \text{ } : \text{ } i u \sigma^+_{j^{\prime}} \sigma^-_{j^{\prime}} \in \mathrm{support} ( A ( u^{\prime})) \}}  \big[ \sigma^+_{j^{\prime}}  i u^{\prime} \sigma^+_{j^{\prime}+1} \sigma^-_{j^{\prime}+1}         + i u^{\prime}     \sigma^+_{j^{\prime}} \sigma^-_{j^{\prime}} \sigma^-_{j^{\prime}+1}        \big]    \bigg]   \\ ,  \bigg[ \underset{0 \leq j < j+1 < N-1}{\prod}  \sigma^-_{j} i u^{\prime} \sigma^+_{j+1} \sigma^-_{j+1}      \textbf{1}_{\{j^{\prime}> j+1 \text{ } : \text{ } i u^{\prime} \sigma^-_{j^{\prime}} \sigma^+_{j^{\prime}} \in \mathrm{support} ( A(u^{\prime}))\}}  \big[ 
       \sigma^-_{j^{\prime}} +  i u^{\prime} \sigma^+_{j^{\prime}} \sigma^-_{j^{\prime}}      \big]  \bigg]     \bigg\} \bigg[ \underset{j \equiv N}{\prod} i u^{\prime} \sigma^+_j \sigma^-_j  \bigg]       \\ \times   \bigg[       \underset{j  \text{ } \mathrm{mod}3 \equiv 0}{\underset{3 \leq j \leq N-1}{\prod}}      i u^{-1} \sigma^+_j \sigma^-_j              \bigg] \bigg[ \underset{j \equiv N}{\prod}  \sigma^+_j \bigg] \end{align*}
       
       \noindent An application of (LR) yields,

       \begin{align*}
              -       \bigg[ \underset{0 \leq j < j+1 < N-1}{\prod}  \sigma^-_{j} i u \sigma^+_{j+1} \sigma^-_{j+1}      \textbf{1}_{\{j^{\prime}> j+1 \text{ } : \text{ } i u \sigma^-_{j^{\prime}} \sigma^+_{j^{\prime}} \in \mathrm{support} ( A(u))\}}  \big[ 
       \sigma^-_{j^{\prime}} +  i u \sigma^+_{j^{\prime}} \sigma^-_{j^{\prime}}      \big]  \bigg]          \bigg\{      \bigg[ \underset{1 \leq j \leq N-1}{\prod} \big[ \sigma^+_j  \\ + i u \sigma^+_j \big]  \textbf{1}_{\{ j^{\prime} >j \text{ } : \text{ } i u \sigma^+_{j^{\prime}} \sigma^-_{j^{\prime}} \in \mathrm{support} ( A ( u^{\prime})) \}} \\  \times   \big[ \sigma^+_{j^{\prime}} i u \sigma^+_{j^{\prime}+1} \sigma^-_{j^{\prime}+1}        + i u     \sigma^+_{j^{\prime}} \sigma^-_{j^{\prime}} \sigma^-_{j^{\prime}+1}        \big]    \bigg] , \bigg[       \underset{j  \text{ } \mathrm{mod}3 \equiv 0}{\underset{3 \leq j \leq N-1}{\prod}}      i \big( u^{\prime} \big)^{-1} \sigma^+_j \sigma^-_j              \bigg]      \bigg\}  \bigg[ \underset{j \equiv N}{\prod} i u^{\prime} \sigma^+_j \sigma^-_j  \bigg]    \bigg[ \underset{j \equiv N}{\prod}  \sigma^+_j \bigg]  \text{, }  \end{align*}

       \noindent corresponding to the first term,

       \begin{align*} -       \bigg[ \underset{0 \leq j < j+1 < N-1}{\prod}  \sigma^-_{j} i u \sigma^+_{j+1} \sigma^-_{j+1}      \textbf{1}_{\{j^{\prime}> j+1 \text{ } : \text{ } i u \sigma^-_{j^{\prime}} \sigma^+_{j^{\prime}} \in \mathrm{support} ( A(u))\}}  \big[ 
       \sigma^-_{j^{\prime}} +  i u \sigma^+_{j^{\prime}} \sigma^-_{j^{\prime}}      \big]  \bigg]       \bigg[ \underset{1 \leq j \leq N-1}{\prod} \big[ \sigma^+_j + i u \sigma^+_j \big] \end{align*}

              \begin{align*}   \times \textbf{1}_{\{ j^{\prime} >j \text{ } : \text{ } i u \sigma^+_{j^{\prime}} \sigma^-_{j^{\prime}} \in \mathrm{support} ( A ( u^{\prime})) \}}  \big[ \sigma^+_{j^{\prime}} i u^{\prime} \sigma^+_{j^{\prime}+1} \sigma^-_{j^{\prime}+1}        + i u^{\prime}     \sigma^+_{j^{\prime}} \sigma^-_{j^{\prime}} \sigma^-_{j^{\prime}+1}        \big]    \bigg]   \bigg\{        \bigg[ \underset{j \equiv N}{\prod} i u^{\prime} \sigma^+_j \sigma^-_j  \bigg]  \\ , \bigg[       \underset{j  \text{ } \mathrm{mod}3 \equiv 0}{\underset{3 \leq j \leq N-1}{\prod}}      i \big( u^{\prime} \big)^{-1} \sigma^+_j \sigma^-_j              \bigg]      \bigg\} \bigg[ \underset{j \equiv N}{\prod}  \sigma^+_j \bigg]   \text{, }  \end{align*}

       \noindent corresponding to the second term,

       \begin{align*} -     \bigg[ \underset{0 \leq j < j+1 < N-1}{\prod}  \sigma^-_{j} i u \sigma^+_{j+1} \sigma^-_{j+1}      \textbf{1}_{\{j^{\prime}> j+1 \text{ } : \text{ } i u \sigma^-_{j^{\prime}} \sigma^+_{j^{\prime}} \in \mathrm{support} ( A(u))\}}  \big[ 
       \sigma^-_{j^{\prime}} +  i u \sigma^+_{j^{\prime}} \sigma^-_{j^{\prime}}      \big]  \bigg]   \bigg[       \underset{j  \text{ } \mathrm{mod}3 \equiv 0}{\underset{3 \leq j \leq N-1}{\prod}}      i u^{-1} \sigma^+_j \sigma^-_j              \bigg]               \end{align*}

       \begin{align*}
       \times  \bigg\{     \bigg[ \underset{1 \leq j \leq N-1}{\prod} \big[ \sigma^+_j + i u \sigma^+_j \big] \textbf{1}_{\{ j^{\prime} >j \text{ } : \text{ } i u \sigma^+_{j^{\prime}} \sigma^-_{j^{\prime}} \in \mathrm{support} ( A ( u^{\prime})) \}}  \big[ \sigma^+_{j^{\prime}} i u^{\prime} \sigma^+_{j^{\prime}+1} \sigma^-_{j^{\prime}+1}        + i u^{\prime}     \sigma^+_{j^{\prime}} \sigma^-_{j^{\prime}} \sigma^-_{j^{\prime}+1}        \big]    \bigg] \\ ,  \bigg[ \underset{j \equiv N}{\prod}  \sigma^+_j \bigg]     \bigg\}  \bigg[ \underset{j \equiv N}{\prod} i u^{\prime} \sigma^+_j \sigma^-_j  \bigg]  \text{, }  \end{align*}

       \noindent corresponding to the third term,

       \begin{align*}  -     \bigg[ \underset{0 \leq j < j+1 < N-1}{\prod}  \sigma^-_{j} i u \sigma^+_{j+1} \sigma^-_{j+1}      \textbf{1}_{\{j^{\prime}> j+1 \text{ } : \text{ } i u \sigma^-_{j^{\prime}} \sigma^+_{j^{\prime}} \in \mathrm{support} ( A(u))\}}  \big[ 
       \sigma^-_{j^{\prime}} +  i u \sigma^+_{j^{\prime}} \sigma^-_{j^{\prime}}      \big]  \bigg]   \bigg[       \underset{j  \text{ } \mathrm{mod}3 \equiv 0}{\underset{3 \leq j \leq N-1}{\prod}}      i u^{-1} \sigma^+_j \sigma^-_j              \bigg]                \\  \times      \bigg[ \underset{1 \leq j \leq N-1}{\prod} \big[ \sigma^+_j + i u \sigma^+_j \big] \textbf{1}_{\{ j^{\prime} >j \text{ } : \text{ } i u \sigma^+_{j^{\prime}} \sigma^-_{j^{\prime}} \in \mathrm{support} ( A ( u^{\prime})) \}}  \big[ \sigma^+_{j^{\prime}} i u^{\prime} \sigma^+_{j^{\prime}+1} \sigma^-_{j^{\prime}+1}        + i u^{\prime}     \sigma^+_{j^{\prime}} \sigma^-_{j^{\prime}} \sigma^-_{j^{\prime}+1}        \big]    \bigg] \\ \times  \bigg\{   \bigg[ \underset{j \equiv N}{\prod} i u \sigma^+_j \sigma^-_j  \bigg] ,  \bigg[ \underset{j \equiv N}{\prod}  \sigma^+_j \bigg]     \bigg\}   \\ -    \bigg[ \underset{1 \leq j \leq N-1}{\prod} \big[ \sigma^+_j + i u \sigma^+_j \big]   \textbf{1}_{\{ j^{\prime} >j \text{ } : \text{ } i u \sigma^+_{j^{\prime}} \sigma^-_{j^{\prime}} \in \mathrm{support} ( A ( u^{\prime})) \}}  \big[ \sigma^+_{j^{\prime}}  i u^{\prime} \sigma^+_{j^{\prime}+1} \sigma^-_{j^{\prime}+1}         + i u^{\prime}     \sigma^+_{j^{\prime}} \sigma^-_{j^{\prime}} \sigma^-_{j^{\prime}+1}        \big]    \bigg]   \bigg\{     \bigg[ \underset{j \equiv N}{\prod} i u^{\prime} \sigma^+_j \sigma^-_j  \bigg]  \\ ,  \bigg[ \underset{0 \leq j < j+1 < N-1}{\prod}  \sigma^-_{j} i u^{\prime} \sigma^+_{j+1} \sigma^-_{j+1}      \textbf{1}_{\{j^{\prime}> j+1 \text{ } : \text{ } i u^{\prime} \sigma^-_{j^{\prime}} \sigma^+_{j^{\prime}} \in \mathrm{support} ( A(u^{\prime}))\}}  \big[ 
       \sigma^-_{j^{\prime}} +  i u^{\prime} \sigma^+_{j^{\prime}} \sigma^-_{j^{\prime}}      \big]  \bigg]     \bigg\}  \\ \times       \bigg[       \underset{j  \text{ } \mathrm{mod}3 \equiv 0}{\underset{3 \leq j \leq N-1}{\prod}}      i u^{-1} \sigma^+_j \sigma^-_j              \bigg]  \bigg[ \underset{j \equiv N}{\prod}  \sigma^+_j \bigg] \\    -  \bigg\{       \bigg[ \underset{1 \leq j \leq N-1}{\prod} \big[ \sigma^+_j + i u \sigma^+_j \big]   \textbf{1}_{\{ j^{\prime} >j \text{ } : \text{ } i u \sigma^+_{j^{\prime}} \sigma^-_{j^{\prime}} \in \mathrm{support} ( A ( u^{\prime})) \}}  \big[ \sigma^+_{j^{\prime}}  i u^{\prime} \sigma^+_{j^{\prime}+1} \sigma^-_{j^{\prime}+1}         + i u^{\prime}     \sigma^+_{j^{\prime}} \sigma^-_{j^{\prime}} \sigma^-_{j^{\prime}+1}        \big]    \bigg]  \\ ,  \bigg[ \underset{0 \leq j < j+1 < N-1}{\prod}  \sigma^-_{j} i u^{\prime} \sigma^+_{j+1} \sigma^-_{j+1}      \textbf{1}_{\{j^{\prime}> j+1 \text{ } : \text{ } i u^{\prime} \sigma^-_{j^{\prime}} \sigma^+_{j^{\prime}} \in \mathrm{support} ( A(u^{\prime}))\}}  \big[ 
       \sigma^-_{j^{\prime}} +  i u^{\prime} \sigma^+_{j^{\prime}} \sigma^-_{j^{\prime}}      \big]  \bigg]     \bigg\} \\ \times   \bigg[ \underset{j \equiv N}{\prod} i u \sigma^+_j \sigma^-_j  \bigg]       \bigg[       \underset{j  \text{ } \mathrm{mod}3 \equiv 0}{\underset{3 \leq j \leq N-1}{\prod}}      i u^{-1} \sigma^+_j \sigma^-_j              \bigg]  \bigg[ \underset{j \equiv N}{\prod}  \sigma^+_j \bigg]      \text{, } 
\end{align*}

\noindent which can be approximated with,

\begin{align*}
        \big( \mathscr{C}^1_1 \big)_{12}        \text{. }
\end{align*}

\noindent For the remaining computations with the Poisson bracket, we make use of the decomposition for $\underline{\mathcal{I}_4}$. In comparison to previous computations from the decompositions of $\underline{\mathcal{I}_1},\underline{\mathcal{I}_2}$ and ,$\underline{\mathcal{I}_3}$, different combinations of Pauli operators appear. The corresponding brackets take the form, after one application of (BL),

\begin{align*}
 \bigg\{   i u \sigma^+_0 \sigma^-_0 \bigg[  \text{ } \bigg[   \underset{j \text{ odd}}{\underset{1 \leq  j \leq N-1}{\prod}}       i u \sigma^+_j \sigma^-_j \textbf{1}_{\{ j^{\prime}> j \text{ } : \text{ } i u \sigma^+_{j^{\prime}} \sigma^-_{j^{\prime}} \in \mathrm{support} ( A ( u ) )\} }       \big[ i u \sigma^+_{j^{\prime}} \sigma^-_{j^{\prime}} i u \sigma^+_{j^{\prime}+1} \sigma^-_{j^{\prime}+1} + \sigma^-_{j^{\prime}} \sigma^+_{j^{\prime}} \sigma^-_{j^{\prime}+1} \big]        \bigg]   \bigg[    \underset{j \equiv N}{\prod}     i u \sigma^+_j \sigma^-_j \bigg]     \\    +   i u \sigma^+_1 \sigma^-_1 \sigma^+_2          \bigg[        \underset{i \text{ odd}}{\underset{1 \leq j \leq N}{\prod}}      i u \sigma^+_j \sigma^-_j             \bigg]     \bigg[ \underset{j \equiv N}{\prod} \sigma^+_j \sigma^-_j  \bigg]                 +  \bigg[   \underset{j \text{ even}}{\underset{1 \leq j \leq N-1}{\prod}}   i u \sigma^+_j \sigma^-_j  \bigg]  \bigg[       \underset{j \text{ odd}}{\underset{3 \leq j \leq N-1}{\prod}}      i u^{-1} \sigma^+_j \sigma^-_j              \bigg]   \bigg[ \underset{j \equiv N}{\prod} i u \sigma^+_j \sigma^-_j   \bigg] \\ ,  \sigma^-_0 \bigg[   i u^{\prime} \sigma^+_1 \sigma^-_1            \bigg[ \underset{0 \leq j \leq N-1}{\prod}            \textbf{1}_{\{ j^{\prime} > j \text{ } : \text{ } i u^{\prime} \sigma^+_{j^{\prime}} \sigma^-_{j^{\prime}} \in \mathrm{support} ( D ( u^{\prime} ) ) \}}  \bigg[  \sigma^+_{j^{\prime}} + i u^{\prime} \sigma^+_{j^{\prime}} \sigma^-_{j^{\prime}} \bigg] \text{ }                   \bigg]  \bigg[ \underset{j \equiv N}{\prod} i \big( u^{\prime}\big)^{-1} \sigma^+_{j-1} \sigma^-_{j+1} i \big( u^{\prime}\big)^{-1} \sigma^+_j \sigma^-_j  \bigg]                          \\ +           \sigma^+_1 \bigg[  \underset{1 \leq j \leq N-1}{\prod}     \textbf{1}_{\{j^{\prime} > j \text{ } : \text{ }      i u^{\prime} \sigma^+_{j^{\prime}} \sigma^-_{j^{\prime}} \in \mathrm{support} ( D ( u^{\prime} ))         \}}    \bigg[         i u^{\prime} \sigma^+_{j^{\prime}} \sigma^-_{j^{\prime}} + \sigma^-_{j^{\prime}} \bigg]  \text{ }      \bigg]     \bigg[          \underset{j \equiv N}{\prod}        i \big( u^{\prime}\big)^{-1} \sigma^+_{j-1} \sigma^-_{j-1} i \big( u^{\prime}\big)^{-1} \sigma^+_j \sigma^-_j     \bigg] \text{ }        \bigg]    \bigg\}     \end{align*}

 \begin{align*}  \overset{(\mathrm{BL})}{=}     \bigg\{   i u \sigma^+_0 \sigma^-_0 \bigg[  \text{ } \bigg[   \underset{j \text{ odd}}{\underset{1 \leq  j \leq N-1}{\prod}}       i u \sigma^+_j \sigma^-_j \textbf{1}_{\{ j^{\prime}> j \text{ } : \text{ } i u \sigma^+_{j^{\prime}} \sigma^-_{j^{\prime}} \in \mathrm{support} ( A ( u ) )\} }       \big[ i u \sigma^+_{j^{\prime}} \sigma^-_{j^{\prime}} i u \sigma^+_{j^{\prime}+1} \sigma^-_{j^{\prime}+1} + \sigma^-_{j^{\prime}} \sigma^+_{j^{\prime}} \sigma^-_{j^{\prime}+1} \big]        \bigg]   \bigg[    \underset{j \equiv N}{\prod}     i u \sigma^+_j \sigma^-_j \bigg]     \\ 
 ,  \sigma^-_0 \bigg[   i u^{\prime} \sigma^+_1 \sigma^-_1            \bigg[ \underset{0 \leq j \leq N-1}{\prod}            \textbf{1}_{\{ j^{\prime} > j \text{ } : \text{ } i u^{\prime} \sigma^+_{j^{\prime}} \sigma^-_{j^{\prime}} \in \mathrm{support} ( D ( u^{\prime} ) ) \}}  \bigg[  \sigma^+_{j^{\prime}} + i u^{\prime} \sigma^+_{j^{\prime}} \sigma^-_{j^{\prime}} \bigg] \text{ }                   \bigg]  \bigg[ \underset{j \equiv N}{\prod} i \big( u^{\prime}\big)^{-1}  \sigma^+_{j-1} \sigma^-_{j+1} i \big( u^{\prime}\big)^{-1} \sigma^+_j \sigma^-_j  \bigg]                          \\ +           \sigma^+_1 \bigg[  \underset{1 \leq j \leq N-1}{\prod}     \textbf{1}_{\{j^{\prime} > j \text{ } : \text{ }      i u^{\prime} \sigma^+_{j^{\prime}} \sigma^-_{j^{\prime}} \in \mathrm{support} ( D ( u^{\prime} ))         \}}    \bigg[         i u^{\prime} \sigma^+_{j^{\prime}} \sigma^-_{j^{\prime}} + \sigma^-_{j^{\prime}} \bigg]  \text{ }      \bigg]     \bigg[          \underset{j \equiv N}{\prod}        i \big( u^{\prime}\big)^{-1} \sigma^+_{j-1} \sigma^-_{j-1} i \big( u^{\prime}\big)^{-1} \sigma^+_j \sigma^-_j     \bigg] \text{ }        \bigg]    \bigg\}    \\ + \bigg\{        i u \sigma^+_1 \sigma^-_1 \sigma^+_2          \bigg[        \underset{i \text{ odd}}{\underset{1 \leq j \leq N}{\prod}}      i u \sigma^+_j \sigma^-_j             \bigg]     \bigg[ \underset{j \equiv N}{\prod} \sigma^+_j \sigma^-_j  \bigg]                 +  \bigg[   \underset{j \text{ even}}{\underset{1 \leq j \leq N-1}{\prod}}   i u \sigma^+_j \sigma^-_j  \bigg]  \bigg[       \underset{j \text{ odd}}{\underset{3 \leq j \leq N-1}{\prod}}      i u^{-1} \sigma^+_j \sigma^-_j              \bigg]   \bigg[ \underset{j \equiv N}{\prod} i u \sigma^+_j \sigma^-_j   \bigg] \\ ,  \sigma^-_0 \bigg[   i u^{\prime} \sigma^+_1 \sigma^-_1            \bigg[ \underset{0 \leq j \leq N-1}{\prod}            \textbf{1}_{\{ j^{\prime} > j \text{ } : \text{ } i u^{\prime} \sigma^+_{j^{\prime}} \sigma^-_{j^{\prime}} \in \mathrm{support} ( D ( u^{\prime} ) ) \}}  \bigg[  \sigma^+_{j^{\prime}} + i u^{\prime} \sigma^+_{j^{\prime}} \sigma^-_{j^{\prime}} \bigg] \text{ }                   \bigg]  \bigg[ \underset{j \equiv N}{\prod} i \big( u^{\prime}\big)^{-1} \sigma^+_{j-1} \sigma^-_{j+1} i \big( u^{\prime}\big)^{-1} \sigma^+_j \sigma^-_j  \bigg]                          \\ +           \sigma^+_1 \bigg[  \underset{1 \leq j \leq N-1}{\prod}     \textbf{1}_{\{j^{\prime} > j \text{ } : \text{ }      i u^{\prime} \sigma^+_{j^{\prime}} \sigma^-_{j^{\prime}} \in \mathrm{support} ( D ( u^{\prime} ))         \}}    \bigg[         i u^{\prime} \sigma^+_{j^{\prime}} \sigma^-_{j^{\prime}} + \sigma^-_{j^{\prime}} \bigg]  \text{ }      \bigg]     \bigg[          \underset{j \equiv N}{\prod}        i \big( u^{\prime}\big)^{-1} \sigma^+_{j-1} \sigma^-_{j-1} i \big( u^{\prime}\big)^{-1} \sigma^+_j \sigma^-_j     \bigg] \text{ }        \bigg]    \bigg\} \text{. } \end{align*}

 \noindent An application of (LR), followed by (BL), yields,

 \begin{align*}
   i u \sigma^+_0 \sigma^-_0   \bigg\{  \bigg[  \text{ } \bigg[   \underset{j \text{ odd}}{\underset{1 \leq  j \leq N-1}{\prod}}       i u \sigma^+_j \sigma^-_j \textbf{1}_{\{ j^{\prime}> j \text{ } : \text{ } i u \sigma^+_{j^{\prime}} \sigma^-_{j^{\prime}} \in \mathrm{support} ( A ( u ) )\} }       \big[ i u \sigma^+_{j^{\prime}} \sigma^-_{j^{\prime}} i u \sigma^+_{j^{\prime}+1} \sigma^-_{j^{\prime}+1} + \sigma^-_{j^{\prime}} \sigma^+_{j^{\prime}} \sigma^-_{j^{\prime}+1} \big]        \bigg]   \bigg[    \underset{j \equiv N}{\prod}     i u \sigma^+_j \sigma^-_j \bigg]     \\ ,  \sigma^-_0 \bigg[   i u^{\prime} \sigma^+_1 \sigma^-_1            \bigg[ \underset{0 \leq j \leq N-1}{\prod}            \textbf{1}_{\{ j^{\prime} > j \text{ } : \text{ } i u^{\prime} \sigma^+_{j^{\prime}} \sigma^-_{j^{\prime}} \in \mathrm{support} ( D ( u^{\prime} ) ) \}}  \bigg[  \sigma^+_{j^{\prime}} + i u^{\prime} \sigma^+_{j^{\prime}} \sigma^-_{j^{\prime}} \bigg] \text{ }                   \bigg]  \bigg[ \underset{j \equiv N}{\prod} i u^{-1} \sigma^+_{j-1} \sigma^-_{j+1} i u^{-1} \sigma^+_j \sigma^-_j  \bigg]                     \\ + 
          \sigma^+_1 \bigg[  \underset{1 \leq j \leq N-1}{\prod}     \textbf{1}_{\{j^{\prime} > j \text{ } : \text{ }      i u^{\prime} \sigma^+_{j^{\prime}} \sigma^-_{j^{\prime}} \in \mathrm{support} ( D ( u^{\prime} ))         \}}    \bigg[         i u^{\prime} \sigma^+_{j^{\prime}} \sigma^-_{j^{\prime}} + \sigma^-_{j^{\prime}} \bigg]  \text{ }      \bigg]     \bigg[          \underset{j \equiv N}{\prod}        i \big( u^{\prime} \big)^{-1} \sigma^+_{j-1} \sigma^-_{j-1} i \big( u^{\prime} \big)^{-1} \sigma^+_j \sigma^-_j     \bigg] \text{ }        \bigg]    \bigg\}   \text{, }     \end{align*}

   \noindent corresponding to the first term, 
   \begin{align*}
    \bigg\{  i u \sigma^+_0 \sigma^-_0      ,  \sigma^-_0 \bigg[   i u^{\prime} \sigma^+_1 \sigma^-_1            \bigg[ \underset{0 \leq j \leq N-1}{\prod}            \textbf{1}_{\{ j^{\prime} > j \text{ } : \text{ } i u^{\prime} \sigma^+_{j^{\prime}} \sigma^-_{j^{\prime}} \in \mathrm{support} ( D ( u^{\prime} ) ) \}}  \bigg[  \sigma^+_{j^{\prime}} + i u^{\prime} \sigma^+_{j^{\prime}} \sigma^-_{j^{\prime}} \bigg] \text{ }                   \bigg]  \bigg[ \underset{j \equiv N}{\prod} i \big( u^{\prime}\big)^{-1} \sigma^+_{j-1} \sigma^-_{j+1}  \\ \times i \big (u^{\prime}\big)^{-1} \sigma^+_j \sigma^-_j  \bigg]        \\ +          \sigma^+_1 \bigg[  \underset{1 \leq j \leq N-1}{\prod}     \textbf{1}_{\{j^{\prime} > j \text{ } : \text{ }      i u^{\prime} \sigma^+_{j^{\prime}} \sigma^-_{j^{\prime}} \in \mathrm{support} ( D ( u^{\prime} ))         \}}    \bigg[         i u^{\prime} \sigma^+_{j^{\prime}} \sigma^-_{j^{\prime}} + \sigma^-_{j^{\prime}} \bigg]  \text{ }      \bigg]     \bigg[          \underset{j \equiv N}{\prod}        i \big( u^{\prime}\big)^{-1} \sigma^+_{j-1} \sigma^-_{j-1} i \big( u^{\prime}\big)^{-1} \sigma^+_j \sigma^-_j     \bigg] \text{ }        \bigg]    \bigg\}  \\   \times  \bigg[  \text{ } \bigg[   \underset{j \text{ odd}}{\underset{1 \leq  j \leq N-1}{\prod}}       i u \sigma^+_j \sigma^-_j \textbf{1}_{\{ j^{\prime}> j \text{ } : \text{ } i u \sigma^+_{j^{\prime}} \sigma^-_{j^{\prime}} \in \mathrm{support} ( A ( u ) )\} }       \big[ i u \sigma^+_{j^{\prime}} \sigma^-_{j^{\prime}} i u \sigma^+_{j^{\prime}+1} \sigma^-_{j^{\prime}+1} + \sigma^-_{j^{\prime}} \sigma^+_{j^{\prime}} \sigma^-_{j^{\prime}+1} \big]        \bigg]   \bigg[    \underset{j \equiv N}{\prod}     i u \sigma^+_j \sigma^-_j \bigg]    \text{, }     \end{align*}

   \noindent corresponding to the second term, 
   \begin{align*}       \bigg\{        i u \sigma^+_1 \sigma^-_1 \sigma^+_2          \bigg[        \underset{i \text{ odd}}{\underset{1 \leq j \leq N}{\prod}}      i u \sigma^+_j \sigma^-_j             \bigg]     \bigg[ \underset{j \equiv N}{\prod} \sigma^+_j \sigma^-_j  \bigg]                  ,  \sigma^-_0 \bigg[   i u^{\prime} \sigma^+_1 \sigma^-_1            \bigg[ \underset{0 \leq j \leq N-1}{\prod}            \textbf{1}_{\{ j^{\prime} > j \text{ } : \text{ } i u^{\prime} \sigma^+_{j^{\prime}} \sigma^-_{j^{\prime}} \in \mathrm{support} ( D ( u^{\prime} ) ) \}}  \\ \times \bigg[  \sigma^+_{j^{\prime}} + i u^{\prime} \sigma^+_{j^{\prime}} \sigma^-_{j^{\prime}} \bigg] \text{ }                   \bigg]  \bigg[ \underset{j \equiv N}{\prod} i \big( u^{\prime} \big)^{-1} \sigma^+_{j-1} \sigma^-_{j+1} i \big( u^{\prime} \big)^{-1} \sigma^+_j \sigma^-_j  \bigg]                             \sigma^+_1 \bigg[  \underset{1 \leq j \leq N-1}{\prod}     \textbf{1}_{\{j^{\prime} > j \text{ } : \text{ }      i u^{\prime} \sigma^+_{j^{\prime}} \sigma^-_{j^{\prime}} \in \mathrm{support} ( D ( u^{\prime} ))         \}} \\ \times    \bigg[         i u^{\prime} \sigma^+_{j^{\prime}} \sigma^-_{j^{\prime}} + \sigma^-_{j^{\prime}} \bigg]  \text{ }      \bigg]     \bigg[          \underset{j \equiv N}{\prod}        i \big( u^{\prime} \big)^{-1} \sigma^+_{j-1} \sigma^-_{j-1} i \big( u^{\prime} \big)^{-1} \sigma^+_j \sigma^-_j     \bigg] \text{ }        \bigg]    \bigg\}           \text{, }     \end{align*}

   \noindent corresponding to the third term, 
   \begin{align*}         \bigg\{       \bigg[   \underset{j \text{ even}}{\underset{1 \leq j \leq N-1}{\prod}}   i u \sigma^+_j \sigma^-_j  \bigg]  \bigg[       \underset{j \text{ odd}}{\underset{3 \leq j \leq N-1}{\prod}}      i u^{-1} \sigma^+_j \sigma^-_j              \bigg]   \bigg[ \underset{j \equiv N}{\prod} i u \sigma^+_j \sigma^-_j   \bigg]  ,  \sigma^-_0 \bigg[   i u^{\prime} \sigma^+_1 \sigma^-_1            \bigg[ \underset{0 \leq j \leq N-1}{\prod}            \textbf{1}_{\{ j^{\prime} > j \text{ } : \text{ } i u^{\prime} \sigma^+_{j^{\prime}} \sigma^-_{j^{\prime}} \in \mathrm{support} ( D ( u^{\prime} ) ) \}}  \\ \times \bigg[  \sigma^+_{j^{\prime}} + i u^{\prime} \sigma^+_{j^{\prime}} \sigma^-_{j^{\prime}} \bigg] \text{ }                   \bigg]  \bigg[ \underset{j \equiv N}{\prod} i u^{-1} \sigma^+_{j-1} \sigma^-_{j+1} i \big( u^{\prime} \big)^{-1} \sigma^+_j \sigma^-_j  \bigg]                           +           \sigma^+_1 \bigg[  \underset{1 \leq j \leq N-1}{\prod}     \textbf{1}_{\{j^{\prime} > j \text{ } : \text{ }      i u^{\prime} \sigma^+_{j^{\prime}} \sigma^-_{j^{\prime}} \in \mathrm{support} ( D ( u^{\prime} ))         \}}   \\ \times  \bigg[         i u^{\prime} \sigma^+_{j^{\prime}} \sigma^-_{j^{\prime}} + \sigma^-_{j^{\prime}} \bigg]  \text{ }      \bigg]     \bigg[          \underset{j \equiv N}{\prod}        i \big( u^{\prime} \big)^{-1} \sigma^+_{j-1} \sigma^-_{j-1} i \big( u^{\prime} \big)^{-1} \sigma^+_j \sigma^-_j     \bigg] \text{ }        \bigg]    \bigg\}    \text{, }     \end{align*}

   \noindent corresponding to the fourth term. An application of (LR), followed by (AC), yields,

   \begin{align*}     i u \sigma^+_0 \sigma^-_0   \bigg\{ \bigg[   \underset{j \text{ odd}}{\underset{1 \leq  j \leq N-1}{\prod}}       i u \sigma^+_j \sigma^-_j \textbf{1}_{\{ j^{\prime}> j \text{ } : \text{ } i u \sigma^+_{j^{\prime}} \sigma^-_{j^{\prime}} \in \mathrm{support} ( A ( u ) )\} }       \big[ i u \sigma^+_{j^{\prime}} \sigma^-_{j^{\prime}} i u \sigma^+_{j^{\prime}+1} \sigma^-_{j^{\prime}+1} + \sigma^-_{j^{\prime}} \sigma^+_{j^{\prime}} \sigma^-_{j^{\prime}+1} \big]        \bigg]       \\ ,  \sigma^-_0 \bigg[   i u^{\prime} \sigma^+_1 \sigma^-_1            \bigg[ \underset{0 \leq j \leq N-1}{\prod}            \textbf{1}_{\{ j^{\prime} > j \text{ } : \text{ } i u^{\prime} \sigma^+_{j^{\prime}} \sigma^-_{j^{\prime}} \in \mathrm{support} ( D ( u^{\prime} ) ) \}}  \bigg[  \sigma^+_{j^{\prime}} + i u^{\prime} \sigma^+_{j^{\prime}} \sigma^-_{j^{\prime}} \bigg] \text{ }                   \bigg]  \bigg[ \underset{j \equiv N}{\prod} i u^{-1} \sigma^+_{j-1} \sigma^-_{j+1} i \big( u^{\prime} \big)^{-1} \sigma^+_j \sigma^-_j  \bigg]        \\ ,          \sigma^+_1 \bigg[  \underset{1 \leq j \leq N-1}{\prod}     \textbf{1}_{\{j^{\prime} > j \text{ } : \text{ }      i u^{\prime} \sigma^+_{j^{\prime}} \sigma^-_{j^{\prime}} \in \mathrm{support} ( D ( u^{\prime} ))         \}}    \bigg[         i u^{\prime} \sigma^+_{j^{\prime}} \sigma^-_{j^{\prime}} + \sigma^-_{j^{\prime}} \bigg]  \text{ }      \bigg]    \bigg[          \underset{j \equiv N}{\prod}        i \big( u^{\prime} \big)^{-1} \sigma^+_{j-1} \sigma^-_{j-1} i u^{-1} \sigma^+_j \sigma^-_j     \bigg] \text{ }        \bigg]    \bigg\} \\ \times  \bigg[    \underset{j \equiv N}{\prod}     i u \sigma^+_j \sigma^-_j \bigg]     \text{, }     \end{align*}

   \noindent corresponding to the first term, 
   \begin{align*}      i u \sigma^+_0 \sigma^-_0  \bigg[   \underset{j \text{ odd}}{\underset{1 \leq  j \leq N-1}{\prod}}       i u \sigma^+_j \sigma^-_j \textbf{1}_{\{ j^{\prime}> j \text{ } : \text{ } i u \sigma^+_{j^{\prime}} \sigma^-_{j^{\prime}} \in \mathrm{support} ( A ( u ) )\} }       \big[ i u \sigma^+_{j^{\prime}} \sigma^-_{j^{\prime}} i u \sigma^+_{j^{\prime}+1} \sigma^-_{j^{\prime}+1} + \sigma^-_{j^{\prime}} \sigma^+_{j^{\prime}} \sigma^-_{j^{\prime}+1} \big]        \bigg]     \bigg\{    \bigg[    \underset{j \equiv N}{\prod}     i u \sigma^+_j \sigma^-_j \bigg]     \\ ,  \sigma^-_0 \bigg[   i u^{\prime} \sigma^+_1 \sigma^-_1            \bigg[ \underset{0 \leq j \leq N-1}{\prod}            \textbf{1}_{\{ j^{\prime} > j \text{ } : \text{ } i u^{\prime} \sigma^+_{j^{\prime}} \sigma^-_{j^{\prime}} \in \mathrm{support} ( D ( u^{\prime} ) ) \}}  \bigg[  \sigma^+_{j^{\prime}} + i u^{\prime} \sigma^+_{j^{\prime}} \sigma^-_{j^{\prime}} \bigg] \text{ }                   \bigg]  \bigg[ \underset{j \equiv N}{\prod} i \big( u^{\prime}\big)^{-1} \sigma^+_{j-1} \sigma^-_{j+1} i u^{-1} \sigma^+_j \sigma^-_j  \bigg]                             \\           \sigma^+_1 \bigg[  \underset{1 \leq j \leq N-1}{\prod}     \textbf{1}_{\{j^{\prime} > j \text{ } : \text{ }      i u^{\prime} \sigma^+_{j^{\prime}} \sigma^-_{j^{\prime}} \in \mathrm{support} ( D ( u^{\prime} ))         \}}    \bigg[         i u^{\prime} \sigma^+_{j^{\prime}} \sigma^-_{j^{\prime}} + \sigma^-_{j^{\prime}} \bigg]  \text{ }      \bigg]     \bigg[          \underset{j \equiv N}{\prod}        i  \big( u^{\prime} \big)^{-1} \sigma^+_{j-1} \sigma^-_{j-1} i \big( u^{\prime} \big)^{-1} \sigma^+_j \sigma^-_j     \bigg] \text{ }        \bigg]    \bigg\}            \end{align*}

      \noindent corresponding to the second term, 
   
   \begin{align*}
   -         \bigg\{     \sigma^-_0 \bigg[   i u \sigma^+_1 \sigma^-_1            \bigg[ \underset{0 \leq j \leq N-1}{\prod}            \textbf{1}_{\{ j^{\prime} > j \text{ } : \text{ } i u \sigma^+_{j^{\prime}} \sigma^-_{j^{\prime}} \in \mathrm{support} ( D ( u ) ) \}}  \bigg[  \sigma^+_{j^{\prime}} + i u \sigma^+_{j^{\prime}} \sigma^-_{j^{\prime}} \bigg] \text{ }                   \bigg]  \bigg[ \underset{j \equiv N}{\prod} i u^{-1} \sigma^+_{j-1} \sigma^-_{j+1} i u^{-1} \sigma^+_j \sigma^-_j  \bigg]                  ,  \\         \sigma^+_1 \bigg[  \underset{1 \leq j \leq N-1}{\prod}     \textbf{1}_{\{j^{\prime} > j \text{ } : \text{ }      i u \sigma^+_{j^{\prime}} \sigma^-_{j^{\prime}} \in \mathrm{support} ( D ( u ))         \}}    \bigg[         i u \sigma^+_{j^{\prime}} \sigma^-_{j^{\prime}} + \sigma^-_{j^{\prime}} \bigg]  \text{ }      \bigg]     \bigg[          \underset{j \equiv N}{\prod}        i u^{-1} \sigma^+_{j-1} \sigma^-_{j-1} i u^{-1} \sigma^+_j \sigma^-_j     \bigg] \text{ }        \bigg] , i u^{\prime} \sigma^+_0 \sigma^-_0    \bigg\}  \\   \times  \bigg[   \underset{j \text{ odd}}{\underset{1 \leq  j \leq N-1}{\prod}}       i u \sigma^+_j \sigma^-_j \textbf{1}_{\{ j^{\prime}> j \text{ } : \text{ } i u \sigma^+_{j^{\prime}} \sigma^-_{j^{\prime}} \in \mathrm{support} ( A ( u ) )\} }       \big[ i u \sigma^+_{j^{\prime}} \sigma^-_{j^{\prime}} i u \sigma^+_{j^{\prime}+1} \sigma^-_{j^{\prime}+1} + \sigma^-_{j^{\prime}} \sigma^+_{j^{\prime}} \sigma^-_{j^{\prime}+1} \big]        \bigg]   \bigg[    \underset{j \equiv N}{\prod}     i u \sigma^+_j \sigma^-_j \bigg]  
     \text{, }     \end{align*}

   \noindent corresponding to the third term, 
   \begin{align*}    
           \bigg\{        i u \sigma^+_1 \sigma^-_1 \sigma^+_2                          ,  \sigma^-_0 \bigg[   i u^{\prime} \sigma^+_1 \sigma^-_1            \bigg[ \underset{0 \leq j \leq N-1}{\prod}            \textbf{1}_{\{ j^{\prime} > j \text{ } : \text{ } i u^{\prime} \sigma^+_{j^{\prime}} \sigma^-_{j^{\prime}} \in \mathrm{support} ( D ( u^{\prime} ) ) \}} \bigg[  \sigma^+_{j^{\prime}} + i u^{\prime} \sigma^+_{j^{\prime}} \sigma^-_{j^{\prime}} \bigg] \text{ }                   \bigg]  \bigg[ \underset{j \equiv N}{\prod} i \big( u^{\prime} \big)^{-1} \sigma^+_{j-1} \sigma^-_{j+1} i \big( u^{\prime} \big)^{-1} \\ \times  \sigma^+_j \sigma^-_j  \bigg]                        +           \sigma^+_1 \bigg[  \underset{1 \leq j \leq N-1}{\prod}     \textbf{1}_{\{j^{\prime} > j \text{ } : \text{ }      i u^{\prime} \sigma^+_{j^{\prime}} \sigma^-_{j^{\prime}} \in \mathrm{support} ( D ( u^{\prime} ))         \}}  \bigg[         i u^{\prime} \sigma^+_{j^{\prime}} \sigma^-_{j^{\prime}} + \sigma^-_{j^{\prime}} \bigg]  \text{ }      \bigg]     \bigg[          \underset{j \equiv N}{\prod}        i \big( u^{\prime} \big)^{-1} \sigma^+_{j-1} \sigma^-_{j-1} i \big( u^{\prime} \big)^{-1} \sigma^+_j \sigma^-_j     \bigg] \text{ }        \bigg]    \bigg\}  \\ \times \bigg[        \underset{i \text{ odd}}{\underset{1 \leq j \leq N}{\prod}}      i u \sigma^+_j \sigma^-_j             \bigg]     \bigg[ \underset{j \equiv N}{\prod} \sigma^+_j \sigma^-_j  \bigg]           \text{, }     \end{align*}

   \noindent corresponding to the fourth term, 
   \begin{align*}        i u \sigma^+_1 \sigma^-_1 \sigma^+_2           \bigg[        \underset{i \text{ odd}}{\underset{1 \leq j \leq N}{\prod}}      i u \sigma^+_j \sigma^-_j             \bigg]    \bigg\{       \bigg[ \underset{j \equiv N}{\prod} \sigma^+_j \sigma^-_j  \bigg]                  ,  \sigma^-_0 \bigg[   i u^{\prime} \sigma^+_1 \sigma^-_1            \bigg[ \underset{0 \leq j \leq N-1}{\prod}            \textbf{1}_{\{ j^{\prime} > j \text{ } : \text{ } i u^{\prime} \sigma^+_{j^{\prime}} \sigma^-_{j^{\prime}} \in \mathrm{support} ( D ( u^{\prime} ) ) \}}  \\ \times \bigg[  \sigma^+_{j^{\prime}} + i u^{\prime} \sigma^+_{j^{\prime}} \sigma^-_{j^{\prime}} \bigg] \text{ }                   \bigg]  \bigg[ \underset{j \equiv N}{\prod} i \big( u^{\prime}\big)^{-1} \sigma^+_{j-1} \sigma^-_{j+1} i u^{-1} \sigma^+_j \sigma^-_j  \bigg]                           +           \sigma^+_1 \bigg[  \underset{1 \leq j \leq N-1}{\prod}     \textbf{1}_{\{j^{\prime} > j \text{ } : \text{ }      i u^{\prime} \sigma^+_{j^{\prime}} \sigma^-_{j^{\prime}} \in \mathrm{support} ( D ( u^{\prime} ))         \}}   \\ \times  \bigg[         i u^{\prime} \sigma^+_{j^{\prime}} \sigma^-_{j^{\prime}} + \sigma^-_{j^{\prime}} \bigg]  \text{ }      \bigg]     \bigg[          \underset{j \equiv N}{\prod}        i \big( u^{\prime} \big)^{-1} \sigma^+_{j-1} \sigma^-_{j-1} i \big( u^{\prime} \big)^{-1} \sigma^+_j \sigma^-_j     \bigg] \text{ }        \bigg]    \bigg\}            \text{, }     \end{align*}

   \noindent corresponding to the fifth term, 
   \begin{align*}       \bigg\{       \bigg[   \underset{j \text{ even}}{\underset{1 \leq j \leq N-1}{\prod}}   i u \sigma^+_j \sigma^-_j  \bigg]  \bigg[       \underset{j \text{ odd}}{\underset{3 \leq j \leq N-1}{\prod}}      i u^{-1} \sigma^+_j \sigma^-_j              \bigg]     ,  \sigma^-_0 \bigg[   i u^{\prime} \sigma^+_1 \sigma^-_1            \bigg[ \underset{0 \leq j \leq N-1}{\prod}            \textbf{1}_{\{ j^{\prime} > j \text{ } : \text{ } i u^{\prime} \sigma^+_{j^{\prime}} \sigma^-_{j^{\prime}} \in \mathrm{support} ( D ( u^{\prime} ) ) \}}  \\ \times \bigg[  \sigma^+_{j^{\prime}} + i u^{\prime} \sigma^+_{j^{\prime}} \sigma^-_{j^{\prime}} \bigg] \text{ }                   \bigg]  \bigg[ \underset{j \equiv N}{\prod} i \big( u^{\prime} \big)^{-1} \sigma^+_{j-1} \sigma^-_{j+1} i \big( u^{\prime} \big)^{-1} \sigma^+_j \sigma^-_j  \bigg]                           +           \sigma^+_1 \bigg[  \underset{1 \leq j \leq N-1}{\prod}     \textbf{1}_{\{j^{\prime} > j \text{ } : \text{ }      i u^{\prime} \sigma^+_{j^{\prime}} \sigma^-_{j^{\prime}} \in \mathrm{support} ( D ( u^{\prime} ))         \}}   \\ \times  \bigg[         i u \sigma^+_{j^{\prime}} \sigma^-_{j^{\prime}} + \sigma^-_{j^{\prime}} \bigg]  \text{ }      \bigg]     \bigg[          \underset{j \equiv N}{\prod}        i \big( u^{\prime} \big)^{-1} \sigma^+_{j-1} \sigma^-_{j-1} i \big( u^{\prime}\big)^{-1} \sigma^+_j \sigma^-_j     \bigg] \text{ }        \bigg]    \bigg\} \bigg[ \underset{j \equiv N}{\prod} i u \sigma^+_j \sigma^-_j   \bigg]      \text{, }     \end{align*}

   \noindent corresponding to the sixth term, 
   \begin{align*}      \bigg[   \underset{j \text{ even}}{\underset{1 \leq j \leq N-1}{\prod}}   i u \sigma^+_j \sigma^-_j  \bigg]  \bigg[       \underset{j \text{ odd}}{\underset{3 \leq j \leq N-1}{\prod}}      i u^{-1} \sigma^+_j \sigma^-_j              \bigg]  \bigg\{         \bigg[ \underset{j \equiv N}{\prod} i u \sigma^+_j \sigma^-_j   \bigg]  ,  \sigma^-_0 \bigg[   i u^{\prime} \sigma^+_1 \sigma^-_1            \bigg[ \underset{0 \leq j \leq N-1}{\prod}            \textbf{1}_{\{ j^{\prime} > j \text{ } : \text{ } i u^{\prime} \sigma^+_{j^{\prime}} \sigma^-_{j^{\prime}} \in \mathrm{support} ( D ( u^{\prime} ) ) \}}  \\ \times \bigg[  \sigma^+_{j^{\prime}} + i u^{\prime} \sigma^+_{j^{\prime}} \sigma^-_{j^{\prime}} \bigg] \text{ }                   \bigg]  \bigg[ \underset{j \equiv N}{\prod} i u^{-1} \sigma^+_{j-1} \sigma^-_{j+1} i \big( u^{\prime} \big)^{-1} \sigma^+_j \sigma^-_j  \bigg]                           +           \sigma^+_1 \bigg[  \underset{1 \leq j \leq N-1}{\prod}     \textbf{1}_{\{j^{\prime} > j \text{ } : \text{ }      i u^{\prime} \sigma^+_{j^{\prime}} \sigma^-_{j^{\prime}} \in \mathrm{support} ( D ( u^{\prime} ))         \}}   \\ \times  \bigg[         i u^{\prime} \sigma^+_{j^{\prime}} \sigma^-_{j^{\prime}} + \sigma^-_{j^{\prime}} \bigg]  \text{ }      \bigg]     \bigg[          \underset{j \equiv N}{\prod}        i \big( u^{\prime} \big)^{-1} \sigma^+_{j-1} \sigma^-_{j-1} i \big( u^{\prime} \big)^{-1} \sigma^+_j \sigma^-_j     \bigg] \text{ }        \bigg]    \bigg\} \text{. } \end{align*}

   \noindent An application of (BL), and (AC), to the superposition above yields,
   
   \begin{align*}
       -          i u \sigma^+_0 \sigma^-_0   \bigg\{  \sigma^-_0 \bigg[   i u \sigma^+_1 \sigma^-_1            \bigg[ \underset{0 \leq j \leq N-1}{\prod}            \textbf{1}_{\{ j^{\prime} > j \text{ } : \text{ } i u \sigma^+_{j^{\prime}} \sigma^-_{j^{\prime}} \in \mathrm{support} ( D ( u ) ) \}}  \bigg[  \sigma^+_{j^{\prime}}  + i u \sigma^+_{j^{\prime}} \sigma^-_{j^{\prime}} \bigg] \text{ }                   \bigg]  \bigg[ \underset{j \equiv N}{\prod} i u^{-1} \sigma^+_{j-1} \sigma^-_{j+1} \\ \times  i u^{-1} \sigma^+_j \sigma^-_j  \bigg]      ,       \bigg[   \underset{j \text{ odd}}{\underset{1 \leq  j \leq N-1}{\prod}}       i u^{\prime} \sigma^+_j \sigma^-_j \textbf{1}_{\{ j^{\prime}> j \text{ } : \text{ } i u^{\prime} \sigma^+_{j^{\prime}} \sigma^-_{j^{\prime}} \in \mathrm{support} ( A ( u^{\prime} ) )\} }       \big[ i u^{\prime} \sigma^+_{j^{\prime}} \sigma^-_{j^{\prime}} i u^{\prime} \sigma^+_{j^{\prime}+1} \sigma^-_{j^{\prime}+1} \\ \times  \sigma^-_{j^{\prime}} \sigma^+_{j^{\prime}} \sigma^-_{j^{\prime}+1} \big]        \bigg]                       \bigg\}   \bigg[  \underset{j \equiv N}{\prod}  i u \sigma^+_j \sigma^-_j    \bigg]      \text{, }     \end{align*}

   \noindent corresponding to the first term, 
   \begin{align*}             -    
 i u \sigma^+_0 \sigma^-_0 \bigg\{            \sigma^+_1 \bigg[  \underset{1 \leq j \leq N-1}{\prod}     \textbf{1}_{\{j^{\prime} > j \text{ } : \text{ }      i u \sigma^+_{j^{\prime}} \sigma^-_{j^{\prime}} \in \mathrm{support} ( D ( u ))         \}}    \bigg[         i u \sigma^+_{j^{\prime}} \sigma^-_{j^{\prime}} + \sigma^-_{j^{\prime}} \bigg]  \text{ }      \bigg]     \bigg[          \underset{j \equiv N}{\prod}        i u^{-1} \sigma^+_{j-1} \sigma^-_{j-1} i u^{-1} \sigma^+_j \sigma^-_j     \bigg] \text{ }        \bigg]  \\ ,             \bigg[   \underset{j \text{ odd}}{\underset{1 \leq  j \leq N-1}{\prod}}       i u^{\prime} \sigma^+_j \sigma^-_j \textbf{1}_{\{ j^{\prime}> j \text{ } : \text{ } i u^{\prime} \sigma^+_{j^{\prime}} \sigma^-_{j^{\prime}} \in \mathrm{support} ( A ( u^{\prime} ) )\} }       \big[ i u^{\prime} \sigma^+_{j^{\prime}} \sigma^-_{j^{\prime}} i u^{\prime} \sigma^+_{j^{\prime}+1} \sigma^-_{j^{\prime}+1}  + \sigma^-_{j^{\prime}} \sigma^+_{j^{\prime}} \sigma^-_{j^{\prime}+1} \big]        \bigg]                    \bigg\}    \bigg[  \underset{j \equiv N}{\prod}  i u \sigma^+_j \sigma^-_j    \bigg]         \text{, }     \end{align*}

   \noindent corresponding to the second term, 
   \begin{align*}                -  i u \sigma^+_0 \sigma^-_0  \bigg[   \underset{j \text{ odd}}{\underset{1 \leq  j \leq N-1}{\prod}}       i u \sigma^+_j \sigma^-_j \textbf{1}_{\{ j^{\prime}> j \text{ } : \text{ } i u \sigma^+_{j^{\prime}} \sigma^-_{j^{\prime}} \in \mathrm{support} ( A ( u ) )\} }       \big[ i u \sigma^+_{j^{\prime}} \sigma^-_{j^{\prime}} i u \sigma^+_{j^{\prime}+1} \sigma^-_{j^{\prime}+1} + \sigma^-_{j^{\prime}} \sigma^+_{j^{\prime}} \sigma^-_{j^{\prime}+1} \big]        \bigg]     \bigg\{        \sigma^-_0 \bigg[   i u \sigma^+_1  \\ \times \sigma^-_1            \bigg[ \underset{0 \leq j \leq N-1}{\prod}            \textbf{1}_{\{ j^{\prime} > j \text{ } : \text{ } i u \sigma^+_{j^{\prime}} \sigma^-_{j^{\prime}} \in \mathrm{support} ( D ( u ) ) \}}  \bigg[  \sigma^+_{j^{\prime}} + i u \sigma^+_{j^{\prime}} \sigma^-_{j^{\prime}} \bigg] \text{ }                   \bigg]  \bigg[ \underset{j \equiv N}{\prod} i u^{-1} \sigma^+_{j-1} \sigma^-_{j+1} i u^{-1} \sigma^+_j \sigma^-_j  \bigg]                         \end{align*}

 \begin{align*}
  ,    \bigg[    \underset{j \equiv N}{\prod}     i u^{\prime} \sigma^+_j \sigma^-_j \bigg]  \bigg\}     \text{, }     \end{align*}

   \noindent corresponding to the third term, 
   \begin{align*}        -  i u \sigma^+_0 \sigma^-_0  \bigg[   \underset{j \text{ odd}}{\underset{1 \leq  j \leq N-1}{\prod}}       i u \sigma^+_j \sigma^-_j \textbf{1}_{\{ j^{\prime}> j \text{ } : \text{ } i u \sigma^+_{j^{\prime}} \sigma^-_{j^{\prime}} \in \mathrm{support} ( A ( u ) )\} }       \big[ i u \sigma^+_{j^{\prime}} \sigma^-_{j^{\prime}} i u \sigma^+_{j^{\prime}+1} \sigma^-_{j^{\prime}+1} + \sigma^-_{j^{\prime}} \sigma^+_{j^{\prime}} \sigma^-_{j^{\prime}+1} \big]        \bigg]     \\ \times   \bigg\{  
         \sigma^+_1  \bigg[  \underset{1 \leq j \leq N-1}{\prod}     \textbf{1}_{\{j^{\prime} > j \text{ } : \text{ }      i u \sigma^+_{j^{\prime}} \sigma^-_{j^{\prime}} \in \mathrm{support} ( D ( u ))         \}}    \bigg[         i u \sigma^+_{j^{\prime}} \sigma^-_{j^{\prime}} + \sigma^-_{j^{\prime}} \bigg]  \text{ }      \bigg]     \bigg[          \underset{j \equiv N}{\prod}        i  u^{-1} \sigma^+_{j-1} \sigma^-_{j-1} i  u^{-1} \sigma^+_j \sigma^-_j     \bigg] \text{ }        \bigg]   \\ ,    \bigg[    \underset{j \equiv N}{\prod}     i u^{\prime} \sigma^+_j \sigma^-_j \bigg]  \bigg\}                     \text{, }     \end{align*}

   \noindent corresponding to the fourth term, 
   \begin{align*}         \bigg\{    \sigma^+_1 \bigg[  \underset{1 \leq j \leq N-1}{\prod}     \textbf{1}_{\{j^{\prime} > j \text{ } : \text{ }      i u \sigma^+_{j^{\prime}} \sigma^-_{j^{\prime}} \in \mathrm{support} ( D ( u ))         \}}    \bigg[         i u \sigma^+_{j^{\prime}} \sigma^-_{j^{\prime}} + \sigma^-_{j^{\prime}} \bigg]  \text{ }      \bigg]   ,  \bigg[          \underset{j \equiv N}{\prod}        i \big( u^{\prime} \big)^{-1} \sigma^+_{j-1} \sigma^-_{j-1} i u^{-1} \sigma^+_j \sigma^-_j     \bigg] \text{ }        \bigg]    \bigg\}    \text{, } \end{align*}

   \noindent corresponding to the fifth term,
   
   \begin{align*}
   \bigg\{        i u \sigma^+_1 \sigma^-_1 \sigma^+_2                          ,  \sigma^-_0    i u^{\prime} \sigma^+_1 \sigma^-_1            \bigg[ \underset{0 \leq j \leq N-1}{\prod}            \textbf{1}_{\{ j^{\prime} > j \text{ } : \text{ } i u^{\prime} \sigma^+_{j^{\prime}} \sigma^-_{j^{\prime}} \in \mathrm{support} ( D ( u^{\prime} ) ) \}} \bigg[  \sigma^+_{j^{\prime}} + i u^{\prime} \sigma^+_{j^{\prime}} \sigma^-_{j^{\prime}} \bigg] \text{ }                   \bigg]  \bigg[ \underset{j \equiv N}{\prod} i  \big( u^{\prime} \big)^{-1} \sigma^+_{j-1} \sigma^-_{j+1}  \\ \times i \big( u^{\prime} \big)^{-1} \sigma^+_j \sigma^-_j  \bigg]                          \bigg\}   \text{, }     \end{align*}

   \noindent corresponding to the sixth term, 
   \begin{align*}           \bigg\{        i u \sigma^+_1 \sigma^-_1 \sigma^+_2                          ,    \sigma^-_0   \sigma^+_1 \bigg[  \underset{1 \leq j \leq N-1}{\prod}     \textbf{1}_{\{j^{\prime} > j \text{ } : \text{ }      i u^{\prime} \sigma^+_{j^{\prime}} \sigma^-_{j^{\prime}} \in \mathrm{support} ( D ( u^{\prime} ))         \}}  \bigg[         i u^{\prime} \sigma^+_{j^{\prime}} \sigma^-_{j^{\prime}} + \sigma^-_{j^{\prime}} \bigg]  \text{ }      \bigg]    \\ \times  \bigg[          \underset{j \equiv N}{\prod}        i \big( u^{\prime} \big)^{-1} \sigma^+_{j-1} \sigma^-_{j-1} i  \big( u^{\prime} \big)^{-1} \sigma^+_j \sigma^-_j     \bigg] \text{ }        \bigg]    \bigg\}  \text{, }     \end{align*}

   \noindent corresponding to the seventh term, 
   \begin{align*}         -  i u \sigma^+_1 \sigma^-_1 \sigma^+_2       \bigg[        \underset{i \text{ odd}}{\underset{1 \leq j \leq N}{\prod}}      i u \sigma^+_j \sigma^-_j             \bigg]    \bigg\{      \sigma^-_0 \bigg[   i u \sigma^+_1 \sigma^-_1            \bigg[ \underset{0 \leq j \leq N-1}{\prod}            \textbf{1}_{\{ j^{\prime} > j \text{ } : \text{ } i u \sigma^+_{j^{\prime}} \sigma^-_{j^{\prime}} \in \mathrm{support} ( D ( u ) ) \}}  \\ \times \bigg[  \sigma^+_{j^{\prime}} + i u \sigma^+_{j^{\prime}} \sigma^-_{j^{\prime}} \bigg] \text{ }                   \bigg]  \bigg[ \underset{j \equiv N}{\prod} i u^{-1} \sigma^+_{j-1} \sigma^-_{j+1} i u^{-1} \sigma^+_j \sigma^-_j  \bigg]                       ,    \bigg[ \underset{j \equiv N}{\prod} \sigma^+_j \sigma^-_j  \bigg]                 \bigg\}     - i u \sigma^+_1 \sigma^-_1 \sigma^+_2   \bigg\{           \sigma^+_1  \\ \times \bigg[  \underset{1 \leq j \leq N-1}{\prod}     \textbf{1}_{\{j^{\prime} > j \text{ } : \text{ }      i u^{\prime} \sigma^+_{j^{\prime}} \sigma^-_{j^{\prime}} \in \mathrm{support} ( D ( u^{\prime} ))         \}}   \bigg[         i u^{\prime} \sigma^+_{j^{\prime}} \sigma^-_{j^{\prime}} + \sigma^-_{j^{\prime}} \bigg]  \text{ }      \bigg]     \bigg[          \underset{j \equiv N}{\prod}        i \big( u^{\prime} \big)^{-1} \sigma^+_{j-1} \sigma^-_{j-1} i \big( u^{\prime} \big)^{-1} \sigma^+_j \sigma^-_j     \bigg] \text{ }        \bigg]    \bigg\}        \text{, }     \end{align*}

   \noindent corresponding to the eighth term, 
   \begin{align*}   \bigg\{       \bigg[   \underset{j \text{ even}}{\underset{1 \leq j \leq N-1}{\prod}}   i u \sigma^+_j \sigma^-_j  \bigg]  \bigg[       \underset{j \text{ odd}}{\underset{3 \leq j \leq N-1}{\prod}}      i u^{-1} \sigma^+_j \sigma^-_j              \bigg]     ,  \sigma^-_0 \bigg[   i u^{\prime} \sigma^+_1 \sigma^-_1            \bigg[ \underset{0 \leq j \leq N-1}{\prod}            \textbf{1}_{\{ j^{\prime} > j \text{ } : \text{ } i u^{\prime} \sigma^+_{j^{\prime}} \sigma^-_{j^{\prime}} \in \mathrm{support} ( D ( u^{\prime} ) ) \}}  \\ \times \bigg[  \sigma^+_{j^{\prime}} + i u^{\prime} \sigma^+_{j^{\prime}} \sigma^-_{j^{\prime}} \bigg] \text{ }                   \bigg]  \bigg[ \underset{j \equiv N}{\prod} i \big( u^{\prime} \big)^{-1} \sigma^+_{j-1} \sigma^-_{j+1} i u^{-1} \sigma^+_j \sigma^-_j  \bigg]                           +           \sigma^+_1 \bigg[  \underset{1 \leq j \leq N-1}{\prod}     \textbf{1}_{\{j^{\prime} > j \text{ } : \text{ }      i u^{\prime} \sigma^+_{j^{\prime}} \sigma^-_{j^{\prime}} \in \mathrm{support} ( D ( u^{\prime} ))         \}}   \\ \times  \bigg[         i u^{\prime} \sigma^+_{j^{\prime}} \sigma^-_{j^{\prime}} + \sigma^-_{j^{\prime}} \bigg]  \text{ }      \bigg]     \bigg[          \underset{j \equiv N}{\prod}        i \big( u^{\prime} \big)^{-1} \sigma^+_{j-1} \sigma^-_{j-1} i \big( u^{\prime} \big)^{-1} \sigma^+_j \sigma^-_j     \bigg] \text{ }        \bigg]    \bigg\} \bigg[ \underset{j \equiv N}{\prod} i u^{\prime} \sigma^+_j \sigma^-_j   \bigg]  \text{, }     \end{align*}

   \noindent corresponding to the ninth term, 
   \begin{align*}   \bigg[   \underset{j \text{ even}}{\underset{1 \leq j \leq N-1}{\prod}}   i u \sigma^+_j \sigma^-_j  \bigg]  \bigg[       \underset{j \text{ odd}}{\underset{3 \leq j \leq N-1}{\prod}}      i u^{-1} \sigma^+_j \sigma^-_j              \bigg]  \bigg\{         \bigg[ \underset{j \equiv N}{\prod} i u \sigma^+_j \sigma^-_j   \bigg]  ,  \sigma^-_0 \bigg[   i u^{\prime} \sigma^+_1 \sigma^-_1            \bigg[ \underset{0 \leq j \leq N-1}{\prod}            \textbf{1}_{\{ j^{\prime} > j \text{ } : \text{ } i u^{\prime} \sigma^+_{j^{\prime}} \sigma^-_{j^{\prime}} \in \mathrm{support} ( D ( u^{\prime} ) ) \}}  \end{align*}

   \begin{align*}
   \times \bigg[  \sigma^+_{j^{\prime}} + i u^{\prime} \sigma^+_{j^{\prime}} \sigma^-_{j^{\prime}} \bigg] \text{ }                   \bigg]  \bigg[ \underset{j \equiv N}{\prod} i \big( u^{\prime} \big)^{-1} \sigma^+_{j-1} \sigma^-_{j+1} i \big( u^{\prime} \big)^{-1} \sigma^+_j \sigma^-_j  \bigg]                           +           \sigma^+_1 \bigg[  \underset{1 \leq j \leq N-1}{\prod}     \textbf{1}_{\{j^{\prime} > j \text{ } : \text{ }      i u^{\prime} \sigma^+_{j^{\prime}} \sigma^-_{j^{\prime}} \in \mathrm{support} ( D ( u^{\prime} ))         \}}   \\ \times  \bigg[         i u^{\prime} \sigma^+_{j^{\prime}} \sigma^-_{j^{\prime}} + \sigma^-_{j^{\prime}} \bigg]  \text{ }      \bigg]     \bigg[          \underset{j \equiv N}{\prod}        i  \big( u^{\prime} \big)^{-1} \sigma^+_{j-1} \sigma^-_{j-1} i \big( u^{\prime} \big)^{-1} \sigma^+_j \sigma^-_j     \bigg] \text{ }        \bigg]    \bigg\}        
  \text{, }
\end{align*}

   \noindent corresponding to the tenth term.

\bigskip

\noindent The final superposition of Poisson brackets above can be approximated through a final application of Leibniz' rule to isolate products of terms appearing in the first argument of fourth bracket, the second argument of the fifth bracket, the first argument of the sixth and seventh brackets, in addition to the second argument of the eighth bracket, with $\big( \mathscr{C}^1_1 \big)_{13}$. From such computations with combinations of $(\mathrm{AC})$, $(\mathrm{LR})$, and $(\mathrm{BL})$, the remaining brackets appearing within the Poisson structure with $\underline{\mathcal{I}_4}$ appearing in the second argument of the bracket can be approximated.

           \begin{align*}     \big\{ \underline{\mathcal{I}^{\prime\cdots\prime}_2 \big( u , u^{-1} \big)}, \underline{\mathcal{I}^{\prime\cdots\prime}_1 \big( u^{\prime} , \big( u^{\prime} \big)^{-1} \big)}\big\}  \text{, }  \\ \\ \big\{ \underline{\mathcal{I}^{\prime\cdots\prime}_2 \big( u , u^{-1} \big)}, \underline{\mathcal{I}^{\prime\cdots\prime}_2 \big( u^{\prime} , \big( u^{\prime} \big)^{-1} \big)}\big\} \text{, } \\ \\ 
           \big\{ \underline{\mathcal{I}^{\prime\cdots\prime}_2 \big( u , u^{-1} \big)}, \underline{\mathcal{I}^{\prime\cdots\prime}_3 \big( u^{\prime} , \big( u^{\prime} \big)^{-1} \big)}\big\} \text{, } \\ \vdots \text{, }    \end{align*}

\bigskip

\noindent can be readily approximated with previous computations using the Poisson bracket, namely by either permutting the order in which the terms appear in the first or second argument, or through similar applications of $(\mathrm{BL})$. Altogether, throught the decomposition of the first constant, $\mathscr{C}_1$, of the Poisson structure, approximations for the remaining constant, $\mathscr{C}_2, \mathscr{C}_3, \mathscr{C}_4$, can be obtained.

\bigskip

\noindent The remaining constants for $\mathcal{C}^{XXX}$ can be obtained by applying $\phi$. By construction, the fact that $\phi$ maps the Pauli basis elements of the 4-vertex L-operator into those of the higher-spin XXX chain implies that each Poisson bracket appearing for each entry of the product representation for the transfer matrix can be approximated with combinations of (LR), and (AC), where appropriate, from which we conclude the argument. \boxed{}

\subsection{Yang-Baxter algebras}

The Yang-Baxter algebra is of central interest for adaptations of the quantum inverse scattering approach. As alluded to in the first section of the Introduction, several previous works of the author have investigated aspects of such algebras, {\color{blue}[42},{\color{blue}46},{\color{blue}47]}, which as adaptations of seminal work, {\color{blue}[17]}, put forth in {\color{blue}[25]} raise several implications for other vertex models, including the number of relations with the Poisson structure, as well as geometric, and combinatorial, perspectives which share connections with the rank of Lie Algebras. For the Yang-Baxter algebra of the 4-vertex model, in comparison to that of the 20-vertex model, the relations that are obtained by multiplying operators together from the product representation of the transfer matrix provide expressions for relations of the algebra which depend upon Pauli basis elements, and spectral parameters $u,u^{\prime}$. By making use of previous representations for operators of the transfer matrix, with the two-dimensional representation,

\[\begin{bmatrix}
        A^{4V} \big( \underline{u} \big) & B^{4V} \big( \underline{u} \big) \\ C^{4V} \big( \underline{u} \big) & D^{4V} \big( \underline{u} \big)  
    \end{bmatrix} \equiv \begin{bmatrix}
        A \big( \underline{u} \big) & B \big( \underline{u} \big) \\ C \big( \underline{u} \big) & D \big( \underline{u} \big) 
    \end{bmatrix} \text{, } \]

\noindent generators of the 4-vertex Yang-Baxter algebra can be obtained. In previous aforementioned applications of the quantum inverse scattering method for the 6-vertex, and 20-vertex, models, in the presence of inhomogeneities components of the Yang-Baxter algebra for the 6-vertex model, in large finite volume, can be captured with,

\begin{align*}
  A^{20V} \big( \underline{u} \big)  A^{20V} \big( \underline{u^{\prime}} \big)   \text{, } \\  A^{20V} \big( \underline{u} \big)  B^{20V} \big( \underline{u^{\prime}} \big)   \text{, } \\ \vdots \text{, }
\end{align*}

\noindent which in the case of the 4-vertex model are stated below.

\bigskip

\noindent \textbf{Lemma YB} (\textit{the Yang-Baxter algebra for the 4-vertex model}). For the 4-vertex model, the Yang-Baxter algebra consists of the set of operator relations,

\[
\left\{\!\begin{array}{ll@{}>{{}}l}  A \big( \underline{u} \big) A \big( \underline{u^{\prime}} \big) \text{, } \\   A \big( \underline{u} \big) B \big( \underline{u^{\prime}} \big) \text{, }  \\   A \big( \underline{u} \big) C \big( \underline{u^{\prime}} \big)  \text{, }  \\     A \big( \underline{u} \big) D \big( \underline{u^{\prime}} \big)  \text{, }      \\  B \big( \underline{u} \big) A \big( \underline{u^{\prime}} \big)  \text{, }      \\  B \big( \underline{u} \big) B \big( \underline{u^{\prime}} \big)  \text{, } \\  B \big( \underline{u} \big) C \big( \underline{u^{\prime}} \big)  \text{, }     \\  B \big( \underline{u} \big) D \big( \underline{u^{\prime}} \big)  \text{, } \\   C \big( \underline{u} \big) A \big( \underline{u^{\prime}} \big)  \text{, } \\      C \big( \underline{u} \big) B \big( \underline{u^{\prime}} \big)  \text{, }   \\   C \big( \underline{u} \big) C \big( \underline{u^{\prime}} \big)  \text{, }    \\   C \big( \underline{u} \big) D \big( \underline{u^{\prime}} \big)  \text{, }    \\                       D \big( \underline{u} \big) A \big( \underline{u^{\prime}} \big)  \text{, }  \\    D \big( \underline{u} \big) B \big( \underline{u^{\prime}} \big)  \text{, }      \\   D \big( \underline{u} \big) C \big( \underline{u^{\prime}} \big)  \text{, }  \\   D \big( \underline{u} \big) D \big( \underline{u^{\prime}} \big)  \text{, }  \end{array}\right. \]
  
  \noindent each of which can respectively be approximated with $C_1, \cdots, C_{16}$.

  \bigskip

  \noindent \textit{Proof of Lemma YB}. The computations of each operator product above are included in the next section, the Appendix, from which we conclude the argument. \boxed{}

  \bigskip

  \noindent We state the last item below, which immediately follows from $\phi$.

  \bigskip

  \noindent \textbf{Lemma XXX} (\textit{the Yang-Baxter algebra for the higher-spin XXX chain}).  For the higher-spin XXX chain, the Yang-Baxter algebra consists of the set of operator relations,

  \[
\left\{\!\begin{array}{ll@{}>{{}}l}  A^{XXX} \big( \underline{u} \big) A^{XXX} \big( \underline{u^{\prime}} \big) \text{, } \\   A^{XXX} \big( \underline{u} \big) B^{XXX} \big( \underline{u^{\prime}} \big) \text{, }  \\   A^{XXX} \big( \underline{u} \big) C^{XXX} \big( \underline{u^{\prime}} \big)  \text{, }  \\     A^{XXX} \big( \underline{u} \big) D^{XXX} \big( \underline{u^{\prime}} \big)  \text{, }      \\  B^{XXX} \big( \underline{u} \big) A^{XXX} \big( \underline{u^{\prime}} \big)  \text{, }      \\  B^{XXX} \big( \underline{u} \big) B^{XXX} \big( \underline{u^{\prime}} \big)  \text{, } \\  B^{XXX} \big( \underline{u} \big) C^{XXX} \big( \underline{u^{\prime}} \big)  \text{, }     \\  B^{XXX} \big( \underline{u} \big) D^{XXX} \big( \underline{u^{\prime}} \big)  \text{, } \\   C^{XXX} \big( \underline{u} \big) A^{XXX} \big( \underline{u^{\prime}} \big)  \text{, } \\      C^{XXX} \big( \underline{u} \big) B^{XXX} \big( \underline{u^{\prime}} \big)  \text{, } \\  C^{XXX} \big( \underline{u} \big) C^{XXX} \big( \underline{u^{\prime}} \big)  \text{, }    \\   C^{XXX} \big( \underline{u} \big) D^{XXX} \big( \underline{u^{\prime}} \big)  \text{, }    \\                       D^{XXX} \big( \underline{u} \big) A^{XXX} \big( \underline{u^{\prime}} \big)  \text{, }  \\    D^{XXX} \big( \underline{u} \big) B^{XXX} \big( \underline{u^{\prime}} \big)  \text{, }      \\   D^{XXX} \big( \underline{u} \big) C^{XXX} \big( \underline{u^{\prime}} \big)  \text{, }  \\   D^{XXX} \big( \underline{u} \big) D^{XXX} \big( \underline{u^{\prime}} \big)  \text{, }  \end{array}\right. \]

\noindent each of which can be respectively approximated with $\mathcal{C}_1, \cdots, \mathcal{C}_{16}$.

\bigskip

\noindent \textit{Proof of Lemma XXX}. The result can be straightforwardly obtained from the following images of each $C_i$ in the previous result under the change of basis mapping $\phi$, through,

\[
\left\{\!\begin{array}{ll@{}>{{}}l}  \phi \bigg[ A^{XXX} \big( \underline{u} \big) A^{XXX} \big( \underline{u^{\prime}} \big) \bigg] \approx \phi \big[ C_1 \big] \approx \mathcal{C}_1 \text{, } \\ \phi\bigg[   A^{XXX} \big( \underline{u} \big) B^{XXX} \big( \underline{u^{\prime}} \big) \bigg] \approx \phi \big[ C_2 \big] \approx \mathcal{C}_2  \text{, }  \\   \phi \bigg[ A^{XXX} \big( \underline{u} \big) C^{XXX} \big( \underline{u^{\prime}} \big) \bigg] \approx \phi \big[ C_3 \big] \approx \mathcal{C}_3  \text{, }  \\     \phi \bigg[ A^{XXX} \big( \underline{u} \big) D^{XXX} \big( \underline{u^{\prime}} \big) \bigg] \approx \phi \big[ C_4 \big] \approx  \mathcal{C}_4  \text{, }      \\  \phi \bigg[ B^{XXX} \big( \underline{u} \big) A^{XXX} \big( \underline{u^{\prime}} \big)  \bigg] \approx \phi \big[ C_5 \big] \approx \mathcal{C}_5 \text{, }      \\  \phi \bigg[ B^{XXX} \big( \underline{u} \big) B^{XXX} \big( \underline{u^{\prime}} \big) \bigg] \approx \phi \big[ C_6 \big] \approx \mathcal{C}_6   \text{, } \\         \phi \bigg[ B^{XXX} \big( \underline{u} \big) C^{XXX} \big( \underline{u^{\prime}} \big) \bigg] \approx \phi \big[ C_7 \big] \approx \mathcal{C}_7   \text{, }  \\ \phi \bigg[ B^{XXX} \big( \underline{u} \big) D^{XXX} \big( \underline{u^{\prime}} \big) \bigg] \approx \phi \big[ C_8 \big] \approx \mathcal{C}_8 \text{, }  \\  \phi \bigg[  C^{XXX} \big( \underline{u} \big) A^{XXX} \big( \underline{u^{\prime}} \big) \bigg] \approx \phi \big[ C_9 \big] \approx \mathcal{C}_9   \text{, } \\               \phi \bigg[ C^{XXX} \big( \underline{u} \big) B^{XXX} \big( \underline{u^{\prime}} \big) \bigg] \approx \phi \big[ C_{10} \big] \approx \mathcal{C}_{10} \text{, } \\   \phi \bigg[ C^{XXX} \big( \underline{u} \big) C^{XXX} \big( \underline{u^{\prime}} \big) \bigg] \approx \phi \big[ C_{11} \big] \approx \mathcal{C}_{11} \text{, }    \\  \phi \bigg[  C^{XXX} \big( \underline{u} \big) D^{XXX} \big( \underline{u^{\prime}} \big) \bigg] \approx \phi \big[ C_{12} \big] \approx \mathcal{C}_{12} \text{, }       \\      \phi \bigg[   D^{XXX} \big( \underline{u} \big) A^{XXX} \big( \underline{u^{\prime}} \big) \bigg] \approx \phi \big[  C_{13} 
 \big] \approx \mathcal{C}_{13} \text{, }  \\  \phi \bigg[   D^{XXX} \big( \underline{u} \big) B^{XXX} \big( \underline{u^{\prime}} \big) \bigg] \approx \phi \big[ C_{14} \big] \approx \mathcal{C}_{14} \text{, }      \\ \phi \bigg[   D^{XXX} \big( \underline{u} \big) C^{XXX} \big( \underline{u^{\prime}} \big) \bigg] \approx \phi \big[ C_{15} \big] \approx \mathcal{C}_{15} \text{, }  \\  \phi \bigg[  D^{XXX} \big( \underline{u} \big) D^{XXX} \big( \underline{u^{\prime}} \big) \bigg] \approx \phi \big[ C_{16} \big] \approx \mathcal{C}_{16}  \text{, }  \end{array}\right. \]

\noindent from which we conclude the argument. \boxed{}

  \subsection{Approximations of Operators from the Transfer Matrix}

\noindent We perform the computations for the result of the item above. To avoid having to repeat very similar computations for all of the constants, WLOG we demonstrate how the first operator product is obtained below, from which the remaining fifteen constants can be obtained with very similar computations.

\bigskip

\noindent \underline{\textit{Operator Product} $\#$ 1}

\begin{align*}
C_1 \approx                                       i u \sigma^+_0 \sigma^-_0 i u^{\prime} \sigma^+_0 \sigma^-_0     \bigg[         \underset{1 \leq  j\leq N-1}{\prod}     i \big( u + u^{\prime} \big) \sigma^+_j \sigma^-_j     \textbf{1}_{\{ j^{\prime} > j \text{ }: \text{ } i u \sigma^+_{j^{\prime}} \sigma^-_{j^{\prime} } \in \mathrm{support} ( A  ( u + u^{\prime} ) )    \}}   \bigg[    i \big( u + u^{\prime} \big) \sigma^+_{j^{\prime} } \sigma^-_j i \big( u + u^{\prime} \big)  \\ \times     \sigma^+_{j^{\prime}+1} \sigma^-_{j^{\prime}+1}      + \sigma^-_{j^{\prime}} \sigma^+_{j^{\prime}}    \sigma^+_{j^{\prime}+1}                        \bigg]  \text{ } \bigg] \bigg[   \underset{j \equiv N}{\prod}    i \big( u + u^{\prime} \big) \sigma^+_j \sigma^-_j             \bigg]    +       \sigma^-_1 \bigg[ \underset{1 \leq j \leq N-1}{\prod} \sigma^+_j \sigma^-_{j+1} + i \big( u + u^{\prime} \big) \sigma^+_{j} \sigma^-_j i \big( u + u^{-1} \big) \\ \times \sigma^+_{j+1} \sigma^-_{j+1} \bigg]  \bigg[           \underset{3 \leq j \leq N-1}{\prod}   i \big( u^{-1} + \big( u^{\prime} \big)^{-1} \big) \sigma^+_j \sigma^-_j    \bigg]  \bigg[ \underset{j \equiv N}{\prod}   \sigma^+_j     \bigg]^2    +   i u \sigma^+_0 \sigma^-_0 i u^{\prime} \sigma^+_1 \sigma^-_1 \bigg[                \underset{1 \leq j \leq N-1}{\prod}  i u \sigma^+_j \sigma^-_j  \\ \times   \textbf{1}_{\{ j^{\prime} > j \text{ } : \text{ } i u \sigma^+_{j^{\prime}} \sigma^-_{j^{\prime}}      \in \mathrm{support} ( A ( u + u^{\prime} ) )          \}}   \bigg[ i u \sigma^+_{j^{\prime}} \sigma^-_{j^{\prime}}  i u \sigma^+_{j^{\prime}+1} \sigma^-_{j^{\prime}+1} + \sigma^-_{j^{\prime}} \sigma^+_{j^{\prime}} \sigma^-_{j^{\prime}+1}                 \bigg] \bigg[      i u^{\prime} \sigma^+_j \sigma^-_j       \bigg] \text{ }  \bigg]  +        i u \sigma^+_0 \sigma^-_0 i u^{\prime} \sigma^+_1 \sigma^-_1 \\ 
\times \bigg[ \underset{1 \leq j \leq 2}{\prod}                 i u \sigma^+_j \sigma^-_j \textbf{1}_{ \{ j^{\prime} > j \text{ } : \text{ }  i u \sigma^+_{j^{\prime}} \sigma^-_{j^{\prime}} \in \mathrm{support} ( A ( u + u^{\prime} ) )        \}  }          \bigg[            i u \sigma^+_{j^{\prime}} \sigma^-_{j^{\prime}} i u \sigma^+_{j^{\prime}+1} \sigma^+_{j+1} \sigma^-_{j^{\prime}}   \sigma^+_{j^{\prime}} \sigma^-_{j^{\prime}+1}                \bigg]   \bigg[   \sigma^-_j + \sigma^+_j      \bigg]     \end{align*}

\begin{align*} \times \bigg[            \underset{3 \leq j \leq N-1}{\prod}                 i u \sigma^+_j \sigma^-_j             \textbf{1}_{ \{ j^{\prime} > j \text{ } : \text{ } i u \sigma^+_{j^{\prime}}    \in \mathrm{support} ( A ( u + u^{\prime} ) )  \} } \bigg[   i u\sigma^+_{j^{\prime}} \sigma^-_{j^{\prime}} + \sigma^-_{j^{\prime}} \sigma^+_{j^{\prime}} \sigma^-_{j^{\prime}+1} \bigg] \text{ }   \bigg]   \bigg[ \underset{3 \leq j \leq N-2}{\prod}      \big[ \sigma^-_j + \sigma^+_j  \big]          \bigg]                     \text{ }      \bigg]  \\  \times  \bigg[     \underset{j \equiv N}{\prod}       i u \sigma^+_j \sigma^-_j       \bigg]                 \bigg[     \underset{3 \leq j < j+1 \leq N}{\prod}  \sigma^+_j \sigma^+_{j+1}  \bigg]       + i u \sigma^+_0 \sigma^-_0               i  u^{\prime} \sigma^+_1  \sigma^-_1 \bigg[     \underset{j^{\prime} \mathrm{mod} 4 \equiv 0}{\underset{j,j^{\prime} \equiv N}{\prod}}   i u \sigma^+_j \sigma^-_j \sigma^+_{j^{\prime}} \bigg]  \bigg[               \underset{j \text{ } \mathrm{odd}}{\underset{1 \leq j \neq j^{\prime} \leq 2}{\prod}}  i u\sigma^+_j \sigma^-_j                       \\  \times \textbf{1}_{\{ j^{\prime\prime}> j^{\prime} \text{ } : \text{ } i u \sigma^+_{j^{\prime\prime}} \sigma^-_{j^{\prime\prime}}  \in \mathrm{support} ( A ( u + u^{\prime} ) )  \}}    \bigg[ i u \sigma^+_{j^{\prime}} \sigma^-_{j^{\prime}} i u \sigma^+_{j^{\prime}+1} \sigma^-_{j^{\prime}+1}       \sigma^-_{j^{\prime\prime}} + \sigma^-_{j^{\prime}} \sigma^+_{j^{\prime}}  \sigma^-_{j^{\prime}+1} i u^{\prime} \sigma^+_{j^{\prime\prime}} \sigma^-_{j^{\prime\prime} }    \bigg]  \text{ }                            \bigg]\\ \times \bigg[ \underset{j \text{ } \mathrm{mod}3 \equiv 0}{\underset{3 \leq j \leq N-1}{\prod}}             i u^{-1}  \sigma^+_j \sigma^-_j           \bigg]  + i u \sigma^+_0 \sigma^-_0         \bigg[    \underset{j{\prime} \text{ } \mathrm{odd}}{\underset{1 \leq j \neq j^{\prime} \leq N-1}{\prod}}   i u \sigma^+_{j^{\prime}} \sigma^-_{j^{\prime}} \textbf{1}_{ \{ j^{\prime\prime} > j^{\prime} \text{ } : \text{ }   i. u\sigma^+_{j^{\prime\prime} } \sigma^-_{j^{\prime\prime}} \in   \mathrm{support} ( A ( u +   u^{\prime} )    \} } \\ \times \bigg[             i u \sigma^+_{j^{\prime\prime} }                  \sigma^-_{j^{\prime\prime}}           i u \sigma^+_{j^{\prime\prime}+1} \sigma^-_{j^{\prime\prime}+1} i u  \sigma^+_j \sigma^-_j i u \sigma^+_{j+1} \sigma^-_{j+1} +  \sigma^-_{j^{\prime\prime}} \sigma^+_{j^{\prime\prime}}    \bigg] \bigg[ i u^{\prime} \sigma^+_j \sigma^-_j   i u^{\prime} \sigma^+_{j+1} \sigma^-_{j+1}    \bigg]   \bigg]  \\ \times \bigg[ \underset{j \mathrm{mod} 3 \equiv 0}{\underset{3 \leq j\leq N-1}{\prod}}           i \big( u^{\prime} \big)^{-1} \sigma^+_j \sigma^-_j     \bigg]     \bigg[ \underset{j \equiv N}{\prod} i u \sigma^+_j \sigma^-_j \sigma^+_j  \bigg] + i u \sigma^+_0 \sigma^-_0 \sigma^-_1 \bigg[   \underset{j \mathrm{odd}}{\underset{1 \leq j^{\prime\prime} \leq N-1}{\underset{1 \leq j \leq N-1}{\prod}} }    \big[ i u \sigma^+_{j^{\prime}} \sigma^-_{j^{\prime}} i u \sigma^+_{j^{\prime}+1} \sigma^-_{j^{\prime}+1} + \sigma^-_{j^{\prime}} \\ \times    \sigma^+_{j^{\prime}} \sigma^-_{j^{\prime}+1}  \big]   \big[ \sigma^+_{j^{\prime\prime}}    \sigma^-_{j^{\prime\prime}+1} + i u^{\prime} \sigma^+_{j^{\prime\prime}} \sigma^-_{j^{\prime\prime}} i u^{\prime} \sigma^+_{j^{\prime\prime}+1} \sigma^-_{j^{\prime\prime}+1} i u^{\prime} \sigma^+_{j^{\prime\prime}+1} \sigma^-_{j^{\prime\prime}+1}   \big]      \bigg]      \bigg[ \underset{j \text{ } \mathrm{mod} 3 \equiv 0}{\underset{3 \leq j \leq N-1}{\prod}}    i u^{-1} \sigma^+_j \sigma^-_j       \bigg]  \\ \times   \bigg[   \underset{ j \equiv N}{\prod}   i u \sigma^+_j \sigma^-_j \sigma^+_j       \bigg]     +    i u \sigma^+_1 \sigma^-_1 \sigma^+_2 i u^{\prime}  \sigma^+_1 \sigma^-_1 \sigma^+_2 \bigg[      \underset{1 \leq. j\leq N}{\prod}   i u \sigma^+_j \sigma^-_j i u^{\prime} \sigma^+_j \sigma^-_j  \bigg]   \bigg[ \underset{ j \equiv N}{\prod} \sigma^+_j \sigma^-_j  \bigg]^2     \\ +     i u \sigma^+_1 \sigma^-_1 \sigma^+_2 \bigg[ \underset{j \text{ } \mathrm{even}}{\underset{1 \leq j \leq N-1}{\prod}}       i u^{\prime}    \sigma^+_j \sigma^-_j      \bigg]    \bigg[               \underset{ j\text{ } \mathrm{odd}}{\underset{1 \leq j \leq N}{\prod}}  i u^{\prime} \sigma^+_j \sigma^-_j     \bigg]   \bigg[ \underset{j \text{ } \mathrm{odd}}{\underset{3 \leq j \leq N-1}{\prod}}     i \big( u^{\prime} \big)^{-1}  \sigma^+_j \sigma^-_j       \bigg]            \bigg[ \underset{j \equiv N}{\prod}     \sigma^+_j \sigma^-_j i u^{\prime}  \sigma^+_j \sigma^-_j       \bigg]     \\ + i u \sigma^+_1 \sigma^-_1 \sigma^+_2 i u^{\prime} \sigma^+_1 \sigma^-_1 \bigg[      \underset{j \text{ } \mathrm{odd}}{\underset{1 \leq j \leq N}{\prod}}             i u \sigma+_j \sigma^-_j               \bigg]  \bigg[ \underset{2 \leq j \leq N-2}{\prod} \big[ \sigma^-_j  +  \sigma^+_j  \big] \bigg]  \bigg[ \underset{3 \leq j < j+1 \leq N+1}{\prod}  \sigma^+_j \sigma^+_{j+1} \bigg] \\ \times  \bigg[ \underset{j \equiv N} {\prod}        \sigma^+_j \sigma^-_j    \bigg]  + i u \sigma^+_1 \sigma^-_1 \sigma^+_2  i u^{\prime} \sigma^+_1 \sigma^-_1 \bigg[ \underset{j^{\prime} \mathrm{mod} 4 \equiv 0}{\underset{j\neq j^{\prime}, j,j^{\prime} \equiv N}{\prod}}               i u \sigma^+_j \sigma^-_j \sigma^+_{j^{\prime}} \bigg] \bigg[ \underset{j \mathrm{mod}3 \equiv 0}{\underset{3 \leq j \leq N-1}{\prod}}     i \big( u^{\prime} \big)^{-1} \sigma^+_j \sigma^-_j       \bigg]         \\ \times \bigg[     \underset{j \mathrm{mod}2 \equiv 0, \text{ } j \mathrm{mod} 4 \neq 0}{\underset{2 \leq j \leq N-1}{\prod}} \big[ \sigma^-_j + i u^{\prime} \sigma^+_j \sigma^-_j  \big]       \bigg]     \bigg[ \underset{j \text{ } \mathrm{odd}}{\underset{1 \leq j \leq N-1}{\prod}}     i u \sigma^+_j \sigma^-_j     \bigg]  \bigg[           \underset{j \mathrm{mod} 3 \equiv 0}{\underset{3 \leq j \leq N-1}{\prod}}              i \big( u^{\prime} \big)^{-1} \sigma^+_j \sigma^-_j        \bigg]   \\ + i u \sigma^+_1 \sigma^-_1 \sigma^+_2 \bigg[               \underset{j \text{ } \mathrm{odd}}{\underset{1 \leq j \leq N-1}{\prod}}   i u \sigma^+_j \sigma^-_j      i u^{\prime} \sigma^+_j \sigma^-_j   i u^{\prime} \sigma^+_{j+1} \sigma^-_{j+1}   \bigg]   \bigg[ \underset{j \mathrm{mod} 3 \equiv 0}{\underset{3 \leq j \leq N-1}{\prod}}    i \big( u^{\prime} \big)^{-1} \sigma^+_j \sigma^-_j       \bigg]   \bigg[ \underset{ j \equiv N}{\prod} \sigma^+_j \sigma^-_j \sigma^+_j  \bigg] \\ + i u \sigma^+_1 \sigma^-_1 \sigma^+_2     \bigg[ \underset{j \text{ } \mathrm{odd}}{\underset{1 \leq j^{\prime\prime} \leq N-1}{\underset{1 \leq j \leq N-1}{\prod}}}    i u \sigma^+_j \sigma^-_j \textbf{1}_{\{ j^{\prime} > j \text{ } : \text{ } i u \sigma^+_{j^{\prime}} \sigma^-_{j^{\prime}} \in \mathrm{support} ( A ( u + u^{\prime} ) ) \}}   \bigg[         i u \sigma^+_{j^{\prime} }  \sigma^-_{j^{\prime}} i u \sigma^+_{j^{\prime}+1} + \sigma^-_{j^{\prime}} \sigma^+_{j^{\prime}} \sigma^-_{j^{\prime}+1}  \bigg]  \\ \times \bigg[ \sigma^+_{j^{\prime\prime}} \sigma^-_{j^{\prime\prime}+1} + i u^{\prime} \sigma^+_{j^{\prime\prime}} \sigma^-_{j^{\prime\prime}} i u^{\prime} \sigma^+_{j^{\prime\prime}+1}        \sigma^-_{j^{\prime\prime}+1} \bigg]   \text{ }   \bigg]     \bigg[ \underset{j \mathrm{mod} 3 \equiv 0}{\underset{3 \leq j \leq N-1}{\prod}}              i \big( u^{\prime} \big)^{-1} \sigma^+_j \sigma^-_j       \bigg]   \bigg[ \underset{ j \equiv N}{\prod} i u \sigma^+_j \sigma^-_j \sigma^+_j  \bigg] \\ + \bigg[ \underset{j \text{ } \mathrm{even}}{\underset{1 \leq j \leq N-1}{\prod}}     i \big( u + u^{\prime} \big) \sigma^+_j \sigma^-_j         \bigg]      \bigg[ \underset{j \text{ } \mathrm{odd}}{\underset{3 \leq j \leq N-1}{\prod}} i \big( u^{-1} + \big( u^{\prime} \big)^{-1} \big) \sigma^+_j \sigma^-_j \bigg]  \bigg[          \underset{ j \equiv N}{\prod}            i \big( u + u^{\prime} \big) \sigma^+_j \sigma^-_j      \bigg]   \\ + i u \sigma^+_1 \sigma^-_1 \bigg[     \underset{j \text{ } \mathrm{even}}{\underset{1 \leq j^{\prime\prime} \leq 2}{\underset{1 \leq j \leq N -1}{\prod}}}   i u \sigma^+_j \sigma^-_j \big[ \sigma^-_{j^{\prime\prime}} + \sigma^+_{j^{\prime\prime}} \big]   \bigg]  \bigg[ \underset{j \equiv N}{\prod} i u \sigma^+_j \sigma^-_j  \bigg]  \bigg[ \underset{3 \leq j < j +1 \leq N+1}{\prod} \sigma^+_j \sigma^+_{j+1} \bigg]    
\end{align*}

\begin{align*}
 +                  i u \sigma^+_1 \sigma^-_1 \sigma^+_2 i u^{\prime} \sigma^+_0 \sigma^-_0 \bigg[ \underset{j \text{ } \mathrm{odd}}{\underset{1 \leq j \leq N}{\prod}} i u \sigma^+_j \sigma^-_j  \bigg]  \bigg[    \underset{ j \text{ } \mathrm{odd}}{\underset{1 \leq j \leq N-1}{\prod}}     i u^{\prime} \sigma^+_j \sigma^-_j \textbf{1}_{\{ j^{\prime} > j \text{ } : \text{ } i u \sigma^+_{j^{\prime}} \sigma^-_{j^{\prime}} \in \mathrm{support} ( A ( u + u^{\prime} ) )  \}}            \\  \times  \bigg[ i u^{\prime} \sigma^+_{j^{\prime}} \sigma^-_{j^{\prime}} i u^{\prime} \sigma^+_{j^{\prime}+1}  \sigma^-_{j^{\prime}+1}  + \sigma^-_{j^{\prime}}   \sigma^+_{j^{\prime}} \sigma^-_{j^{\prime}+1 }          \bigg] \text{ } \bigg]  \bigg[ \underset{j \equiv N}{\prod}       \sigma^+_j \sigma^-_j i u^{\prime} \sigma^+_j \sigma^-_j  \bigg]  + i u^{\prime} \sigma^+_1 \sigma^-_1 \bigg[ \underset{j \text{ }\mathrm{even}}{\underset{1 \leq j\leq N-1}{\prod}}              i u \sigma^+_j \sigma^-_j       \bigg]   \\ \times     \bigg[  \underset{j^{\prime} \mathrm{mood} 3 \equiv 0}{\underset{j \neq j^{\prime} , \text{ } j \text{ } \mathrm{odd}}{\underset{3 \leq j^{\prime} \leq N-1}{\underset{3 \leq j \leq N-1}{\prod}   } }}   i \big( u^{-1} + \big( u^{\prime} \big)^{-1} \big) \sigma^+_j \sigma^-_j       \bigg]        \bigg[ \underset{j \text{. }\mathrm{odd}}{\underset{2 \leq j \leq N-1}{\prod}}     i u^{-1} \sigma^+_j \sigma^-_j       \bigg]                 \bigg[  \underset{ j^{\prime} \mathrm{mod} 4 \equiv 0 }{\underset{j , j^{\prime} \equiv N}{\prod}}  i u \sigma^+_j \sigma^-_j \sigma^+_{j^{\prime}}   \bigg] \\ + \bigg[     \underset{j \text{ } \mathrm{even}}{\underset{1 \leq j^{\prime} < j^{\prime}+1 \leq N-1}{\underset{1 \leq j \leq N-1}{\prod}}}    i u \sigma^+_j \sigma^-_j \big[        i u^{\prime}  \sigma^+_{j^{\prime}} \sigma^-_{j^{\prime}} i u \sigma^+_{j^{\prime}+1} \sigma^-_{j^{\prime}+1}      \big]             \bigg] \bigg[   \underset{j \mathrm{mod}3 \equiv 0}{\underset{j \text{ } \mathrm{odd}}{\underset{3 \leq j^{\prime} \leq N-1}{\underset{3 \leq j \leq N-1}{\prod}}}}     i \big( u + \big( u^{\prime} \big)^{-1} \big) \sigma^+_j \sigma^-_j   \bigg]    \bigg[ \underset{j \equiv N}{\prod}  i u \sigma^+_j \sigma^-_j \sigma^+_j \bigg]   \\ + \sigma^-_1 \bigg[ \underset{1 \leq j^{\prime} \leq N-1}{\underset{j \text{ } \mathrm{even}}{\underset{1 \leq j \leq N-1}{\prod}}}    i u \sigma^+_j \sigma^-_j \big[ \sigma^+_{j^{\prime}} \sigma^-_{j^{\prime}+1} + i u \sigma^+_{j^{\prime}} \sigma^-_{j^{\prime}} i u \sigma^+_{j^{\prime}+1} \sigma^-_{j^{\prime}+1} \big]        \bigg] \bigg[ \underset{j^{\prime} \mathrm{mod} 3 \equiv 0}{\underset{j \text{ } \mathrm{odd}}{\underset{3 \leq j^{\prime} \leq N-1}{\underset{3 \leq j \leq N-1}{\prod}}}}   i \big( u + \big( u^{\prime} \big)^{-1} \big) \sigma^+_j \sigma^-_j  \bigg]  \\ \times \bigg[   \underset{j \equiv N}{\prod} i u \sigma^+_j \sigma^-_j \sigma^+_j      \bigg] + i \big( u + u^{\prime} \big) \sigma^+_1 \sigma^-_1 \bigg[ \underset{2 \leq j \leq N-2}{\prod} \big( \sigma^-_j \big)^2 + \big( \sigma^+_j \big)^2  \bigg]  \bigg[            \underset{3 \leq j < j+1 \leq N+1}{\prod}   \sigma^+_j \sigma^+_{j+1}      \bigg]^2 \\ +   i u^{\prime} \sigma^+_1 \sigma^-_1 \bigg[         \underset{j^{\prime} \text{ } \mathrm{odd}}{\underset{1 \leq j^{\prime} \leq N-1}{\underset{2 \leq j \leq N-2}{\prod}}}                    \big[ \sigma^-_j i u \sigma^+_{j^{\prime}} \sigma^-_{j^{\prime}} + \sigma^+_j i u \sigma^+_{j^{\prime}} \sigma^-_{j^{\prime}} \big]   \textbf{1}_{ \{ j^{\prime\prime} > j^{\prime} \text{ } : \text{ } i u \sigma^+_{j^{\prime\prime}} \sigma^-_{j^{\prime\prime}} \in \mathrm{support} ( A ( u + u^{\prime} )) \}}   \\ \times \big[ i u \sigma^+_{j^{\prime\prime}} \sigma^-_{j^{\prime\prime}} i u \sigma^+_{j^{\prime\prime}+1} \sigma^-_{j^{\prime\prime}+1}  + \sigma^-_{j^{\prime\prime}} \sigma^+_{j^{\prime\prime}} \sigma^-_{j^{\prime\prime}+1} \big]       \bigg]     \bigg[ \underset{3 \leq j < j+1 \leq N+1}{\prod} \sigma^+_j \sigma^+_{j+1} \bigg] \bigg[ \underset{j \equiv N}{\prod}     i u \sigma^+_j \sigma^-_j \bigg]   \\ + i u^{\prime} \sigma^+_1 \sigma^-_1 \bigg[ \underset{2 \leq j \leq N-2}{\prod} \big[ \sigma^-_j + \sigma^+_j \big] \bigg]    \bigg[  \underset{j \text{ } \mathrm{odd}}{\underset{1 \leq j \leq N}{\prod}} i u^{\prime} \sigma^+_j \sigma^-_j   \bigg] \bigg[ \underset{3 \leq j < j+1 \leq N+1}{\prod}     \sigma^+_j \sigma^+_{j+1} \bigg]  \bigg[ \underset{j \equiv N}{\prod} \sigma^+_j \sigma^-_j \bigg]  \\ + i u^{\prime} \sigma^+_1 \sigma^-_1 \bigg[   \underset{j \equiv N}{\prod} i u^{\prime}  \sigma^+_j \sigma^-_j   \bigg]   \bigg[ \underset{j \text{ } \mathrm{odd}}{\underset{3 \leq j \leq N-1}{\prod}} i \big( u^{\prime} \big)^{-1} \sigma^+_j \sigma^-_j  \bigg]    \bigg[   \underset{1 \leq j \leq N-1}{\prod}  i u^{\prime} \sigma^+_j \sigma^-_j              \bigg]  \bigg[ \underset{2 \leq j \leq N-2}{\prod}  \big( \sigma^-_j + \sigma^+_j  \big) \bigg]  \\ \times \bigg[ \underset{3 \leq j < j+1 \leq N+1}{\prod}  \sigma^+_j \sigma^+_{j+1} \bigg]    + i u \sigma^+_1 \sigma^-_1 i u^{\prime} \sigma^+_1 \sigma^-_1     \bigg[ \underset{j \equiv N}{\prod} \sigma^+_j  \bigg] \bigg[ \underset{j \mathrm{mod} 3 \equiv 0}{\underset{3 \leq j \leq N-1}{\prod}}  i \big( u^{\prime} \big)^{-1} \sigma^+_j \sigma^-_j      \bigg]  \\ \times \bigg[   \underset{j \mathrm{mod} 4 \neq 0}{\underset{j \mathrm{mod} 2 \equiv 0 }{\underset{2 \leq j \leq N-1}{\prod}}}  \big[ \sigma^-_j + i u^{\prime} \sigma^+_j \sigma^-_j  \big]  \bigg]  \bigg[ \underset{2 \leq j \leq N-2}{\prod}     \big[ \sigma^-_j + \sigma^+_j \big]       \bigg]   \bigg[ \underset{3 \leq j < j +1 \leq N+1}{\prod} \sigma^+_j \sigma^+_{j+1}  \bigg]    \bigg[        \underset{2 \leq j \leq N-2}{\prod}  \big[ \sigma^-_j + \sigma^+_j \big]     \bigg] \\ + i u \sigma^+_1 \sigma^-_1 \bigg[ \underset{j \equiv N}{\prod} \sigma^+_j  \bigg]      \bigg[ \underset{j \mathrm{mod} 3 \equiv 0}{\underset{2 \leq j \leq N-1}{\prod}}    i \big( u^{\prime} \big)^{-1} \sigma^+_j \sigma^-_j    \bigg]  \bigg[ \underset{1 \leq j \leq 2}{\prod}  i u \sigma^+_j \sigma^-_j i u \sigma^+_{j+1}\sigma^-_{j+1}   \bigg]  \bigg[ \underset{2 \leq j^{\prime} \leq N-2}{\underset{2 \leq j < j+1 \leq N-2}{\prod}} \big[ \sigma^-_{j^{\prime}} + \sigma^+_{j^{\prime}} \big]  \\ \times \big[ i u \sigma^+_j \sigma^-_j i u \sigma^+_{j+1} \sigma^-_{j+1 } \big]   \bigg]  \bigg[ \underset{3 \leq j < j+1 \leq N+1}{\prod} \sigma^+_j \sigma^+_{j+1} \bigg] \bigg[  \underset{N-2 \leq j < j+1}{\prod} i u \sigma^+_j \sigma^-_j i u \sigma^+_{j+1} \sigma^-_{j+1}   \bigg]  \\ + i u \sigma^+_1 \big( \sigma^-_1 \big)^2 \bigg[        \underset{j \equiv N}{\prod}  \sigma^+_j  \bigg]   \bigg[    \underset{j \mathrm{mod} 3 \equiv 0}{\underset{3 \leq j \leq N-1}{\prod}}   i \big( u^{\prime} \big)^{-1} \sigma^+_j \sigma^-_j   \bigg]    \bigg[  \underset{1 \leq j \leq N-1}{\prod} \sigma^+_j \sigma^-_{j+1} + i u^{\prime} \sigma^+_j \sigma^-_j i u^{\prime} \sigma^+_{j+1} \sigma^-_{j-1} \bigg]  \\ \times \bigg[ \underset{3 \leq j < j+1 \leq N+1}{\prod} \sigma^+_j \sigma^+_{j+1} \bigg]  \bigg[ \underset{2 \leq j \leq N-2}{\prod}  \big[ \sigma^-_j + \sigma^+_j  \big] \bigg]   + \sigma^-_0 \sigma^+_1 \bigg[ \underset{1 \leq j \leq N-1}{\prod} i \big( u + u^{\prime} \big) \sigma^+_j \sigma^-_j  \textbf{1}_{\{ j^{\prime} > j \text{ } : \text{ } i u \sigma^+_{j^{\prime}} \sigma^-_{j^{\prime}} \in \mathrm{support} ( A ( u + u^{\prime} ) )  \}}  
\end{align*}

\begin{align*}
  \times         \bigg[  \big( \sigma^+_{j^{\prime}} \big)^2 + i \big( u^{-1} + \big( u^{\prime} \big)^{-1} \big) \sigma^+_{j^{\prime}} \sigma^-_{j^{\prime}} \bigg] \text{ }              \bigg] \bigg[ \underset{j \equiv N}{\prod} \sigma^+_j  \bigg]^2  + \sigma^-_0 \sigma^+_1 \bigg[ \underset{1 \leq j \leq N-1}{\prod} i u \sigma^+_j \sigma^-_j \textbf{1}_{\{ j^{\prime} > j \text{ } : \text{ } i u \sigma^+_{j^{\prime}} \sigma^-_{j^{\prime}} \in \mathrm{support} ( A ( u + u^{\prime}  ) ) \}}  \\  \times    \big[ \big( \sigma^+_j \big)^2 \sigma^-_{j^{\prime}+1}  + i \big( u^{-1} + u^{\prime} \big) \sigma^+_{j^{\prime}} \sigma^-_{j^{\prime}}            \big]  \bigg]  \bigg[  \underset{j \mathrm{mod} 3 \equiv 0 }{\underset{3 \leq j \leq N-1}{\prod}} i \big( u^{\prime} \big)^{-1} \sigma^+_j \sigma^-_j \bigg] \bigg[ \underset{j \equiv N}{\prod}  \sigma^+_j \bigg]^2   + \big( \sigma^-_1 \big)^2 \\ \times \bigg[ \underset{1 \leq j \leq N}{\prod} \textbf{1}_{\{  j^{\prime} > j \text{ } : \text{ } i u \sigma^+_{j^{\prime}} 
 \sigma^-_{j^{\prime}} \in \mathrm{support} ( A ( u+ u^{\prime})) \}} \big[   \sigma^+_{j^{\prime}} \sigma^-_{j^{\prime}+1} + i \big( u + u^{\prime} \big) \sigma^+_{j^{\prime}}  \sigma^-_{j^{\prime}} \big] \bigg[ \underset{j \mathrm{mod} 3 \equiv 0}{\underset{3 \leq j \leq N-1}{\prod}} i \big( u^{-1} + \big( u^{\prime} \big)^{-1} \big) \sigma^+_j \sigma^-_j  \bigg] \text{ }  \bigg]  \\ \times \bigg[ \underset{j \equiv N}{\prod}  \sigma^+_j \bigg]^2  \text{. }
\end{align*}

\section{References}

\noindent [1] Amico, L., Frahm, H., Osterloh, A., Ribeiro, G.A.P. Integrable spin-boson models descending from rational six-vetex models. \textit{Nucl. Phys. B.} \textbf{787}: 283-300 (2007).

\bigskip

\noindent [2] Alcaraz, F.C., Lazo, M.J. Exactly solvable interacting vertex models. \textit{J. Stat. Mech.} P08008 (2007).

\bigskip

\noindent [3] Batchelor, M.T., Baxter, R.J., O'Rourke, M.J., Yung, C.M. Exact solution and interfacial tension of the six-vertex model with anti-periodic boundary conditions. \textit{J. Phys. A: Math. Gen.} \textbf{28}: 2759 (1995).

\bigskip

\noindent [4] Bogoliubov, N.M. Four-Vertex Model and Random Tilings. \textit{Theor. and Math. Phys.}, \textbf{155}: 523-535 (2008).

\bigskip

\noindent [5] Baxter, R.J. Exactly Solved Models in Statistical Mechanics. Dover Publishing.

\bigskip

\noindent [6] Boos, H., et al. Universal R-matrix and functional relations. \textit{Rev. Math. Phys.} \textbf{26}: 143005 (2014).

\bigskip

\noindent [7] Boos, H., Gohmann, F., Klumper, A., Nirov, K.S., Razumov, A.V.. Exercises with the universal R-matrix. \textit{J. Phys. A.} \textbf{43} (2010).

\bigskip

\noindent [8] Colomo, F., Giulio, G.D., Pronko, A.G. Six-vertex model on a finite lattice: Integral representations for nonlocal correlation functions. \textit{Nuclear Physics B} \textbf{972}: 115535 (2021).

\bigskip

\noindent [9] Colomo, F., Pronko, A.G. The Arctic Circle Revisited. \textit{Contemp. Math.} \textbf{458}: 361-376 (2008).

\bigskip

\noindent [10] Deguchi, T. Introduction to solvable lattice models in statistical and mathematical physics. \textit{CRC Press}: 9780429137891 (2003).

\bigskip

\noindent  [11] Duminil-Copin, H., Karrila, A., Manolescu, I.,  Oulamara, M. Delocalization of the height function
of the six-vertex model. J. Eur. Math. Soc. (2024).

\bigskip

\noindent [12] Duminil-Copin, H., Kozlowski, K.K., Krachun, D. et al. On the Six-Vertex Model’s Free Energy.
Commun. Math. Phys. 395, 1383–1430 (2022).

\bigskip

\noindent [13] de Vega, H.J. Bethe Ansatz and Quantum Groups. \textit{LPTHE} \textbf{93/17} (1992).

\bigskip

\noindent [14] de Vega, H.J. Boundary K-matrices for the XYZ, XXZ and XXX spin chains. \textit{LPTHE-PAR} \textbf{93/29} (1993).

\bigskip

\noindent [15] de Vega, H.J., Ruiz, A.G. Boundary K-matrices for the six vertex and the $n \big( 2n-1 \big)$ $A_{n-1}$ vertex models. \textit{J. Phys. A: Math. Gen.} \textbf{26} (1993).

\bigskip

\noindent [16] Di Francesco, Philippe. Twenty Vertex model and domino
tilings of the Aztec triangle. the electronic journal of combinatorics \textbf{28}(4), (2021).

\bigskip

\noindent [17] Faddeev, L.D., Takhtajan, L. A. Hamiltonian Methods in the Theory of Solitons. \textit{Springer}, 978-3540698432.

\bigskip

\noindent [18] Frahm, H., Seel, A. The staggered six-vertex model: Conformal invariance and corrections to scaling. \textit{Nucl. Phys. B.} \textbf{879}: 382-406 (2014).

\bigskip

\noindent [19] Garbali, A., de Gier, J., Mead, W., Wheeler, M. Symmetric functions from the six-vertex model in half-space. \textit{arXiv: 2312.14348} (2023).

\bigskip

\noindent [20] Gier, J.D., Korepin, V. Six-vertex model with domain wall boundary conditions: variable inhomogeneities. \textit{J. Phys. A: Math. Gen.} \textbf{34} (2001).

\bigskip

\noindent [21] Gohmann, F. Bethe ansatz. \textit{arXiv: 2309.02008} (2023).

\bigskip

\noindent [22] Gorbounov, V., Korff, C. Quantm integrability and generalised quantum Schubert calculus. \textit{Adv. Math.} \textbf{313}: 282-356 (2017).

\bigskip

\noindent [23] Gorbounov, V., Korff, C., Stroppel, C. Yang-Baxter Algebras as Convolution Algebras: The Grassmannian case. \textit{Russian Mathematical Surveys} \textbf{75} (2020).

\bigskip

\noindent [24] Ikhlef, Y., Jacobsen, J., Saleur, H. A staggered six-vertex model with non-compact continuum limit. \textit{Nucl. Phys. B.} \textbf{789}(3): 483-524 (2008).

\bigskip

\noindent [25] Keating, D., Reshetikhin, N., Sridhar, A. Integrability of Limit Shapes of the Inhomogeneous Six Vertex Model. Commun. Math. Phys. 391, 1181–1207 (2022). https : //doi.org/10.1

\bigskip

\noindent [26] Kitanine, N., et al. Thermodynamic limit of particle-hole-form factors in the massless XXZ Heisenberg chain. \textit{J. Stat. Mech. P05028} (2011).

\bigskip

\noindent [27] Kitanine, N., et al. On correlation functions of integrable models associated with the six-vertex R-matrix. \textit{J. Stat. Mech.} P01022 (2007).

\bigskip

\noindent [28] Kitanine, N., et al. On the spin-spin correlation functions of the XXZ spin-$\frac{1}{2}$ infinite chain. \textit{J. Phys. A: Math. Gen.} \textbf{38} (2005).

\bigskip

\noindent [29] Kozlowski, K.K. Riemann-Hilbert approach to the time-dependent generalized sine kernel. \textit{Adv. Theor. Math. Phys.} \textbf{15}: 1655-1743 (2011).

\bigskip

\noindent [30] Korff, C., McCoy, B.M. Loop symmetry of integrable vertex models at roots of unity. \textit{Nucl. Phys. B.} \textbf{618}: 551-569 (2001).

\bigskip

\noindent [31]  Lieb, E. H. The Residual Entropy of Square Ice, Phys Rev. \textbf{162} 162-172 (1967).

\bigskip

\noindent [32] Lamers, J. Introduction to quantum integrability. \textit{10th Modave Sumemr School in Mathematical Physics} (2015).

\bigskip

\noindent [33] Motegi, K. Symmetric functions and wavefunctions of the XXZ-type six-vertex models and elliptic Felderhof models by Izergin-Korepin analysis. \textit{J. Math. Phys.} \textbf{59}: 053505 (2018).

\bigskip

\noindent [34] Naprienko, S. Free fermionic Schur functions. \textit{Adv. Math.} \textbf{436}: 109413 (2024).

\bigskip

\noindent [35] Pauling, L. J. Am. Chem. Soc. \textbf{57}, 2680 (1935)

\bigskip

\noindent [36] Reshetikhin, N., Sridhar, A. Integrability of limits shapes of the six-vertex model. \textit{Comm. Math. Phys.} \textbf{356}: 535-565 (2017).

\bigskip

\noindent [37]  Rigas, P. From logarithmic delocalization of the six-vertex height function under sloped boundary
conditions to weakened crossing probability estimates for the Ashkin-Teller, generalized random-cluster,
and $\big(q_{\sigma} , q_{\tau} \big)$-cubic models, \textit{arXiv:2211.14934} (2022).

\bigskip

\noindent [38] Rigas, P. Renormalization of crossing probabilities in the dilute Potts model. \textit{arXiv: 2211.10979 v2} (2022).

\bigskip

\noindent [39] Rigas, P. The phase transition for the Gaussian free field is sharp. \textit{arXiv: 2307.12925} (2023).

\bigskip

\noindent [40] Rigas, P. Phase transition of the long range Ising model in lower dimensions, for $d < \alpha \leq d+1$, with a Peierls' argument. \textit{arXiv: 2309.07943} (2023).

\bigskip

\noindent [41] Rigas, P. Operator formalism for discretely holomorphic parafermions of the two-color Ashkin-Teller, loop $\mathrm{O} \big( 1 \big)$, staggered eight-vertex, odd eight-vertex, and abelian sandpile models. \textit{arXiv: 2310.08212} (2023).

\bigskip

\noindent [42] Rigas, P. Poisson structure and Integrability of a Hamiltonian flow for the inhomogeneous six-vertex model. \textit{arXiv: 2310.15181} (2023).

\bigskip

\noindent [43] Rigas, P. Open boundary conditions of the $D^{(2)}_3$ spin chain and sectors of conformal field theories. \textit{arXiv: 2310.18499} (2023).

\bigskip

\noindent [44] Rigas, P. Scaling limit of the triangular prudent walk. \textit{arXiv: 2312.16236} (2023).

\bigskip

\noindent [45] Rigas, P. Eigenvalue attraction in open quantum systems, biophysical systems, and Parity-time symmetric materials. \textit{arXiv: 2309.07943} (2023).

\bigskip

\noindent [46] Rigas, P. Quantum-inverse scattering for the 20-vertex model up to Dynkin automorphism: crossing probabilities, 3D Poisson structure, triangular height functions, weak integrability. \textit{arXiv 2407.11066} (2024).

\bigskip

\noindent [47] Rigas, P. The emptiness formation probability, and nonlocal correlation functions, of the 20-vertex model. \textit{arXiv: 2409.05309} (2024).

\end{document}